\documentclass{thesisclass}
\begin{document}

\begin{titlepage}
\centering
\vspace*{2.5cm}
\bfseries\Large
Fine structures of solar radio bursts: \\ origins and radio-wave propagation effects

\vspace{1.5cm}
\normalfont\large
Nicolina Chrysaphi \\ BSc, MSc
\vspace{1.5cm}

Submitted in fulfilment of the requirements for the\\
Degree of Doctor of Philosophy\\
\vspace{1.5cm}
School of Physics and Astronomy\\
College of Science and Engineering\\
University of Glasgow\\
\vspace{1.5cm}
\includegraphics[scale=0.6]{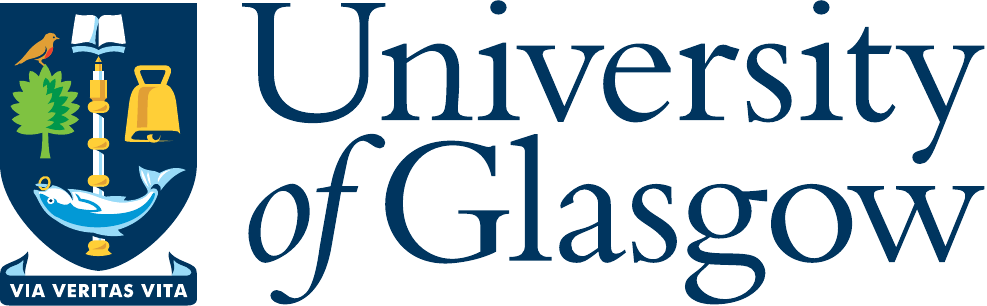}
\vspace{1.5cm}

February 2021
\end{titlepage}
\cleardoublepage

\frontmatter

\addtocounter{page}{1}
\thispagestyle{empty}

\newpage\markboth{}{}

\vspace*{7cm}
\begin{flushright}
\parbox{130mm}{
\hrulefill

This thesis is my own composition except where indicated in
the text. No part of this thesis has been submitted elsewhere for any other degree
or qualification.
\vspace*{1cm}

{\bf Copyright \copyright ~\the\year \ Nicolina Chrysaphi}
\vspace*{0.4cm}

15 February 2021

\hrulefill  }
\end{flushright}

\cleardoublepage
\pagestyle{empty}
\vspace*{\fill}
\begin{flushright}

	\vspace{-7cm}
	
	\begin{otherlanguage}{greek}
		Εγώ έχω μάθει σα παιδί να λαχταρώ να 'ρθούν τα καλοκαίρια.
	\end{otherlanguage}

\end{flushright}

\vspace*{\fill}

\cleardoublepage
\chapter*{Abstract}
Solar eruptive events are associated with radio emissions that appear as impulsive increases in intensity, known as solar radio bursts.  Turbulence in the solar corona impacts the propagation of radio waves, obscuring the intrinsic emission properties.
Here, anisotropic scattering on small-scale density fluctuations is investigated using novel 3D radio-wave propagation simulations.
Several observed radio properties are simultaneously reproduced for the first time, verifying the necessity to consider anisotropic scattering.
The sub-second evolution of fine radio burst properties at a single frequency is also investigated, enabled by conducting observations that utilise the unprecedented imaging capabilities of the LOw-Frequency ARray (LOFAR).
The fundamental and harmonic sources of a Type IIIb burst are quantitatively compared, demonstrating that harmonic emissions arise from an intrinsic source with a finite size and finite emission duration.  Drift-pair burst observations are successfully described by the radio echo hypothesis.  It is shown that the radio echo, which produces the second Drift-pair component, is detected only when the anisotropy is strong.  A dependence of the observed properties on the source's intrinsic location and on the assumed emission-to-plasma frequency ratio is inferred.  Moreover, the subbands of a split-band Type II burst are simultaneously imaged for the first time.  Despite the large separations observed between subband sources, it is shown that once scattering is quantitatively accounted for, the sources become co-spatial.  Corrections on the observed source locations also allude to lower coronal densities.
Additionally, the first observation of a Type II burst that transitions between a stationary and drifting state---termed as a transitioning Type II burst---is reported.
The radio emissions are related to a jet eruption that drives a streamer-puff CME.
Overall, state-of-the-art simulations and radio observations are combined and compared.  The importance of accounting for radio-wave propagation effects---primarily anisotropic scattering---and the consequence of neglecting to do so on any subsequent interpretations is illustrated.

\cleardoublepage
\pagestyle{fancy}

\begingroup
\renewcommand{\leftline}{}
\tableofcontents
\endgroup
\cleardoublepage

\listoffigures
\cleardoublepage
\phantomsection
\addcontentsline{toc}{chapter}{Preface}
\clearpage \markboth{PREFACE}{PREFACE}
\chapter*{Preface}

Chapter~\ref{chap:intro} provides a brief introduction to the topics and theory relevant to this thesis.  It includes a description of solar eruptive events, the types of radio bursts observed, the plasma emission mechanism, Type II bursts and the debated band-splitting models, the observing capabilities of the LOw-Frequency Array (LOFAR), and radio-wave propagation effects.

In Chapter~\ref{chap:scattering} the recently-developed 3D radio-wave propagation simulations that take into account the anisotropy of the density fluctuations are presented.  A large collection of observed Type III source sizes and decay times is used to demonstrate that, without anisotropic scattering, the observed radio source properties cannot be simultaneously reproduced.  A successful description of multiple radio burst properties enables the estimation of the level of density fluctuations and anisotropy in the corona.  The dependence of the observed properties on the projected location of the source is also investigated.  It is found that anisotropic scattering can produce large apparent sources while maintaining a highly-directional emission, something not possible when isotropic scattering is invoked.
This chapter is based on work published in \cite{2019ApJ...884..122K}.

In Chapter~\ref{chap:observation_simulations}, the sub-second evolution of Type IIIb and Drift-pair solar radio bursts, observed across a single frequency with LOFAR, is quantitatively reproduced by implementing the anisotropic scattering simulations.  The fundamental and harmonic properties of the Type IIIb burst are investigated, demonstrating that the harmonic emissions arise from an intrinsic source with a finite size and finite emission duration.  The simulations also indicate that the second component of Drift-pair bursts is the result of reflected rays reaching the observer when strong anisotropic fluctuations are present, validating the radio echo hypothesis.  The effects of varying the emission-to-plasma frequency ratio are also investigated, showing that the delay between the Drift-pair components is significantly affected by the value this ratio.  The level of density fluctuations, the anisotropy, and the source-polar angle of the observations are inferred.
The work presented in this chapter is published in \cite{2020ApJ...898...94K} and \cite{2020ApJ...905...43C}.

Chapter~\ref{chap:split-band_typeII} presents the first simultaneous imaging of split-band Type II subband sources.  A large separation between the upper- and lower-frequency subband sources is observed, but is shown to be consistent with radio-wave scattering effects.  The impact of the scattering correction on other inferred properties like the local coronal density is also discussed.  An analytical expression for estimating the scattering-induced radial shift is derived and applied on this observation.  A model for estimating the out-of-plane locations of the radio sources, as long as simultaneous imaging of the subbands is possible, is also presented.  This chapter is based on work published in \cite{2018ApJ...868...79C}.

Chapter~\ref{chap:transitioning_typeII} presents the first observation of a transitioning Type II burst---a new sub-class of Type II bursts---where the emissions transition from a stationary to a drifting state.  Double band splitting and intriguing fine structures during the stationary Type II part are also reported.  The observed emissions are related to a jet eruption which led to a streamer-puff CME.  The work presented in this chapter is published in \cite{2020ApJ...893..115C}.

Chapter~\ref{chap:conclusions} gives a brief summary of the main outcomes and conclusions of this thesis, as well as a short discussion on the current understanding in light of the presented results.

\cleardoublepage
\chapter*{Acknowledgements}
I must start by thanking my supervisor, Prof. Eduard P. Kontar, for the continuing guidance, support, and constructive criticism he has provided me over the past years.  I feel privileged to have been able to conduct research under your supervision; thank you.

I want to thank my family, without whom I would not be who I am or where I am today.  I owe everything to my mother, my aunt, both of my grandmothers, and my father.  To my sister and my brothers, I've missed you.  Thank you all for endlessly supporting me no matter where I decided to be or what I decided to do.

I should also thank everyone in room 604 and the 6th floor of the Kelvin Building.  To the friends I made here, thank you for making Glasgow feel as close to home as possible.

I also wish to thank Dr Gordon D. Holman who was constantly by my side for 4 months during my research visit at NASA's GSFC.  I've made some great memories during my PhD travels, whether I was in Washington, D.C., in Pune, India, or short conferences around the globe.

I also need to thank my collaborators and co-authors of my publications, who contributed to the work comprising this thesis.  Last but not least, I would like to thank my funder, the Science and Technology Facilities Council (STFC), whose financial support enabled me to undertake my PhD in Glasgow.

\addcontentsline{toc}{chapter}{Acknowledgements}
\cleardoublepage

\mainmatter

\titleformat{\chapter}[display]
{\filleft \bfseries \color{myblue} \raggedleft}
{\raggedleft \filleft \chapnumfont \textcolor{myblue!100} {\thechapter}}
{-40pt}
{\Huge}
\titlespacing{\chapter}{0pt}{-20pt}{60pt}[0pt]

\chapter{Introduction} \label{chap:intro}

\section{Solar Atmosphere and Activity} \label{sec:theSun}
Even though what the human brain can interpret is limited to optical wavelengths (i.e. visible light), the Sun is an active emitter of the entire spectrum of electromagnetic radiation---from the lowest radio frequencies to the highest energy gamma rays.  Besides radiation, the Sun is also constantly releasing plasma, populating the interplanetary space and forming the heliosphere.  Although the Earth constantly moves through this solar plasma, there are many aspects of the interplanetary environment that we do not yet fully understand.

Strong solar eruptions often excite bright emissions across the electromagnetic spectrum which reflect the local behaviour and structure of the Sun and its atmosphere.  This thesis focuses on the study of radio emissions, which can be used as a diagnostic tool of both their exciters and the properties of the interplanetary medium through which they propagate.  The advantage of using radio observations over other wavelengths to probe the interplanetary environment, is that many of the observed radio emissions are emitted near the local plasma frequency.  As a consequence, they can be directly related to the fundamental behaviour of the ambient particles.

\subsection{From the photosphere to the solar wind} \label{sec:solar_layers}

Solar radio emissions can be excited throughout the solar atmosphere which is divided into five different layers \citep{1985srph.book.....M}.  The density and temperature of each layer varies, thus measurements across different wavelengths are required to probe the emissions originating from each layer.
The innermost layer and the one visible with the human eye is the \textit{photosphere}, often thought of as the solar ``surface'', as it is opaque to visible light.
It is also the layer used to estimate the radius of the Sun, given as $\Rs \approx 6.96\times10^{5}$~km.
Its temperature is often taken to be $\sim$5780~K, although it should be noted that no solar layer has a uniform temperature throughout.  Just like density, the temperature of the solar atmosphere varies with heliocentric distance.  In the photosphere, the density and temperature decrease with distance.

The $\sim$2000~km following the photosphere define the \textit{chromosphere}, where the density continues to fall \citep{1985srph.book.....M}.
A temperature minimum is reached at the boundary between the photosphere and chromosphere, after which the temperature of the chromosphere rises; at first slowly and then very rapidly as the outer chromospheric boundary is approached.
At this outer edge---where a temperature of $\sim$25,000~K is reached---lies a narrow layer that is merely $\sim$100~km wide, known as the \textit{transition region} \citep{1985srph.book.....M}.
The temperature within the transition region increases steeply by two orders of magnitude (up to $\sim$\vphantom{}$10^{6}$~K), whereas the density decreases by roughly two orders of magnitude \citep{2004psci.book.....A}.  Both the chromosphere and transition region are highly-inhomogeneous layers.

Past the transition region lies the \textit{corona}, another highly-inhomogeneous medium.  This is the largest layer of the Sun and it permeates the interplanetary space.  While the underlying density of the corona is very low (lower than any other layer) and gradually decreases with increasing distance, the underlying temperature is very high (between $\sim$1--2$\times 10^6$~K) and gradually increases with distance \citep{2004psci.book.....A}.  This behaviour is a long-standing mystery known as ``the heating of the solar corona''.

The constant, steady outflow of solar plasma makes the corona a time-varying medium, although some large-scale structures can exist over longer time-scales and may not always show significant variations during their lifetime (see Section~\ref{sec:solar_activities}).  Notably, radio emissions are excited in the solar corona.  Hence, any coronal structures and interactions which are associated to radio excitations are of relevance to this thesis.
Observations of radio emissions have been used (among others) to estimate the density of the solar corona and its structures (as discussed in Section~\ref{sec:f_vs_R_relation}).

The extended parts of the corona (i.e. the outer corona) are often referred to as the \textit{solar wind} \citep{1985srph.book.....M}.  Over the past few decades, the properties of the solar wind have been explored in-situ using several space-based instruments.  Some examples include the \textit{WIND} spacecraft \citep{1997AdSpR..20..559O}, the \textit{Cluster} mission \citep{1997SSRv...79...11E}, as well as the more recent \textit{Parker Solar Probe} \citep[\textit{PSP}; e.g.,][]{2016SSRv..204....7F} and \textit{Solar Orbiter} missions \citep{2020A&A...642A...1M}.
When the corona and solar wind are distinguished as two separate layers, the assumed region of evolution from one into the other normally depends on the interpretation of the specific study, or---more commonly---on the convention used in the specific field (e.g., studies focused on ground-based versus space-based instrumentation).
When it comes to describing large heliocentric distances in this thesis, like those near the Earth (i.e. at 1~au $\simeq$ 215~$\Rs$), the two terms are used interchangeably as no boundary is defined between the corona and solar wind.

\subsection{Signatures of the active Sun} \label{sec:solar_activities}

Besides the constant outflow of solar plasma comprising the basal coronal environment, sporadic solar activities can transiently alter the ambient coronal conditions.  This section offers a brief description of such sporadic phenomena that have been related to the acceleration of electrons and to the subsequent excitation of radio emissions (discussed in Section~\ref{sec:radio_obs}).  It is also common for several of the activities described in this section to occur in sequence, where one is often the driver of the other.

\subsubsection{Active regions and sunspots}
An \textit{active region} (AR) is a compact (and complex) area on the Sun comprised of strong, dynamic magnetic fields which emerge through the photosphere into the corona, and are associated with solar emissions across a broad range of wavelengths \citep{2015LRSP...12....1V}.  These dense concentrations of strong magnetic fields make active regions the brightest structures on the Sun when observed in ultra violet (UV), extreme UV (EUV), and X-ray wavelengths.  At these wavelengths, large assemblies of coronal loops (magnetic arcades whose footpoints are on opposite magnetic polarities) in the low corona are strongly illuminated, contributing to the straightforward identification of active regions.

\textit{Sunspots} are also manifestations of the strong magnetic fields emerging through the solar surface.  They are distinct dark "spots" commonly visible in optical wavelengths, although observed beyond the optical range as well.  The central, darkest region of a sunspot is called the \textit{umbra}, whereas the surrounding, lighter region is the \textit{penumbra} \citep{2004psci.book.....A}.  The magnetic polarity of sunspots is classified according to the number of sunspots (or groups of) that have the same polarity---a classification scheme known as the Hale class.  For example, the two simplest cases are sunspots of Hale class $\alpha$, implying there is only a single polarity, and sunspots of Hale class $\beta$, implying that two opposite polarities exist \citep{1919ApJ....49..153H}.  Occasionally, a bright feature that appears to emanate from two sides of the penumbra and splits the sunspot in two parts is observed.  This is known as a \textit{light bridge}.

Historically, areas defined as active regions were distinguished in terms of the presence of sunspots in optical-wavelength observations, rather than collections of strong magnetic fields resulting to a multitude of emissions.  Although---scientifically---this relation no longer constrains the definition of an active region, the National Oceanic and Atmospheric Administration (NOAA) has continued to assign numbers to active regions based on whether they are associated to at least one sunspot observed in the visible-light range \citep{2015LRSP...12....1V}.

In this thesis, active regions related to both radio and X-ray emissions are presented.  They were imaged in UV and EUV wavelengths by the Atmospheric Imaging Assembly \citep[AIA;][]{2012SoPh..275...17L} instrument onboard the \textit{Solar Dynamics Observatory} \citep[\textit{SDO};][]{2012SoPh..275....3P}.
The AIA instrument, which consists of four telescopes, captures the entire solar disk up to $\sim$0.5~$\Rs$.  Images are taken at nine different wavelengths (94, 131, 171, 193, 211, 304, 335, 1600, and 1700~\AA), with very high temporal and spatial resolutions ($\sim$12~s and $\sim$1.5~arcsec, respectively).

\subsubsection{Solar flares}
\textit{Solar flares} are the most prolific and violent particle accelerators---in comparison to other solar activities---capable of exciting large numbers of semi-relativistic electrons.  They are sudden releases of energy ($\gtrsim 10^{30}$~erg) triggered by instabilities causing rapid re-configurations of the magnetic field, usually within active regions \citep{2004psci.book.....A}.  These localised explosive increases in brightness can be observed on the Sun across several wavelengths \citep{2011SSRv..159...19F}, but are commonly observed in (E)UV and X-ray spectra.  They are classified according to the maximum observed X-ray flux density, categorised (from weakest to strongest) as A, B, C, M, or X flares, and then into sub-divisions denoted by numbers.  An instrument that is often used to examine the X-ray emissions from flares is the X-Ray Sensor (XRS) \citep{1985SoPh...95..323T, 1994SoPh..154..275G} on board the \textit{Geostationary Operational Environmental Satellite} (\textit{GOES}).  Data from GOES/XRS is presented in this thesis.

\subsubsection{Coronal mass ejections}
\textit{Coronal Mass Ejections} (CMEs) are the second most violent solar phenomena observed.  Unlike solar flares, they are ejections of large-scale solar material and frozen-in magnetic flux that propagate away from the Sun.
For the frozen-in approximation to hold, the convection needs to dominate the diffusion (such that the magnetic Reynolds number $R_M \gg 1$), a condition easily satisfied in naturally occurring plasmas.  In the solar corona (over large scales), the magnetic Reynolds number $R_M = 10^{12}$, thus, coronal magnetic fields follow the motion of the plasma, including that carried by CMEs (see, e.g., \cite{1969pldy.book.....B}).
During and after their passage, they tend to transiently but strongly disturb the coronal environment, inducing density enhancements and observable changes in the coronal structure which can last from several minutes to several hours \citep{2006LRSP....3....2S}.
They are often related to solar flares \citep{2008ApJ...673L..95T, 2010ApJ...712.1410T}---in terms of timing and region of origin---although the nature of their relation is subject of debate as one is not always accompanied by the other \citep{2004psci.book.....A}.
Some CMEs appear to accelerate before they decelerate (impulsive CMEs) and others decelerate and then accelerate at larger distances (gradual CMEs; \cite{1999JGR...10424739S}).  Those that retain their structure up to large distances (i.e. become interplanetary CMEs) can also cross the Earth's orbit \citep{2003AnGeo..21..847V}.  Even though several morphologies and varieties of CMEs exist (see, e.g., \cite{2001ApJ...550.1093G, 2005ApJ...635L.189B, 2010ApJ...712.1410T, 2018ApJ...861..103V}), they are generally very dynamic structures that expand with increasing distance, appearing as massive clouds.  The bright leading edge of a CME is referred to as the "front", whereas its sides are known as the "flanks".

Such emissions are imaged with the use of \textit{coronagraphs}, which (artificially) eclipse the solar surface in order to emphasise the fainter surrounding coronal structures.  An example of a space-based coronagraphic instrument utilised in this thesis is the Large Angle Spectroscopic Coronagraph \citep[LASCO;][]{1995SoPh..162..357B} onboard the \textit{Solar and Heliospheric Observatory} \citep[\textit{SOHO};][]{1995SSRv...72...81D}.  LASCO was built with three white-light coronagaphic cameras tasked with imaging the corona constantly, each covering a different spatial range but with overlaps to ensure continuous coverage.  The first one, known as the C1 coronagraph, was designed to mask the solar surface up to a distance of $\sim$1.1~$\Rs$ and image up to $\sim$3~$\Rs$.  However, the C1 camera failed to restart after contact with \textit{SOHO} was temporarily lost in 1998.  The C2 camera covers plane-of-sky distances from $\sim$1.5--6~$\Rs$, although, light diffracted from the occulting disk limits the practical lower limit of the C2 field of view (FoV) to $\sim$2.2~$\Rs$ \citep{1995SoPh..162..357B}.  The C3 coronagraph covers the largest range of distances, from $\sim$3.7--30~$\Rs$.  LASCO's imaging cadence is of the order of tens of minutes, depending on the data rate.

Data obtained at different times is often combined to enhance certain coronal structures.  For example, subtracting two consecutive images from each other outputs what is known as a "running-difference" image \citep{1995SoPh..162..357B}.  On the other hand, taking the ratio of the data from two consecutive images produces a "running-ratio" image.  So-called ``base difference'' images are sometimes utilised too, where data obtained at a specific starting (or base) time is subtracted from all subsequent images.  Such images, obtained using C2 data, were utilised in the upcoming analysis.

\subsubsection{Coronal dimming}
Small regions of the solar surface near (or within) active regions often show a rapid and dramatic decrease in brightness following solar eruptive events.
This darkening is known as \textit{coronal dimming} and is interpreted as the signature of large density depletions caused by ejected solar mass \citep{2004psci.book.....A}.  Notably, their durations were found to be too short to be explained by mere radiative cooling of the corona (which occurs over larger time scales; \cite{1996ApJ...470..629H}).  Coronal dimmings are observed in (E)UV or soft X-ray (SXR) wavelengths and are often used to identify the launch site of CMEs \citep{2018ApJ...863..169D}.  A coronal dimming event associated with a CME eruption is presented in this thesis (Chapter~\ref{chap:split-band_typeII}), observed using \textit{SDO}/AIA.  Similar to white-light coronagraphic observations, running-difference and running-ratio images can be constructed to emphasise the change in intensity at the region of interest.

\subsubsection{Shocks}
A \textit{shock} is formed when the main parameters of a wave---such as the fluid density, temperature (pressure), and velocity---suffer from an abrupt discontinuity \citep{2014masu.book.....P}.  In other words, there is an abrupt transition between the undisturbed (\textit{upstream}) and disturbed (\textit{downstream}) parts of the medium.

The properties of the upstream and downstream regions of the shock front are related using the Rankine-Hugoniot conditions (or ``jump conditions'' across the shock), which describe the momentum, mass, and energy conservation within the shock \citep{1969pldy.book.....B, 2014masu.book.....P}.

Shock waves in the magnetised and ionised coronal medium are collisionless, since the shock front is significantly thinner than the mean-free path of particles \citep{1969pldy.book.....B, 2014masu.book.....P}.  However, for simplicity, their basic properties are often approximated using the magnetohydrodynamic (MHD) wave modes.
There are three main propagating MHD waves modes: the fast shock, the slow shock, and intermediate shocks.  Each of these is characterised by a different set of Rankine-Hugoniot conditions and a different Mach number \citep{Oliveira_2017}.  The Mach number is defined as the ratio between the speed of the shock wave and the characteristic speed of the medium.  One of the characteristic speeds that can be used as a proxy for the (magnetised) coronal medium is the Alfv\'en speed $V_A$, which will subsequently define the Alfv\'en Mach number $M_A$ \citep{2014masu.book.....P, Oliveira_2017}.  Another characteristic speed is the magnetosonic speed $V_{MS}$, which corresponds to the magnetosonic Mach number $M_{MS}$ \citep{Oliveira_2017}, but in this thesis, only the Alfv\'en speed $V_A$ (and Alfv\'en Mach number $M_A$) will be considered.  Shock fronts form when the Alfv\'en Mach number $M_A$ is other than unity.  The further the Alfv\'en Mach number is from unity, the stronger the shock.  In this thesis, only shocks characterised by Alfv\'en Mach numbers $M_A > 1$ will be considered, as well as only those whose downstream region has a higher density than the upstream region (although the reverse is also possible; \cite{Oliveira_2017}).

The behaviour of the shock also depends on the relative geometry of the magnetic field.  For shocks parallel to the magnetic field, the field plays no significant role.  On the other hand, shocks that are perpendicular to the magnetic field have a minimum speed that is set by the speed of compressional waves perpendicular to the magnetic field, effectively reducing the shock strength \citep{1969pldy.book.....B}.

Coronal shocks can be observed using either remote-sensing or in-situ instruments.  The remote-sensing identification of shocks is most-commonly (but not solely) based on their association with radio emissions \citep{2008A&ARv..16....1P}, whereas in-situ instruments have recorded the sharp discontinuity in the local coronal properties when crossed by the shock \citep{1999GeoRL..26.1573B, 2008ApJ...676.1330P}.  Both flares and CMEs are known to drive shock fronts \citep{1999SoPh..187...89C}.  Two radio bursts related to CME-driven shocks are analysed in this thesis.

\subsubsection{Streamers}
\textit{Streamers} are long-lived, physically long, and approximately radially-orientated structures that seem to be rooted in the solar surface but can extend over several solar radii away from the Sun \citep{1985srph.book.....M, 2004psci.book.....A}.  They are physical manifestations of open magnetic fields that extend into the corona, through which solar material escapes.  In white-light coronagraphic images, the denser plasma regions appear as bright stripes overlaid on the background coronal medium.  Just like with CMEs, these features are usually presented in running-difference or running-ratio images, enhancing the fainter, finer regions.  Many of them are associated to active regions, emanate after a solar eruptive event, and sometimes appear to be confining the expansion of solar ejections like CMEs (\cite{2005ApJ...635L.189B}), while other times they are torn apart by the ejections \citep{2018ApJ...861..103V}.  This difference in the streamers' reaction is used to identify the type of CME event, rather than the streamer itself (as is discussed in Chapter~\ref{chap:transitioning_typeII}).  Similarly, streamers can play an integral part in the transport of electron beams far into the heliosphere, which lead to radio emissions, as detailed in this thesis.

\subsubsection{Jets}
Solar \textit{jets} are transient and narrow bright features observed on the solar surface that tend to be associated with active regions \citep{2018JPhCS1100a2024S}.  They result from plasma flowing along open magnetic fields, although they are not as long and do not extend as far into the corona as streamers \citep{2004psci.book.....A}.  Instead, they look like a sharp-edged structure whose footpoint is located on a bright spot, and are observed across different wavelengths, from UV to X-rays \citep{2016A&A...589A..79M}.  This sharp-edged structure---or the body of the jet---is referred to as the \textit{spire}.  Occasionally, the jet spire can split into two components, a process known as \textit{bifurcation} \citep{2012ApJ...745..164S}.  A jet with a bifurcated spire observed by \textit{SDO}/AIA in (E)UV wavelengths is presented in Chapter~\ref{chap:transitioning_typeII}.

\subsection{Space weather} \label{sec:space_weather}
\begin{figure}[ht!]
    \centering
	\includegraphics[width=0.65\textwidth, keepaspectratio=true]{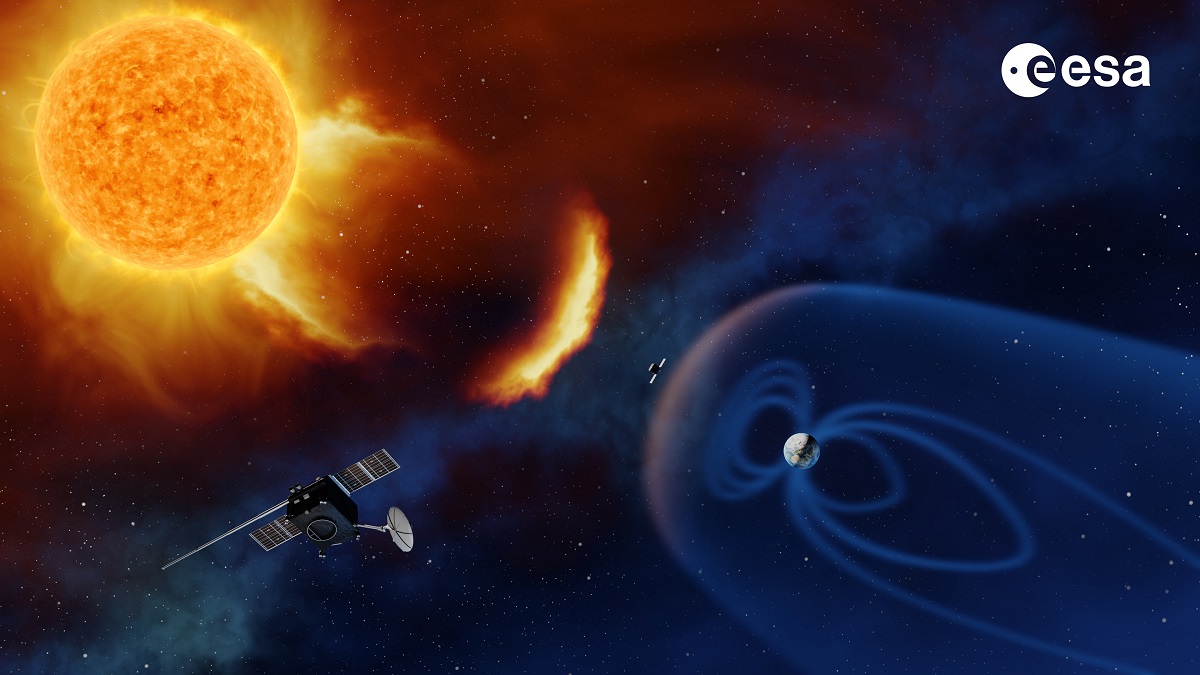}
    \caption[Graphical illustration of the solar-terrestrial relation.]
    {Artistic illustration of the solar-terrestrial relation and space weather.  A coronal mass ejection propagates through the interplanetary space, directed towards the Earth's magnetosphere and surrounding spacecraft.
    Figure credit: ESA/A. Baker, \href{https://creativecommons.org/licenses/by-sa/3.0/igo/}{\color{black}{\textsl{CC BY-SA 3.0 IGO}}}.
	}
    \label{fig:space_weather_cartoon}
\end{figure}

It should be mentioned that particles and magnetic fields propelled towards Earth by the aforementioned solar eruptive events (Section~\ref{sec:solar_activities}) can have an impact on:
\begin{enumerate*}[label=(\roman*)]
	\item the near-Earth environment (e.g., drive interplanetary shocks and damage satellite electronics),
	\item atmospheric events (e.g., excite auroras and interfere with telecommunication systems), and
	\item even ground-based activities susceptible to the induced currents (e.g., transmission and railway networks; \cite{2007LRSP....4....1P}).
\end{enumerate*}
Commonly referred to as \textit{space weather}, understanding the impact and predicting the triggers of such disturbances can be crucial for the successful shielding of the electronics dominating our modern-day functions.  Figure~\ref{fig:space_weather_cartoon} is an artistic illustration of the solar-terrestrial relation, depicting a CME propagating from the Sun towards the Earth's magnetosphere and orbiting spacecraft.

Radio emissions are particularly useful as they are early signatures of such eruptive events, especially for strong flares and CMEs that tend to pose the greatest threat \citep{2006LRSP....3....2S}.  Unlike other wavelengths, the Sun is constantly observed over a large range of frequencies within the radio domain---thanks to the plethora of inexpensive ground-based antennas installed internationally (e.g., \cite{2009EM&P..104..277B})---enabling the monitoring of heights from the solar surface until 1~au.  Moreover, due to the relation between radio emissions and the local plasma frequency (see Sections~\ref{sec:plasma_emmission} and \ref{sec:f_vs_R_relation}), radio bursts can be used to infer information on their local coronal environment, as well as their exciter.  An example of such bursts are Type II radio bursts (discussed in Section~\ref{sec:typeIIs}) which are excited by shock waves and trace the propagation of the shocks through the corona.  They are thought to be the most reliable and direct diagnostic tool of coronal shocks and their drivers, especially in the upper corona which cannot be probed in situ \citep{2008SoPh..253....3N, 2010ApJ...712..188R}.

Understanding the evolution of solar eruptions and constructing a complete picture of the sequence of events from the Sun to the Earth requires the combination of multi-wavelength observations, usually from both space-based and ground-based instruments.  Such approach is often key to identifying the generation mechanisms of specific emissions, as illustrated in Chapters~\ref{chap:split-band_typeII} and \ref{chap:transitioning_typeII}.

\section{Solar Radio Bursts} \label{sec:all_bursts}

\begin{figure}[t!]
    \centering
	\includegraphics[width=0.7\textwidth, keepaspectratio=true]{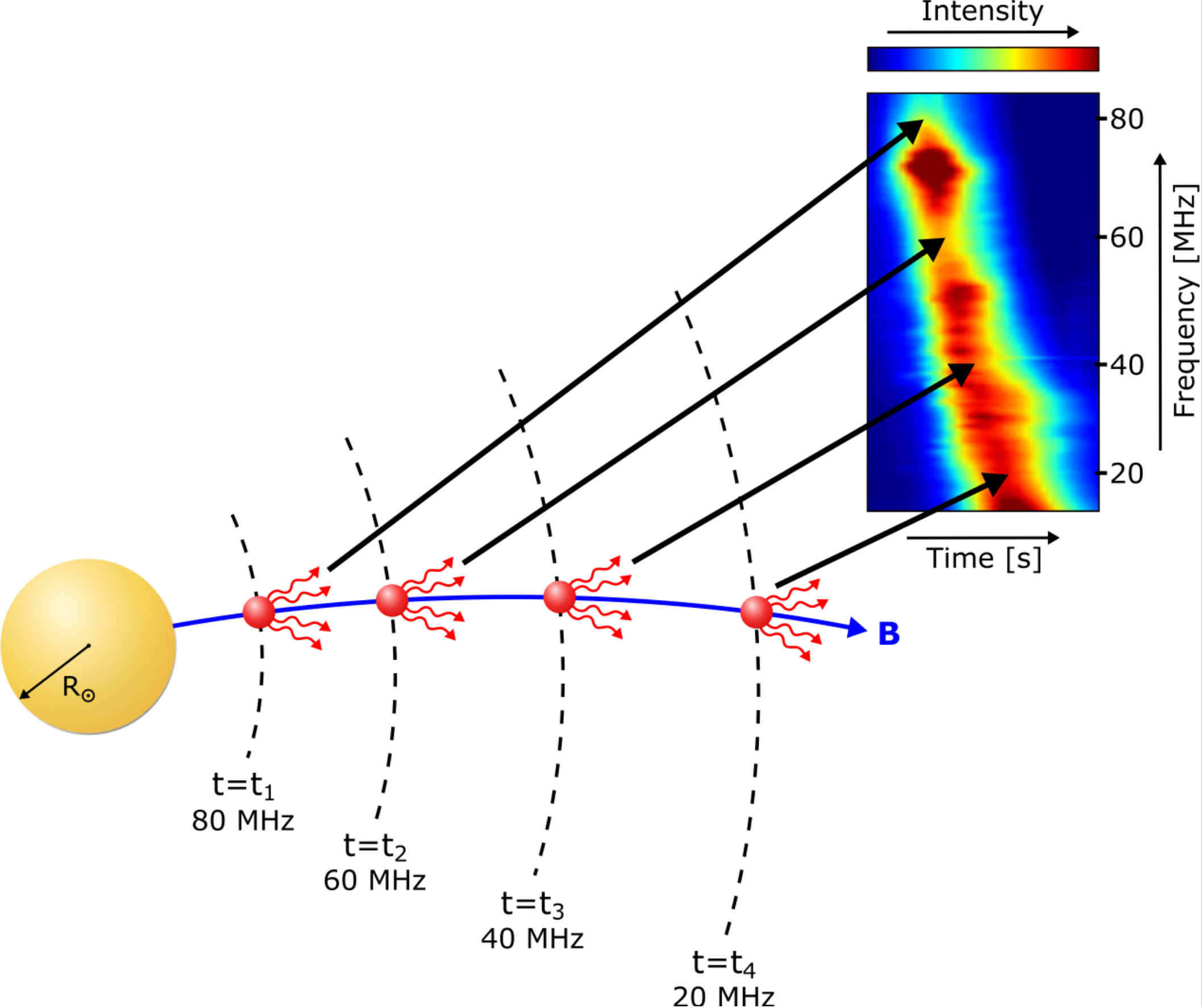}
    \caption[Cartoon of a dynamic spectrum.]
    {Schematic illustration of what dynamic spectra represent.  A radio source propagating away from the Sun along an open magnetic field is shown (i.e. a Type III source).  As it moves, the source emits radio radiation whose frequency decreases with increasing distance from the Sun (see Section~\ref{sec:f_vs_R_relation} for details), meaning that lower frequencies are emitted at a later time $t$ than higher frequencies.  Mapping the intensity of these emissions as a function of frequency and time produces a so-called ``dynamic spectrum''.  The recorded emissions reflect that the source drifts from high to low frequency over a certain period of time.
    }
    \label{fig:dyn_spec_cartoon}
\end{figure}

Solar radio astronomy was born during World War II when solar emissions interfered with the signals from metre-wavelength radars used to monitor the aerial space for aircraft \citep{1985srph.book.....M}.  These strong emissions sparked the interest of several physicists and engineers who then linked the observed interference to regions of high activity on the Sun \citep{1946PMag...37...73A}.  The effect of solar activities on Earth had thereafter become a field of interest and led to major technological developments dedicated to its study. Explosive increases in intensity during solar radio emission measurements had been termed \textit{radio bursts} \citep{1947Natur.160..256P}.  Studies by \cite{1947Natur.160..256P} emphasised the need for a radiospectrograph - a device that can record the intensities of solar emissions over continuous frequency and time steps \citep{1985srph.book.....M}.  Mapping the intensity of emissions as a function of frequency and time produces what is known as a \textit{dynamic spectrum}, shown in Figure~\ref{fig:dyn_spec_cartoon}.  Emissions that appear at higher frequencies in dynamic spectra are caused by radio sources which are located closer to the Sun in comparison to their lower-frequency counterparts, as detailed in Section~\ref{sec:f_vs_R_relation}.

\begin{figure}[t!]
    \centering
	\includegraphics[width=0.7\textwidth, keepaspectratio=true]{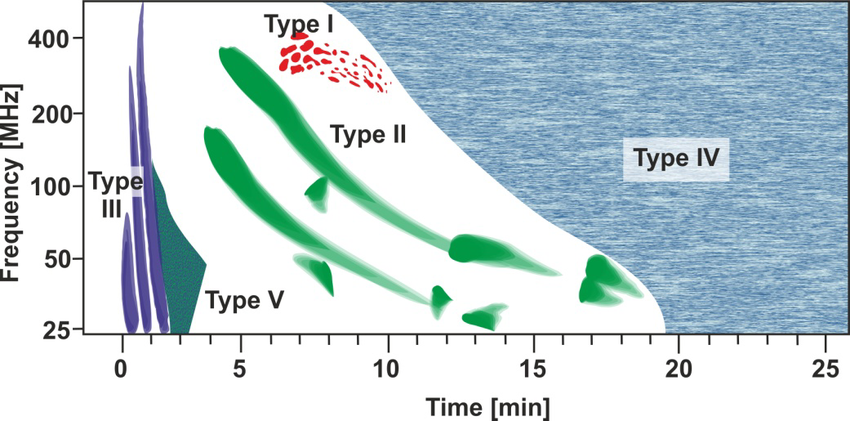}
    \caption[Morphology of the five classical solar radio bursts.]
    {Graphical representation of the distinct morphologies of the five classical types of radio bursts (Type I, II, III, IV, and V), as observed on dynamic spectra.
	Figure taken from \cite{2016AcGeo..64..825D} and reproduced under \href{https://creativecommons.org/licenses/by/4.0/}{\color{black}{\textsl{CC BY 4.0}}}.
    }
    \label{fig:all_bursts}
\end{figure}

The first radiospectrograph, referred to as the ``Aerial'', was built in Penrith in New South Wales, Australia \citep{1950AuSRA...3..387W}.  What followed was the first identification and classification of solar radio bursts from dynamic spectra---as used today---presented by \cite{1950AuSRA...3..387W}, who distinguished between the features of Type~I, Type II, and Type III bursts.  Shortly after, other categories of radio bursts were also identified, including Type IV and Type V bursts.  Collectively, these bursts are considered as the classical types of radio bursts \citep{2004psci.book.....A}.  They have distinct morphologies that can be identified in dynamic spectra, as illustrated in Figure~\ref{fig:all_bursts}.  The frequency-drift rate ($df/dt$), the duration ($\Delta t$), and the (total) bandwidth ($\Delta f_t$) of the emissions is used to characterise and categorise the different radio bursts.  Their appearance is strongly affected by the process that has excited the radio emissions, allowing for the extraction of information on the exciter mechanism, and even on the local coronal conditions (see, e.g., Sections~\ref{sec:f_vs_R_relation} and \ref{sec:bandsplitting_models}).

\begin{figure}[ht!]
    \centering
	\includegraphics[width=0.35\textwidth, keepaspectratio=true]{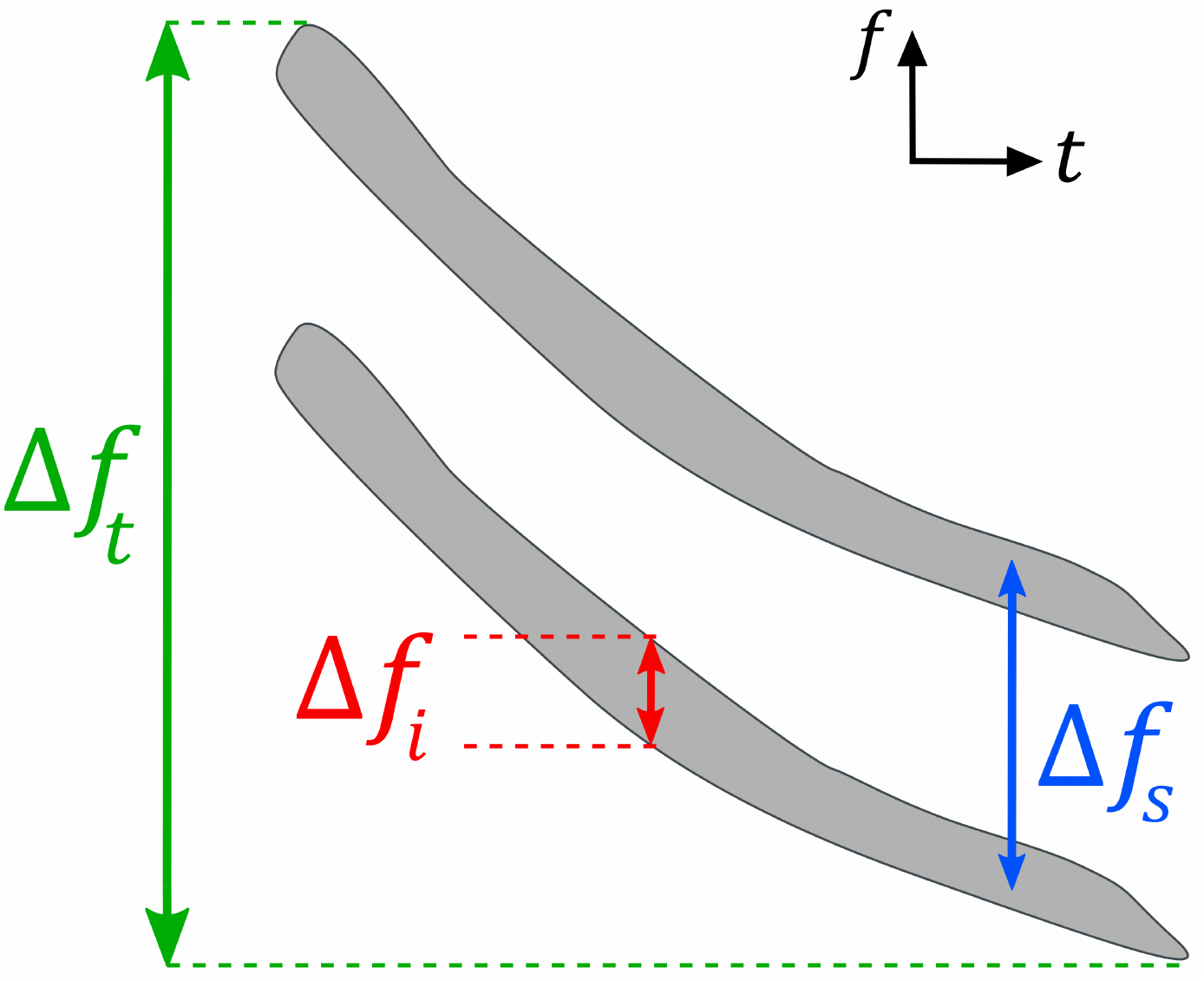}
    \caption[Schematic illustration of the different measured bandwidths.]
    {Schematic illustration of the different bandwidths mentioned in this thesis, depicted on a Type II burst.  The total bandwidth of the burst is annotated as $\Delta f_t$ (green), the instantaneous bandwidth describing the spectral width of a specific structure is annotated as $\Delta f_i$ (red), and the frequency split between two structures is denoted with $\Delta f_s$ (blue).
    }
    \label{fig:bandwidths_cartoon}
\end{figure}

The different bandwidths discussed in this thesis are depicted in Figure~\ref{fig:bandwidths_cartoon}, for clarity, where a Type II burst is used for the demonstration.  The total bandwidth of a burst (i.e. the entire range of frequencies for which it appears on the dynamic spectrum) is annotated as $\Delta f_t$, the bandwidth used to define the frequency separation (or split) between two structures is annotated as $\Delta f_s$, and the instantaneous bandwidth characterising the spectral width of a specific (fine) structure is denoted as $\Delta f_i$.

\textit{Type I} solar radio bursts are short-lived emissions with durations of $\sim$1~s and narrow bandwidths $\Delta f_t/f \simeq 0.025$ \citep{1985srph.book.....M}.  They can appear in groups, superimposed on a slowly-varying background continuum, forming a noise storm which can last for hours or days, known as a ``Type I storm'' \citep{1947RSPSA.190..357M}.

\textit{Type II} radio bursts appear as slowly-drifting lanes that tend to last for several minutes, believed to be driven by shocks \citep{1974IAUS...57..301M}.  Section~\ref{sec:typeIIs} provides an in-depth description of their characteristics and their exciting mechanism.

\textit{Type III} bursts are very spiky, short-lived emissions with higher frequency-drift rates than any other burst.  They are believed to be the manifestation of electrons accelerated by flares along open magnetic fields (see, e.g., \cite{2014RAA....14..773R}).  Consecutive Type III bursts can be observed quasi-continuously over a period of hours or days, forming a ``Type III storm'' \citep{1957IAUS....4..321W, 1985srph.book.....M}.  A detailed description of their characteristics and exciter is given in Section~\ref{sec:typeIIIs}.

\textit{Type IV} radio bursts are described as broadband emission continua that tend to be associated with solar flares and CMEs, and have durations that can vary from $\gtrsim$~$10$~min up to a few hours \citep{1985srph.book.....M}.  They often appear after Type II bursts, although this is not always the case \citep{2008A&ARv..16....1P}.  Two classifications of Type IV bursts exist: stationary Type IV bursts (often referred to as ``flare continua'') and moving Type IV bursts.  The most reliable way to distinguish moving Type IV bursts from stationary ones is by studying the imaged positions of the emission sources \citep{1974IAUS...57..301M}.  Unlike stationary Type IV sources, the emission sources of moving Type IV bursts will appear to move away from the Sun.

First identified by \cite{1959IAUS....9..176W}, \textit{Type V} bursts are short-lived continua that have durations between $\sim$10~s and a few minutes.  They appear shortly after Type III bursts (or groups of them), and due to this temporal relation, they are believed to be a by-product of Type III bursts \citep{1985srph.book.....M, 2004psci.book.....A}.

Notably, variations in the morphology of solar radio bursts and further fine structures can be observed.  It is also common for different types of radio bursts to overlap in dynamic spectra, creating complex radio emissions that do not represent their idealised morphology depicted in Figure~\ref{fig:all_bursts} (see, e.g., \cite{2001JGR...10629989R, 2007SoPh..241..145C, 2008SoPh..253....3N, 2008A&ARv..16....1P}).  Distinct variations in the idealised form of radio bursts have prompted a more detailed categorisation of emissions into sub-classes, some of which are discussed in this thesis (see Section~\ref{sec:typeIIIs} and Chapter~\ref{chap:transitioning_typeII}).  In addition, other types of radio bursts---besides these five classical ones---have been identified over the years.  One such example are the \textit{Drift-pair} bursts which are presented and analysed in Section~\ref{sec:drift_pairs}.

\subsubsection{Fine structures of radio bursts}
Fine radio burst structures are commonly observed in dynamic spectra, whether at near-Sun or near-Earth frequencies \citep{1982srs..work..182M, 2007AstL...33..192C, 2014Ge&Ae..54..406C, 2019A&A...624A..76A}.  These fine structures are sub-second emissions with narrow bandwidths that tend to be identified in dynamic spectra when sufficient temporal and spectral resolutions are available.  Radio bursts often appear as fragmented emissions---or a collection of fine structures---instead of smooth, continuous emissions.  Short-lived and narrow emissions that are not associated with broader bursts are also observed.  Fine emission patterns can generally be considered as: (i) stand-alone fine-structure bursts, or (ii) fine structures observed within a broader emission structure (i.e. ``sub-bursts'').

An example of stand-alone fine structures are the Drift-pair bursts, discussed in detail in Section~\ref{sec:drift_pairs}.  A well-known example of sub-burst emissions are the striations observed within Type III bursts, which define a sub-class known as \textit{Type IIIb} bursts (discussed in Section~\ref{sec:typeIIIs} and, e.g., \cite{2017NatCo...8.1515K}).  Sub-bursts similar to Type IIIb striae have also been observed within Type II bursts \citep{2015SoPh..290.2031D}.
Attempts to categorise the morphology of fragments and fine structures of bursts related to CME-driven shocks (Type II and Type IV bursts) into certain groups have been made \citep{2006ApJ...642L..77M, 2020ApJ...897L..15M}.  This suggests that the morphology of fragments is not unique to a single event, or, perhaps, to the specific exciting mechanism.

Fine radio burst structures are particularly interesting as they can provide a unique insight into what excites radio waves and when such excitations can occur---conditions that, in some cases, have been debated for decades (see, e.g., Section~\ref{sec:bandsplitting_models}).  The short duration and spectral bandwidth ($\Delta f_i$) of fine structures imply that they are associated with a single emission source, something that cannot always be (confidently) stated for smooth emissions (e.g., broad lower-frequency Type III bursts might be the result of several overlapping Type III bursts).  It is also interesting to explore whether fine structures arise due to the same mechanism as the broader, smooth bursts.  In other words, as long as fine structures can be resolved and fully imaged, they can be used to identify the nature of the exciter and the coronal conditions necessary for the production of radio emissions.

\subsection{Plasma emission mechanism} \label{sec:plasma_emmission}
The mechanisms causing radio emissions can be classified into two broad categories: coherent emissions and incoherent emissions.  Coherent emissions are generated by particles that emit in phase with each other after a kinetic instability affects the existing unstable particle distribution, whereas incoherent emissions result from continuum processes \citep{2004psci.book.....A, 2017RvMPP...1....5M}.  Coherent emission mechanisms are characterised by radio radiation with brightness temperatures too high ($T_B \approx 10^8 - 10^{12}$~K) to be accounted for by incoherent emissions \citep{2004psci.book.....A}.  The brightness temperature is defined as the temperature that a black body would need to have in order to produce an equal intensity to the one observed, at the given frequency.  Given that the particles emit in phase, the brightness temperature exceeds the mean energy of the emitting particles (i.e. a non-thermal brightness temperature; \cite{2008SoPh..253....3N}).

The solar radio bursts described in Section~\ref{sec:all_bursts} are excited via the plasma emission mechanism, which is a coherent emission mechanism that dominates other mechanisms at frequencies $\lesssim 1$~GHz \citep{2004psci.book.....A}.  Plasma emission arises due to the presence of electrons of varying energies in a quasi-collisionless plasma, whose velocity dispersion will be characterised by a distribution function $f(\vec{V})$.

\begin{figure}[hb!]
    \centering
	\includegraphics[width=0.5\textwidth, keepaspectratio=true]{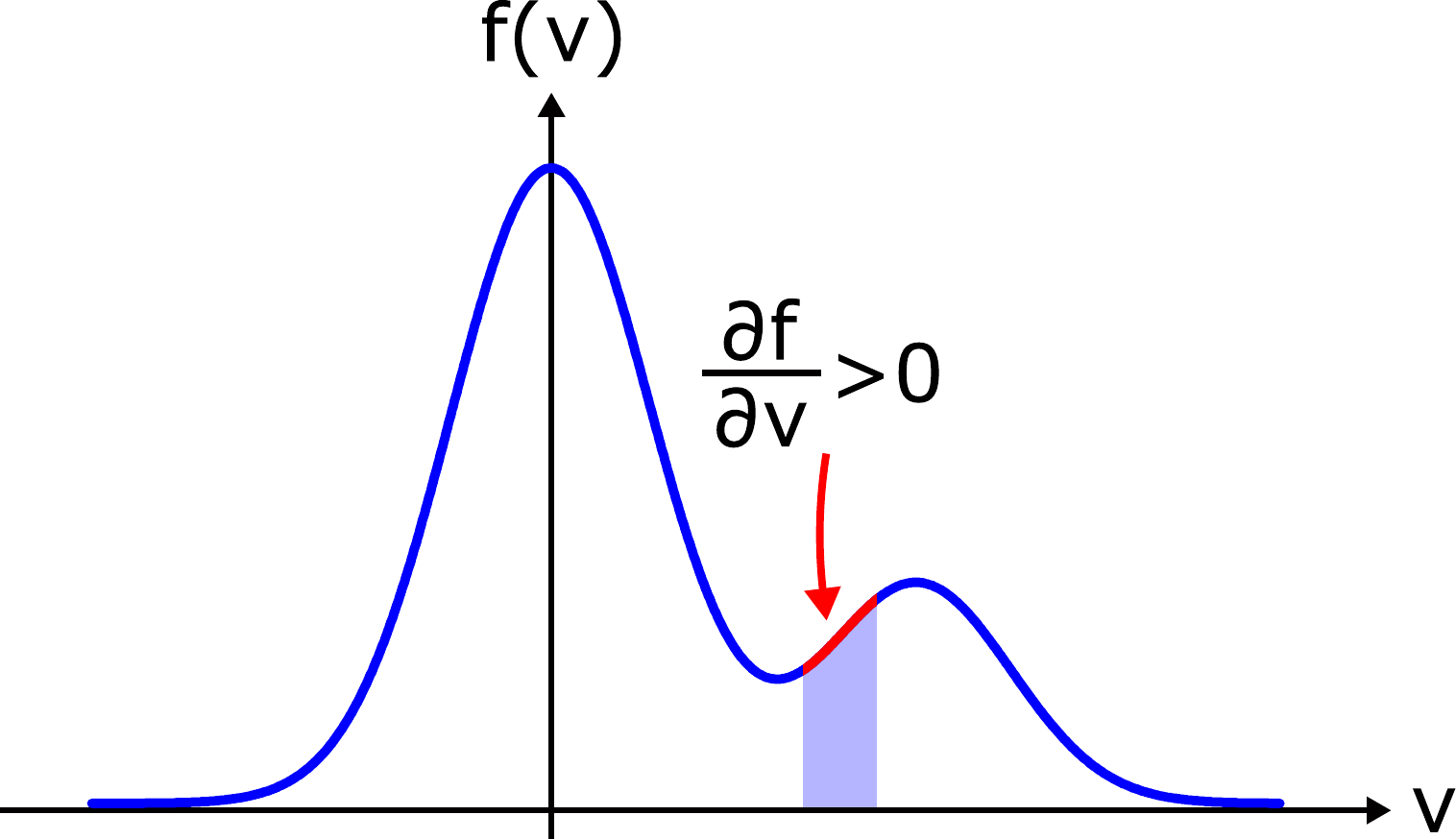}
    \caption[Velocity distribution with an electron beam.]
    {One-dimensional thermal velocity distribution $f(\vec{V})$ of electrons.  Electrons with sufficiently-high energies produce a secondary positive slope in the distribution (i.e. a ``bump''; $\partial f / \partial V > 0$), known as an electron beam.  This beam is unstable and thus susceptible to the bump-in-tail instability which can excite Langmuir waves.
	}
    \label{fig:bump_in_tail}
\end{figure}

When the higher-energy electrons are sufficiently faster than the lower-energy electrons (such that they have velocities $V \gtrsim 3 \, V_{th}$, where $V_{th}$ is the thermal speed of electrons; i.e. they are non-thermal), a positive slope ($\partial f / \partial V > 0$)---or ``bump''---forms at the high-velocity tail of the distribution \citep{2004psci.book.....A}, as shown in Figure~\ref{fig:bump_in_tail}.  If this bump occurs in the component of velocity parallel to the magnetic field ($\partial f / \partial V_\parallel > 0$), it is referred to as a ``beam'' and it is susceptible to the ``bump-in-tail'' instability, a type of streaming instability \citep{1985srph.book.....M}.  Electrons are known to be accelerated to such non-thermal speeds by solar flares and shock waves (e.g., \cite{1981PASAu...4..139M}).

When the bump-in-tail instability is triggered, electrons in the unstable beam can (on average) lose energy (through Landau damping) which is transferred to the electric field and excites electron plasma oscillations known as \textit{Langmuir waves}.  These are electrostatic waves (i.e. longitudinal) with a wavevector $\vec{k}$ parallel to the magnetic field $\vec{B}$ (since the electron beam occurs parallel to the magnetic field), meaning that the magnetic field does not affect their oscillations \citep{1985srph.book.....M}.  Therefore, the dispersion relation of Langmuir waves
\begin{equation} \label{eqn:Langmuir_disp_rel}
	\omega_L^2 = \omega_{pe}^2 + 3 \, k_L^2 \, V_{th}^2
\end{equation}
is the same in both magnetised and unmagnetised plasmas.
Here, $\omega_L$ is the (angular) frequency of the Langmuir waves, $k_L = 2\pi / \lambda$ is the (angular) wavenumber of Langmuir waves, where $\lambda$ is the wavelength, and $V_{th} = \sqrt{k_B T_e/m_e}$ is the thermal speed of the electrons, where $k_B$ is the Boltzmann constant, $T_e$ is the electron temperature, and $m_e$ is the electron mass.  The local (angular) electron plasma frequency $\omega_{pe}$ is defined as:
\begin{equation} \label{eqn:omega_pe}
	\omega_{pe} = 2\pi f_{pe} = \sqrt{\dfrac{4 \pi e^2}{m_e} \, n_e(r)} \, ,
\end{equation}
where $f_{pe}$ is the local electron plasma frequency, $e$ is the electron charge, and $n_e$ is the electron plasma density which is an inverse function of the heliocentric distance $r$ (see Section~\ref{sec:f_vs_R_relation}).  The wavenumber $k_L$ of Langmuir waves can then be expressed as
\begin{equation} \label{eqn:Langmuir_wavenum}
	k_L = \sqrt{\dfrac{m_ep(\omega_L^2 - \omega_{pe}^2)}{3 \, k_B \, T_e}} \, ,
\end{equation}
indicating that a cut-off occurs at $\omega_L = \omega_{pe}$, such that the condition $\omega_L \gtrsim \omega_{pe}$ needs to be satisfied in order for a physical wavelength $\lambda$ to exist.  Therefore, the frequency of Langmuir waves is just above the local electron plasma frequency of the corona (see, e.g., \cite{1981PASAu...4..139M}).

Plasma emission is the process through which part of the energy from Langmuir turbulence is converted into escaping (electromagnetic) radiation \citep{1985srph.book.....M}.  Langmuir waves can convert into transverse waves (i.e. radio waves) by undergoing non-linear interactions with other waves in their vicinity.  Considering the presence of Langmuir waves ($L$) generated directly by the beam, secondary (scattered) Langmuir waves ($L'$), ion-sound waves ($S$), and transverse (electromagnetic) waves ($T$), the following interactions can occur \citep{1985srph.book.....M, 2004psci.book.....A, 2017RvMPP...1....5M}:
\begin{subequations} \label{eqn:wave_interactions}
	\begin{align}
		& L \rightarrow S+L' \label{eqn:L=S+L'} \\
		& L \rightarrow T+S \label{eqn:L=T+S} \\
		& L+S \rightarrow L' \label{eqn:L+S=L'} \\		
		& L+S \rightarrow T \label{eqn:L+S=T} \\
		& T+S \rightarrow T \label{eqn:T+S=T} \\
		& T+S \rightarrow L \label{eqn:T+S=L} \\
		& T \rightarrow L+L' \label{eqn:T=L+L'} \\
		& L+L' \rightarrow T \label{eqn:L+L'=T} \, .
	\end{align}
\end{subequations}
These three-wave interactions are possible because they satisfy two conditions, known as the Manley-Rowe (or ``beat'') conditions, which can be thought of as expressions of momentum and energy conservation \citep{1985srph.book.....M, 2017RvMPP...1....5M}.  For example, in the case of two waves coalescing into a third, the Manley-Rowe conditions are given as
\begin{equation}\label{eqn:3wave_interaction_conditions}
	\vec{k}_1 + \vec{k}_2 = \vec{k}_3	
	\qquad \text{and} \qquad
	\omega(\vec{k}_1) + \omega(\vec{k}_2) = \omega(\vec{k}_3) \, .
\end{equation}

It can be seen that Langmuir waves can either decay into ion-acoustic waves and secondary Langmuir waves (Equation~(\ref{eqn:L=S+L'})), or decay into ion-acoustic and transverse waves (Equation~(\ref{eqn:L=T+S})).  They can also coalesce with ion-acoustic waves to form either secondary Langmuir waves (Equation~(\ref{eqn:L+S=L'})) or transverse waves (Equation~(\ref{eqn:L+S=T})).  The generation of transverse waves during these interactions results in radio emissions with frequencies $f_t$ close to those of the Langmuir waves---i.e. near the local plasma frequency ($f_t \gtrsim f_{pe}$)---referred to as \textit{fundamental} plasma emissions \citep{1985srph.book.....M, 2017RvMPP...1....5M}.

The proximity of the emission frequency of electromagnetic (radio) waves to the local plasma frequency can be illustrated using the dispersion relation of transverse waves in an unmagnetised plasma \citep{2017RvMPP...1....5M}:
\begin{equation} \label{eqn:EM_disp_rel}
	\omega_t^2 = \omega_{pe}^2 + k_t^2 \, c^2 \, .
\end{equation}
Here, $\omega_t$ is the (angular) frequency of transverse waves, $c$ is the speed of light, and $k_t$ is the wavenumber of transverse waves.  The wavenumber $k_t$ can therefore be expressed as
\begin{equation} \label{eqn:EM_wavenum}
	k_t = \dfrac{1}{c} \, \sqrt{\omega_t^2 - \omega_{pe}^2} \, .
\end{equation}
Similar to Langmuir waves (Equation~(\ref{eqn:Langmuir_disp_rel})), a cut-off occurs at $\omega_t = \omega_{pe}$.  Thus, transverse waves can only propagate when $\omega_t \gtrsim \omega_{pe}$, i.e. $f_t \gtrsim f_{pe}$.

These transverse waves can couple with ion-acoustic waves to further produce transverse waves (Equation~(\ref{eqn:T+S=T})) or Langmuir waves (Equation~(\ref{eqn:T+S=L})).
Alternatively, they can decay into Langmuir and secondary Langmuir waves (Equation~(\ref{eqn:T=L+L'})).
The production of secondary Langmuir waves is important for the generation of Langmuir turbulence \citep{2004psci.book.....A}.

Langmuir waves can also coalesce with secondary Langmuir waves and convert into transverse waves (Equation~(\ref{eqn:L+L'=T})).  This interaction results in radiation emitted close to the second-harmonic of the local plasma frequency ($f_t \approx 2 f_{pe}$), i.e. \textit{harmonic} plasma emission (see, e.g., \cite{1985srph.book.....M}, \cite{2004psci.book.....A}, or \cite{2017RvMPP...1....5M} for a review).

As shown, beam-driven plasma emissions have frequencies that are a strict function of the coronal electron density $n_e$ \citep{2004psci.book.....A}.

\subsection{The frequency-distance relation} \label{sec:f_vs_R_relation}
The electron plasma density $n_e$ is a function of the heliocentric distance $r$.  While it is known that the basal coronal density decreases with increasing distance from the Sun, the exact relation is unknown and can vary depending on the solar activity and associated disturbances (see Section~\ref{sec:solar_activities}).  Several studies have utilised a statistically-significant number of observations to derive empirical relationships between the coronal density and the heliocentric distance.  An example of such an empirical relationship was deduced by \cite{1961ApJ...133..983N} using K-coronameter observations of the upper corona ($< 3 \, \Rs$) during a sunspot maximum (i.e. solar maximum) period:
\begin{equation} \label{eqn:n_Newkirk}
	n_e = N \cdot n_0 \cdot 10^{4.32 R_\sol / r} \quad \mathrm{[cm^{-3}]} \, ,
\end{equation}
where $N$ is a constant (such that N=1 for the ``one-fold'' Newkirk model; or ``1$\times$Newkirk'') and $n_0 = 4.2\times10^4 \, \mathrm{cm^{-3}}$.  It should be emphasised that this Newkirk model (which is spherically symmetric) assumes a radial evolution of the density, i.e. it is a one-dimensional (1D) model.

Equation~(\ref{eqn:omega_pe}) can be written as:
\begin{equation} \label{eqn:fpe_vs_ne}
	f_{pe} = \kappa \, \sqrt{n_e} \, ,
\end{equation}
where the constant $\kappa = \sqrt{e^2 / \pi m_e}$, with the density $n_e$ given in $\mathrm{cm^{-3}}$ and the plasma frequency $f_{pe}$ given in Hz.
Therefore, by combining the density model in Equation~(\ref{eqn:n_Newkirk}) with Equation~(\ref{eqn:fpe_vs_ne}), the heliocentric distance $r$ can be expressed as a function of the plasma frequency $f_{pe}$:
\begin{equation} \label{eqn:r_Newkrik}
	\dfrac{r}{R_\sun} = \dfrac{2.16}{\log_{10}(f_{pe})-\log_{10}(\kappa \, \sqrt{n_{0}N})} \,.
\end{equation}
The frequency $f$ at which radio waves are emitted (and observed) is just above the local plasma frequency ($f \gtrsim f_{pe}$), as described in Section~\ref{sec:plasma_emmission}.  Since the exact ratio between the observed and the plasma frequency is unknown, but is close to 1, it is often convenient to assume that $f = f_{pe}$.  In other words, the distance of the radio source away from the Sun is directly related to the observed frequency of radio emissions.

Dynamic spectra provide the frequency-drift rate $df/dt$ of radio bursts (see Figures~\ref{fig:dyn_spec_cartoon} and~\ref{fig:all_bursts}).  The speed $V_{exc}$ of the exciter of the radio bursts can be inferred from the observed frequency-drift rate (taken, here, in \si{\hertz \per \second}) using the chain rule:
\begin{equation*}
	\dfrac{df}{dt} = \dfrac{df}{dr} \, \dfrac{dr}{dt} = \dfrac{df}{dr} \, V_{exc} \, .
\end{equation*}
By differentiating $df/dr$, where $f=\kappa \sqrt{n_e}$, the following expression is obtained:
\begin{equation} \label{eqn:dfdt_all_models}
	\dfrac{df}{dt} = \dfrac{1}{2} \, \dfrac{f}{n_e} \, \dfrac{dn_e}{dr} \, V_{exc} \, ,
\end{equation}
and thus
\begin{equation} \label{eqn:v_exc}
	V_{exc} = \dfrac{2 n_e}{f} \, \dfrac{df}{dt} \left( \dfrac{dn_e}{dr} \right)^{-1} \, ,
\end{equation}
where
\begin{equation*}
	\dfrac{dn_e}{dr} = n_e \, \dfrac{d}{dr} \ln(n_{e}) \, ,
\end{equation*}
such that \citep{2017NatCo...8.1515K}:
\begin{equation} \label{eqn:v_exc_Newkirk}
	V_{exc} = \dfrac{2}{f} \, \dfrac{df}{dt} \left(\dfrac{d}{dr} \ln(n_{e})\right)^{-1} \, .
\end{equation}

Solving for the Newkirk density model (Equation~(\ref{eqn:n_Newkirk})) gives:
\begin{equation*}
	\left(\dfrac{d}{dr} \ln(n_{e})\right)^{-1} = \dfrac{-r^2}{4.32 R_\sun \ln(10)} \, .
\end{equation*}

It is evident from Equation~(\ref{eqn:v_exc}) that by combining the observed drift rate---as obtained from dynamic spectra---with a coronal density model, the exciter speed can be estimated.  If the density model characterises the radial evolution in the corona (like the Newkirk model does), then the inferred exciter speed can only be interpreted as the radial speed (and may, thus, not be representative of the true exciter propagation).

\subsection{Type II solar radio bursts} \label{sec:typeIIs}
\begin{figure}[t!]
    \centering
	\includegraphics[width=0.9\textwidth, keepaspectratio=true]{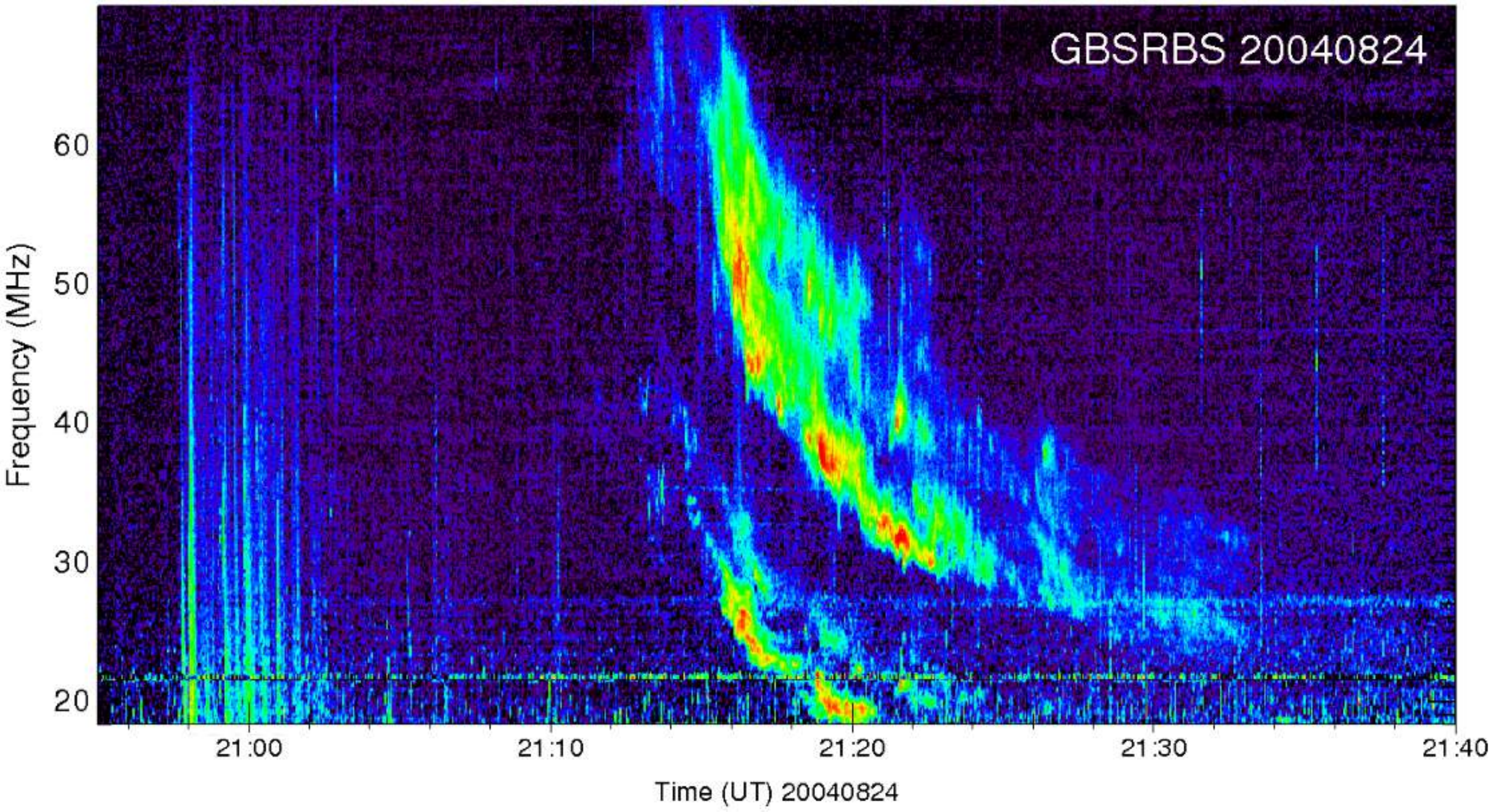}
    \caption[Split-band Type II burst with fundamental and harmonic emissions.]{Dynamic spectrum depicting a group of Type III bursts (around 21:00~UT) followed by a Type II burst with fundamental and harmonic bands, both of which experience band splitting.
    Figure courtesy of Stephen M. White; reproduced and adapted for this thesis with permission.
    }
    \label{fig:typeII_bursts}
\end{figure}

Radio emissions that slowly drift from high to low frequencies at rates of $\lesssim -1$~$\MHzs$ are referred to as Type II solar radio bursts (Figure~\ref{fig:typeII_bursts}; \cite{1950AuSRA...3..399W, 1985srph.book.....M}).  A negative frequency drift rate implies that the emitter progressively encounters regions of lower densities, a behaviour reflecting its increasing distance from the Sun (given that $f \propto \sqrt{n_e(r)}$; see Section~\ref{sec:f_vs_R_relation}).
Due to the typical exciter speeds inferred from the observed drift rates (Section~\ref{sec:f_vs_R_relation}), Type II bursts are believed to be the manifestations of radio emissions excited by shock waves which are driven by solar eruptive events like flares and CMEs (see Section~\ref{sec:solar_activities}; \cite{1962ApJ...135..138M, 1999SoPh..187...89C, 2000JGR...10518225L, 2008SoPh..253....3N, 2014SoPh..289.2123K}).

However, the intrinsic location of Type II radio sources on the shock front has been disputed.  Some studies imply that radio sources are excited near the nose of the CME (i.e. the leading edge; see, e.g., \cite{2012ApJ...752..107R, 2012A&A...547A...6Z}), and others suggest excitation at the flanks of the CME (see, e.g., \cite{2007A&A...461.1121C, 2018A&A...615A..89Z}).  As a result, current observational evidence is inconclusive as both cases are supported (e.g., \cite{2014SoPh..289.2123K}).  A statistical analysis of interplanetary Type II bursts has suggested that excitation near the flanks is more likely (see, e.g., \cite{2019ApJ...882...92K}).  The flanks of the CME are often considered to be a more probable location as this is where a compression between the CME-driven shock and regions of enhanced density---like coronal streamers---is likely to occur, creating conditions thought to be favourable for Type II emission excitation \citep{2003ApJ...590..533R}.

As seen from Figure~\ref{fig:typeII_bursts}, Type II bursts tend to last over several minutes, have a narrow instantaneous bandwidth $\Delta f_i$ (cf. Figure~\ref{fig:bandwidths_cartoon}), and can appear in pairs with a frequency ratio close to 1:2.  These pairs of Type II bands are the result of plasma emissions excited at both the local plasma frequency $f_{pe}$ (forming the fundamental band) and its second harmonic $2f_{pe}$ (forming the harmonic band), as described in Section~\ref{sec:plasma_emmission}.  Each of these bands can split into thinner lanes, or ``subbands'', a phenomenon known as \textit{band splitting} \citep{1950AuSRA...3..399W, 1959AuJPh..12..327R}.  Another common feature is that the emissions that outline the Type II shape do not always appear to be continuous with time, but instead can be fragmented or patchy (e.g., \cite{1959AuJPh..12..327R, 2001JGR...10629989R, 2001A&A...377..321V}).

Type II bursts that experience band splitting---referred to as \textit{split-band Type II} bursts---are identified via a number of typical characteristics.  First, a 1:2 frequency ratio between the subbands is not observed, meaning that they are not harmonically related.  Both the upper-frequency ($f_U$) and lower-frequency ($f_L$) subbands evolve in a synchronised manner in frequency and time---appearing as quasi-parallel lanes---and the intensity fluctuations across the two subbands are similar \citep{1985srph.book.....M, 2001A&A...377..321V}.
The similarity between the subbands suggests that their emission sources may propagate through the same coronal density region simultaneously (see, e.g., \cite{1974IAUS...57..389S, 1975ApL....16R..23S, 2001A&A...377..321V}).
The relative frequency split
\begin{equation} \label{eqn:deltaf_over_f}
	\dfrac{\Delta f_s}{f} = \dfrac{f_U - f_L}{f_L}
\end{equation}
between the subbands (see Figure~\ref{fig:bandwidths_cartoon}) is found to be approximately constant within a single event, but also varies very little from one event to another, ranging between 0.1 and 0.5 \citep{2001A&A...377..321V, 2015ApJ...812...52D}.  The physical reason behind this narrow range of observed $\Delta f_s/f$ values between split-band Type II bursts is unknown.  Some studies suggested that there is a link between the amount of frequency split $\Delta f_s/f$ and the emission frequency, but the exact relation seems to be ambiguous.  \cite{1974IAUS...57..389S, 1975ApL....16R..23S} found that the split increases with increasing frequency ($\Delta f_s \simeq 0.27 f - 3.0$ and thus $\Delta f_s/f \simeq 0.27 - 3.0 f^{-1}$), whereas \cite{2004A&A...413..753V} who repeated the analysis for a larger range of observed frequencies found that the average $\Delta f_s/f$ values increase with decreasing frequency ($\Delta f_s/f = 0.37 f^{-0.061}$), although considerable variation in the data was present.

Spiky, short-lived emissions are sometimes seen to emanate from the Type II band (referred to as the ``backbone'') which have much higher drift rates than the backbone itself, but (normally) somewhat lower than that of Type III bursts (e.g., \cite{1987SoPh..111..365C, 2004SoPh..222..151M, 2005A&A...441..319M, 2015A&A...581A.100C, 2019NatAs...3..452M}).  The spikes that appear on the higher-frequency side of the backbone show positive frequency-drift rates (implying motion towards the Sun), whereas the ones on the lower-frequency side have negative frequency-drift rates (implying motion away from the Sun).  These structures are known as ``herringbones'' and are believed to be caused by shock-accelerated electrons escaping along open magnetic fields, similar to a Type III burst \citep{1959AuJPh..12..327R}.  Herringbones do not always appear to emanate from both sides of the backbone and are sometimes observed without the presence of a backbone (see, e.g., \cite{1985srph.book.....M, 1987SoPh..111..365C}).

Another interesting aspect of Type II bursts is that, occasionally, multiple Type II lanes that are neither harmonically related nor can be classified as split bands have been observed \citep{1985srph.book.....M, 2015SoPh..290.1195F, 2015AdSpR..56.2811Z}.  Moreover, what are thought to be Type II bands emitted at the third harmonic (i.e. at $3 f_{pe}$) have also been reported \citep{1994ESASP.373...95A, 1996A&AS..119..489M, 1998A&A...331.1087Z}.

Occasionally, shock-related narrow-band emissions that show little or no drift with frequency are also observed.  These emissions have been termed as \textit{stationary} (or quasi-stationary) Type II bursts \citep{2002A&A...384..273A}, differentiating them from the classical \textit{drifting} Type II bursts (see Figure~\ref{fig:typeII_bursts}).  Stationary Type II bursts have been interpreted as the signatures of standing shocks that are related to solar flares, known as termination shocks \citep{2002A&A...384..273A, 2004ApJ...615..526A, 2009A&A...494..669M, 2019ApJ...884...63C}.  No drift with frequency implies that the radio source does \textit{not} propagate into a region where the ratio of the emitter's density ($n_{s}$) to the local background coronal density ($n_{bg}$) is different than that of its previous location.  In other words, $n_{s}/n_{bg}$ remains constant, thus the source continues to emit at the same frequency over time.

\subsection{Band-splitting models} \label{sec:bandsplitting_models}
The mechanisms causing band splitting in Type II radio bursts have long been debated.  Several models have been proposed over the decades in an attempt to describe the characteristics of split-band Type II bursts, but none of them has dominated over the rest.  Instead, there currently are two interpretations with opposite predictions that have been widely-accepted (e.g., \cite{2012A&A...547A...6Z, 2018A&A...609A..41M}).  These are the \cite{1974IAUS...57..389S, 1975ApL....16R..23S} and \cite{1983ApJ...267..837H} band-splitting models.  As described in this section, the major difference between these models---and the one relevant to this thesis---is the origin of the subband sources with respect to the shock front.

Specifically, it is unclear whether the two subband sources are located on a single side of the shock front, or both sides.  The \cite{1974IAUS...57..389S, 1975ApL....16R..23S} model is based on the assumption that one source is located in the \textit{upstream} region (the undisturbed region ahead of the shock) and the other is located in the \textit{downstream} region (the disturbed region behind the shock).  The \cite{1983ApJ...267..837H} model, on the other hand, requires both subband sources to be located upstream of the shock front.

In some cases (like in the \cite{1983ApJ...267..837H} model) the orientation of the shock at the location where the subband sources originate is also crucial.  The shock is described as \textit{quasi-perpendicular} when the upstream magnetic field is approximately perpendicular to the vector normal to the plane of the given emission location on the shock.  When, however, the magnetic field is approximately parallel to the shock's normal vector, the shock is referred to as \textit{quasi-parallel}.

\subsubsection{Upstream shock emissions}
The \cite{1983ApJ...267..837H} model attributes band splitting to radiation produced by electrons reflected (and accelerated) at two different locations upstream of the shock front, where the curvature of the shock front is quasi-perpendicular to the local magnetic field (acting as a magnetic mirror).  In this case, the two subband sources are expected to be physically \textit{separated}.  The relative frequency split ($\Delta f_s/f$) characterising split-band Type II bursts is explained by the different locations along the curved shock front which are found at different heliocentric heights.
Since the coronal density decreases with increasing distance from the Sun, at any given time, the subband sources encounter different densities to each other, meaning that they emit at different frequencies (Section~\ref{sec:f_vs_R_relation}).

\subsubsection{Upstream and downstream shock emissions}
The \cite{1974IAUS...57..389S, 1975ApL....16R..23S} model attributes band splitting to the simultaneous emission of radio waves from the upstream (ahead, undisturbed) and downstream (behind, disturbed) regions of a shock front.  Given that the thickness of the shock is negligible (see Section~\ref{sec:solar_activities}), the sources of each subband are emitted from the same source region and are thus expected to be virtually \textit{co-spatial}.  The observed frequency split between the subbands results from the density jump between the upstream and downstream regions of the shock forming the discontinuity.  Therefore, the source on the upstream region emits at a lower frequency than that of the downstream region.

Due to the presumed relation of the frequency split to the density jump across the shock front, the Rankine-Hugoniot jump conditions can be invoked \citep{1974IAUS...57..389S, 1975ApL....16R..23S, 2014masu.book.....P}.  The inferred shock speed (see Section~\ref{sec:f_vs_R_relation}) and the inferred density jump (Equation~(\ref{eqn:shock_density_jump_n2/n1})) allow for the estimation of the Alfv\'en Mach number, the Alfv\'en speed, and the local coronal magnetic field, as illustrated below \citep{1974IAUS...57..389S, 1975ApL....16R..23S, 1995A&A...295..775M, 2002A&A...396..673V}.

The relative frequency split $\Delta f_s/f$ (Equation~(\ref{eqn:deltaf_over_f})) can be related (using $f = \kappa \sqrt{n} \,$; see Equation~(\ref{eqn:fpe_vs_ne})) to the density jump $n_U/n_L$ across the shock front via
\begin{equation} \label{eqn:shock_density_jump_deltaf/f}
	\dfrac{\Delta f_s}{f} = \dfrac{f_U - f_L}{f_L} = \dfrac{f_U}{f_L} - 1 = \sqrt{\dfrac{n_U}{n_L}} - 1 \, .
\end{equation}
Here, $n_U$ and $n_L$ represent the densities corresponding to the upper-frequency source (emitted downstream of the shock front) and the lower-frequency source (emitted upstream of the shock front), respectively.
The density jump $n_U/n_L$ is also known as the electron density ``compression ratio'' $X$ \citep{1995A&A...295..775M, 2014masu.book.....P} and can be expressed as
\begin{equation} \label{eqn:shock_density_jump_n2/n1}
	X \equiv \dfrac{n_U}{n_L} = \left( \dfrac{f_U}{f_L} \right)^2 = \left( \dfrac{\Delta f_s}{f} + 1 \right)^2 \, .
\end{equation}
For example, in the case of a perpendicular shock ($\theta = 90\degr$)---i.e. one whose normal is at $90\degr$ to the upstream magnetic field---and an adiabatic index $\gamma_\alpha$ taken to be $5/3$, the relationship between the compression ratio $X$ and the Alfv\'en Mach number $M_A$ is given by \citep{2002A&A...396..673V}:
\begin{equation} \label{eqn:Alfven_Mach_number}
	M_A = \sqrt{ \dfrac{X (X + 5 + 5\beta)}{2 (4-X)} } \, ,
\end{equation}
where $\beta$ is the plasma beta (i.e. the ratio between the plasma and magnetic pressures).  Therefore, assuming a value of $\beta$ (e.g., 0.5) allows for the estimation of the Alfv\'en Mach number $M_A$.
The Alfv\'en speed $V_A$ can in turn be calculated as
\begin{equation} \label{eqn:Alfven_speed_vs_M_A}
	V_A = \dfrac{V_{exc}}{M_A} \, ,
\end{equation}
where $V_{exc}$ is the exciter speed (in this case the shock speed) estimated using Equation~(\ref{eqn:v_exc}), obtained from the observed frequency-drift rate of the Type II burst in dynamic spectra.
The Alfv\'en speed $V_A$ depends on the magnetic field $B$ as:
\begin{equation} \label{eqn:Alfven_speed_vs_Bfield}
	V_A = \dfrac{B}{\sqrt{\mu_0 \, \rho}} = \dfrac{B}{\sqrt{\mu_0 \, m_p \, n_e}} \, ,
\end{equation}
where $\mu_0$ is the permeability of free space, $m_p$ is the proton mass, and $n_e$ is the electron density, such that $\sqrt{n_e} = \kappa / f_{pe}$ (Equation~(\ref{eqn:fpe_vs_ne})).  Therefore, substituting for all constants,
the (upstream) local magnetic field $B$ can be approximated as \citep{1974IAUS...57..389S, 1975ApL....16R..23S}:
\begin{equation} \label{eqn:local_Bfield}
	B \approx 5.1 \times 10^{-5} \, V_A \, f_{pe} \, ,
\end{equation}
where $V_A$ is given in $\kms$, $f_{pe}$ is given in MHz, and the magnetic field is given in gauss (where $10^{4} \, \mathrm{G} = 1 \, \mathrm{T}$).

The ability to obtain the local coronal conditions at such a large range of distances through observations of split-band Type II radio bursts is what makes the \cite{1974IAUS...57..389S, 1975ApL....16R..23S} model attractive.  Hence, it is often applied in the literature, but without particular evidence that it is the mechanism at play for the studied event.  Moreover, the assumptions made in order to obtain the presented expressions (Equations~(\ref{eqn:Alfven_Mach_number})--(\ref{eqn:local_Bfield})) may not represent the true coronal conditions.  To add to that, the physical mechanism causing band splitting is contested (as explained in this section).  Thus, any values describing the coronal conditions deduced by applying this model ought to be used conservatively, until robust evidence for this band-splitting interpretation becomes available (see Chapter~\ref{chap:split-band_typeII}).

\subsection{Type III solar radio bursts} \label{sec:typeIIIs}
Some of the most commonly-observed and intense radio emissions are Type III bursts, easily distinguished by their spiky morphology, broad frequency bandwidths ($\sim$100~MHz), and very short duration of no more than a few seconds (see Figures~\ref{fig:all_bursts} and \ref{fig:typeII_bursts}; \cite{1950AuSRA...3..387W, 1985srph.book.....M}).  Due to their high drift rates, they are believed to be excited by energetic (semi-relativistic) electron beams that trace open magnetic fields, leading to their spiky appearance (as shown in Figure~\ref{fig:dyn_spec_cartoon}; \cite{2018A&A...611A..57M}).  The derived exciter speeds (Equation~(\ref{eqn:v_exc})) tend to be a fraction of the speed of light $c$, ranging between $\sim 0.1 c - 0.6 c$, with the fastest ones observed at higher frequencies where the background density changes faster with distance, corresponding to a higher frequency-drift rate \citep{1985srph.book.....M, 2014RAA....14..773R}.  They are often temporally and spatially associated to solar flares, which are thought to be the accelerators of such energetic electrons.  They have been observed over a very large range of frequencies, corresponding to distances from the low corona up to (and beyond) 1~au.  When they appear at frequencies below $\sim$1~MHz, they are referred to as ``interplanetary'' Type III bursts \citep{2014RAA....14..773R}.

The configuration of the magnetic field dictates the motion of Type III sources. If, for example, the magnetic field is not open, variants of Type III bursts known as Type J or Type U bursts can form, named after their respective morphology in dynamic spectra (e.g., \cite{2017A&A...606A.141R}).  These are simply Type III bursts whose electron beam is confined by, and thus follows, the curvature of the magnetic field, eventually propagating towards the Sun.

\begin{figure}[t!]
\centering

\captionsetup[subfigure]{position=top, aboveskip=0em, belowskip=0em}

\begin{subfigure}[t]{0.5\textwidth}
    \centering
    \includegraphics[width=1\textwidth, keepaspectratio=true]{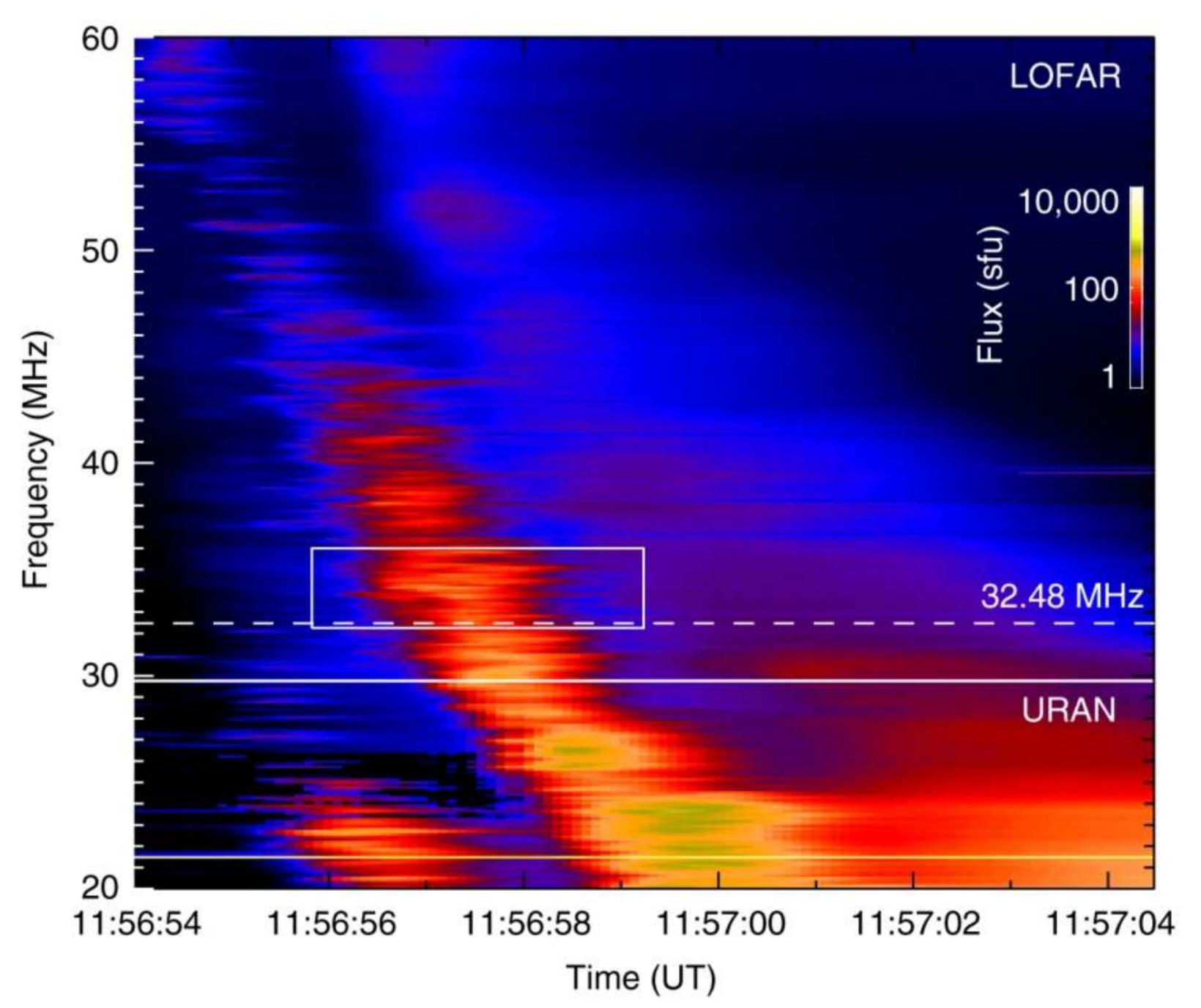}
    \caption{}
    \label{fig:TypeIIIb_F_H}
\end{subfigure}%
\begin{subfigure}[t]{0.5\textwidth}
    \centering
    \includegraphics[width=1\textwidth, keepaspectratio=true]{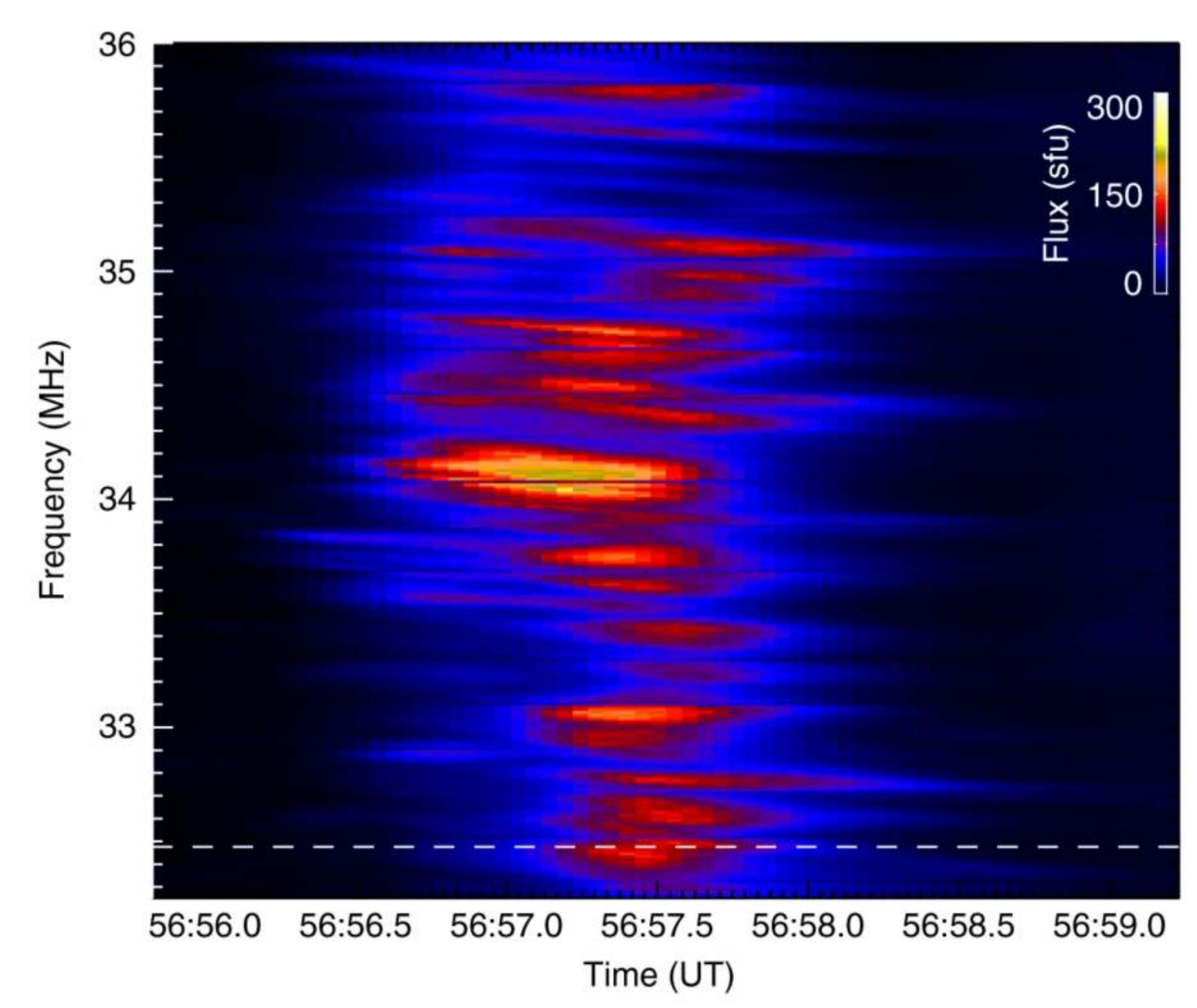}
    \caption{}
    \label{fig:TypeIIIb_F_striae}
\end{subfigure}

\caption[Type IIIb burst striae.]
{
(a)~Fundamental and harmonic band of a Type IIIb burst observed on 16 April 2015.
(b)~Expanded section of the fundamental Type IIIb band highlighting the fine structures known as striae.  This section is indicated by the white box in panel (a).
Figure taken from \cite{2017NatCo...8.1515K} and then adapted (under \href{https://creativecommons.org/licenses/by/4.0/}{\color{black}{\textsl{CC BY 4.0}}}).
}
\label{fig:TypeIIIb_bursts}
\end{figure}

Type III bursts that occasionally display multiple highly-elliptical fine structures whose duration corresponds to the major axis---known as \textit{striae}---have also been observed in dynamic spectra \citep{1967AuJPh..20..583E}.  The striae show no, or very little, drift with frequency, ranging from approximately 0 to $\sim$\nolinebreak$-$0.3~$\MHzs$ \citep{2018SoPh..293..115S}.  When striations are present, the bursts are referred to as \textit{Type IIIb} bursts, a sub-class of Type III bursts \citep{1972A&A....20...55D}.
Striae are believed to be the result of small-scale density inhomogeneities in the corona that modulate the emitted radiation \citep{1975SoPh...40..421T, 2018ApJ...856...73C}.  Figure~\ref{fig:TypeIIIb_bursts} shows an example of such a Type IIIb burst, where both fundamental and harmonic emissions were detected.  As can be seen, the fundamental band has well-defined striae (highlighted in Figure~\ref{fig:TypeIIIb_F_striae}), whereas the harmonic striae are broader and cannot be easily distinguished, forming a smoother band.

Type IIIb bursts are of particular importance as they allow for an (order-of-magnitude) estimation of the characteristic size of the region emitting the radio waves, i.e. the intrinsic source size.  This is possible thanks to the very narrow instantaneous frequency bandwidth $\Delta f_i$ of the individual striae (cf. Figure~\ref{fig:bandwidths_cartoon}).  Specifically, since they are observed as distinguished fine structures, one can reasonably assume that they result from a single source, and that the radial extent $dr$ of that emitting source is reflected by the spectral extent $\Delta f_i$ of the emissions on dynamic spectra (since $f$ is a function of $r$; see Section~\ref{sec:f_vs_R_relation}).  For example, a fundamental Type IIIb stria near $\sim$32~MHz has a ful width at half maximum (FWHM) width $\Delta f_i \lesssim 0.3$~MHz, measured at the peak-flux time \citep{2017NatCo...8.1515K, 2018SoPh..293..115S}.
Therefore, it can be assumed that the radio source emits from $f=32.15$~MHz to $f=31.85$~MHz.  The spatial extent can then be approximated by assuming a density model.  For example, the 1$\times$Newkirk coronal density model (Equation~(\ref{eqn:n_Newkirk}); $N=1$) suggests that there is a heliocentric separation of $dr \approx 6$~arcsec between the emission locations of the two frequencies, implying that the intrinsic FWHM size of the source is approximately 0.1$\arcmin$.  In other words, the fundamental source can be considered as a point source which subtends a small solid angle $\Omega \lesssim 10^{-2}$~$\mathrm{arcmin^{2}}$ (see, e.g., \cite{2017NatCo...8.1515K}).

\section{Observing Solar Radio Emissions} \label{sec:radio_obs}

The very first instruments capable of detecting sporadic solar radio emissions were limited to observing frequencies above $\sim$10~MHz, meaning that only emissions near the Sun's surface could be identified and studied.
This restriction---known as the ``ionospheric cut-off''---is imposed by the density of the Earth's ionosphere which reflects radio waves below $\sim$10 MHz and does not allow them to reach ground-based detectors \citep{1986psun....2.....S, 2013A&A...556A...2V}.  As technology and the interest in solar radio bursts advanced, spacecraft that were capable of detecting radio emissions without being confined by the ionospheric cut-off were employed, exploring emissions below $\sim$10~MHz and thus larger heliocentric heights (including the near-Earth environment; \cite{1986psun....2.....S}).  Space-based instruments can combine in-situ and remote-sensing data to probe the radio emissions and the ambient coronal conditions (e.g., \cite{2020A&A...642A..12M}).
On the other hand, many commonly-observed solar radio bursts are excited by solar eruptive events like solar flares and CMEs, so they are more likely to be observed at higher frequencies due to the spatial proximity to the solar surface and origin of these eruptive events.  Hence, ground-based instruments that capture emissions $\gtrsim 10$~MHz still allow for the examination of a significant number of bursts and any relevant fine structures.
Moreover, ground-based instruments maintain a significant advantage over any space-based instruments: they can provide a much higher sensitivity and angular resolution, crucial for resolving the fine structures of bursts (see Section~\ref{sec:all_bursts} and Figure~\ref{fig:TypeIIIb_bursts}).

Radio telescopes with larger effective collecting areas (i.e. apertures)---whether single large disks or interferometers---will observe with a higher spatial resolution than "smaller" telescopes (see Section~\ref{sec:lofar}).  Observations at higher frequencies require larger collecting areas $D$ than lower frequencies in order to be resolved to the same degree, since a detector's resolution is proportional to $\lambda/D$, where $\lambda$ is the wavelength of the received radiation (see Equation~(\ref{eqn:lofar_res})).  The largest possible collecting areas are achieved using interferometers: collections of multiple radio antennas that are strategically placed and spaced so that the signals measured at each receiver can be merged into a single signal.
The signal recorded at two antennas is delayed (due to the geometry) by
\begin{eqnarray} \label{eqn:interf_delay}
	\tau_{g} = \dfrac{D}{c} \, \sin(\theta_{s}) \, ,
\end{eqnarray}
where $D$ is the distance between the two antennas (known as  the ``baseline''), $c$ is the speed of light, and $\theta_{s}$ is the angle between the normal to the baseline vector and the vector pointing towards the source \citep{2017isra.book.....T}.
By knowing the position of the antennas and the delay between the signal received at different antennas, the location of the emitting source can be determined.

Data products of radio instruments can be classified as either spectroscopic or imaging.  Instruments with purely spectroscopic outputs are sometimes referred to as ``radiospectrographs'', whereas those that conduct imaging observations are sometimes referred to as ``radioheliographs'' \citep{1985srph.book.....M}.  Radiospectrographs measure the intensity of radio emissions as a function of frequency and time.  Those capable of simultaneously recording multiple intensity profiles at consecutive frequencies produce the so-called dynamic spectra (see Section~\ref{sec:all_bursts}).  Radioheliographs, on the other hand, simply image the regions from which radio waves are excited (i.e. intensity versus position) with respect to the Sun.

For the purposes of conducting an in-depth analysis of any radio emissions, both the dynamic spectra and images are required.  Dynamic spectra are used to identify the type of radio bursts captured, but images are necessary to examine the characteristics and evolution of the emission sources.  It is therefore desirable to have an instrument that can produce both dynamic spectra and images.

In order to have a one-to-one correlation between the spectroscopic structures and the emission sources (especially for fine structures), it is crucial to be able to record both the spectra and images with the same temporal and spectral resolutions.  This, however, proved to be technologically challenging to achieve (until recently), due to the large computational power demanded.
As such, radio telescopes resorted to imaging the emissions only at selected frequencies.
For example, the Culgoora radioheliograph produced 2D emission images at a few fixed frequencies.
Images were initially taken at 80~MHz and later at 160~MHz, having the potential to capture both the fundamental and harmonic emissions of a burst.  Eventually, 43 and 327~MHz were also added to the imaging capabilities \citep{1967PASAu...1...38W, 1973IEEEP..61.1312S, 1980A&A....88..203D, 1985srph.book.....M}.
Some of the most advanced low-frequency radio observations were conducted at the Nan\c{c}ay Radio Observatory in France.  Two spectrographs and one radioheliograph were dedicated to solar observations.  The radioheliograph, known as the Nan\c{c}ay Radioheliograph \citep[NRH;][]{1997LNP...483..192K}, covered frequencies from 150--450~MHz and was, until recently (see Section~\ref{sec:lofar}), the instrument with the highest imaging capability for low-frequency radio observations.  Nevertheless, NRH could only image at a maximum of 10 fixed frequencies which---even though an improvement from previous telescopes--was very limiting and did not allow the examination of fine radio structures.  It is worth pointing out the NRH is still operational.  Although observations were interrupted in early 2015, observations recommenced in November 2020 following an upgrade, and the radioheliograph is expected to become fully-operational in March 2021.  However, the upgrade did not affect the number of frequencies imaged by NRH, which is still limited to a maximum of ten.

\subsection{LOFAR: the LOw-Frequency ARray} \label{sec:lofar}
The LOw-Frequency ARray \citep[LOFAR;][]{2013A&A...556A...2V} is a ground-based radio interferometer that commenced operations in late 2010.  It consists of two main types of antennas, the Low-Band Antenna (LBA) composed of dipoles, and the High-Band Antenna (HBA) composed of tiles (where each tile is a group of 16 dipoles).  They collectively cover the largely-unexplored frequency range of 10--240~MHz, corresponding to wavelengths from 30 to 1.25~m (i.e. within the decametric and metric domain; \cite{2011A&A...530A..80S, 2013A&A...556A...2V}).  Specifically, the LBA is designed to operate between 10--90~MHz (starting from the ionospheric cut-off), and the HBA between 110--240~MHz.  The gap from 90--110~MHz exists because this frequency range is reserved for the purposes of commercial FM radio broadcasting \citep{2013A&A...549A..11O}, as per the United Nations Geneva Agreement of 1984.

LOFAR has an innovative phased-array design constructed with low-cost, fixed antennas, meaning that no part of the telescope has to physically move.  Instead, in order to ``point'' the telescope's beams at the desired location, a powerful computer uses the phase delays recorded at each antenna to digitally re-construct the signal coming from the direction of interest.
Additionally, LOFAR is a powerful tool as it allows for observations with very long baselines---thanks to its stations being distributed over several countries---whilst not limiting users to a specific baseline size.  As of this date, the International LOFAR Telescope (ILT) consists of 24 core stations (located in Exloo, the Netherlands), 14 remote stations (spread across the Netherlands), and 14 stations spread across 7 other European countries, with more stations under way.  The 12 inner-most core stations are located in a compact area known as the Superterp \citep{2013A&A...556A...2V}.  The number of antennas and layouts differ between core, remote, and international stations.  Core and remote stations consist of 48 HBA and 96 LBA antennas, and each of the core HBA stations is split into two sub-stations (2$\times$24 antennas each).  The international HBA and LBA stations consist of 96 antennas each.

LOFAR has unprecedented observing capabilities with very high temporal, spectral, and spatial resolutions, as well as sensitivity \citep{2013A&A...556A...2V}.  However, what makes it ground-breaking is its ability to \textit{simultaneously} output both spectroscopic data (presented in the form of dynamic spectra) and imaging data.  Crucially, it is the first radio telescope to produce images at the same temporal and spectral increments (i.e. same resolution) as for the dynamic spectra.  In other words, for every pixel on the dynamic spectrum, LOFAR produces a corresponding 2D image of the radio emissions.  This means that the radio source of every emission structure that appears on the dynamic spectrum at a given time and frequency can be studied.  This makes LOFAR the prime radio telescope for studying fine structures of solar radio bursts and their (sub-second) evolution both in time and frequency.

Full-Stokes (Stokes I, Q, U, and V) polarisation measurements can also be conducted \citep{2013A&A...556A...2V}.  The Stokes I parameter describes the total intensity of the radiation, the Stokes Q and U parameters are used to define the linear polarisation, and Stokes V is used to define the circular polarisation of the radiation.  However, the opportunity cost of choosing to record all four Stokes parameters is that the computational power needs to be diverted from other processes.  For this reason, it is sometimes preferred to record only the Stokes I parameter (total intensity) in order to preserve a higher processing power for other parameters like, for example, the number of formed (synthesised) beams whose upper limit is defined by the computational power available.  This approach was adapted for the observations presented in this thesis, since (in this case) the polarisation information would not provide any essential diagnostic information.  Instead, it was deemed vital to maintain the most beams and highest temporal, spectral, and spatial resolutions possible; crucial for an adequate analysis of the fine, sub-second radio burst structures.

Observations can be conducted with a single antenna, a single station, or a combination of antennas and stations.
The FoV of a single antenna is referred to as an ``element beam'', a summation of the signals from all station elements produces a ``station beam'', and a summation of the signals received at multiple stations produces an ``array beam'' \citep{2011A&A...530A..80S, 2013A&A...556A...2V}.  Due to LOFAR's signal processing power, a station can be split into several beams and each beam can be individually pointed at a different direction (within the combined FoV), making LOFAR a very flexible instrument \citep{2011A&A...530A..80S, 2013A&A...556A...2V}.

The higher the number of antennas used, the higher the angular resolution of the synthesised beam, given that the baseline becomes longer.  The nominal FWHM angular resolution $\theta_{res}$ (or ``spatial resolution'') of synthesised LOFAR beams is given (in radians) by:
\begin{equation} \label{eqn:lofar_res}
	\theta_{res} = \alpha \, \dfrac{\lambda}{D} = \alpha \, \dfrac{c}{fD} \, ,
\end{equation}
where $\alpha$ is a constant ($\sim 1$) that depends on the imaging weighting scheme and the chosen array configuration, $\lambda = c/f$ is the wavelength of the observed signal (with $f$ and $c$ being the observed frequency and speed of light, respectively), and $D$ is the largest (projected) separation between the outermost antennas in use (i.e. the maximum baseline; \cite{2013A&A...556A...2V}).  The FoV of the beam---i.e. the FWHM beam area $A_{beam}$---is therefore defined as:
\begin{equation} \label{eqn:lofar_fov}
	A_{beam} = \pi \left( \dfrac{\theta_{res}}{2} \right)^2 \, .
\end{equation}
It is important to note that Equation~(\ref{eqn:lofar_res}) (and consequently Equation~(\ref{eqn:lofar_fov})) represents an ideal (spherical beam) scenario.  In practise, the shape and size of the beam in the $xy$-plane of the sky depends on---for example---the altitude of the source and is not a perfect circle, meaning that the angular resolution of the beam is somewhat lower \citep{2013A&A...556A...2V}.  The higher the altitude (or declination---if measured from the equator), the less elliptical the shape of the beam, improving the angular resolution.  Higher altitudes also correspond to a lower atmospheric attenuation, i.e. less absorption and scattering of the incoming radiation by the Earth's atmosphere.  Hence, observations around noon (12:00---at the stations' location) are preferred since the Sun is found at its maximum elevation above the horizon.  Moreover, observations near the summer solstice (which is in June for the Northern Hemisphere) are also preferred as the Sun reaches higher altitudes than any other time of the year.  As such, the observations presented in this thesis were conducted around midday and near the summer solstice.
Another factor to consider when attempting to estimate the angular resolution of the synthesised beams for a given observation with higher accuracy, is that the effective collecting area $D$ is the maximum baseline projected perpendicular to the source direction, at any given time (i.e. it is a function of the source's elevation).

It is also worth mentioning that LOFAR's ability to image at any frequency and time step (defined by its temporal resolution) offers several advantages to analyses of emissions, beyond the ability to investigate fine structures.  One of these advantages is that the \textit{evolution} of radio source properties can be examined at a single frequency (and consecutive times), and will therefore not be influenced by any ionospheric scintillations.
Ionospheric effects (like refraction) can alter the apparent absolute position of sources, but crucially, they are frequency-dependent and vary over larger time scales (on the order of minutes) than those investigated in this thesis (i.e. sub-seconds; see, e.g., \cite{2016AcGeo..64..825D} and \cite{2019ApJ...873...48G}).  Therefore, an ionospheric calibration (beyond the one automatically conducted during LOFAR observations; \cite{2013A&A...556A...2V}) is redundant for studies that focus on the evolution of radio source properties (primarily positions), such that the relative (and not absolute) values are of interest (e.g., \cite{2018SoPh..293..115S}), as presented in this thesis.  In other words, if the emissions are impacted by the ionosphere, they are affected in an equal manner, meaning that their apparent evolution is not distorted.

\subsection{LOFAR Low-Band Antenna (LBA)} \label{sec:LBA}
The LBA frequency range (10--90~MHz) starts from the lowest frequency that can be observed from Earth due to the ionospheric cut-off ($\sim$10~MHz).  However, the operational range of the LBA antennas is limited in practise from 30 to 80~MHz \citep{2012A&A...543A..43V, 2013A&A...556A...2V}.  Observations below $\sim$30~MHz are severely polluted by Radio-Frequency Interference (RFI).  This is caused by ionospheric reflections of low-frequency signals from man-made sources---like mobile, broadcasting (AM radio), navigation, and military system transmission---back towards the ground \citep{2011A&A...530A..80S}.  Additionally, there is RFI at higher frequencies ($\gtrsim 80$~MHz) due to FM radio broadcasting.

\begin{figure}[ht!]
    \centering

	\captionsetup[subfigure]{aboveskip=0.5em, belowskip=0em}
	
\begin{subfigure}[t]{0.49\linewidth}
    \centering
    \includegraphics[width=1\textwidth, keepaspectratio=true]{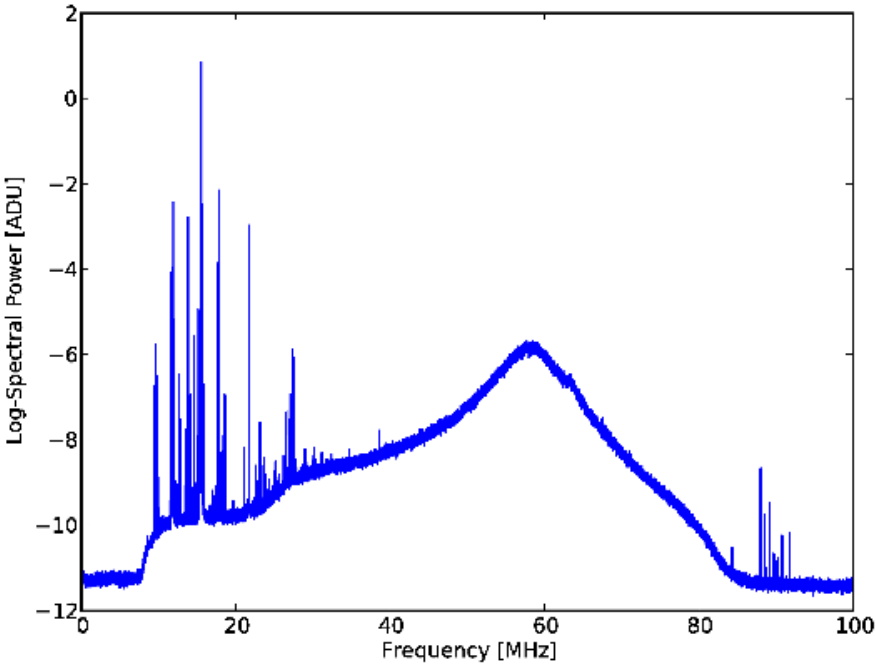}
    \caption{}
    \label{fig:lba_rfi}
\end{subfigure}%
\begin{subfigure}[t]{0.514\linewidth}
    \centering
    \includegraphics[width=1\textwidth, keepaspectratio=true]{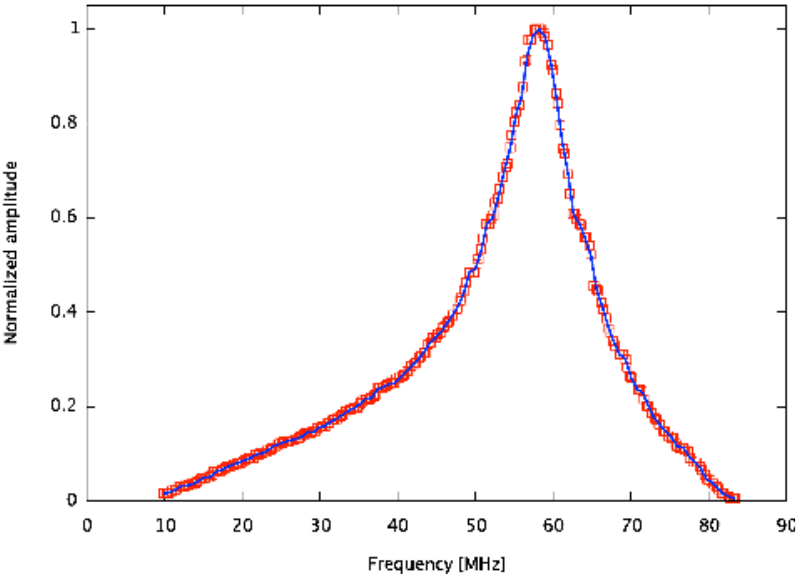}
    \caption{}
    \label{fig:lba_bandpass}
\end{subfigure}

    \caption[LOFAR/LBA spectral power and global bandpass.]
    {
    (a)~Averaged spectral power as a function of frequency for a LOFAR core LBA station.
    (b)~Normalised global bandpass of the LBA antennas.
    Figures taken from \cite{2013A&A...556A...2V}, reproduced with permission \small{\copyright}~ESO.
    }
    \label{fig:lba_curves}
\end{figure}

The presence of RFI from these combined sources within the LBA range is indicated by the sharp peaks in Figure~\ref{fig:lba_rfi}, which depicts the averaged spectral power of a given LBA core station.  It should be noted that RFI is not constant and (to some degree) varies from one observation to another.  During observations, an analogue filter can be applied to suppress frequencies below 30~MHz \citep{2013A&A...549A..11O}, as chosen for the observations presented in this thesis.

Figure~\ref{fig:lba_bandpass} depicts an example of the normalised global bandpass of the LBA antennas \citep{2012A&A...543A..43V, 2013A&A...556A...2V}.  The bandpass is a property defining the sensitivity of a detector to incoming radiation as a function of frequency.  It is clear from the peak in this figure (and Figure~\ref{fig:lba_rfi}) that the LBA antennas are most sensitive to frequencies around 58~MHz---a characteristic property of the LBA.  Due to their physical structure, the dipoles forming the LBA antennas have a resonance frequency around 58~MHz (in dry conditions; \cite{2013A&A...556A...2V}), meaning that the response of the antenna increases at that frequency.  This implies that emissions near $\sim$58~MHz will artificially appear to be brighter relative to emissions at other frequencies.  Thus, flux calibration (see Section~\ref{sec:flux_cal}) is essential for the correct interpretation of the emissions observed in dynamic spectra.

Moreover, the core LBA stations can be divided into several configurations.  One of those is the ``LBA outer'' configuration which uses the 48 outermost core antennas.  Observations using the LBA outer antennas benefit from a maximum baseline of $\sim$3.5~km, which provides an angular resolution of $\sim$10$\arcmin$ at 30 MHz (see Equation~(\ref{eqn:lofar_res}); \cite{2017NatCo...8.1515K}).

\subsection{LOFAR's tied-array observing mode} \label{sec:tied-array}
The versatility of LOFAR is reflected in the different observing modes that can be utilised, each optimised to suit different observational objectives.  Multiple observing modes can be simultaneously run, adding to LOFAR's unprecedented flexibility.  The three major observing modes are: (i) the interferometric imaging mode, (ii) beam-formed modes, and (iii) direct storage modes \citep{2013A&A...556A...2V}.  Beam-formed modes---where weighted additions of the beam signals are performed---can be used to observe with very high temporal resolutions which are desired for capturing transient solar radio bursts and the sub-second evolution of their properties (e.g., \cite{2011arXiv1105.0661M, 2013A&A...556A...2V, 2017A&A...606A.141R}).  There exist three beam-formed sub-modes: (a) Coherent Stokes (which is of relevance to this thesis), (b) Incoherent Stokes, and (c) Fly's Eye.

As mentioned in Section~\ref{sec:lofar}, several beams can be summed together.  The Coherent Stokes sub-mode applies a coherent summation of the beams, forming what is referred to as a ``tied-array beam'' \citep{2011arXiv1105.0661M, 2013A&A...556A...2V}.
In a coherent summation, the phase (and time) delays between signals received at individual antennas (whether geometric, instrumental, or environmental) are corrected and aligned (i.e. made coherent) before the antenna signals are added \citep{2011arXiv1105.0661M}.  This coherent combination of signals results in a sensitivity that is equivalent to that of the total collecting area of all the stations used \citep{2011A&A...530A..80S}.
More than one tied-array beam can be created for each station beam \citep{2011A&A...530A..80S}.  Tied-array observations offer very high sensitivity, temporal, spectral, and spatial resolutions---all necessary for resolving the fine radio burst structures.

As already mentioned, radio emissions from a source arrive at different antennas at different times, introducing a phase delay in the received signal which depends on the antenna's location.  If a coherent summation of the signals is to be achieved, LOFAR needs to align the phase of the signal at each antenna---\textit{in real time}---by correcting for such geometric delays, as well as other known instrumental and environmental delays (like ionospheric delays; \cite{2013A&A...556A...2V}).  To enable the phase alignment without the need for a real-time clock calibration, the same clock signal has been implemented for all 24 LOFAR core stations.  Hence, only core stations are used for tied-array observations.  If larger baselines than those offered by the core stations were to be used, the FoV of the synthesised beams would decrease.  As a result, more beams would be needed to cover the same area around the Sun with the same beam density, at any given frequency.  To achieve this would require additional computational power that would have to be diverted from other processes, potentially impacting the quality of other observing properties like the temporal and spectral resolutions, something which is undesirable \citep{2011arXiv1105.0661M, 2013A&A...556A...2V}.  In addition to that, as will be demonstrated throughout this thesis, the spatial resolution offered by core stations is more than enough to resolve the solar radio emission sources which are broadened by radio-wave scattering effects (described in Section~\ref{sec:prop_effs} and in subsequent chapters; or see, e.g., \cite{2004P&SS...52.1381B}).

\begin{figure}[ht!]
    \centering
	\includegraphics[width=0.7\textwidth, keepaspectratio=true]{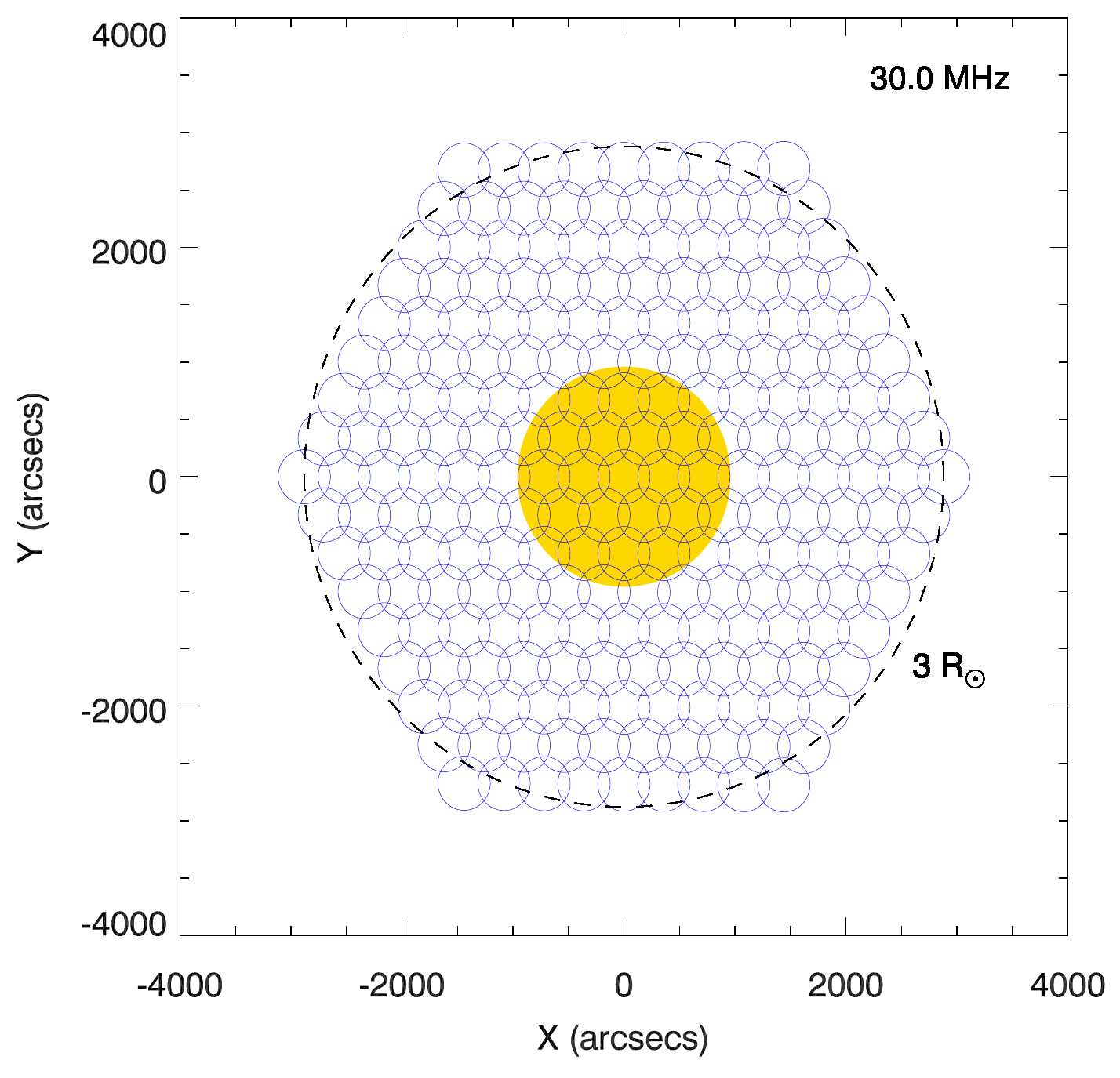}
    \caption[LOFAR tied-array beam mosaic.]
    {LOFAR tied-array beam mosaic formed from 217 individual beams, covering an area up to $\sim$3~$\Rs$ (shown by the black dashed line).  The yellow disk represents the solar surface and the blue circles indicate the theoretical FWHM areas of the synthesised beams at 30~MHz, as given by Equation~(\ref{eqn:lofar_fov}).
    }
    \label{fig:lofar_beams}
\end{figure}

Figure~\ref{fig:lofar_beams} depicts a tied-array beam of 217 individual beams obtained using the LBA outer core stations (i.e. a maximum baseline of $\sim$3.5~km).  At 30~MHz, the resulting mosaic covers a hexagonal area around the Sun which extends up to $\sim$3~$\Rs$.  The theoretical FoV of each synthesised beam is $< 10\arcmin$ (calculated using Equation~(\ref{eqn:lofar_fov})).  For comparison, Type III sources at $\sim$30~MHz have a FWHM of $\sim$20$\arcmin$ \citep{1980A&A....88..203D, 2017NatCo...8.1515K}, therefore the sources are resolvable.  It can also be seen that the small separation between the beams ensures partial overlapping at the lower frequencies.  In this case, each beam centre is separated by $\sim$6$\arcmin$ from the centres of its neighbouring beams.  It is worth pointing out that a 10~MHz source is located at $\lesssim 3 \, \Rs$, according to the 1$\times$Newkirk density model (Equation~(\ref{eqn:r_Newkrik})), meaning that this FoV is adequate for observing emissions within the LBA range.  A source at $\sim$30~MHz---which is the lowest frequency recorded in the presented observations---is found even closer to the solar centre.

\begin{figure}[ht!]
    \centering
	\includegraphics[width=0.6\textwidth, keepaspectratio=true]{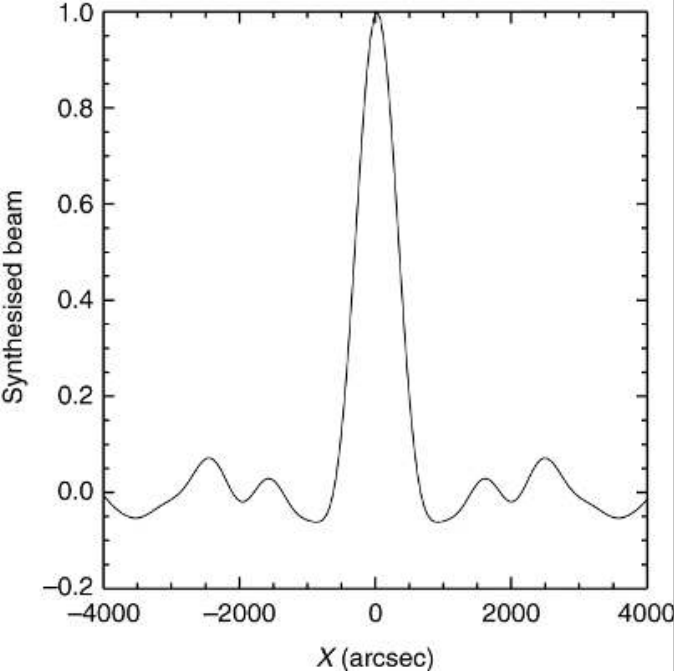}
    \caption[LOFAR tied-array synthesised beam cross-section.]
    {1D cross-section of LOFAR's tied-array synthesised beam at $\sim$32~MHz, assuming central beam separations of $\sim$6$\arcmin$.
    Figure taken from \cite{2017NatCo...8.1515K} and then adapted (under \href{https://creativecommons.org/licenses/by/4.0/}{\color{black}{\textsl{CC BY 4.0}}}).
	}
    \label{fig:lofar_psf}
\end{figure}

Compact tied-array configurations correspond to better coverage of the plane of the sky (and thus a better ``$uv$-coverage''), meaning that they result in synthesised beams with smaller side-lobes (e.g., \cite{1999ASPC..180..537H}).  In other words, increasing the number of beams comprising the tied-array mosaic and decreasing the spacing between them decreases the side-lobes and, therefore, decreases any unwanted emission contributions.  A 1D cross-section (along the $y$-axis) through the centre of one of LOFAR's tied-array beams is shown in Figure~\ref{fig:lofar_psf}.
In this case, 127 beams with a centre-to-centre spacing of $\sim$6$\arcmin$ were used for the observation.  As evident, the main lobe (or ``primary beam'') is centred at $x=0$ and the associated side-lobes do not exceed 10\% of the maximum intensity value, for the given configuration \citep{2017NatCo...8.1515K}.
The negative side-lobes result from the fact that interferometers sample discrete points in the $uv$-plane, due to the minimum possible spacing between two antennas (defined by mechanical considerations; see, e.g., \cite{2013A&A...556A...2V} and \cite{2017isra.book.....T}).  This implies that there is an incomplete $uv$-coverage, leading to a loss of information (destructive interference) on particular angular scales in the sky brightness distribution.

The signal to noise ratio for coherent signal summations increases linearly with an increase in the number of stations used \citep{2013A&A...556A...2V}.  The sensitivity of each tied-array beam is equal to the cumulative sensitivity of the combined stations used for the observation.  For single-polarisation measurements, the sensitivity of each synthesised beam
is estimated as:
\begin{equation} \label{eqn:lofar_sensitivity}
	\Delta S_i = \dfrac{S_{sys}}{\sqrt{N_s(N_s-1) \, dt \, df}} \, ,
\end{equation}
where $dt$ and $df$ are the integration time and bandwidth, respectively (i.e. the resolutions),
$N_s$ is the number of stations, and $S_{sys}$ is the system sensitivity (or System Equivalent Flux Density (SEFD)) given by
\begin{equation} \label{eqn:lofar_SEFD}
	S_{sys} = \dfrac{2 \, \eta \, k_B}{A_{eff}} \, T_{sys} \, ,
\end{equation}
where $\eta$ is the system efficiency factor ($\approx 1$), $k_B$ is the Boltzmann constant, $A_{eff}$ is the effective area, and the system noise temperature $T_{sys} = T_{sky} + T_{instr}$.  Here, $T_{sky} \approx 60 \, \lambda^{2.55}$ (in K; where $\lambda$ is the wavelength of the incoming radiation in metres) is the sky temperature and $T_{instr}$ is the instrumental noise temperature \citep{2013A&A...556A...2V}.
Below $\sim$65~MHz, the LBA system noise temperature is dominated by the sky temperature \citep{2013A&A...556A...2V}.

It should be emphasised that tied-array observations are able to produce both dynamic spectra and images with \textit{equal} sensitivity and temporal and spectral resolutions, an ability sometimes referred to as ``imaging spectroscopy'' (e.g., \cite{2017A&A...606A.141R, 2019A&A...631L...7K}).  Each beam records one dynamic spectrum.  At a given time and frequency, the signals from each beam are combined in order to obtain the dynamic spectrum of the entire tied-array beam.
Similarly, the intensity observed by each beam at each frequency and time can be related to the beam's location.  Combining all beam positions and their respective intensities (at the given time and frequency) produces a 2D image; a snapshot of the radio emissions with respect to the Sun.  If a radio burst is observed, the beams pointing at the source's location will record higher intensities than the rest of the beams, meaning that the source's location can be inferred.  The coordinates of the LOFAR beam centres in the raw data files are supplied in terms of right ascension ($\alpha_b$) and declination ($\delta_b$)---in radians---as this is how the beam directions are defined during the observation.  These values are transformed into Cartesian ($X_b$, $Y_b$) coordinates using their offset from the solar disk centre (which is located at $\alpha_{c}$ and $\delta_{c}$) and a rotation defined by the polar angle $\theta_p$ (the angle from the solar north pole to the celestial north, which accounts for the Earth's precession; \cite{2017A&A...606A.141R}) as follows:
\begin{align*}
	X_b & = - (\alpha_b - \alpha_c) \cos(\delta_c) \cos(\theta_p) + (\delta_b - \delta_c)\sin(\theta_p) \, , \\
	Y_b & = + (\alpha_b - \alpha_c) \cos(\delta_c) \sin(\theta_p) + (\delta_b - \delta_c)\cos(\theta_p) \, .
\end{align*}
Then, the $X_b$ and $Y_b$ locations are translated into arcsecs (as presented in Figure~\ref{fig:lofar_beams}) and the intensity values associated to each beam are interpolated between the beam locations in order to re-construct the radio emission images \citep{2017A&A...606A.141R}.

\subsubsection{Calibrating the flux of tied-array observations} \label{sec:flux_cal}

Besides the emissions from the radio burst source of interest, a variety of unwanted radio emissions can contribute to the observed signal.  Contributions like the radiation from the quite Sun and the background (galactic) radiation are continuous \citep{1985srph.book.....M}.  In order to accurately represent the radio burst emissions, these contributions need to be estimated and subtracted from the aggregate signal received during the radio burst observation.  In addition to that, as mentioned in Section~\ref{sec:LBA}, the response of the LBA antenna peaks at $\sim$58~MHz (Figure~\ref{fig:lba_bandpass}).  This instrumental enhancement in power needs to be corrected, in order to represent the signal received at all frequencies appropriately.

These environmental and instrumental contributions to the observed radio burst emissions can be adjusted through the use of a flux calibrator \citep{2011A&A...530A..80S, 2013A&A...556A...2V}.  To select a calibrator, the following criteria are used \citep{2017isra.book.....T}:
\begin{enumerate*}[label=(\roman*)]
	\item it has a well-defined spectrum (i.e. a well-defined flux density) at the frequencies to be calibrated,
	\item it is non-variable over the observation's time scale,
	\item it is a bright source (so that a good signal-to-noise ratio is obtained in a short time),
	\item it is a point-like source,
	\item its position (and motion) on the sky is well-defined,
	\item its projected position is relatively close to that of the burst but sufficiently separated from the Sun (such that solar contributions are not recorded by the primary beam), and
	\item it is isolated from other radio sources which may also contribute to the observed signal.
\end{enumerate*}
A commonly used calibrator is Taurus~A (or Tau~A), also known as the Crab Nebula \citep{1970A&A.....6..406B, 1985srph.book.....M}.  Along with Tau~A, signals from the ``empty-sky''---a part of the sky not associated with any bright radio sources (at the frequencies of interest)---can also be observed to estimate the background radiation.  This flux calibration method has been applied on the tied-array beam observations presented in this thesis.

The empty-sky contribution is subtracted to remove the background noise, and then the Tau~A spectrum is used to normalise the spectrum obtained when observing the radio burst emissions.  Given that the flux density of the calibrator is known, the flux density of the radio burst source can also be obtained.

The calibrators can be observed in two ways.  The first option is to observe Tau~A and the empty sky over short periods of time, both immediately before and after the solar observations.  The other option is to sacrifice two beams from the total of tied-array beams---one for Tau~A and one for the empty sky---and observe the calibrators during the solar observation \citep{2011A&A...530A..80S, 2017NatCo...8.1515K}.

\subsection{Estimating source locations and sizes from radio images} \label{sec:centroid_calc}

The location of a radio source is given as the centroid position of the function used to describe the shape of the source.  Solar radio burst sources tend to be observed as elliptical, and as a result, emission images are usually (and in this thesis) fitted with a 2D elliptical Gaussian function, allowing for the estimation of the centroid location and size of the source.

The 2D Gaussian function describing an ellipse with an intensity profile $S(x,y)$ centred at ($x_0$, $y_0$) and whose semi-major axis is rotated clockwise with respect to the $x$-axis by an angle $\phi$, is given by (see, e.g., \cite{2017NatCo...8.1515K}):
\begin{equation}
\begin{split}
	S(x,y) = S_0 \cdot \exp \Bigg{[} & - \dfrac{ \left[ (x-x_0)\cos(\phi) - (y-y_0)\sin(\phi) \right]^2 }{ 2 \, \sigma_x^2 } \\
	& - \dfrac{ \left[ (x-x_0)\sin(\phi) + (y-y_0)\cos(\phi) \right]^2 }{ 2 \, \sigma_y^2} \Bigg{]} + \Gamma \, .
\end{split}
\end{equation}
Here, $S_0$ is the peak amplitude, $\sigma_x$ and $\sigma_y$ are the one-standard deviation of the $x$- and $y$-size, respectively, and $\Gamma$ is the offset from $z=0$ (used when a noise floor is defined).  The heliocentric coordinates of the centroid location are $x_0$ and $y_0$, meaning that the heliocentric location of the source is given by $\sqrt{x_0^2 + y_0^2}$.

The parameters are obtained by minimising the $\chi^2$ \citep{2017NatCo...8.1515K}:
\begin{equation}
	\chi^2 = \sum_{i=1}^{N} \dfrac{(F_i - S( x_i , y_i ; S_0, x_0, y_0, \sigma_x, \sigma_y, \phi, \Gamma))^2}{\delta F^2} \, ,
\end{equation}
where $F_i$ represents the independent amplitude measurement under evaluation (located at ($x_i$, $y_i$) of the image) and $\delta F$ is the uncertainty in the flux density measurement, taken to be equal to the background flux level before the burst.

The uncertainties in the centroid estimations are calculated as \citep{1997PASP..109..166C, 2017NatCo...8.1515K}:
\begin{equation}\label{eqn:gauss_centroid_error}
	\delta x_0  = \sqrt{\dfrac{2}{\pi}} \, \dfrac{\sigma_y}{\sigma_x} \, \dfrac{\delta F}{S_0} \, \theta_{res}
	\qquad \text{and} \qquad
	\delta y_0  = \sqrt{\dfrac{2}{\pi}} \, \dfrac{\sigma_x}{\sigma_y} \, \dfrac{\delta F}{S_0} \, \theta_{res} \, ,
\end{equation}
for the $x$- and $y$-coordinates, respectively.  Here, $\theta_{res}$ is the angular resolution (defined in Equation~(\ref{eqn:lofar_res})).

The source size $l$ is defined as the FWHM of the Gaussian distribution, given by
\begin{equation} \label{eqn:gauss_source_size}
	l_{x} = 2 \, \sqrt{2 \ln2} \, \sigma_{x}
	\qquad \text{and} \qquad
	l_{y} = 2 \, \sqrt{2 \ln2} \, \sigma_{y}
\end{equation}
for the $x$- and $y$-size, respectively.
The FWHM source area is therefore given as
\begin{equation} \label{eqn:gauss_area}
	A = \dfrac{\pi}{4} \, l_x \, l_y \, ,
\end{equation}
and the associated error on the FWHM area $A$ is computed using \citep{2017NatCo...8.1515K}
\begin{equation} \label{eqn:gauss_area_error}
	\dfrac{\delta A}{A} = 2 \, \dfrac{\delta F}{S_0} \, \dfrac{h}{\sqrt{A}} \, .
\end{equation}
Higher signal-to-noise ratios $F_i / \delta F$ result in a more accurate determination of the source centroids and areas.

However, it should be noted that the observed source area $A_{obs}$ represents the accumulated contributions of the intrinsic (or true) source area $A_{true}$, the broadened (due to scattering) area $A_{scatt}$, and the beam area $A_{beam}$ (i.e. the beam's FoV; Equation~(\ref{eqn:lofar_fov})):
\begin{equation} \label{eqn:obs_source_contributions}
	A_{obs} = A_{true} + A_{scatt} + A_{beam} \, .
\end{equation}
An interferometer with an infinite baseline $D$ (cf. Equation~(\ref{eqn:lofar_res})) would resolve sources perfectly, i.e. its FoV would be one-dimensional (a point) and thus observe only $A_{obs} = A_{true} + A_{scatt}$.  However, given that baselines have a finite size, the beam has an area that needs to be deconvolved (subtracted; see, e.g., \cite{2013ApJ...762...60S}) from the image in order to represent the actual source: $A_{obs} - A_{beam} = A_{true} + A_{scatt}$.
It is worth highlighting that the intrinsic and scattered areas cannot be distinguished from each other without complete knowledge of the several radio-wave propagation effects at play (as described in this thesis).

\section{Radio-Wave Propagation Effects} \label{sec:prop_effs}
\begin{figure}[b!]
    \centering
	\includegraphics[width=0.8\textwidth, keepaspectratio=true]{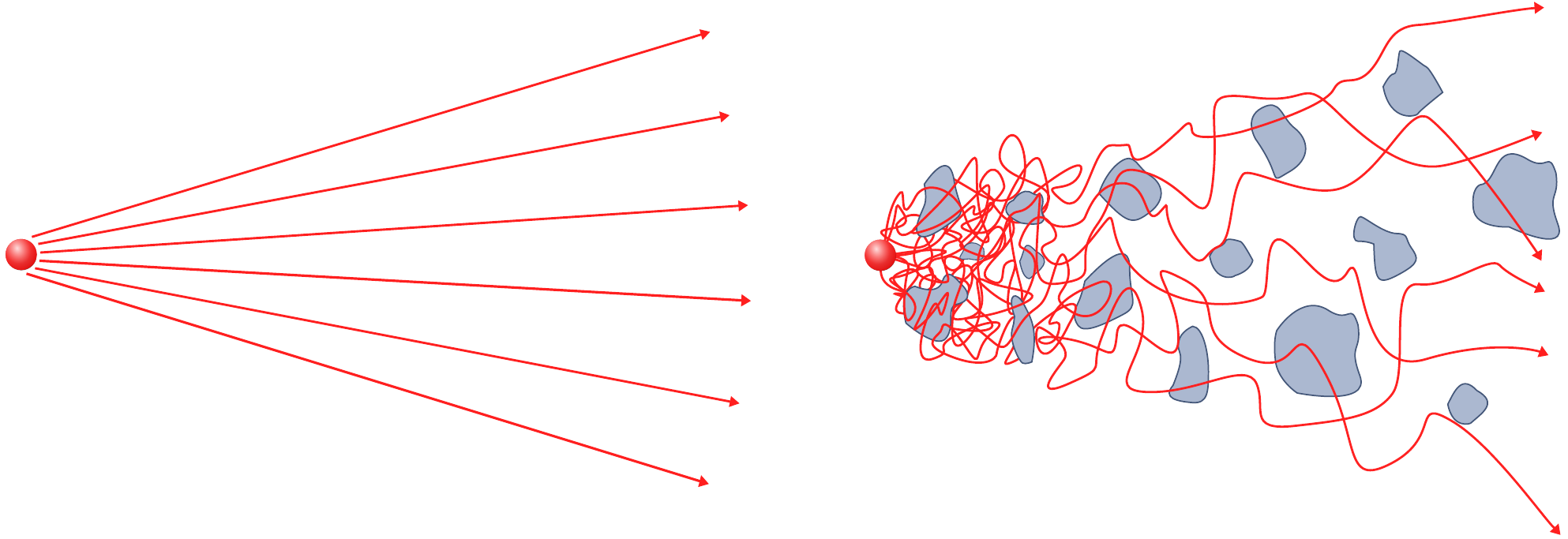}
    \caption[Photon propagation through free space vs density inhomogeneities.]
    {Schematic depiction of the propagation of photons through free space (left) and the propagation of photons through density inhomogeneities (right).  Rays of photons (red arrows) are emitted from a radio source (red sphere).
	}
    \label{fig:inhomogeneities_cartoon}
\end{figure}

Streams of plasma that reflect the high activity and variability of the Sun are continuously ejected into the heliospheric environment.  Consequently, there are random density fluctuations throughout the heliosphere that can cause small-scale turbulence, as well as sporadic large-scale inhomogeneities (like CMEs) which traverse the interplanetary space (or streamers, which are quasi-stationary) and transiently disturb their local coronal environment.
As a result, the trajectory of photons propagating through the heliosphere does not resemble that of propagation in free space (see left panel of Figure~\ref{fig:inhomogeneities_cartoon}); instead, it is influenced by the interactions with encountered density inhomogeneities (right panel of Figure~\ref{fig:inhomogeneities_cartoon}).

The interactions these photons experience can be divided into three categories: (i) scattering, (ii) refraction, and (iii) absorption, each of which impacts the observed radio emissions in a different way.  In the case of radiation at radio frequencies, these interactions are referred to as \textit{radio-wave propagation effects}.

The propagation of electromagnetic waves is characterised via the dispersion relation, as it relates the wavevector $\vec{k}$ (and thus wavelength $\lambda = 2 \pi / k$) to the wave's frequency $\omega$ (where $\omega = 2 \pi f$).  The dispersion relation of electromagnetic (transverse) waves in an unmagnetised plasma is given in Equation~(\ref{eqn:EM_disp_rel}).  Given that the refractive index is defined as $\mu = kc / \omega$ (where $k$ is the wavenumber; \cite{1985srph.book.....M}), the dispersion relation of radio waves can also be expressed in terms of the refractive index as
\begin{equation} \label{eqn:refractive_index}
	\mu^2(r) = 1 - \dfrac{\omega_{pe}^2}{\omega^2} = 1 - \dfrac{f_{pe}^2}{f^2} \, .
\end{equation}
Moreover, the phase and group velocities of radio waves are given by $v_{p} = \omega / k$ and $v_{g} = \partial \omega / \partial k$, respectively.

As radio waves travel away from the Sun into regions where the emission frequency $\omega$ becomes increasingly larger than $\omega_{pe}$ (such that $\mu \rightarrow 1$), the strength of refraction diminishes, implying that the medium encountered by the photons no longer affects their propagation to the same extent.  Therefore, radio-wave propagation effects are most significant near the location of the emission source, where the emitted frequency $\omega \approx \omega_{pe}$, and weaken with increasing distance (as depicted in the left panel of Figure~\ref{fig:inhomogeneities_cartoon}).  By extension, harmonic plasma emissions (for which $\omega \approx 2 \, \omega_{pe}$) will also be less affected by radio-wave propagation effects compared to fundamental emissions ($\omega \approx \omega_{pe}$).
At the cut-off, where $\mu^2=0$ (i.e. $\omega = \omega_{pe}$) and the phase velocity $v_p \rightarrow \infty$, the radio waves will undergo a total reflection \citep{2008SoPh..253....3N}.
On the other hand, at the resonance, where $\mu^2 \rightarrow \infty$ and the phase velocity $v_p = 0$, the waves will be absorbed \citep{1969pldy.book.....B}.

Fluctuations in density $n$ correspond to fluctuations in the refractive index $\mu$ (given that $n \propto \omega^2$; Equation~(\ref{eqn:omega_pe})).  Hence, stronger density fluctuations (i.e. larger $\delta n / n$ values) result in more dramatic changes in the refractive index, and thus a more significant impact on the radio waves \citep{1971A&A....10..362S}.  Since these density fluctuations alter the propagation of photons, it is reasonable to expect that what is detected by an observer (who is located far away from the emission source) does not represent the intrinsic properties of the source or of the interplanetary medium through which the photons propagate.

However, the cumulative contribution of these random perturbations on the direction of the photon's propagation vector (as induced by coronal density fluctuations) is not well understood, igniting long-standing debates, or even suggestions (often without quantitative support) that scattering effects are negligible.  Given the potential of density fluctuations to significantly impact radio observations, it is vital for the understanding of plasma emission processes and the coronal environment to---not only consider---but quantitatively evaluate radio-wave propagation effects, until a satisfactory mathematical description is obtained.  Any propositions must be evaluated on the basis of whether they can successfully reproduce the entirety of observed radio properties.  Recent progress on this matter is presented in Chapters~\ref{chap:scattering} and \ref{chap:observation_simulations}.

\subsection{Scattering dominance} \label{sec:scatt_dom}
The dominance of scattering over other radio-wave propagation effects was recently demonstrated by \cite{2017NatCo...8.1515K}.  A Type IIIb burst observed by LOFAR between 30--80~MHz (shown in Figure~\ref{fig:TypeIIIb_bursts}) with very high temporal and spectral resolutions was analysed.  The properties of striae were investigated as a function of time, the first investigation of its kind.  Specifically, each stria was imaged at a single frequency but for multiple (and consecutive) moments in time, probing the sub-second evolution of the radio source.

As detailed in Section~\ref{sec:typeIIIs}, Type IIIb striae are believed to arise due to small-scale density fluctuations, enabling the estimation of the intrinsic radio source size using the bandwidth $\Delta f_i$ of the striae.  The inferred intrinsic sizes are $\sim$0.1$\arcmin$ for emissions at $\sim$32~MHz.  However, when imaged, sources appear to have sizes $\sim$200 times larger than the expected (i.e. $\sim$20$\arcmin$ at $\sim$32~MHz).  It should be emphasised that the utilisation of LOFAR by \cite{2017NatCo...8.1515K} ensured that the observed source sizes were resolved (i.e. the beam size was smaller than the observed source size).  Therefore, the large source sizes observed cannot be attributed to instrumental limitations (i.e. insufficient spatial resolution), as sometimes suggested (see, e.g., \cite{2011JGRA..116.3104S}).  This dramatic discrepancy between the predicted and observed source sizes can only be explained within the framework of scattering, as none of the other radio-wave propagation effects can justify such significant angular broadening.
For example, refraction on large-scale density inhomogeneities acts as a radio-wave focusing effect, meaning that the large source sizes observed cannot be accounted for.  Consequently, \cite{2017NatCo...8.1515K}, provided strong evidence for the dominance of scattering effects in observations of radio emissions.

\cleardoublepage
\chapter{Radio-Wave Propagation Simulations} \label{chap:scattering}

\textit{The results in this chapter have been published in \cite{2019ApJ...884..122K}. \\
The author of this thesis contributed to the publication by \cite{2019ApJ...884..122K} by creating all the figures (except those labelled as
Figure~\ref{fig:2019_sim_size_and_decay_vs_freq} and \ref{fig:2019_sim_size_and_decay_vs_anis} in this thesis), collecting and analysing the data necessary to create the figures, as well as writing parts of the text.
}

\section{Describing Radio-Wave Propagation Effects} \label{sec:ray-tracing_simulations_motivation}

By the 1980's, the hypothesis that radio-wave scattering effects which result from random small-scale density inhomogeneities define the observed properties of radio sources was becoming less and less popular---``even for those features, such as apparent height and size of a source, for which [they offered] a plausible explanation'' \citep{1985srph.book.....M}.  For example, \cite{1977A&A....61..777B} resorted to a different interpretation of scattering in the corona---from large fibrous structures---after rejecting the idea of \textit{isotropic} scattering from small-scale density inhomogeneities, which could not simultaneously account for both the large source sizes of Type I bursts and their highly-directional emission.  Nevertheless, the contribution of scattering to the observed properties was recognised in the following decades \citep{1999A&A...351.1165A, 2007ApJ...671..894T, 2008A&A...489..419B, 2009SoPh..259..255R, 2011JGRA..116.3104S}.

Recent observations by \cite{2017NatCo...8.1515K}, however, provided strong evidence for a governing role of scattering from small-scale inhomogeneities on the observed emission properties (as described in Section~\ref{sec:scatt_dom}), showing that it dominates other radio-wave propagation effects like refraction, reviving the interest and necessity to understand its effects.  Moreover, the analysed observations suggested that density inhomogeneities in the corona must be \textit{anisotropic}, with the perpendicular component being stronger than the radial.  This result contradicted the favoured assumption of isotropic density fluctuations (and hence isotropic scattering) often used in scattering descriptions (e.g., \cite{1971A&A....10..362S, 2007ApJ...671..894T, 2018ApJ...857...82K}).  Following the outcome of this study, \cite{2019ApJ...884..122K} developed three-dimensional (3D) ray-tracing simulations that can account for anisotropic density fluctuations, as will be discussed in this chapter.  Ray-tracing simulations describe the trajectory of photons (or rays of photons) as they propagate through the coronal medium and experience radio-wave propagation effects.

To examine both the validity of an anisotropic scattering description and the extent of anisotropy required, the simulation outputs need to be compared to observations.  The isotropic scattering description has been used to successfully reproduce individual characteristics of observed radio emissions (e.g., source sizes \citep{1971A&A....10..362S} and decay times \citep{2018ApJ...857...82K}), but a variety of characteristics must be successfully reproduced simultaneously, in order for a robust conclusion to be drawn.  This is the main aim of this chapter: (i) to investigate whether an anisotropic scattering description can simultaneously describe multiple observed source properties, and (ii) to examine which parameters impact---and to what degree---the observed radio emissions.

\section{Anisotropic Radio-Wave Scattering Simulations} \label{sec:anisotropic_simulations_2019}
Ray-tracing simulations that account for anisotropy were developed by \cite{2019ApJ...884..122K}, improving previous (isotropic) descriptions of radio-wave propagation effects in the coronal medium, in light of recent observational results (introduced in Section~\ref{sec:ray-tracing_simulations_motivation}; \cite{2017NatCo...8.1515K}).

The \cite{2019ApJ...884..122K} approach is a 3D stochastic description of radio-wave propagation in a turbulent medium with background density fluctuations, characterised using the kinetic plasma approach and the Fokker-Planck equation, and simulated using a numerical Monte-Carlo ray-tracing technique by solving the Langevin equations.  It utilises the geometric optics approximation which assumes that the scale length of the variations in wavelength $\lambda$ (due to inhomogeneities) is much smaller than the wavelength itself.
The Fokker-Planck equation describes the spectral number density in the geometric optics approximation, and the Hamilton equation gives the dispersion relation (see Appendix~\ref{sec:appendix_fokker_planck_eqn}).
The average plasma density ($n \equiv \langle n \rangle$) is assumed to be a slowly-varying function of position (see Equation~(\ref{eqn:2019_sim_dens_model})).  The adopted mathematical description of scattering is valid only for small-amplitude density fluctuations and unmagnetised plasma environments.  Diffraction effects are therefore ignored, but collisional (free-free) absorption is considered, as well as refraction on large-scale density inhomogeneities (introduced by the gradually-decreasing coronal density with increasing heliocentric distance).
Given that the speed of light is much greater than the velocity of density fluctuations, the density fluctuations are treated as static and only elastic scattering is considered, conserving the wavevector $\vert \vec{k} \vert$ of radio waves during the random changes in the propagation direction (i.e. the frequency $\omega$ of emitted photons is conserved).

The attenuation of the signal (i.e. reduction in intensity) in the collisional coronal medium, resulting from the free-free absorption, is simulated via
\begin{equation} \label{eqn:2019_absorb_attenuation}
	N(t) = N_0 \, e^{- \tau_a} \, .
\end{equation}
Here, $\tau_a$ is the Coulomb collision depth:
\begin{equation} \label{eqn:2019_collis_depth}
	\tau_a = \int \gamma \, \vec{r}(t) \, dt \, ,
\end{equation}
where $\gamma$ is the collisional absorption coefficient of radio waves in a plasma (for details, see Appendix~\ref{sec:appendix_abs_rate}).  In other words, at every simulated time step (i.e. at each interaction), the weight of the wave packet is reduced by $e^{- \tau_a}$.
The effects of absorption are stronger in higher density plasmas, i.e. they affect higher frequencies ($\gtrsim 50$~MHz) the most.  Generally, absorption effects also depend on the strength of scattering, since strong scattering can cause the photons to be trapped near the (intrinsic) source for a time period longer than the free-free absorption time $1/\gamma$, and thus be absorbed.

When density fluctuations are taken to have a Gaussian correlation, a Gaussian autocorrelation function is used to characterise them.  Density inhomogeneities in the corona are described using a spectrum of density fluctuations ($S(\vec{q})$) normalised to the variance of the relative density fluctuations ($\epsilon^2$):
\begin{equation} \label{eqn:2019_S(q)_normalised}
	\epsilon^2 \equiv \dfrac{\langle \delta n^2 \rangle}{n^2} = \int S(\vec{q}) \, \dfrac{d^3 q}{(2\pi)^3} \, ,
\end{equation}
where $\vec{q}$ is the wavevector of the electron density fluctuations.
The (anisotropic) density fluctuations are taken to be axially symmetric, meaning that the spectrum can be parametrised as a spheroid in $\vec{q}$-space:
\begin{equation} \label{eqn:2019_S(q)_spheroid}
	S(\vec{q}) = S \left( \left[ q^2_\perp + \alpha^{-2} \, q^2_\parallel \right]^{1/2} \right) \, ,
\end{equation}
where $q$ is the wavenumber (i.e. size) of density fluctuations and $\alpha$ is their anisotropy, defined as the ratio of the perpendicular and parallel correlation lengths $h$ (also referred to as the characteristic density scale heights):
\begin{equation} \label{eqn:2019_anisotropy_def}
	\alpha = \dfrac{h_\perp}{h_\parallel} \, .
\end{equation}
The perpendicular and parallel directions are defined with respect to the local radial direction from the Sun.
A value of $\alpha = 1$ means that the density fluctuations (and thus the scattering) are isotropic.  If $\alpha < 1$ (i.e. $h_\perp < h_\parallel$) the spectrum of density fluctuations is dominated by fluctuations in the parallel direction, making scattering stronger in the perpendicular direction (and vice versa).
Both levels of density fluctuations $\epsilon$ and anisotropy $\alpha$ are assumed to be independent of the radial distance $r$.

\begin{figure}[ht!]
    \centering
	\includegraphics[width=0.8\textwidth, keepaspectratio=true]{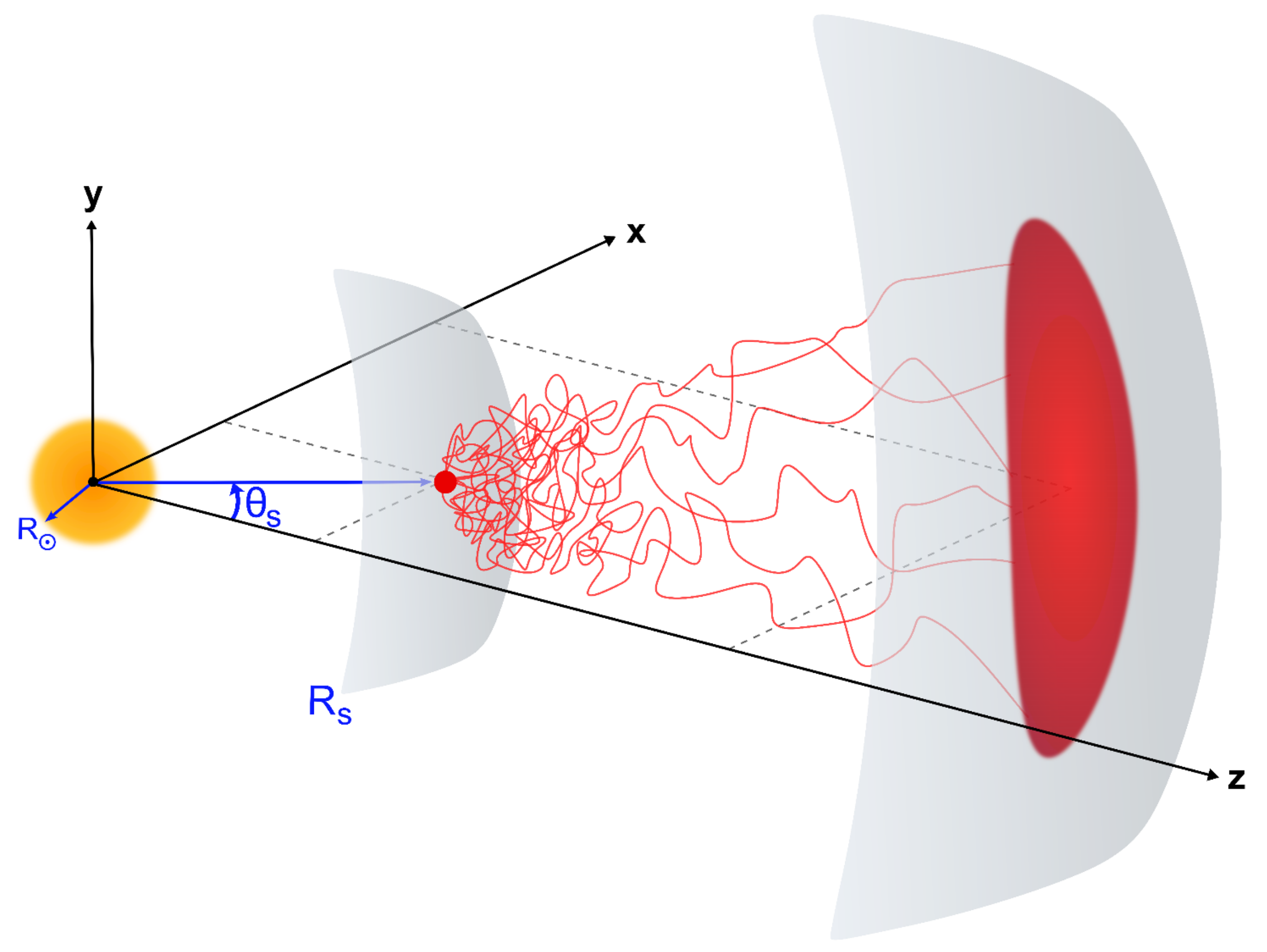}
    \caption[Illustration of the radio-wave propagation effects and simulation set-up.]
    {Schematic demonstration of scattering effects and the Sun-centred Cartesian coordinate system used in the simulations.  The $z$-axis is directed towards the observer.  The intrinsic location of the point source (red disk) is annotated as $R_s$, and the source-polar angle is denoted by $\theta_s$.  Due to the assumed spherical geometry, the azimuthal angle in the ($x$, $y$) plane of the sky is not relevant to the simulations.  Rays of photons emitted from the point source scatter as they propagate away from the Sun, until they reach a sphere at a distance where scattering becomes negligible (i.e. the scattering screen).  The distance of the scattering screen defines the observed source position, as well as the extent to which the perceived source size broadens, as indicated by the large red area.
    Figure taken from \cite{2019ApJ...884..122K}.
	}
    \label{fig:prop_effects_cartoon}
\end{figure}

The solar corona is assumed to be spherically symmetric with a radial magnetic field, such that the parallel component of the anisotropic density fluctuations is aligned with the local radial direction (i.e. $q_{\parallel}$ is parallel to $\vec{r}$).
The assumed spherically-symmetric corona (and the radial magnetic field) is a not realistic assumption for the entire (and vast) range of distances probed by the simulations, but it is consistent with the density model employed (see Equation~(\ref{eqn:2019_sim_dens_model})).
The simulations use a Sun-centred Cartesian coordinate system ($x$, $y$, $z$), where the $z$-axis is always directed towards the observer (see Figure~\ref{fig:prop_effects_cartoon}).  Due to the spherical symmetry, the azimuthal angle in the plane of the sky ($xy$-plane) is not relevant to this description.

The intrinsic source is modelled as an isotropically-emitting point source, i.e. the distribution of (photon) wavevectors $\vec{k}$ is isotropic near the source.  A given source emits only at a single frequency, so all emitted photons have the same absolute value of wavevector $\vec{k}$ (cf. Equation~(\ref{eqn:EM_wavenum})).  The outcome of simulations that assume isotropic emission patterns are applicable to those obtained assuming other emission patterns (like, e.g., the dipole and quadrupole patterns), since the focusing caused by refraction is independent of the emission pattern, and also, the initial emission patterns are annihilated as a result of the scattering of photons \citep{2007ApJ...671..894T}.  In addition to assuming a point source, the simulations consider an instantaneous injection of photons in the corona, such that the time profile of the injected radio pulse is initially characterised by a delta function (i.e. all photons are emitted at the same instance).  As a consequence, if propagation effects were to be ignored, all photons would arrive at the observer at the same time, delayed only by the amount that photons take to propagate through free space, i.e. delayed by $dt = dr / c$, where $dr$ is the radial distance covered from the location of emission and $c$ is the speed of light.

The position of the emission source (i.e. the intrinsic source position) is characterised using a radial distance $R_s$ from the Sun and an angle $\theta_s$ (referred to as the ``source-polar angle''), as depicted in Figure~\ref{fig:prop_effects_cartoon}.
The source-polar angle $\theta_s$ is defined as the heliocentric angle from the $z$-axis (i.e. the observer's LoS and Sun-Earth vector) to the source's centroid position---it is the polar angle of a spherical coordinate system.
In the simulations, positive angles are defined counter-clockwise (with respect to the $z$-axis).  As a result, a source at $\theta_s = 0\degr$ appears to coincide with the solar centre, whereas a source at $\theta_s = \pm 90\degr$ is observed at the solar limb.  As suggested by Figure~\ref{fig:prop_effects_cartoon}, the rate of scattering is higher near the emission source where the photon frequency $\omega$ is close to local plasma frequency $\omega_{pe}$, and decreases at larger distances where $\omega \gg \omega_{pe}$, as described in Section~\ref{sec:prop_effs}.

The emission frequency $\omega_F$ of a fundamental source is defined as $\omega_F \gtrsim \omega_{pe}$ (in this chapter, specifically, $\omega_F = 1.1 \, \omega_{pe}$; see Section~\ref{sec:plasma_emmission}), whereas $\omega_H = 2 \, \omega_{pe}$ is used to define the emission frequency of a harmonic source.
Although it is known that radio waves must have a frequency $\omega$ greater than the local plasma frequency $\omega_{pe}$, in order for propagation to occur (as detailed in Section~\ref{sec:plasma_emmission}), the exact relation between $\omega$ and $\omega_{pe}$ remains an open question.  The effect of varying the ratio between these two frequencies on the simulated radio properties is explored in Section~\ref{sec:driftpairs_frequency_dependence}.
The local electron plasma frequency $\omega_{pe}$ is calculated using a spherically-symmetric Parker density model \citep{1960ApJ...132..821P} where a constant temperature is assumed and the model's constants are chosen in such a way as to agree with satellite measurements adapted from \cite{1999A&A...348..614M}.  The temperature of the (isothermal) corona is taken to be $\sim$1~MK.
The adapted density model, however, lacks a simple analytical form, which is required for solving the differential equations describing the time steps of the stochastic process (see Appendix~\ref{sec:appendix_stoc_eqs}).  The adapted density model is simplified by fitting three power-law functions which result in the following form:
\begin{equation} \label{eqn:2019_sim_dens_model}
	n(r) = 4.8 \times 10^9 \left(\dfrac{\Rs}{r}\right)^{14} + 3 \times 10^8 \left(\dfrac{\Rs}{r}\right)^6 + 1.4 \times 10^6 \left(\dfrac{\Rs}{r}\right)^{2.3} \, ,
\end{equation}
where $r$ is the heliocentric radial distance.
An advantage of using this density model over others is that it can describe the coronal density from distances close to the Sun up to distances close to the Earth,
unlike, e.g., the Newkirk model (see Equation~(\ref{eqn:n_Newkirk})) which is only valid for distances close to the Sun ($< 5 \, \Rs$).
Nevertheless, similar to density models like that of \cite{1961ApJ...133..983N}, the Parker density model is a radial density model (i.e. a 1D model), implying that---like in previous ray-tracing simulations---the characterisation of any shift in the observed source position induced by radio-wave propagation effects is restricted to the radial direction.  In other words, any simulated displacement of the source's centroid in the simulation outputs when $\theta_s \neq 0\degr$ will be portrayed along the $x$-direction (see Figure~\ref{fig:prop_effects_cartoon}).

The simulations are run for $10^4$ photons and consider initial heliocentric emission distances $R_s$ ranging from 1.05 to 57~$\Rs$, which---according to the density model used---correspond to frequencies from $\sim$460 to $\sim$0.1~MHz, respectively.
As will become evident from the simulation outputs presented throughout this thesis, the chosen number of photons ($10^4$) leads to a successful reproduction of several radio-source properties, with reasonable statistical errors.  Moreover, larger numbers of photons increase the run time of these (already) computationally-intensive simulations.
The photons are traced until a distance at which both scattering and refraction become negligible (hereafter referred to as the ``scattering screen'') or until 1~au.  Whether the photons are traced up to the scattering screen (i.e. $< 1$~au) or up to 1~au depends on which of the two heliocentric distances is encountered first (i.e. the smallest distance).  The scattering screen is defined as the distance after which the cumulative change of the angular spread of the photons is $\leq$~1\% of the value it has already reached by that point.  As such, beyond the scattering screen, the simulated properties are not significantly affected and can thus be considered to remain the same at larger distances as what they are at the scattering screen.

The properties and arrival times of photons are recorded when they reach the scattering screen, defining the observed properties (like the time profile) of the simulated source.  To simulate the radio source images, the scattering-screen locations of the photons whose propagation vectors are directed towards the observer (i.e. those with $0.9 < \dfrac{k_z}{k} < 1$) are projected back to the source plane (i.e. the $xy$-plane of the intrinsic source; similar to \cite{2010A&A...513L...2K} and \cite{2011A&A...536A..93J}).  In this way, the intensity map $I(x,y)$ defining how the source will be observed is constructed (depicted by the enlarged red region in Figure~\ref{fig:prop_effects_cartoon}).

The simulations enable the calculation of several source properties such as:
\begin{enumerate*}[label=(\roman*)]
	\item the source intensity map $I(x,y)$,
	\item the intensity-time profile (i.e. the light curve),
	\item the total flux $S = \int \, I(x,y) \, dx \, dy$,
	\item the peak-flux time,
	\item the decay time,
	\item the delay time,
	\item the centroid positions,
	\item the source sizes, and
	\item any associated statistical errors.
\end{enumerate*}
The statistical errors arise due to the finite number of photons used in each simulation run, such that the uncertainty decreases with increasing photon numbers.
The dependence of these properties with respect to the source-polar angle $\theta_s$ can be examined, in addition to their behaviour with changing levels of anisotropy $\alpha$ and density fluctuations $\epsilon$.

\subsection{Source size and centroid location computation} \label{sec:size_and_centr_distribution_calc}

The source sizes and centroids (and associated errors) can be calculated by fitting the simulated intensity map $I(x,y)$ with a 2D elliptical Gaussian, as discussed in Section~\ref{sec:centroid_calc}.
They can also be calculated using the first-normalised moments of the distribution:
\begin{equation} \label{eqn:2019_sim_distr_centroids}
	\bar{x} = \dfrac{\int^{\infty}_{-\infty} \, x \, I(x,y) \, dx \, dy}{\int^{\infty}_{-\infty} I(x,y) \, dx \, dy}
	\qquad \text{and} \qquad
	\bar{y} = \dfrac{\int^{\infty}_{-\infty} \, y \, I(x,y) \, dx \, dy}{\int^{\infty}_{-\infty} I(x,y) \, dx \, dy} \, ,
\end{equation}
where $\bar{x}$ and $\bar{y}$ give the $x$- and $y$-centroid positions, respectively, and the variances $\sigma_x$ and $\sigma_y$ are computed from the second normalised moments
\begin{equation} \label{eqn:2019_sim_distr_variance}
	\sigma_x^2 = \dfrac{\int_{-\infty}^{\infty} \, (x-\bar{x})^2 \, I(x,y) \, dx \, dy}{\int_{-\infty}^{\infty} \, I(x,y) \, dx \, dy}
	\qquad \text{and} \qquad
	\sigma_y^2 = \dfrac{\int_{-\infty}^{\infty} \, (y-\bar{y})^2 \, I(x,y) \, dx \, dy}{\int_{-\infty}^{\infty} \, I(x,y) \, dx \, dy} \, .
\end{equation}
Therefore, the one-standard-deviation uncertainties $\sigma_x$ and $\sigma_y$ are also obtained.
The uncertainty in the respective $x$- and $y$-centroids is given as:
\begin{equation} \label{eqn:2019_sim_distr_centroid_errors}
	\delta \bar{x} \simeq \dfrac{\sigma_x}{\sqrt{N}}
	\qquad \text{and} \qquad
	\delta \bar{y} \simeq \dfrac{\sigma_y}{\sqrt{N}} \, ,
\end{equation}
where $N$ is the number of photons making up the intensity map $I(x,y)$.

Given that radio sources are assumed to have a Gaussian shape, the FWHM $x$- and $y$-size of the sources is calculated using:
\begin{equation} \label{eqn:2019_sim_distr_sizes}
	\mathrm{FWHM}_{x,y} = 2 \, \sqrt{2 \ln2} \, \sigma_{x,y} \, ,
\end{equation}
which is equivalent to Equation~(\ref{eqn:gauss_source_size}) of Section~\ref{sec:centroid_calc}.  The area of the source is therefore estimated as $A = \mathrm{FWHM}_{x} \cdot \mathrm{FWHM}_{y} \cdot \pi /4$.
The associated FWHM size uncertainty is estimated through
\begin{equation} \label{eqn:2019_sim_distr_size_errors}
	\delta \, \mathrm{FWHM}_{x,y} \simeq 2 \, \sqrt{2\ln2} \, \dfrac{\sigma_{x,y}}{\sqrt{2N}} \, .
\end{equation}

\subsection{The spectrum of density fluctuations} \label{sec:dens_fluct_spec}
Radio-wave propagation effects arise due to density fluctuations in the solar corona.  The spectrum of electron density fluctuations in the solar wind (near the Earth) is often obtained either directly through in-situ observations of the electron density, or inferred from in-situ observations of the plasma peak in radio quasi-thermal noise spectra \citep{1987A&A...181..138C, 1995JGR...10019881M, 2012PhRvL.109c5001C, 2020ApJS..246...44M}.
However, it cannot be measured using in-situ observations at distances close to the Sun, due to the subsequent lack of spacecraft at such heights.  Therefore, the spectrum of density fluctuations near the Sun can only be probed with remote-sensing detectors (e.g., \cite{2018ApJ...856...73C}).

Ground-based instruments like LOFAR are limited by the ionospheric cut-off to frequencies above $\sim$10~MHz ($\lesssim 2.5 \, \Rs$), but can allow for very high resolutions and sensitivity (see Section~\ref{sec:radio_obs}).  LOFAR's observing capabilities, specifically, provide a unique ability to record the sub-second evolution of emission sources close to the Sun \citep{2017NatCo...8.1515K, 2018SoPh..293..115S}.  This means that the sub-second behaviour of the radio sources can be examined, providing an insight into the small-scale density fluctuations that define the near-Sun environment.  It it therefore of interest to take advantage of high-resolution data to better understand how radio photons are affected by their local environment and how this environment varies with radial distance.

In order to enable a characterisation of the spectrum of density fluctuations for the range of distances considered in the ray-tracing simulations (from the Sun to the Earth), values obtained empirically can be extrapolated.
The spectrum of density fluctuations $S(q)$---which is a function of distance---is given as a power law of the form:
\begin{equation} \label{eqn:dens_fluc_spec_S(q)}
	S(q) \propto q^{-(p+2)} \, ,
\end{equation}
where $p$ is the exponent.  Assuming an isotropic turbulence, in-situ observations showed that the value of the exponent is often $5/3$ (a Kolmogorov scaling) at distances closer to the Earth (see, e.g., \cite{2013SSRv..178..101A}).  This value was found to hold for broad inertial ranges, specifically, from outer scales $l_o = 2\pi / q_o$ to inner scales $l_i = 2\pi / q_i$ \citep{2013SSRv..178..101A}.  Here, $q_o$ and $q_i$ are the wavenumbers of the electron density fluctuations at the outer and inner scale, respectively.
The outer scale is defined as the point at which the spectral index of the spectrum of density fluctuations decreases from $\sim$\nolinebreak$-1$ to $\sim$\nolinebreak$-5/3$, marking the beginning of the non-linear cascade of turbulence.  The inner scale is defined as the point where the spectral index decreases from $\sim$\nolinebreak$-5/3$ to $\sim$\nolinebreak$-2.5$, after which the density fluctuations dissipate.
In other words, the outer scale corresponds to the largest scales (i.e. smallest wavenumbers $q$) present in the turbulent cascade, whereas the inner scale corresponds to the smallest scales (and largest wavenumbers $q$).

By assuming a large range of wavenumbers so that $q_o \ll q_i$, a simplified model of density fluctuations can be obtained.  Following \cite{2007ApJ...671..894T} and \cite{2018ApJ...857...82K}, $p=5/3$ is taken within this limit ($q_o \ll q_i$), leading to the following simplified model:
\begin{equation} \label{eqn:simple_dens_fluc_model}
	\overline{q \epsilon^2} \simeq 4\pi l_o^{-2/3} l_i^{-1/3} \epsilon^2 \, ,
\end{equation}
where $\bar{q}$ is the spectrum-weighted mean wavenumber (of density fluctuations).
The angular rate of scattering is proportional to $\overline{q \epsilon^2}$, meaning that an increase in either $q$ or $\epsilon$ results in stronger scattering (where larger values of $q$ correspond to smaller density scales).
The inner scale of electron density fluctuations---which is the primary parameter determining the scattering rate---is given (between $r \approx$ 2--70~$\Rs$; \cite{1989ApJ...337.1023C}) in units of $\Rs$ as
\begin{equation} \label{eqn:inner_scale}
	l_i = \dfrac{r}{6.957 \times 10^5} \, ,
\end{equation}
whereas the outer scale (for distances $r$ = 7--80~$\Rs$) is expressed (in units of $\Rs$) using the following empirical formula \citep{2001SSRv...97....9W}:
\begin{equation} \label{eqn:outer_scale}
	l_o = (0.23 \pm 0.11) \times r\,^{0.82 \pm 0.13} \, .	
\end{equation}
The outer and inner scales are poorly known for distances closer to the Sun ($\lesssim 3 \, \Rs$) and so these relations are extrapolated to describe smaller heliocentric heights.
For example, at a distance $r=2$~$\Rs$, the inner and outer scales are approximated as $l_i \approx 2.9 \times 10^{-6}$~$\Rs$ and $l_o \approx 0.4$~$\Rs$, respectively.

It should be emphasised that any subsequent characterisation of the local coronal conditions will depend on the model of density fluctuations used (Equation~(\ref{eqn:simple_dens_fluc_model})).  For example, any inferred value of $\epsilon$ will only be valid for the specific outer scale ($l_o$) model adopted, which might not be representative of the locations resulting to emissions observed by ground-based instruments like LOFAR ($\lesssim 2.5 \, \Rs$, corresponding to $f_{pe} \gtrsim 10$~MHz).  As such, the values of $\epsilon$ inferred using this model may not be directly comparable to in-situ density fluctuation measurements in the corona.

\section{Isotropic vs Anisotropic Scattering Description} \label{sec:isotropic_vs_anisotropic_scatt_sim}

\subsection{Multi-frequency observational data} \label{sec:obs_size_and_decay_vs_freq}
\begin{figure}[htp!]
    \centering

\vspace{-3ex}

	\captionsetup[subfigure]{aboveskip=-2ex, belowskip=-0.5ex, singlelinecheck=off}
	
\begin{subfigure}[t]{1.0\linewidth}
    \centering
    \caption{}
    \includegraphics[width=0.8\textwidth, keepaspectratio=true]{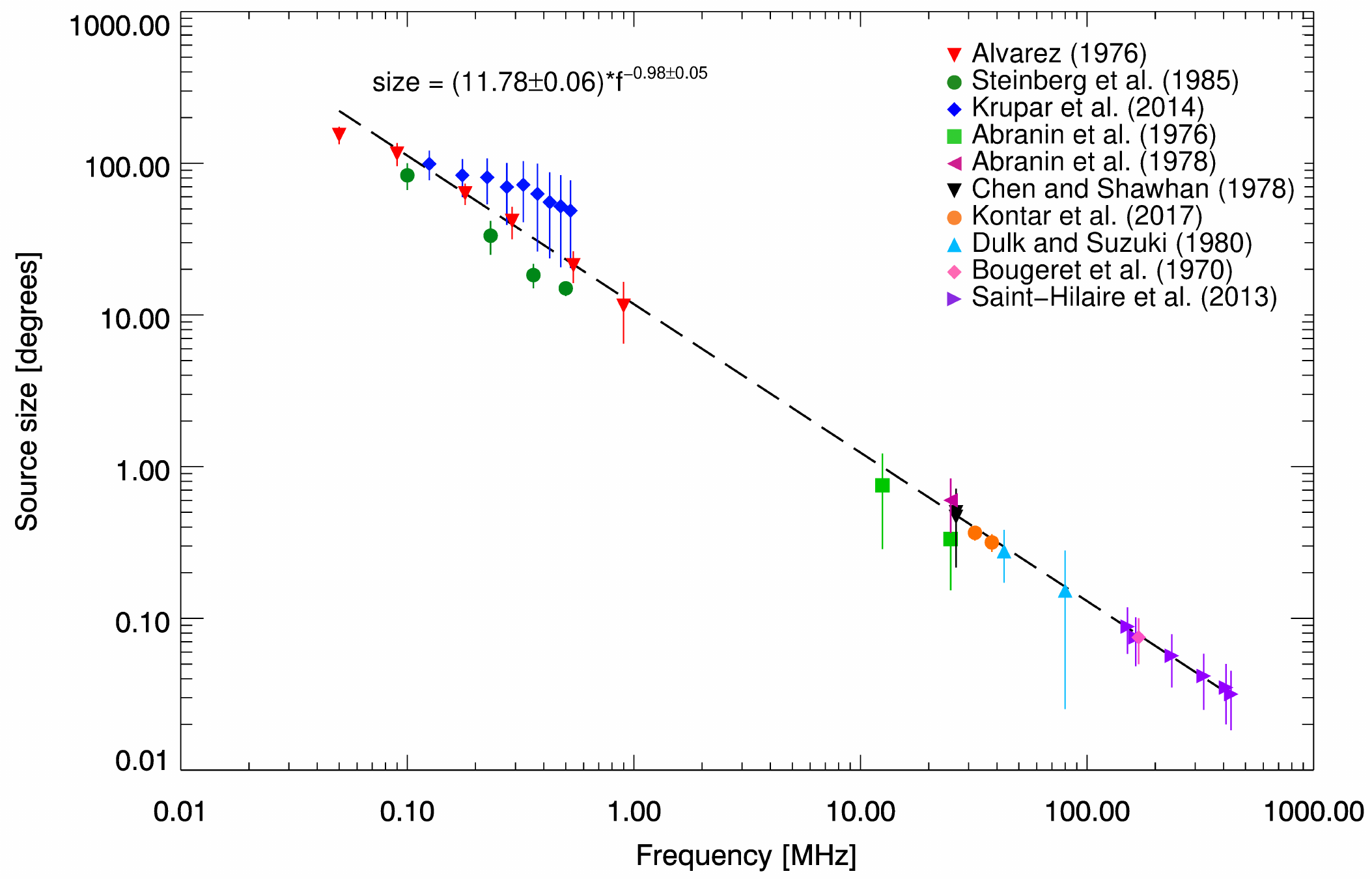}
    \label{fig:source_sizes_fit}
\end{subfigure}
\begin{subfigure}[t]{1.0\linewidth}
    \centering
	\caption{}
    \includegraphics[width=0.8\textwidth, keepaspectratio=true]{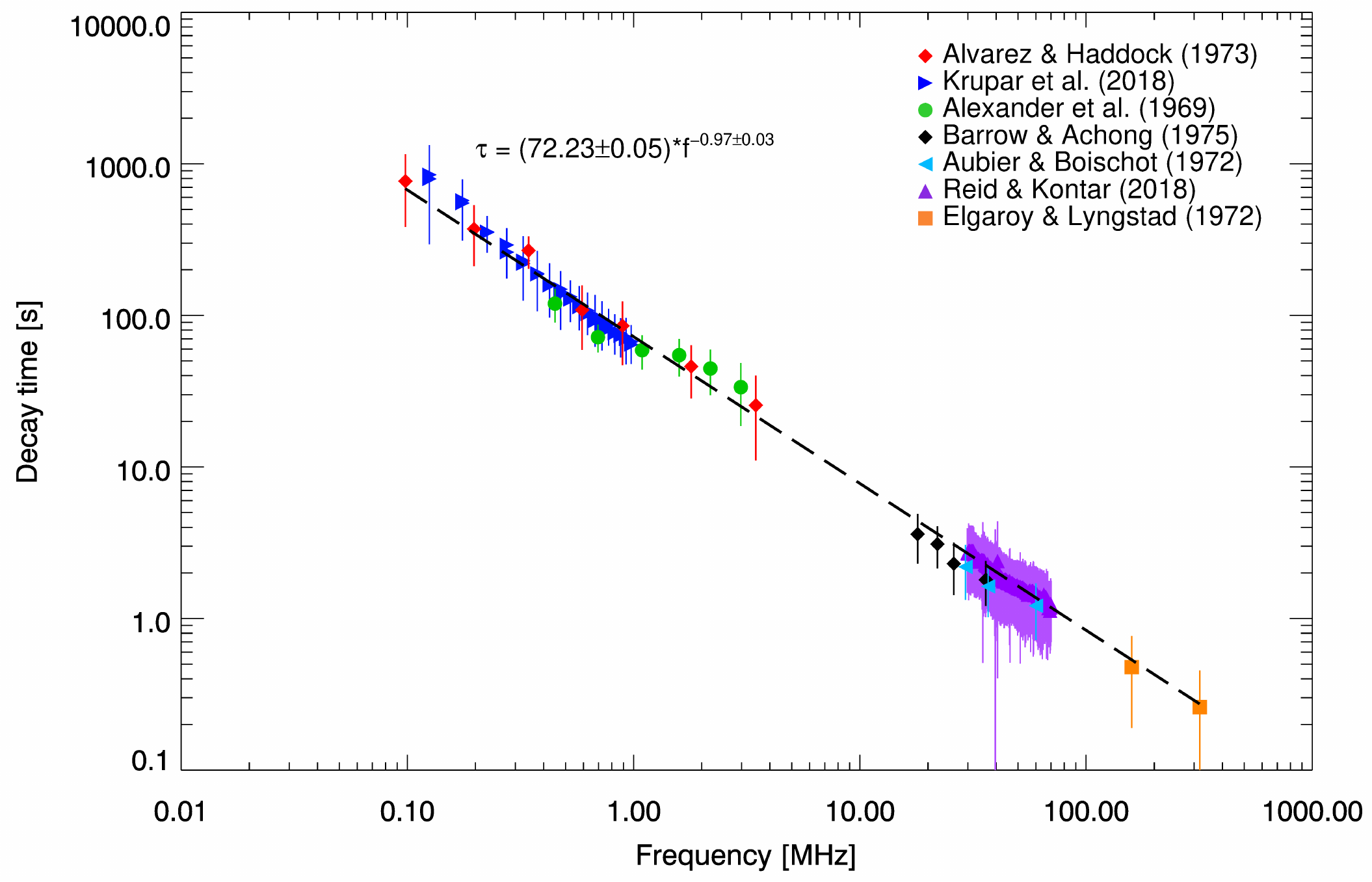}
    \label{fig:decay_times_fit}
\end{subfigure}

    \caption[Observed Type III source sizes and decay times vs frequency.]
    {Collection of observed Type III burst properties from several studies (indicated by the legends) over a large range of frequencies.  The error bars represent the standard deviation (calculated from the statistical distribution of the data) and measurement errors, where reported.  The dashed lines depict the applied weighted linear fit (in log-space) which provided the power-law relation between the observed properties and the emission frequency (as annotated).
    (a)~FWHM source sizes of Type III bursts (given in degrees) spanning frequencies from $\sim$0.05 to 500~MHz.  The power-law relation is given in Equation~(\ref{eqn:obs_size_vs_freq}).
    (b)~Decay times $\tau$ of Type III bursts (defined as the $e$-folding times and given in seconds) spanning a range of frequencies from $\sim$0.1 to 300~MHz.  The power-law relation is given in Equation~(\ref{eqn:obs_decay_vs_freq}).
    Figure taken from \cite{2019ApJ...884..122K}.}
    \label{fig:size_and_decay_fit}
\end{figure}

A collection of Type III source size and decay time data observed over several decades by several instruments was gathered, as illustrated in Figure~\ref{fig:size_and_decay_fit}.  Measurements across a large range of frequencies are displayed, corresponding to emissions excited from distances near the Sun to distances near the Earth.  The top panel shows data of Type III source sizes between $\sim$0.05 and 500~MHz obtained from \cite{1970A&A.....6..406B}, \cite{1976SvA....19..602A}, \cite{1976SoPh...46..483A}, \cite{1978SoPh...57..229A}, \cite{1978SoPh...57..205C}, \cite{1980A&A....88..203D}, \cite{1985A&A...150..205S}, \cite{2013ApJ...762...60S}, \cite{2014SoPh..289.4633K}, and \cite{2017NatCo...8.1515K}.
The angular resolution of the instruments was taken into account, such that observed source sizes were deconvolved (where possible; see Section~\ref{sec:centroid_calc}).
The bottom panel shows decay time data between $\sim$0.1 and 300~MHz obtained from \cite{1969SoPh....8..388A}, \cite{1972A&A....19..343A}, \cite{1972A&A....16....1E}, \cite{1973SoPh...30..175A}, \cite{1975SoPh...45..459B}, \cite{2018ApJ...857...82K}, and \cite{2018A&A...614A..69R}.  A power-law characterisation of the observed Type III decay times and source sizes as a function of emission frequency was obtained by applying a weighted linear fit to the data in logarithmic space.

Prior to fitting, measurements from different studies were recalculated where necessary to present comparable values.  The Type III source sizes are given as the FWHM value in degrees.  Where the source sizes were originally reported as the full width at $1/e$ of the distribution \citep{1980A&A....88..203D}, the values were transformed into FWHM by multiplying by a factor of $\sqrt{\ln 2}$.  It should be noted that data above 1~MHz reported by \cite{2014SoPh..289.4633K} was deemed unreliable and thus not plotted in this study, since ``the analysis above 1~MHz is perhaps distorted by background signals resulting in increased source sizes'' \citep{2014SoPh..289.4633K}.

Type III decay times are given as $1/e$ measurements (or ``$e$-folding times'') in seconds, i.e the time it takes for the flux density to decrease from its peak value to $1/e$ of the peak value.  \cite{1972A&A....16....1E}, however, defined the decay time as the time from the peak of the light curve until the time the intensity reached $1/10$ of its maximum value.  These data were therefore translated into $1/e$ values by multiplying them by a factor of $\ln{(e)}/\ln{(10)}$.  \cite{2018A&A...614A..69R}, on the other hand, fitted the observed light curves with a Gaussian distribution and presented decay times as the half width at half maximum (HWHM) value.  The results from \cite{2018A&A...614A..69R} were multiplied by a factor of $1 / \sqrt{\ln{(2)}}$ in order to make them comparable with the decay times measured at $1/e$ of the peak value.

Where multiple measurements of source size and decay time were made at a given frequency within a single study, the average value of the statistical spread of the single-frequency data was used in the figures presented in this section.  If asymmetric errors were provided in the original studies, the maximum of the two values was assumed and used for the fit.
In some cases where no uncertainties were stated or only the spread of the data was provided (e.g., \cite{2018A&A...614A..69R}), the (sample) standard deviation $\sigma_{sample}$ of the data was calculated via
\begin{equation} \label{eqn:sigma}
        \sigma_{sample} = \sqrt{ \dfrac{1}{M-1} \sum_{i=1}^{M} \left( \chi_i - \bar{\chi} \right)^2  } \,,
\end{equation}
where $M$ is the total number of measurements (or bins, if a histogram was provided), $\chi_i$ is the value of each measurement (or bin), and $\bar{\chi}$ denotes the mean value of the sample.

The best-fit power-law dependence of FWHM source size ($\theta_\mathrm{FWHM}$) on the emission frequency $f$ was found to be
\begin{equation} \label{eqn:obs_size_vs_freq}
    \theta_\mathrm{FWHM} = (11.78 \pm 0.06) \times f^{-0.98 \pm 0.05} \, ,
\end{equation}
where the source size is given in degrees and the frequency in MHz (Figure~\ref{fig:source_sizes_fit}).  The corresponding relation of the decay time ($\tau_{decay}$) to the frequency $f$ was obtained as
\begin{equation} \label{eqn:obs_decay_vs_freq}
    \tau_{decay} = (72.23 \pm 0.05) \times f^{-0.97 \pm 0.03} \, ,
\end{equation}
where the decay time is given in seconds and the frequency in MHz (Figure~\ref{fig:decay_times_fit}).

The inferred dependence of decay time on frequency is consistent with that of previous studies which examined observations within a relatively-narrower frequency range.  \cite{1973SoPh...30..175A} obtained $\tau_{decay} = 51.29 \times f^{-0.95}$ for frequencies from $\sim$0.05--3.5~MHz, \cite{1973SoPh...31..501E} obtained $\tau_{decay} = (2.0 \pm 1.2) \times 100 \times f^{-1.09 \pm 0.05}$ for data between $\sim$0.07--2.8~MHz, and \cite{1950AuSRA...3..541W} obtained $\tau_{decay} = 100 \times f^{-1}$ for the frequency range of $\sim$80--120~MHz.

\subsection{Isotropic scattering simulations vs observations} \label{sec:isotropic_scatt_sim_vs_obs}
The adequacy of the isotropic density fluctuations assumption ($\alpha=1$) is tested by comparing the output of the simulations to the collection of observational data shown in Figure~\ref{fig:size_and_decay_fit}.  The level of density fluctuations $\epsilon$ is varied and the resulting size and (HWHM) decay time values are calculated for 10 different frequencies ranging from 0.1 to 1~MHz.  The simulations are run for a source-polar angle $\theta_s = 0\degr$, meaning that the FWHM $x$- and $y$-sizes are equal (given the isotropic scattering assumption).

\begin{figure}[ht!]
    \centering
	\includegraphics[width=0.475\textwidth, keepaspectratio=true]{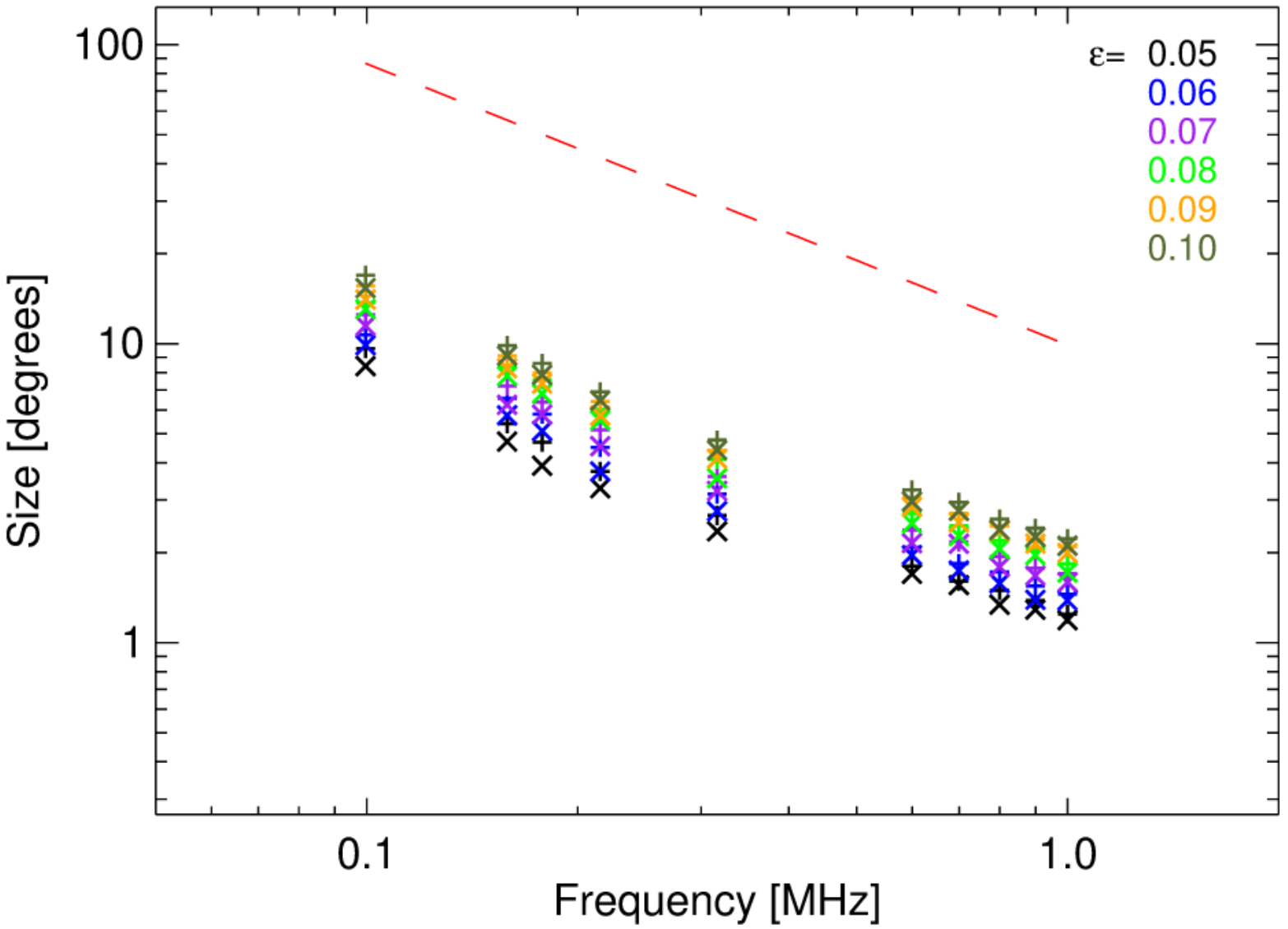}
	\includegraphics[width=0.517\textwidth, keepaspectratio=true]{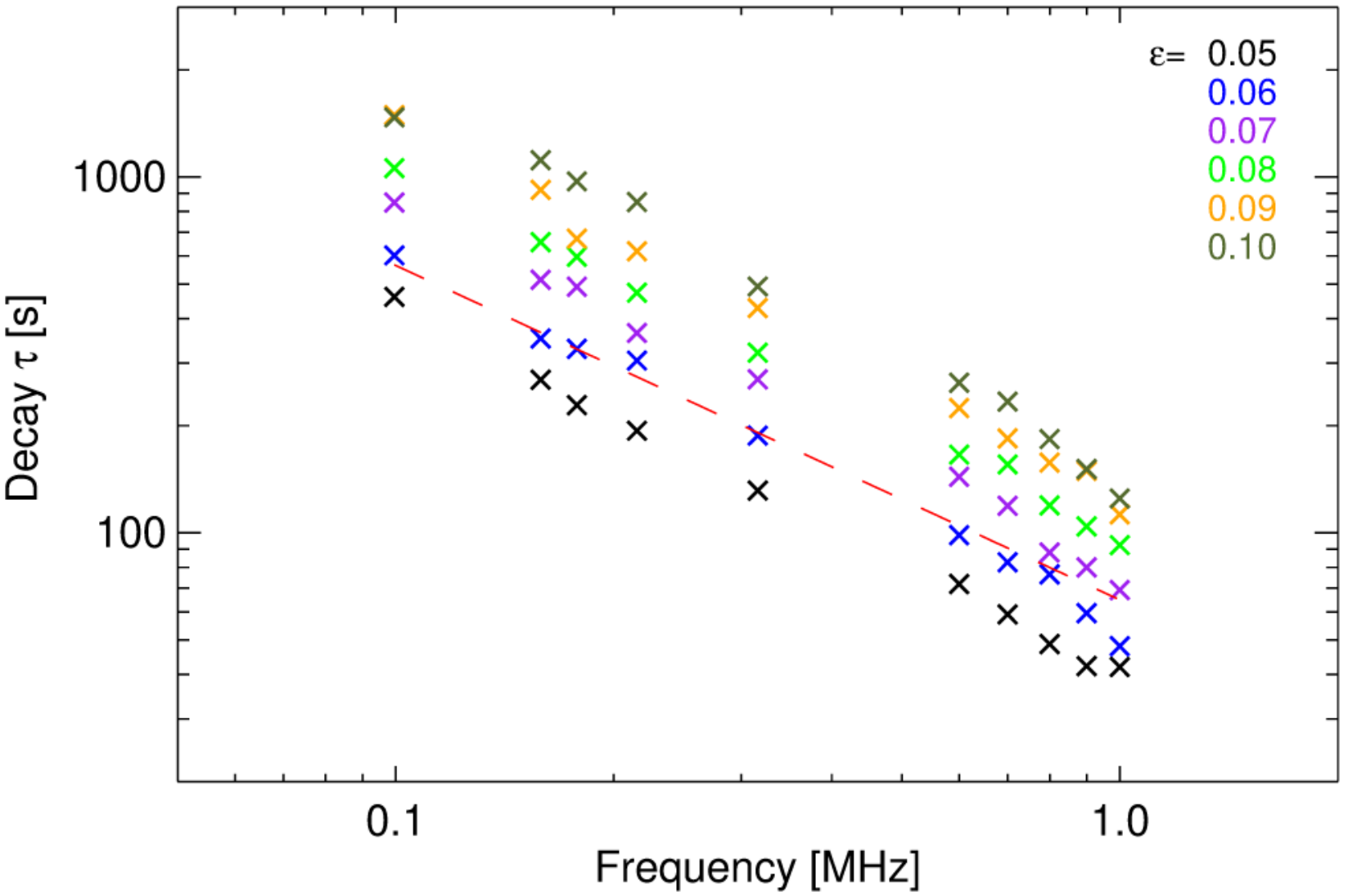}
    \caption[Isotropic scattering simulations vs observed source sizes and decay times.]
    {Simulated FWHM source sizes (left) and HWHM decay times (right) for a range of frequencies (0.1--1~MHz), assuming isotropic scattering ($\alpha = 1$) and sources located as the disk centre (i.e. $\theta_s = 0\degr$, where $\mathrm{FWHM}_{x} = \mathrm{FWHM}_{y}$).  Varying levels of density fluctuations $\epsilon$ (from 0.05--0.1) are used, as indicated by the colour codes and legends.  The simulated properties are compared to the best-fit relationships obtained from observations (see Figure~\ref{fig:size_and_decay_fit}), shown by the red dashed line in each panel.
    Figure taken from \cite{2019ApJ...884..122K}.
	}
    \label{fig:2019_sim_size_and_decay_vs_freq}
\end{figure}

The results are indicated in Figure~\ref{fig:2019_sim_size_and_decay_vs_freq}, where the red dashed lines represent the obtained frequency dependence of the source sizes (left panel) and decay times (right panel) given in Equations~(\ref{eqn:obs_size_vs_freq}) and (\ref{eqn:obs_decay_vs_freq}), respectively.  The crosses represent the simulated size and decay times, colour-coded for values of $\epsilon$ ranging from 0.05--0.1, as indicated by the legends.

It can be seen that the isotropic scattering simulations match the observed decay times, but fail to match the observed source sizes as they consistently produce smaller values.  Increasing, for example, the level of density fluctuations $\epsilon$ in order to produce larger source sizes such that they match the observed sizes, will also lead to the decay times increasing and no longer agreeing with the observed values.  It is thus clear that the assumption of isotropic scattering cannot sufficiently describe the observed source properties, and instead, an anisotropic scattering description must be considered.

\section{Reproducing the Observed Properties of a 35~MHz Type IIIb Radio Source} \label{sec:sim_TypeIIIb_properties}

The coronal and observing conditions vary from event to event, meaning that a single set of simulated parameters is unlikely to be valid for all observations.  It is, therefore, important to evaluate the simulated level of density fluctuations, the level of anisotropy, and the source-polar angle for every individual event, in order to describe the entirety of that event's observed characteristics with confidence.  As such, the developed ray-tracing simulations are compared to the observed properties of a high-resolution LOFAR observation of a Type IIIb burst (observed on 16 April 2015), presented and analysed in several studies \citep{2017NatCo...8.1515K, 2018ApJ...856...73C, 2018ApJ...861...33K, 2018SoPh..293..115S}.  The simulations were run for $f \approx 35$~MHz in order to allow for a direct comparison with the results from these recent observational studies, which analysed the Type IIIb burst at $\sim$35~MHz.  A comparison with Type IIIb bursts is advantageous since their intrinsic source sizes can be estimated (see Section~\ref{sec:prop_effs}) and are found to be very small (with respect to what is observed; \cite{2017NatCo...8.1515K}), consistent with the assumption of intrinsic point sources made in the simulations.

The simulations are compared to the time profile and source size of the fundamental component of the Type IIIb stria observed around 35~MHz.  According to the assumed (fundamental) emission-to-plasma frequency relation ($f = 1.1 \, f_{pe}$; see Section~\ref{sec:anisotropic_simulations_2019}), a source that is observed at $f \approx 35$~MHz is emitted at $f_{pe} \approx 32$~MHz.  Given the adopted density model (Equation~\ref{eqn:2019_sim_dens_model}), the plasma frequency $f_{pe} \approx 32$~MHz is located at $R_s = 1.75~\Rs$, defining the heliocentric distance from which the simulated photons are set to originate.

\cite{2017NatCo...8.1515K} found that for the fundamental component of the Type IIIb burst, sources imaged at $\sim$35~MHz have a size of $\sim$19$\arcmin$, consistent with previous studies (see, e.g., \cite{1980A&A....88..203D}).  The level of density fluctuations $\epsilon$ is set so that the resulting simulated source is equally large ($\sim$19$\arcmin$).  A value of $\epsilon=0.8$ is required to satisfy this condition.

\begin{figure}[htp!]
	\centering

\vspace{-0ex}

	\captionsetup[subfigure]{aboveskip=0.5ex, belowskip=-0.5ex, singlelinecheck=off}
	
\begin{subfigure}[t]{1.0\linewidth}
    \caption{~$\alpha$=0.5 \vspace{0.5ex}}
    \includegraphics[width=0.329\textwidth, keepaspectratio=true]{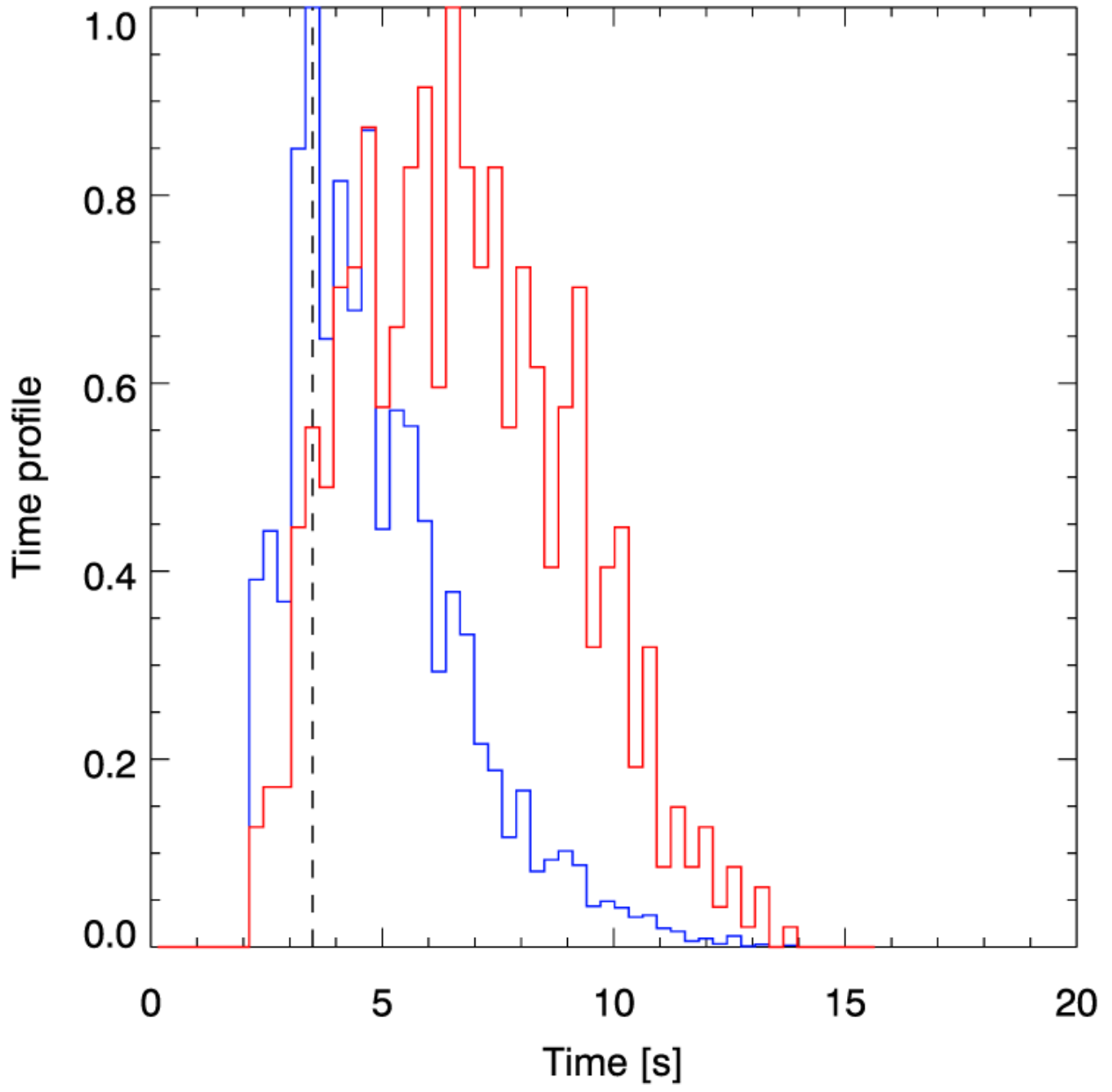}
	\includegraphics[width=0.329\textwidth, keepaspectratio=true]{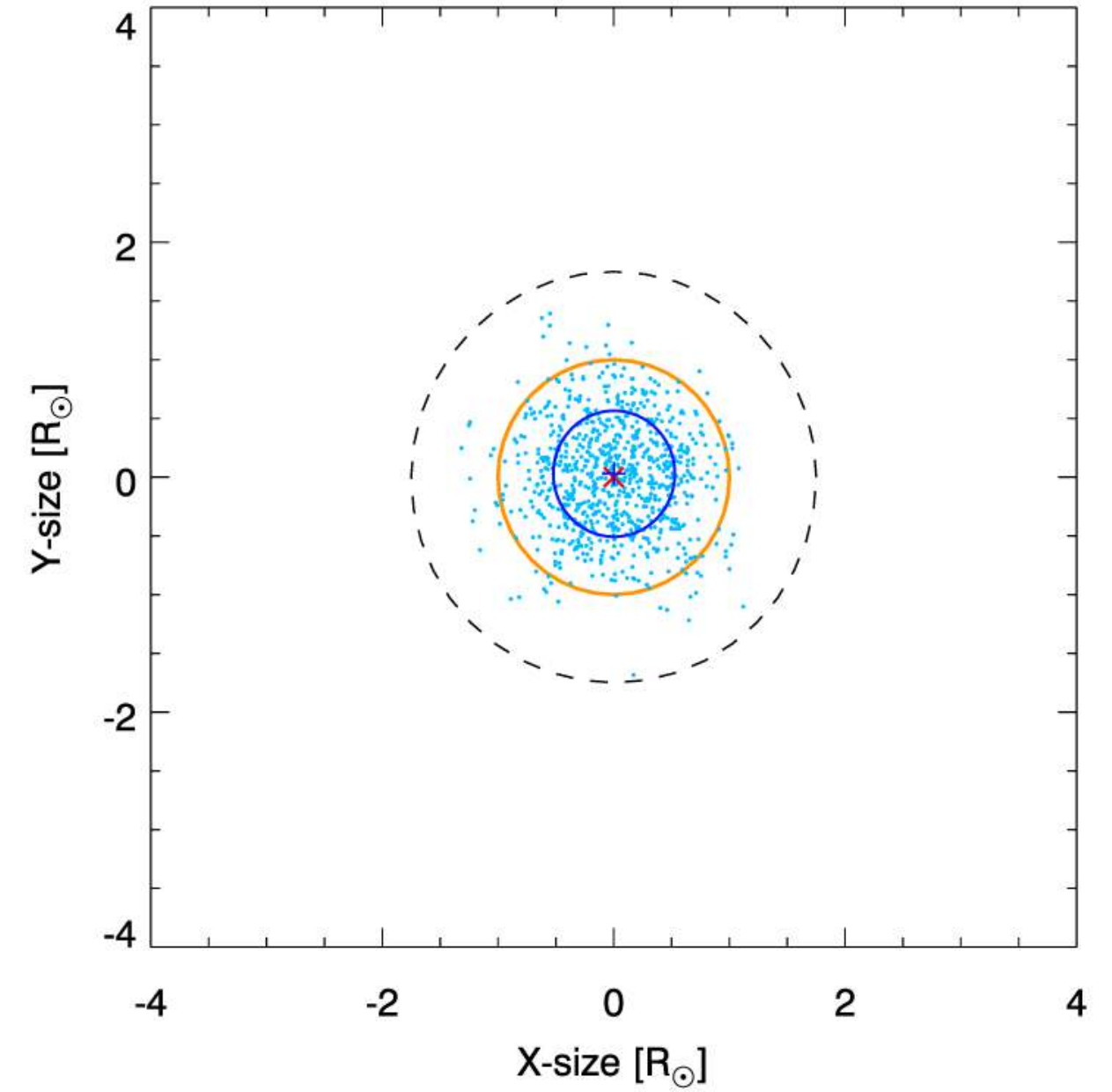}
	\includegraphics[width=0.329\textwidth, keepaspectratio=true]{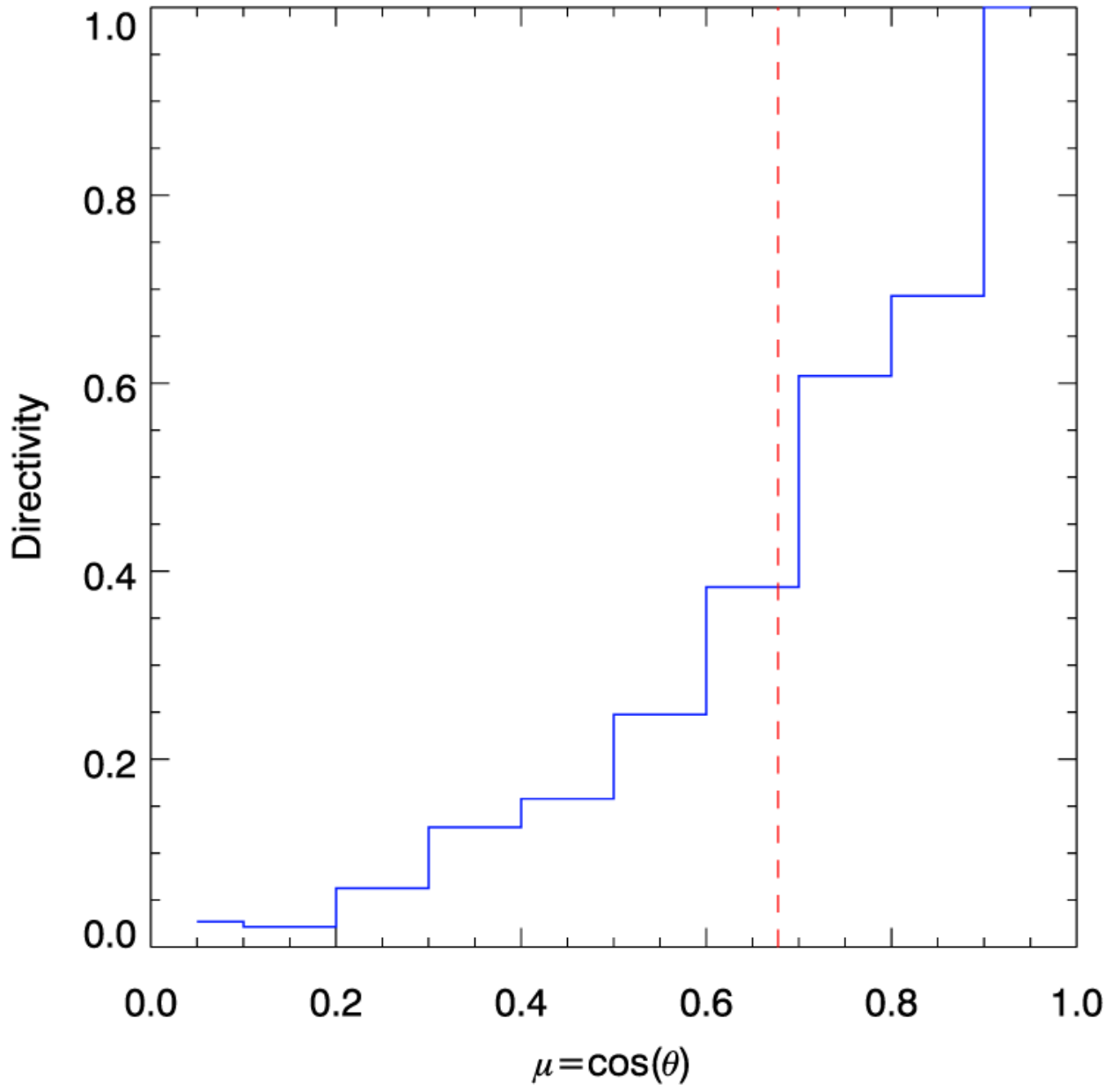}
	\label{fig:2019_sim_prof_source_dir_ans0.5}
\end{subfigure}

\begin{subfigure}[t]{1.0\linewidth}
    \caption{~$\alpha$=0.3 \vspace{0.5ex}}
    \includegraphics[width=0.329\textwidth, keepaspectratio=true]{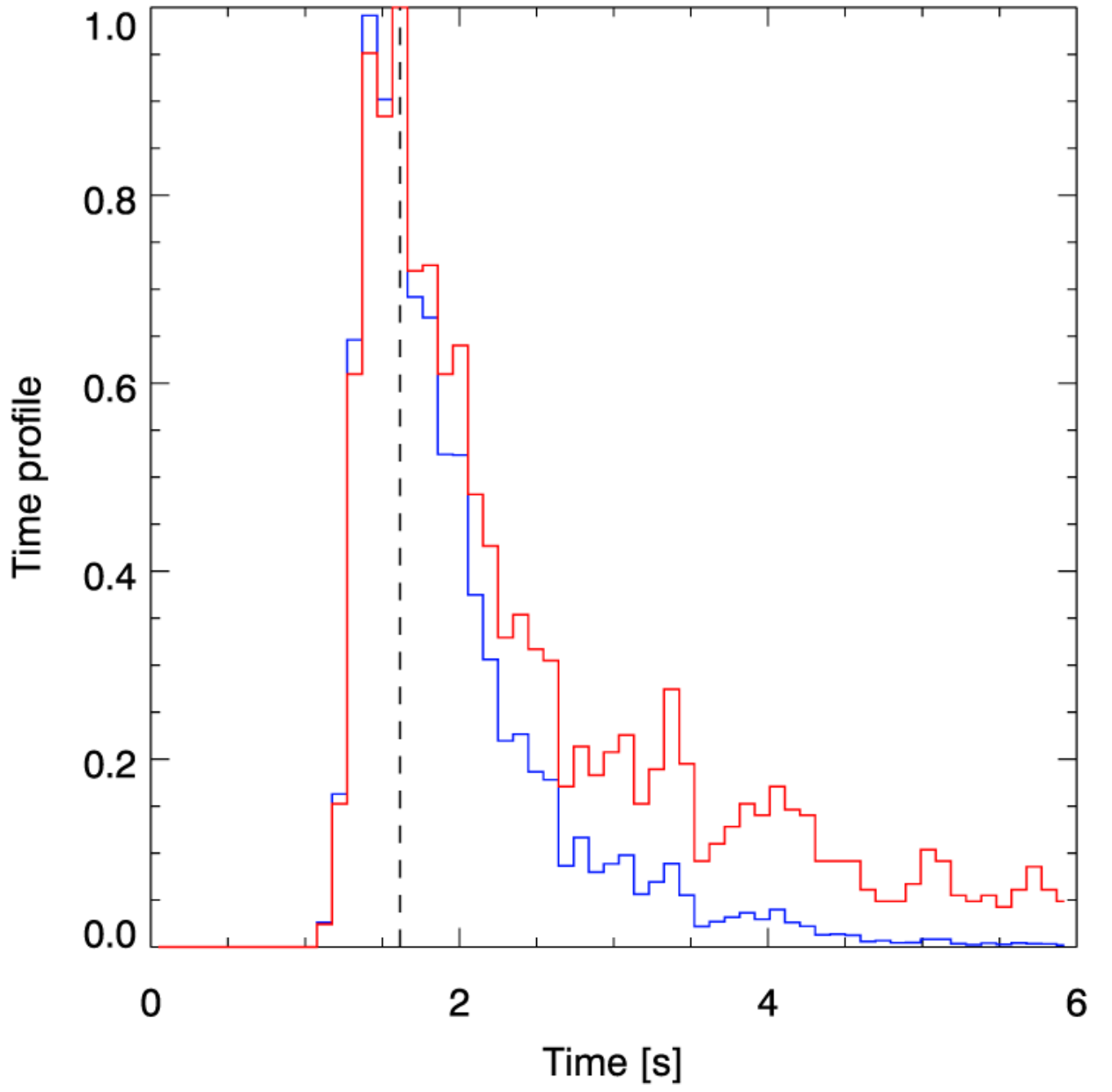}
	\includegraphics[width=0.329\textwidth, keepaspectratio=true]{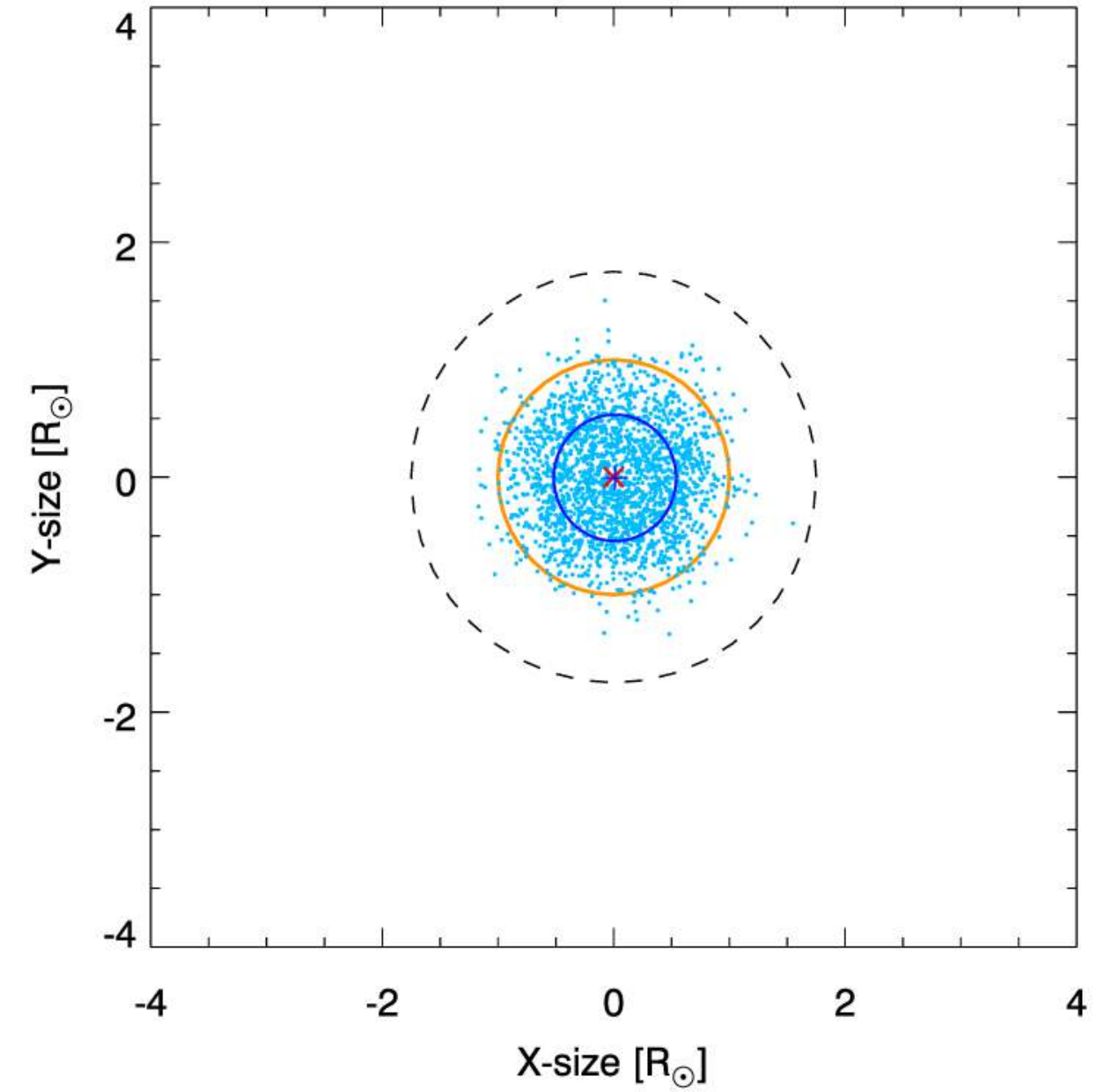}
	\includegraphics[width=0.329\textwidth, keepaspectratio=true]{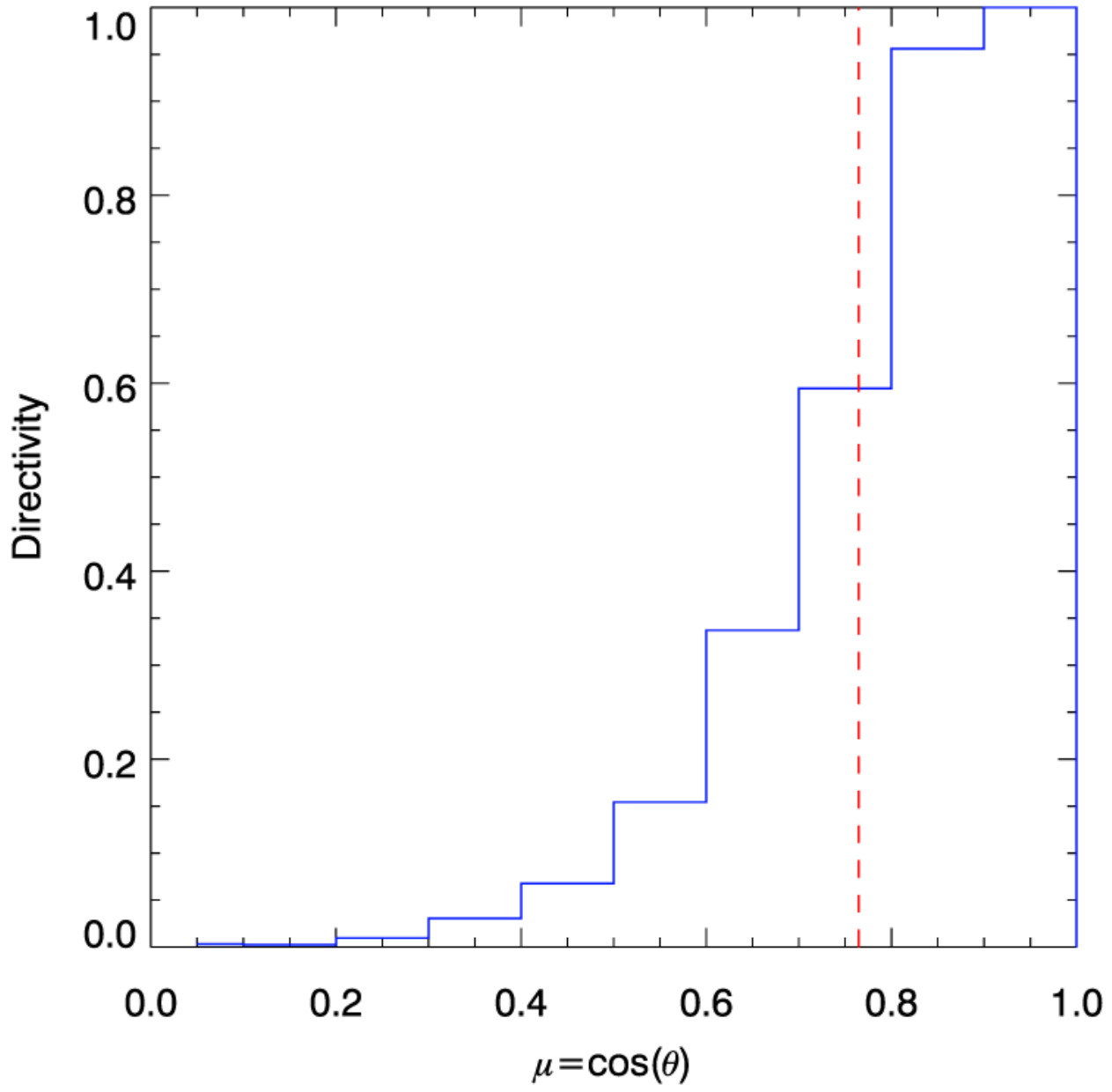}
	\label{fig:2019_sim_prof_source_dir_ans0.3}
\end{subfigure}

\vspace{-2ex}

	\caption[Simulated time profile, source size, and directivity assuming $\epsilon=0.8$ and anisotropies $\alpha=0.3$ and $0.5$.]
    {Simulated properties for a point source located at $R_s = 1.75~\Rs$ (where $f_{pe} \approx 32$~MHz), assuming $\epsilon = 0.8$, $\theta_s = 0\degr$, and two anisotropies $\alpha = 0.5$ (panel (a)) and 0.3 (panel (b)).
    The left panels indicate the simulated (normalised) time profile of the observed photons, both with and without absorption taken into account (blue and red curves, respectively).  The black dashed line indicates the peak location of the time profile that includes absorption.
    The middle panels depict the simulated radio image in Sun-centred coordinates.  Photons are indicated by blue dots, the solar limb is illustrated by the orange curve, the intrinsic heliocentric source distance ($R_s = 1.75~\Rs$) is indicated with the black dashed line, and the source FWHM size is shown by the blue ellipse.  The red cross and blue plus sign represent the source's projected intrinsic and observed positions, respectively, which overlap at the solar centre when $\theta_s = 0\degr$.
    The right panels show the simulated directivity of the observed radio emission, where the red dashed line annotates the width at half maximum.
    Figures taken from \cite{2019ApJ...884..122K}.
	}
	\label{fig:2019_sim_prof_source_dir}
\end{figure}

The anisotropy $\alpha$ is varied in order to examine its effect on the source properties.  Figure~\ref{fig:2019_sim_prof_source_dir} depicts the simulated time profile, source size, and directivity of a fundamental (point) source located at $R_s = 1.75~\Rs$ ($f_{pe} \approx 32$~MHz) and a source-polar angle $\theta_s = 0\degr$, where the assumed anisotropy $\alpha=0.5$ and 0.3, respectively.  The left panels of Figure~\ref{fig:2019_sim_prof_source_dir} show the simulated (normalised) time profiles, with the red curve calculated assuming no free-free absorption and the blue curve including the effects of absorption.  As expected, the time profiles simulated without absorption are longer in duration than those for which absorption is considered, given that all scattered photons eventually make it to the observer.  The middle panels illustrate the simulated observed sources with respect to the Sun (orange circle), where the black dotted circle indicates the radial distance of $R_s = 1.75~\Rs$, the blue ellipse indicates the FWHM size of the sources, the blue plus signs indicate the observed source centroids, and the red crosses indicate the intrinsic position of the sources.  Given that the simulations in Figure~\ref{fig:2019_sim_prof_source_dir} are run for a source-polar angle $\theta_s = 0\degr$ and the scattering-induced shift is radial, no shift is observed between the observed and true source centroids, as expected.  The right panels illustrate the directivity of the observed radio emission, i.e. the number of photons at each angular position.  The red dashed lines mark the (angular) width of the emission at the half maximum level.

The FWHM source size obtained for both the $\alpha=0.5$ (Figure~\ref{fig:2019_sim_prof_source_dir_ans0.5}) and $\alpha=0.3$ (Figure~\ref{fig:2019_sim_prof_source_dir_ans0.3}) cases is approximately 1.15~$\Rs$.
This value is consistent with the observed FWHM sizes of $\sim$19$\arcmin$ (i.e. $\sim$1.19~$\Rs$; \cite{2017NatCo...8.1515K}).
The (HWHM) decay time for $\alpha=0.3$ is found to be $\sim$0.6~s, which agrees with the decay time observed for the fundamental component of Type IIIb bursts at $\sim$35~MHz, as reported by \cite{2018SoPh..293..115S}.  The time profiles, however, for the two anisotropies differ significantly, with the pulse produced assuming anisotropy $\alpha=0.5$ (Figure~\ref{fig:2019_sim_prof_source_dir_ans0.5}) being broader and being observed later (see black dashed line) than that of anisotropy $\alpha=0.3$ (Figure~\ref{fig:2019_sim_prof_source_dir_ans0.3}).  When the level of anisotropy is higher ($\alpha=0.3$), the turbulent density fluctuations have a power that is stronger (by a factor of 3) in the perpendicular direction compared to the radial direction.  In other words, scattering is weaker in the radial direction, which is (in this case) along the observer's LoS (given that $\theta_s = 0\degr$).  Since time profiles reflect the sources' properties along the observer's LoS only, the anisotropy affects both the duration and arrival time of the radio pulse, as well as how much the time profile of the absorbed pulse differs from the one where absorption is ignored (cf. red and blue curves in the left panels of Figure~\ref{fig:2019_sim_prof_source_dir}).  When photons scatter less (in the radial direction; $\alpha=0.3$) they spend less time in the corona before they reach the observer, thus being observed earlier than photons that scatter more in that direction ($\alpha=0.5$).  This also implies that all photons reach the observer faster (compared to $\alpha=0.5$) and the duration of the observed pulse is shorter, corresponding to a shorter decay time.  In addition to that, less scattering in the radial direction ($\alpha=0.3$) corresponds to less absorption (since photons stay less in the collisional coronal medium) and thus more photons reach the observer, which means that the time profile of the absorbed pulse is more similar to its no-absorption time profile than cases where stronger radial scattering occurs ($\alpha=0.5$).

The directivity of the escaping radio emission is determined by the interplay between scattering on small-scale inhomogeneities (which makes the radiation less directional) and refraction on large-scale inhomogeneities (a focusing effect which makes the radiation more directional; \cite{1985srph.book.....M}).
As shown in the left panels of Figure~\ref{fig:2019_sim_prof_source_dir}, the directivity is found to be anisotropic.  The simulated directivity pattern for both levels of anisotropy is primarily in the radial direction, where the HWHM is calculated to be $\simeq 47\degr$ and $40\degr$ for $\alpha=0.5$ and $\alpha=0.3$, respectively.
In other words, anisotropic scattering results in a directional emission, even when an isotropically-emitting point source is assumed in the simulations.  This outcome contradicts previous results suggesting that the directivity due to scattering is isotropic, as reviewed by \cite{1985srph.book.....M}.  Furthermore, it shows that scattering due to small-scale and \textit{anisotropic} density inhomogeneities can lead to sufficiently-large observed source sizes whilst the radiation remains directional (and predominantly along the radial direction).  As such, these results address some of the previous arguments against scattering (see Section~\ref{sec:ray-tracing_simulations_motivation}) which were based on the generation of a less directional emission produced when stronger---but isotropic---scattering was invoked (see, e.g., \cite{1977A&A....61..777B} and \cite{1985srph.book.....M}).

\subsection{Inferring the level of anisotropy and density fluctuations} \label{sec:inferring_eps_and_anis}
\begin{figure}[ht!]
	\centering

	\captionsetup[subfigure]{aboveskip=0.5em, belowskip=-0.5em, singlelinecheck=off}
	
\begin{subfigure}[t]{1.0\linewidth}
    \caption{}
    \includegraphics[width=0.496\textwidth, keepaspectratio=true]{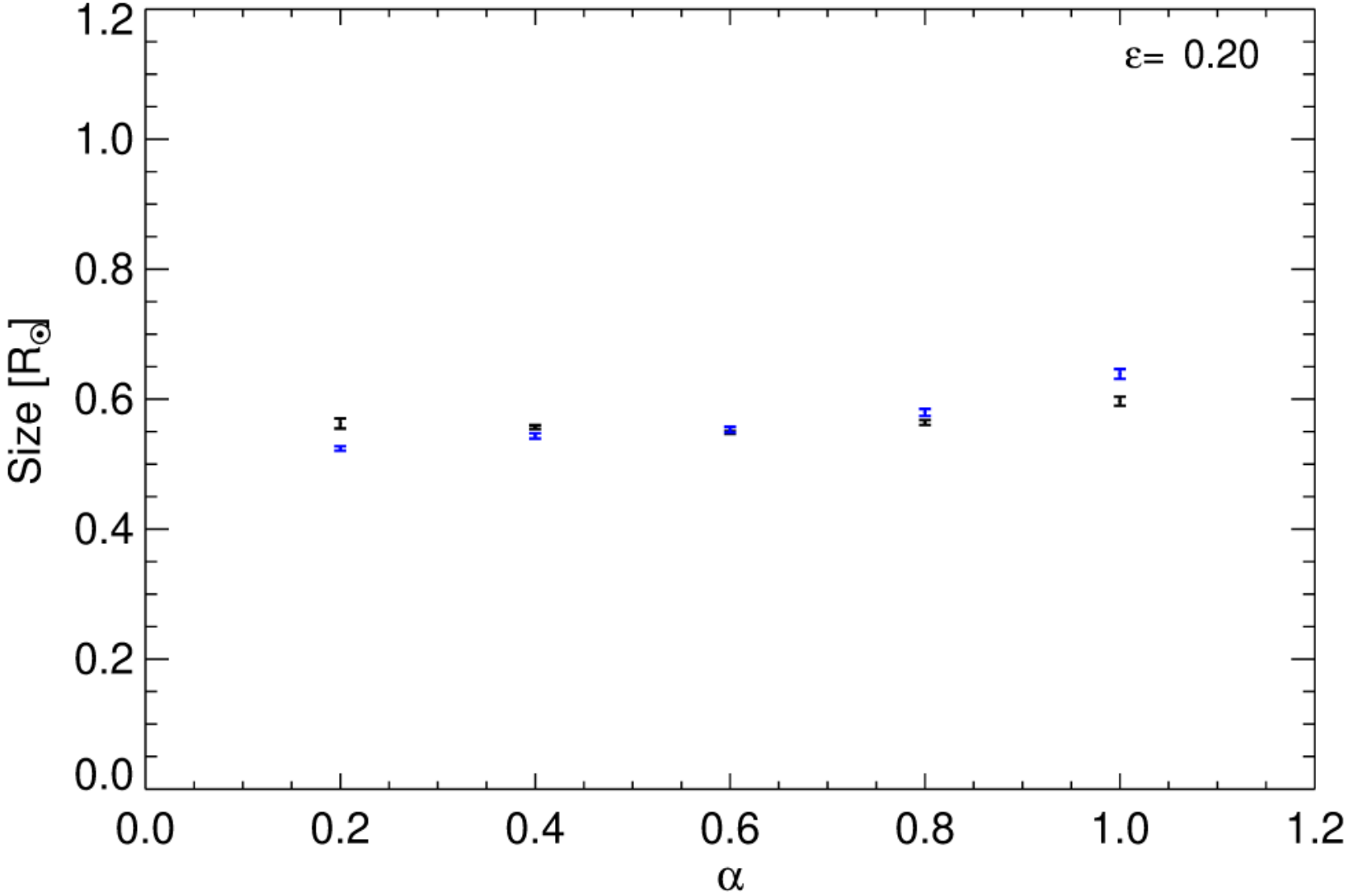}
	\includegraphics[width=0.496\textwidth, keepaspectratio=true]{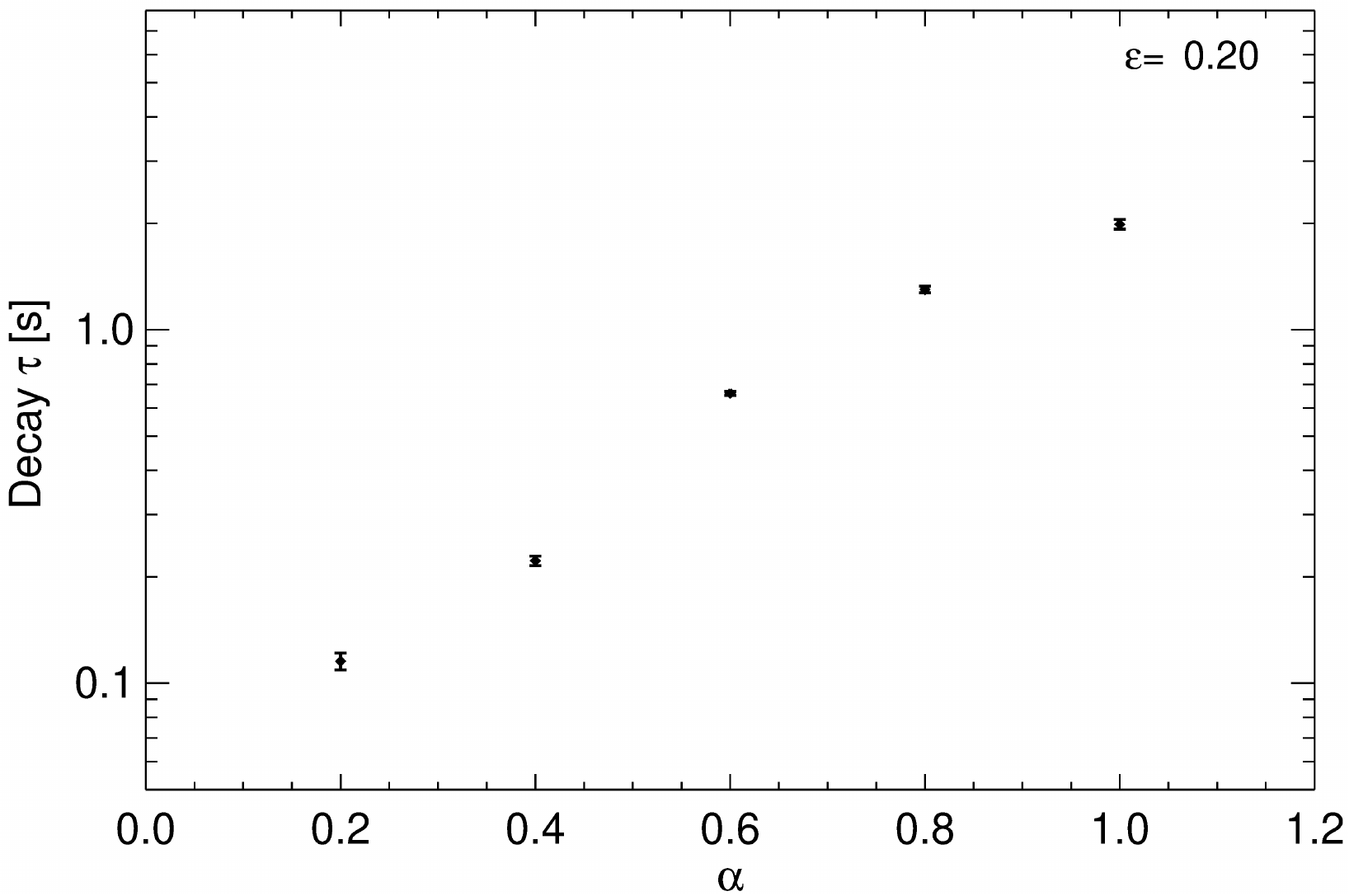}
	\label{fig:2019_sim_size_and_decay_vs_anis_for_eps0.2}
\end{subfigure}

\begin{subfigure}[t]{1.0\linewidth}
    \caption{}
    \includegraphics[width=0.496\textwidth, keepaspectratio=true]{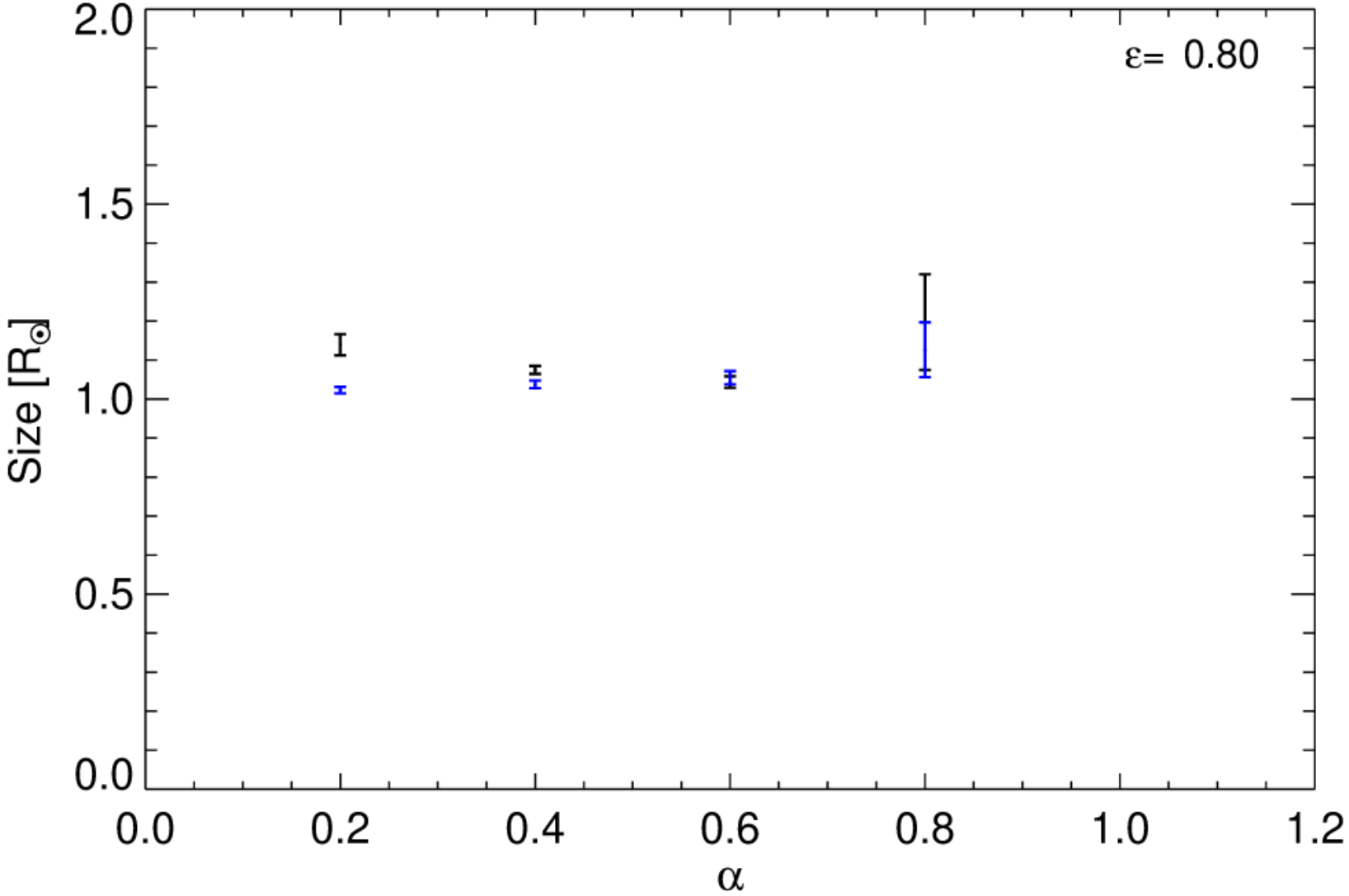}
	\includegraphics[width=0.496\textwidth, keepaspectratio=true]{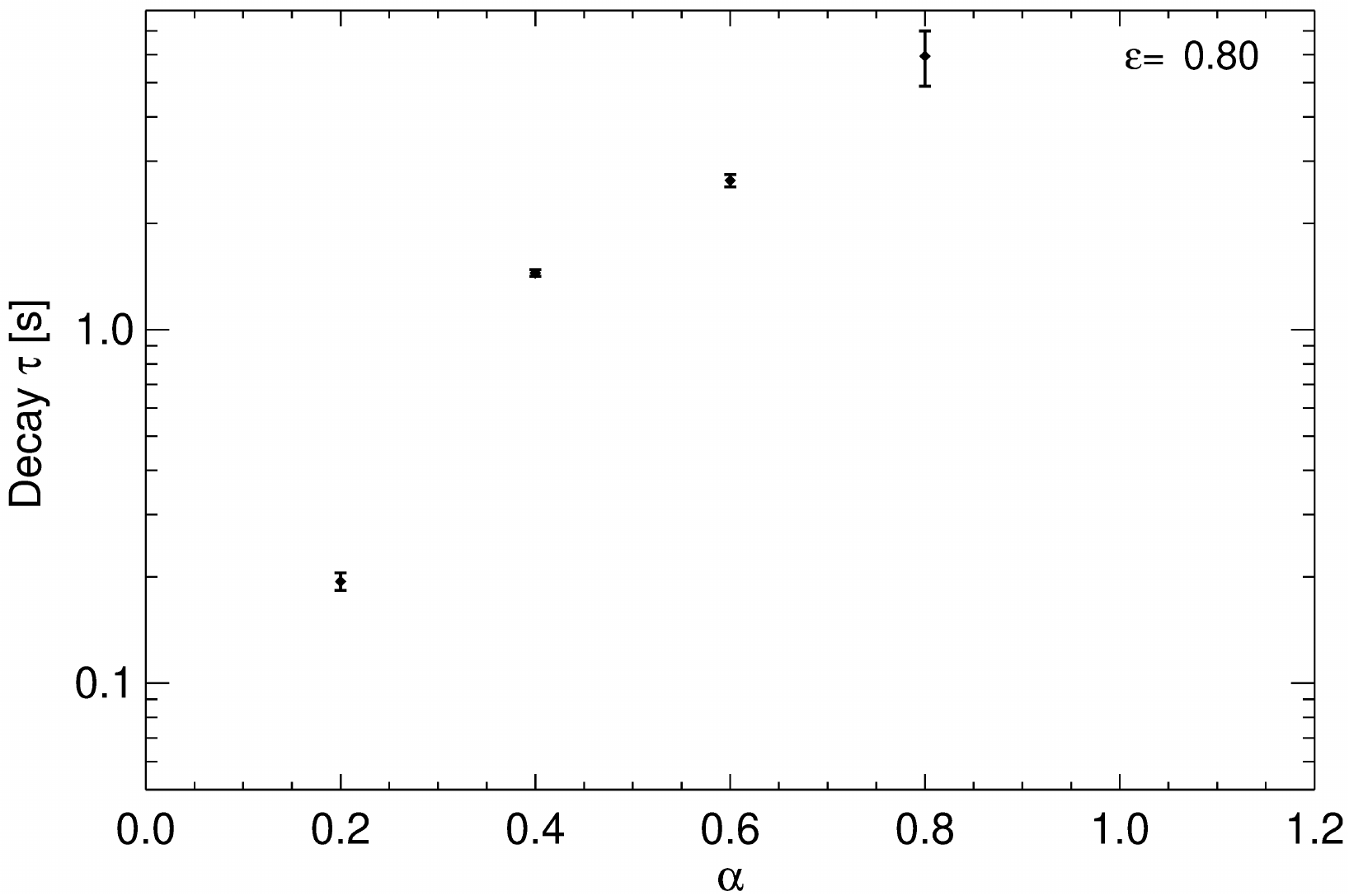}
    \label{fig:2019_sim_size_and_decay_vs_anis_for_eps0.8}
\end{subfigure}

\vspace{-1ex}

	\caption[Simulated source size and decay time as a function of the anisotropy $\alpha$, assuming $\epsilon=0.2$ and $0.8$.]
    {Simulated FWHM source sizes (left panels) and HWHM decay times (right panels) as a function of the anisotropy $\alpha$, assuming $f_{pe} \approx 32$~MHz, $\theta_s = 0\degr$, and $\epsilon = 0.2$ and $0.8$ (panels (a) and (b), respectively).  The source sizes were calculated by fitting the simulated data with a 2D elliptical Gaussian function (black data points; see Section~\ref{sec:centroid_calc}), as well as using the statistical moments of the simulated distribution (blue data points; see Equations~(\ref{eqn:2019_sim_distr_centroids})--(\ref{eqn:2019_sim_distr_size_errors})).  The error bars represent the uncertainties obtained using the one-standard-deviation estimations.
    Figures taken from \cite{2019ApJ...884..122K}.
    }
	\label{fig:2019_sim_size_and_decay_vs_anis}
\end{figure}

The effect that the level of density fluctuations $\epsilon$ and level of anisotropy $\alpha$ have on the FWHM size and decay time of a radio source (at $f_{pe}=32$~MHz and $\theta_s = 0\degr$) is indicated in Figure~\ref{fig:2019_sim_size_and_decay_vs_anis}, where the left panels show the source size against different anisotropy values and the right panels show the decay times against anisotropy.  Figure~\ref{fig:2019_sim_size_and_decay_vs_anis_for_eps0.2} was produced assuming a level of density fluctuations $\epsilon=0.2$, whereas stronger density fluctuations $\epsilon=0.8$ were assumed for Figure~\ref{fig:2019_sim_size_and_decay_vs_anis_for_eps0.8}.  The source sizes were computed by fitting the simulated image with a 2D elliptical Gaussian (black data points; as discussed in Section~\ref{sec:centroid_calc}), as well as using the statistical moments of the simulated distribution (blue data points; as discussed in Section~\ref{sec:size_and_centr_distribution_calc}).

A comparison of Figures~\ref{fig:2019_sim_size_and_decay_vs_anis_for_eps0.2} and \ref{fig:2019_sim_size_and_decay_vs_anis_for_eps0.8} shows that irrespective of the value of anisotropy chosen, the source sizes for $\epsilon=0.2$ are too small to explain the observed sizes of $\sim$1.19~$\Rs$.  The density fluctuations are not strong enough to result in sufficient amounts of scattering and broaden the simulated sources to the degree observed.  However, the obtained source sizes for $\epsilon=0.8$ match the observed source sizes.  Although varying the level of anisotropy does not affect the source sizes significantly (as inferred from Figure~\ref{fig:2019_sim_prof_source_dir}), the decay times are considerably affected (right panel of Figures~\ref{fig:2019_sim_size_and_decay_vs_anis_for_eps0.2} and \ref{fig:2019_sim_size_and_decay_vs_anis_for_eps0.8}).  It is clear that as the anisotropy parameter $\alpha$ increases and approaches unity ($\alpha=1$, i.e. isotropic density fluctuations), the decay times increase beyond the observed value of 0.6~s.  In other words, the stronger the scattering in the perpendicular direction the shorter the observed decay time becomes, given that the extended elongation of the source along the perpendicular direction corresponds to a shorter pulse in the direction parallel to the observer's LoS (as discussed for Figure~\ref{fig:2019_sim_prof_source_dir}).  It is found that when $\epsilon=0.8$, an anisotropy $\alpha=0.3$ is required to produce a characteristic decay time that matches the observations (cf. right panel of Figure~\ref{fig:2019_sim_size_and_decay_vs_anis_for_eps0.8}).

It is therefore deduced that for the adopted model of density fluctuations (i.e. value of $l_o$; Equations~(\ref{eqn:simple_dens_fluc_model}) and (\ref{eqn:outer_scale})), a level of density fluctuations $\epsilon=0.8$ and an anisotropy $\alpha=0.3$ is required in order to obtain a strong agreement between the simulated and observed time profiles and source sizes of the Type IIIb burst near 35~MHz.

\section{Considering the Influence of the Source-Polar Angle} \label{sec:angular_dependence}
\begin{figure}[ht!]
    \centering
	\includegraphics[width=0.3281\textwidth, keepaspectratio=true]{{Kontar_2019_obs_source_fpe32.04MHz_FE35.24MHz_eps0.800_anis0.30_ang0.00deg}.pdf}
	\includegraphics[width=0.3281\textwidth, keepaspectratio=true]{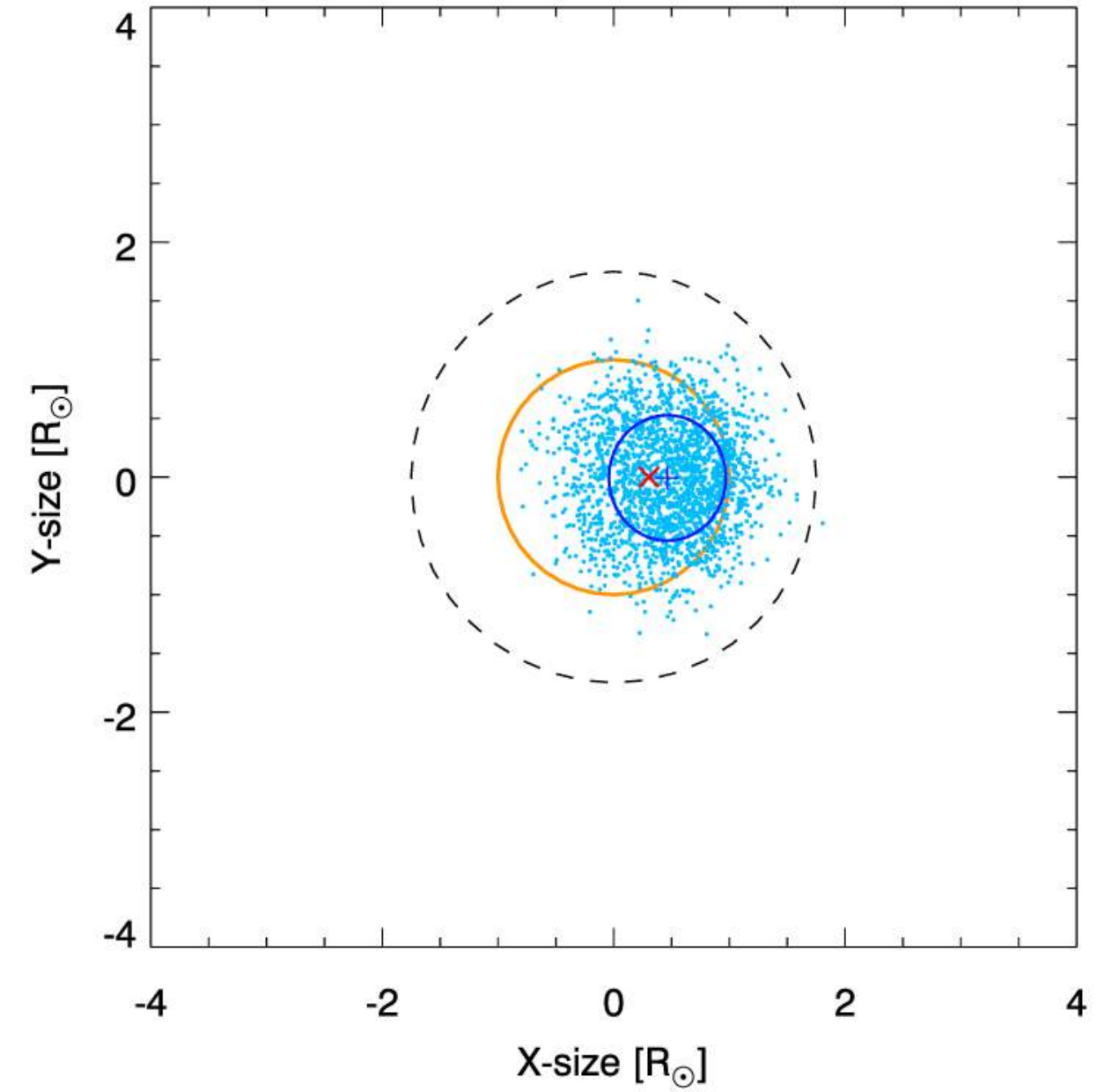}
	\includegraphics[width=0.3281\textwidth, keepaspectratio=true]{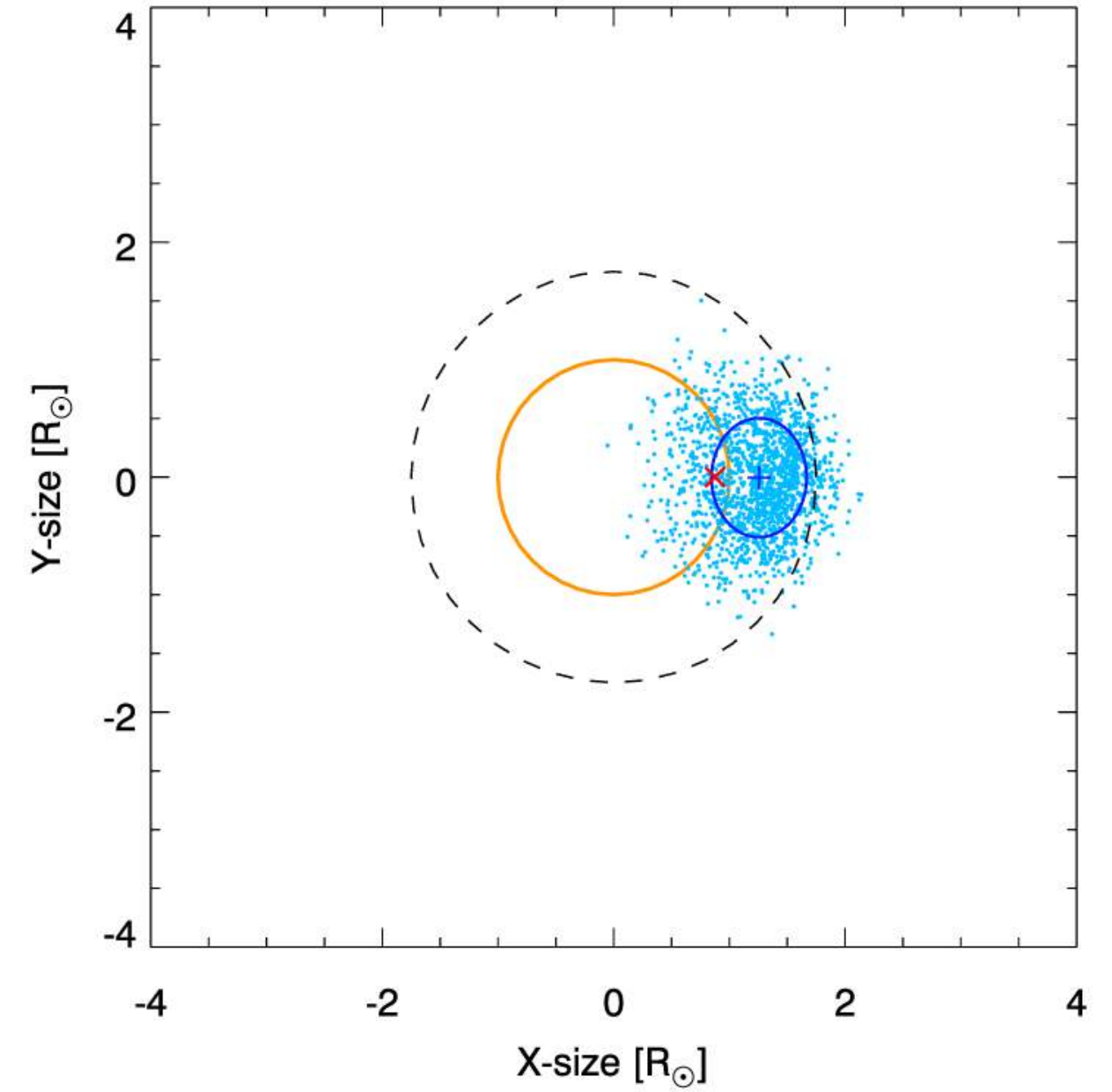}
    \caption[Simulated emission images for source-polar angles $\theta_s = 0\degr$, $10\degr$, and $30\degr$.]
    {Simulated radio images (in Sun-centred coordinates) for a point source located at $R_s = 1.75~\Rs$ (where $f_{pe} \approx 32$~MHz), assuming $\epsilon = 0.8$, $\alpha=0.3$, and for different source-polar angles: $\theta_s = 0\degr$ (left panel), $10\degr$ (middle panel), and $30\degr$ (right panel).  The photons are indicated by blue dots, the solar limb is illustrated by the orange curve, the intrinsic heliocentric source distance ($R_s = 1.75~\Rs$) is indicated with the black dashed line, and the source FWHM size is shown by the blue ellipse.  The red cross represents the source's projected intrinsic location, whereas the blue plus sign represents the projected imaged location.
    Figure taken from \cite{2019ApJ...884..122K}.
	}
    \label{fig:2019_source_vs_angle}
\end{figure}

The effects that the source-polar angle $\theta_s$ (see Figure~\ref{fig:prop_effects_cartoon}) has on the observed properties of radio sources are also investigated.  Figure~\ref{fig:2019_source_vs_angle} depicts the obtained source sizes for a source emitted at $R_s = 1.75$~$\Rs$ ($f_{pe}=32$~MHz), $\alpha=0.3$, and $\epsilon=0.8$ (as in Figure~\ref{fig:2019_sim_prof_source_dir_ans0.3}), but for three different angles: $\theta_s = 0\degr$, $10\degr$, and $30\degr$ (from left to right, respectively).  Similar to Figure~\ref{fig:2019_sim_prof_source_dir}, the red crosses represent the source's intrinsic position (used to define the source-polar angle), whereas the blue plus sign indicates the observed position and the blue ellipse represents the observed FWHM size of the source (as projected in the plane of the sky).

It is shown that when the source is located at the centre of the solar disk ($\theta_s = 0\degr$) the apparent source size is maximised.  As the source approaches the solar limb ($\theta = 90\degr$), its size significantly decreases and appears as more elliptical, as expected, given that scattered radio sources are not perfect spheres.  As evident, the source's observed size can vary significantly depending on the source-polar angle.  This implies that imaging observations may depict radio sources that have a relatively small area, simply due to a large source-polar angle (i.e. a projection effect).  Therefore, it is important to take this effect into account---especially for sources that are observed near or beyond the solar limb---as it can (erroneously) lead to an underestimation of the scattering effects and the notion that scattering plays an insignificant role in the determination of the event's observed properties.

\begin{figure}[htp!]
	\centering

	\captionsetup[subfigure]{aboveskip=0.5ex, belowskip=-0em, singlelinecheck=off}
	
\begin{subfigure}[t]{1.0\linewidth}
    \caption{~$\alpha$=0.5 \vspace{1ex}}
    \includegraphics[width=0.3281\textwidth, keepaspectratio=true]{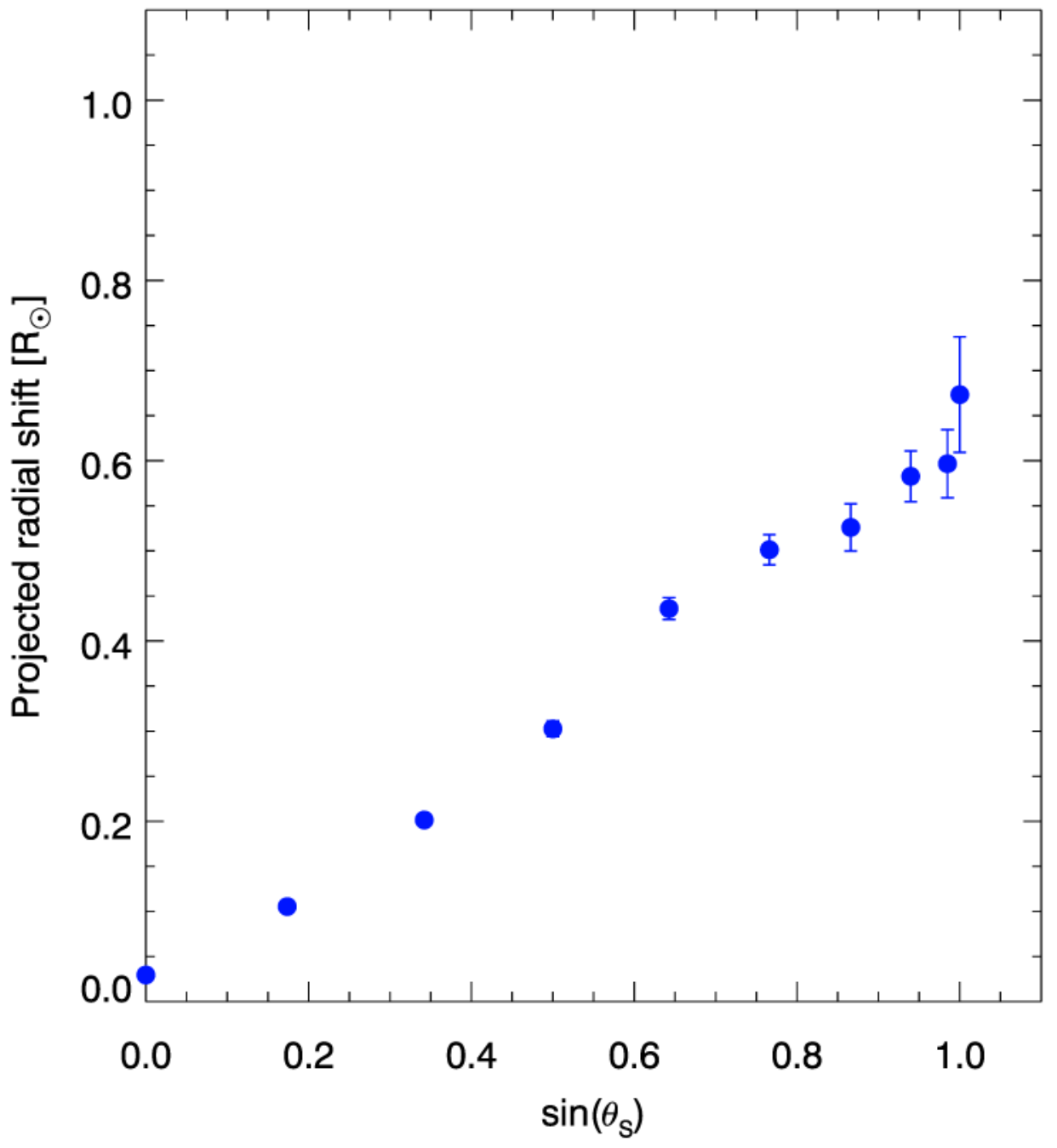}
	\includegraphics[width=0.3281\textwidth, keepaspectratio=true]{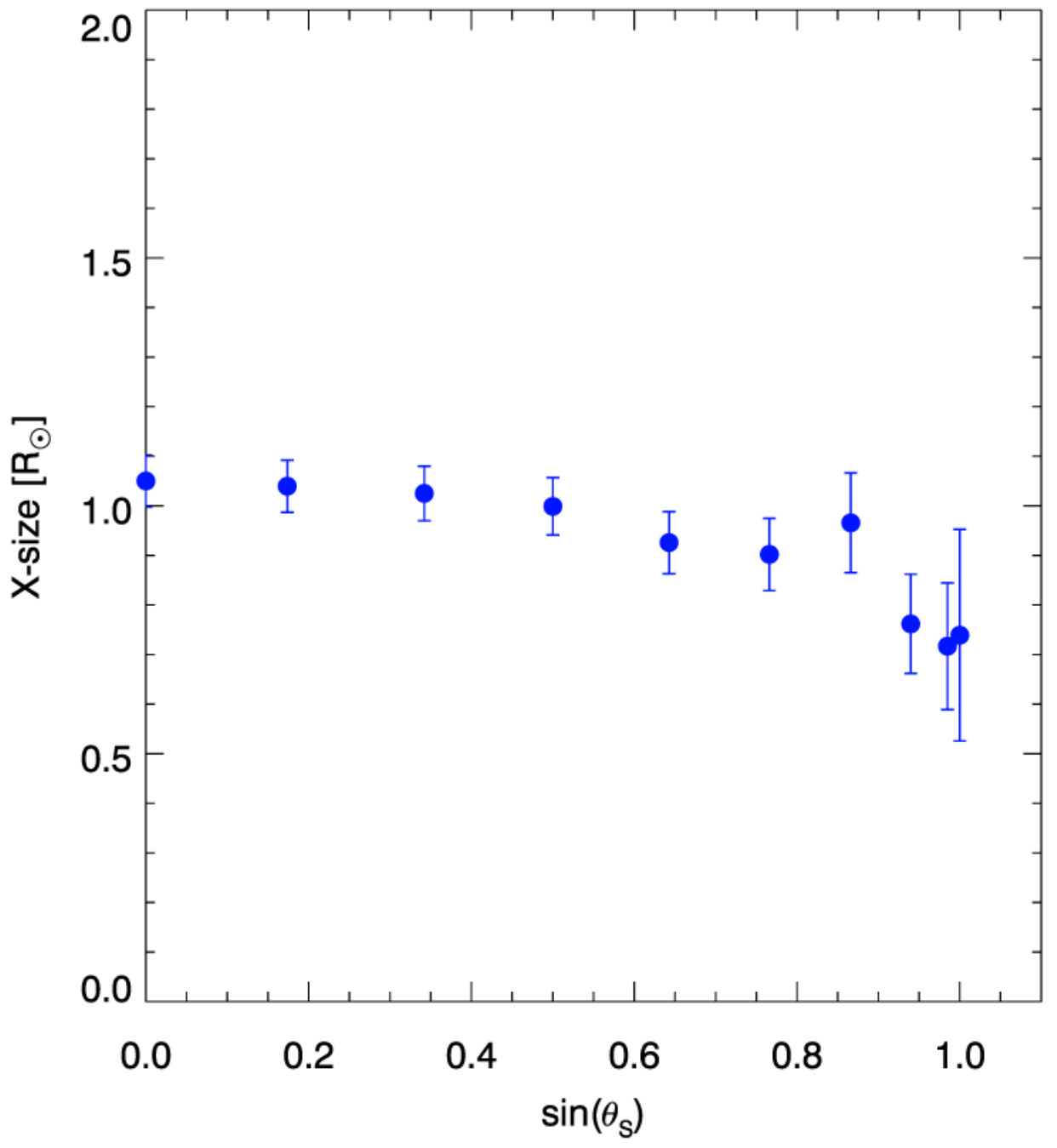}
	\includegraphics[width=0.3281\textwidth, keepaspectratio=true]{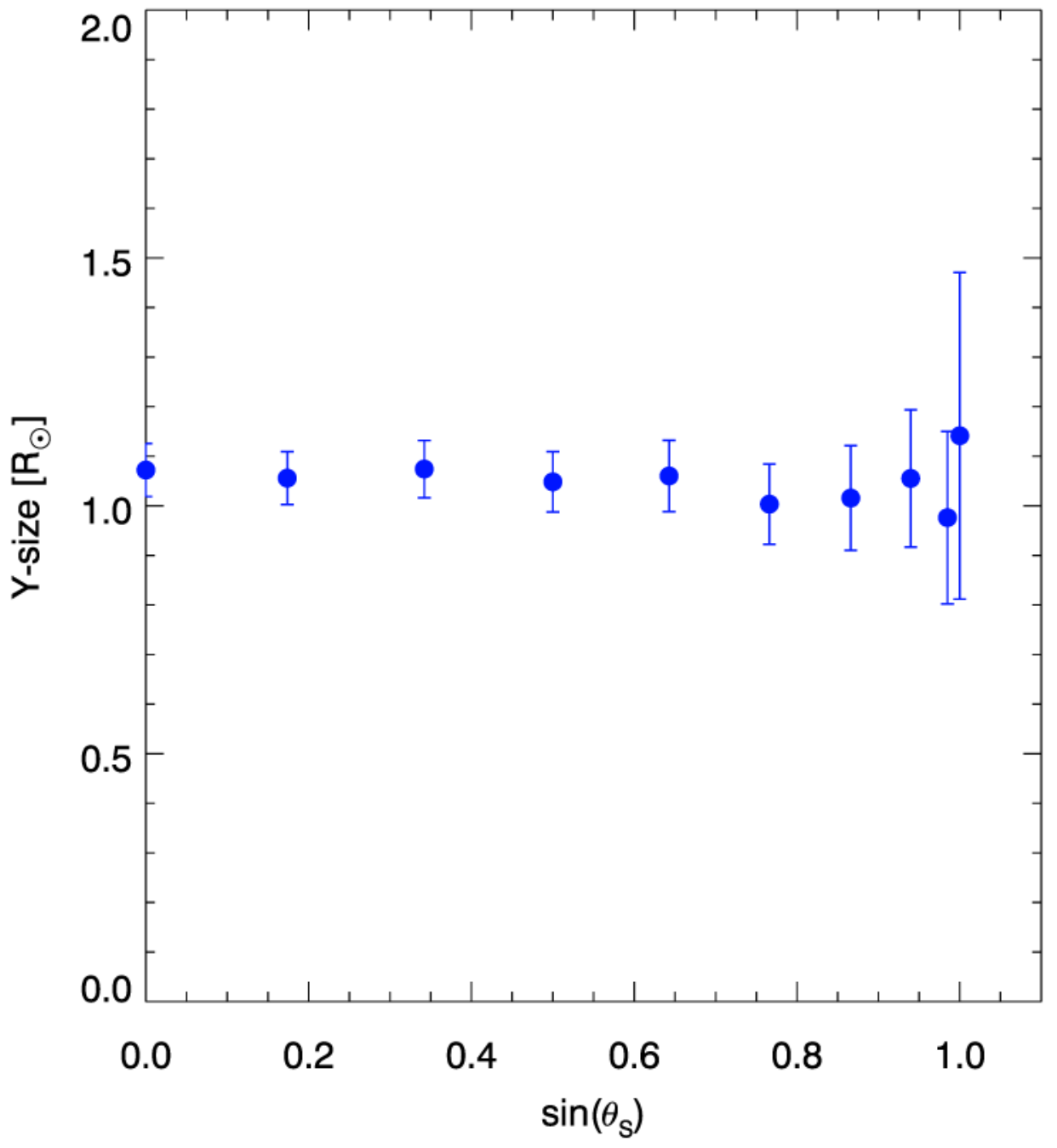}
	\label{fig:2019_sim_angular_dependence_ans0.5}
\end{subfigure}

\begin{subfigure}[t]{1.0\linewidth}
    \caption{~$\alpha$=0.3 \vspace{1ex}}
    \includegraphics[width=0.3281\textwidth, keepaspectratio=true]{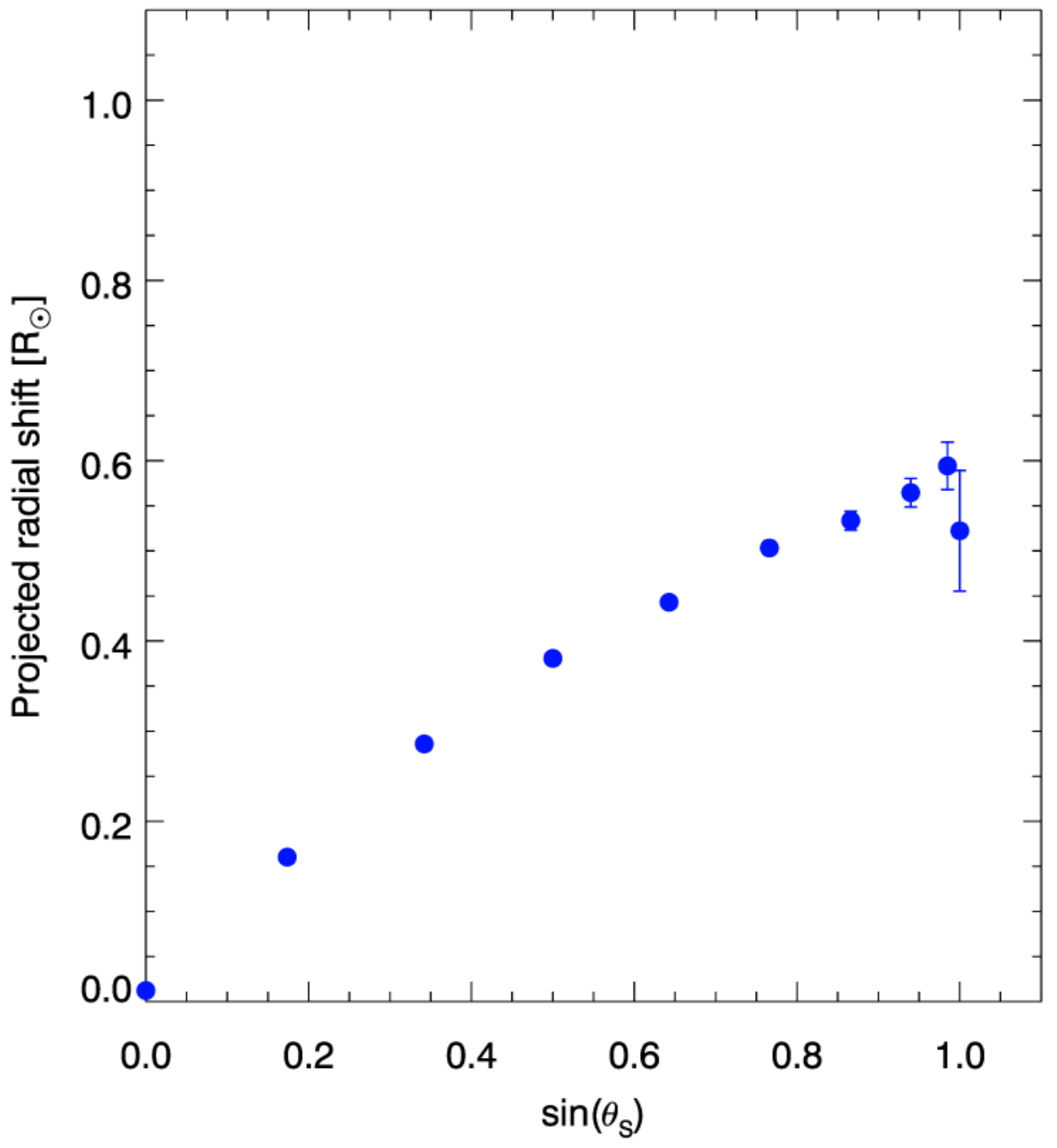}
	\includegraphics[width=0.3281\textwidth, keepaspectratio=true]{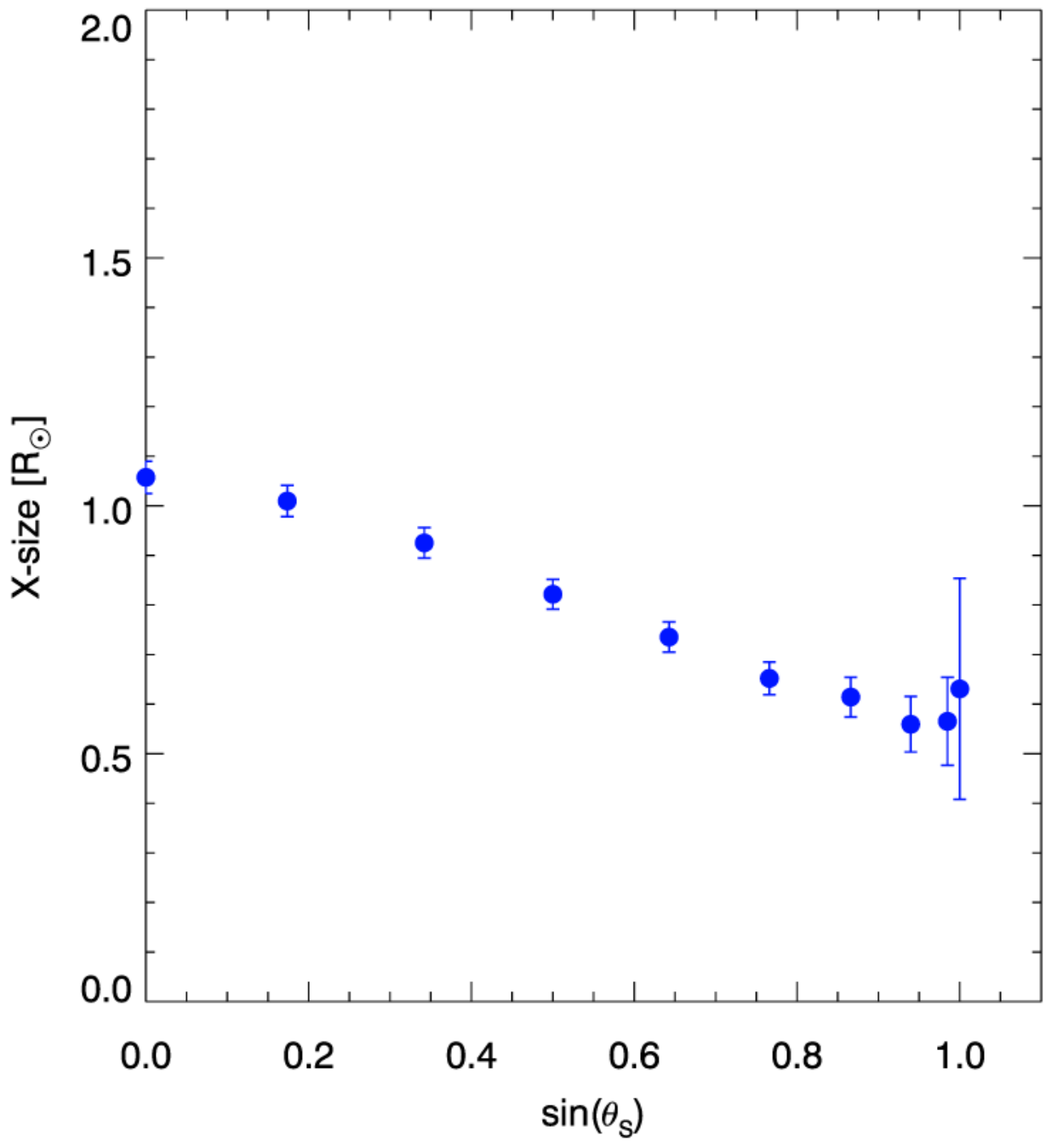}
	\includegraphics[width=0.3281\textwidth, keepaspectratio=true]{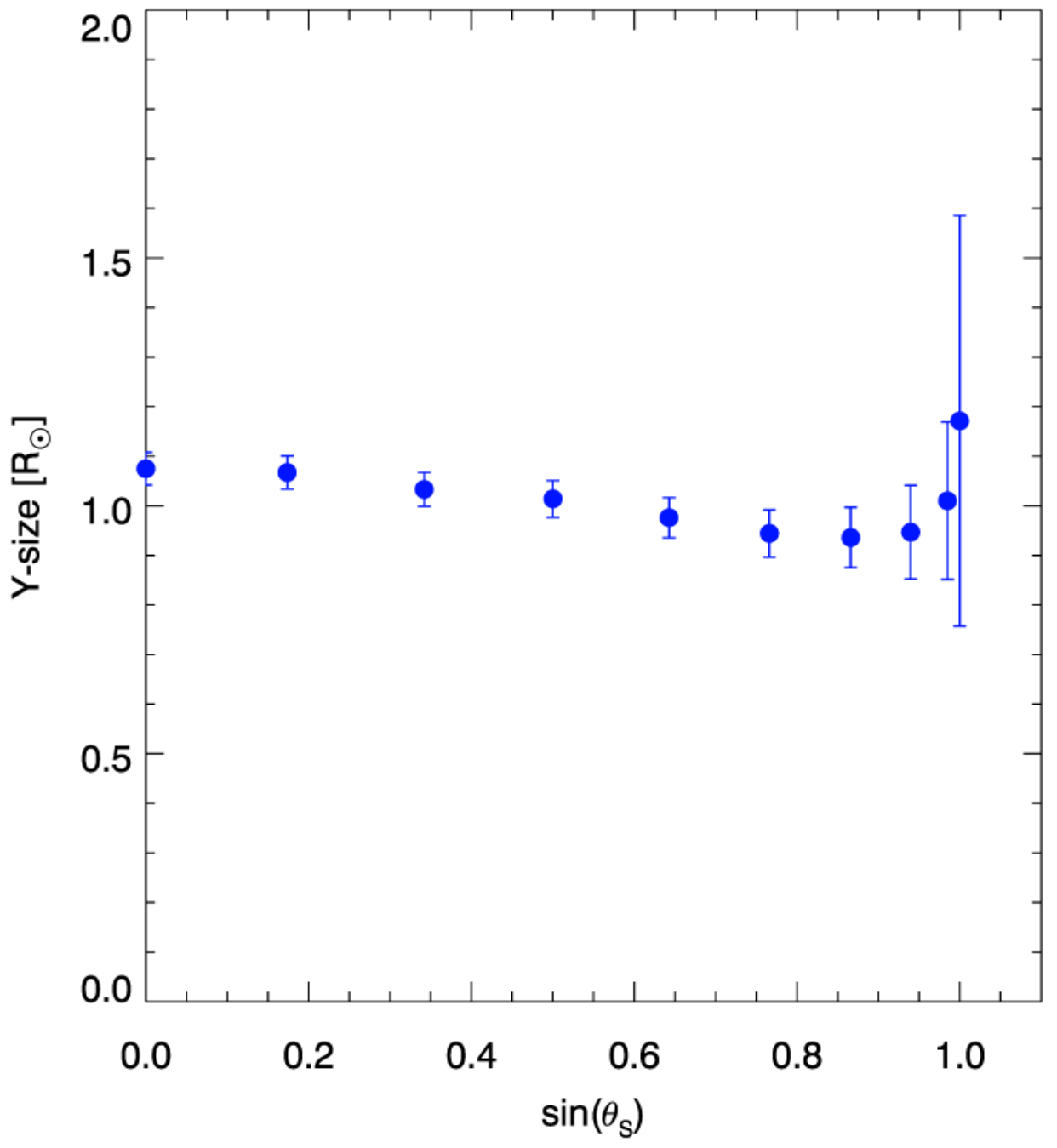}
    \label{fig:2019_sim_angular_dependence_ans0.3}
\end{subfigure}

	\caption[Simulated radial shift, $x$-size, and $y$-size as a function of $\sin (\theta_s)$, assuming $\epsilon = 0.8$ and anisotropies $\alpha = 0.3$ and $0.5$.]
    {Simulated source properties as a function of $\sin (\theta_s)$ for $f_{pe} \approx 32$~MHz, $\epsilon = 0.8$, and $\alpha = 0.5$ (panel (a)) and $0.3$ (panel (b)).  The left panels show the projected radial shift of the observed source centroid position ($\bar{x}$; Equation~(\ref{eqn:2019_sim_distr_centroids})) from its intrinsic position.  The middle and right panels show the source's FWHM $x$-size and $y$-size, respectively, as obtained through Equation~(\ref{eqn:2019_sim_distr_centroid_errors}).  The errors in each panel were calculated using the one-standard-deviation uncertainty obtained from Equations~(\ref{eqn:2019_sim_distr_centroid_errors}) and (\ref{eqn:2019_sim_distr_size_errors}) for the centroids and sizes, respectively.  The error bars become larger with increasing angle (as $\theta_s$ approaches $90\degr$) since the number of photons in the $z$-direction (i.e. towards the observer) is decreasing.
    Figures taken from \cite{2019ApJ...884..122K}.
	}

	\label{fig:2019_sim_angular_dependence}
\end{figure}

Figure~\ref{fig:2019_sim_angular_dependence} illustrates the dependence of the observed source properties for a larger range of source-polar angles ($\theta_s = 0\degr$--90$\degr$).  The source's radial shift from its true location (as projected in the plane of the sky; left panels), the FWHM $x$-size of the source (middle panels), as well as the FWHM $y$-size (right panels) are plotted as a function of $\sin \theta_s$.  Figure~\ref{fig:2019_sim_angular_dependence_ans0.5} was produced assuming an anisotropy $\alpha=0.5$, whereas Figure~\ref{fig:2019_sim_angular_dependence_ans0.3} was produced assuming $\alpha=0.3$.

It is evident that the heliocentric separation between the source's true position and its observed position (red cross and blue plus sign in Figure~\ref{fig:2019_source_vs_angle}, respectively) has a near-linear dependence on $\sin \theta_s$.  As the source approaches the limb and the source-polar angle $\theta_s$ gets closer to $90\degr$ ($\sin \theta_s = 1$), the observed radial separation increases (see Section~\ref{sec:proj_effs_impact}).  Sources that are located away from the disk centre are shifted radially along the $x$-direction in the simulations, therefore, the shift projected on the plane of the sky appears proportional to $\sin \theta_S$.  As mentioned in Section~\ref{sec:sim_TypeIIIb_properties}, when $\theta_s = 0\degr$, the true and apparent source positions coincide in the ($x$, $y$) plane of the sky at the solar centre (i.e. no shift is observed).

Since the sources are shifted radially along the $x$-direction, projection effects also act along the radial $x$-direction (cf. Figure~\ref{fig:2019_source_vs_angle}).  As can be seen by the middle and right panels of Figure~\ref{fig:2019_sim_angular_dependence} ($x$-size and $y$-size, respectively), the $x$-size of the source decreases as $\theta_s$ approaches $90\degr$, whereas the $y$-size remains nearly constant (varying between 1--1.2~$\Rs$).  It is also illustrated that the level of anisotropy makes little difference to the source size when the source is located near the solar centre.  However, as the source approaches the limb ($\theta_s = 90\degr$), its $x$-size decreases less for the reduced anisotropic case ($\alpha=0.5$, Figure~\ref{fig:2019_sim_angular_dependence_ans0.5}) compared to the more rapid and higher degree of decrease observed for stronger anisotropy levels ($\alpha=0.3$, Figure~\ref{fig:2019_sim_angular_dependence_ans0.3}).  This behaviour is consistent with the reported angular broadening of galactic radio sources (like the Crab Nebula) when observed through the corona, which are elongated along the tangential direction to the solar limb (e.g., \cite{1958MNRAS.118..534H, 1972PASAu...2...86D}).  It is also worth noting that the uncertainty in the simulated source properties becomes larger with increasing source-polar angles (as $\theta_s$ approaches $90\degr$, or $\sin \theta_s \rightarrow 1$), since the number of photons in the $z$-direction (i.e. towards the observer) is decreasing.

\section{Discussion and Conclusions} \label{sec:simulation_conclusions}

The need to assess the impact of scattering on solar radio observations was highlighted by \cite{2017NatCo...8.1515K} who---using a high-resolution Type IIIb LOFAR observation---provided strong evidence that scattering (from small-scale density fluctuations) is the dominant radio-wave propagation effect.  Following observations suggesting that density fluctuations in the solar corona are anisotropic, 3D ray-tracing simulations that account for anisotropy were developed in order to examine the impact of radio-wave propagation effects on the observed properties.  The simulations consider
\begin{enumerate*}[label=(\roman*)]
	\item scattering on small-scale density inhomogeneities,
	\item large-scale refraction due to the gradual variation of the ambient coronal density, and
	\item collisional (free-free) absorption.
\end{enumerate*}
The variables affecting the simulation outputs are the level of density fluctuations $\epsilon$, the level of anisotropy $\alpha$, and the source-polar angle $\theta_s$.

In order to test whether an anisotropic scattering description is indeed required, a collection of observational data from several studies and across a large range of frequencies (covering a large distance from the Sun until the Earth) was compared to the simulation outputs.  The objective was to investigate whether the isotropic ($\alpha=1$) scattering simulations could simultaneously describe several observed properties.  Specifically, a collection of Type III source sizes and decay times were considered.  It was found that when the decay times are successfully described, the simulated sizes are smaller than observed.  Increasing the strength of scattering (i.e. value of $\epsilon$) to produce larger source sizes also increased the decay times.  Thus, the source sizes and decay times could not be simultaneously reproduced within the framework of isotropic density fluctuations, affirming the need to consider a degree of anisotropy.

Anisotropic scattering simulations were therefore used to reproduce the observed properties of Type IIIb sources near $\sim$35~MHz, given their recent, detailed examination in several observational studies, where high resolution and high sensitivity LOFAR data was utilised.  The simulated time profiles, emission images, and directivity patterns were compared to the observed ones, suggesting that density fluctuations $\epsilon=0.8$ and an anisotropy $\alpha=0.3$ are required to reproduce the observed characteristics.  It should be emphasised, though, that these values are only valid for the adopted model of outer scales $l_o$ (Equation~(\ref{eqn:outer_scale})) and the specific observation presented here.  They do not represent the universal properties of the (highly-variable) corona at these frequencies.

The impact of varying the level of density fluctuations, level of anisotropy, and source-polar angle was investigated.  All anisotropy values considered ($\alpha < 1$) correspond to scattering that is stronger in the perpendicular (to the radial) direction compared to the parallel one, consistent with the observed elongation of radio sources along the perpendicular direction.

\begin{figure}[h!p]
    \centering
	\includegraphics[width=0.75\textwidth, keepaspectratio=true]{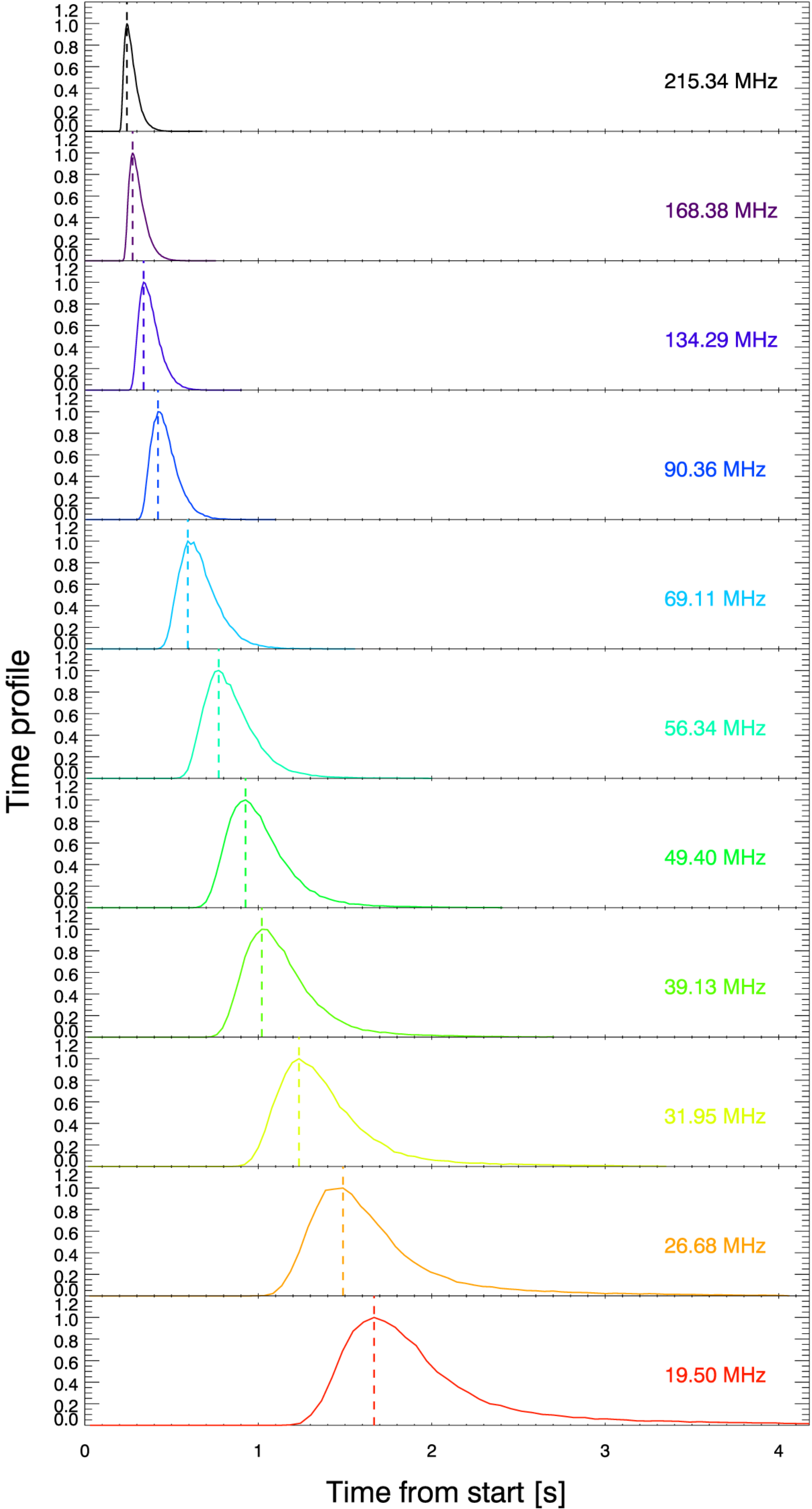}
    \caption[Simulated time profiles for cascading frequencies.]
    {Simulated time profiles (normalised with respect to the peak flux) for different frequencies ranging from $\sim$20--215~MHz, as indicated by the legends.  Dashed lines annotate the peak-flux times.
	}
    \label{fig:drift_rate_time_profs_vs_freq}
\end{figure}

It was demonstrated that the apparent source sizes increase with increasing values of $\epsilon$, whereas the level of anisotropy has a negligible impact.  For a given anisotropy, a larger value of $\epsilon$ produced a longer decay time.  In addition to that, it was shown that due to projection effects, sources which are observed closer to the solar limb ($\theta_s \rightarrow 90\degr$) will display a smaller area compared to when they are near the solar centre ($\theta_s=0\degr$), given that their $x$-size decreases with increasing angle, whilst the $y$-size remains fairly constant.  The projected radial shift of the sources also increases with increasing angle.  Crucially, it was found that anisotropies less than 1 resulted in very directional emissions (along the radial direction), even though the intrinsic source was taken to emit isotropically.  This result addresses some of the previous arguments against scattering from small-scale density inhomogeneities, which were based on the fact that \textit{isotropic} scattering can provide large observed sizes but also results in emissions that are not directional, inconsistent with observations.  It was also found that the level of anisotropy can significantly affect the observed time profile of the radio pulse.  Weaker scattering in the radial direction (i.e. stronger anisotropy levels; $\alpha \rightarrow 0$) corresponds to a reduced level of radio-wave cloud broadening and thus a reduced time-broadening effect.  In other words, strong anisotropies produce time profiles that are characterised by a shorter duration and a shorter decay time than those for weaker anisotropies.  Moreover, the time profiles for stronger anisotropies are less absorbed than their counterparts.  This is due to the fact that photons which scatter more, stay longer in the collisional coronal medium and are thus absorbed more, meaning that fewer photons reach the observer (compared to when weaker radial scattering is at play).

The impact of anisotropy on the observed time profiles is particularly relevant when it comes to understanding the properties obtained from dynamic spectra, which are essentially stack plots of cascading-frequency light curves (cf. Figures~\ref{fig:dyn_spec_cartoon} and \ref{fig:drift_rate_time_profs_vs_freq}).  The frequency-drift rates $df/dt$ of radio bursts are estimated by tracking the peak-flux time of the emissions at consecutive frequencies across the dynamic spectrum.  Drift rates are often used to deduce the speed of the radio emissions' exciter (see Section~\ref{sec:f_vs_R_relation}), as well as other parameters (based on the exciter speed) that describe the local coronal environment (see, e.g., Section~\ref{sec:bandsplitting_models}).  Therefore, understanding whether radio-wave propagation effects disguise the intrinsic drift of the radio source---and the extent to which they might do so under certain conditions---is important.

Lower-frequency photons are more sensitive to scattering, meaning that their path will be altered to a greater extent and they will collectively arrive at the observer later than higher-frequency photons (i.e. a scattering-induced delay is introduced).  Figure~\ref{fig:drift_rate_time_profs_vs_freq} illustrates this delay on radio pulses observed at different frequencies, where scattering, refraction, absorption, and free-space photon propagation were considered in the simulations.  The dashed lines indicate the time at which the peak flux is observed, highlighting the increasing delay in (peak-flux) arrival times with decreasing frequency.  It can also be seen that the decay of the time profiles broadens with decreasing frequency, as expected (see Figure~\ref{fig:decay_times_fit}).

Therefore, the contributions of radio-wave propagation effects to the observed frequency-drift rate merits a detailed investigation.  Simulation outputs for different levels of density fluctuations $\epsilon$, anisotropy $\alpha$, as well as source-polar angles $\theta_s$ should be compared.  The aim is to decouple the intrinsic drift rate of the source from radio-wave and geometric effects, if and when they are found to have a significant impact.

In this chapter, the necessity to consider the anisotropy of density fluctuations in the corona has been demonstrated.  Moreover, the governing role of scattering on the observed radio emissions was confirmed through the simultaneous reproduction of several observed properties.  Consequently, the contribution of scattering on the observed emission features should be acknowledged and accounted for in analyses and interpretations of radio observations.

\cleardoublepage
\chapter{Reproducing Observed Sub-Second Radio Properties} \label{chap:observation_simulations}

\textit{This work was published in \cite{2020ApJ...905...43C} and \cite{2020ApJ...898...94K}.\\
The author of this thesis contributed to the publication by \cite{2020ApJ...905...43C} through discussions on the data analysis and writing parts of the text.
The author of this thesis also contributed to the publication by \cite{2020ApJ...898...94K} by producing all the figures showing results from the simulations, as well as writing parts of the text.
}

\section{Simulating the Temporal Evolution of Type IIIb Source Properties} \label{sec:typeIIIb_sim_setup}

The fascinating aspect of LOFAR observations is the ability to image the time dependence of radio emission properties with extremely high resolutions (see Section~\ref{sec:lofar}).  In Chapter~\ref{chap:scattering}, the observed properties of a Type IIIb burst observed with LOFAR were compared to anisotropic simulations, with the aim to understand the dependence of the observed radio properties on the level of density fluctuations and anisotropy in their local environment, but without probing the temporal evolution of the radio properties.

As mentioned in Chapter~\ref{chap:scattering}, the specific Type IIIb burst has been analysed several times.  Some of these studies examined the time-dependence of its observed properties at sub-second scales \citep{2017NatCo...8.1515K, 2018SoPh..293..115S}, for both its fundamental and harmonic branches.  The observation was conducted on 16 April 2015 with LOFAR's LBA antenna using the tied-array beam configuration which produced a temporal and spectral resolution of $\sim$0.01~s and $\sim$12.1~kHz, respectively, synthesised beams with a FWHM size (Equation~(\ref{eqn:lofar_res})) of $\sim$10$\arcmin$, and a centre-to-centre beam separation of $\sim$6$\arcmin$ at $\sim$32~MHz.  Each Type IIIb stria (whether fundamental or harmonic) lasts for $\sim$1~s and has a short (instantaneous) bandwidth $\Delta f_i$ of $\sim$100~kHz.  This implies that---given LOFAR's resolution---a statistically-sufficient number of data points can be analysed for each stria.

\cite{2017NatCo...8.1515K} examined the time dependence of the well-resolved source properties within the $\sim$32.5~MHz stria, for both the fundamental and harmonic emissions.
It is worth mentioning that stria observed at other (similar) frequencies were also examined and showed nearly-identical properties.  It is of interest to attempt a characterisation of the sub-second temporal evolution of the fundamental and harmonic properties using radio-wave propagation simulations, a comparison not previously performed.
Therefore, in this chapter, the newly-developed Monte Carlo ray-tracing simulations (presented in Chapter~\ref{chap:scattering}) are compared to the time profile, source positions, and source sizes of the $\sim$32.5~MHz fundamental and harmonic striae (as analysed by \cite{2017NatCo...8.1515K}).

For the purposes of this comparison, the model of electron density fluctuations (Equation~(\ref{eqn:simple_dens_fluc_model})) is expressed as
\begin{equation} \label{eqn:sim_Cq_defintion}
	\overline{q \epsilon^2} \simeq 4\pi l_o^{-2/3} l_i^{-1/3} \epsilon^2 = C_q \, r^{-0.88} \, ,
\end{equation}
where $C_q$ is a constant that characterises the level of density fluctuations in units of $1/\Rs$.  Since the adopted outer scale model $l_o$ was obtained empirically for distances from 7--80~$\Rs$ (see Section~\ref{sec:dens_fluct_spec}), it is reasonable to assume that $l_o$ is poorly known for the distances of interest ($< 3$~$\Rs$, covered by LOFAR).  Furthermore, $\epsilon$ is (by definition; see Equation~(\ref{eqn:2019_S(q)_normalised})) the integral over all wavenumbers $\vec{q}$, meaning that radio observations limited to certain frequencies (and thus wavenumbers) cannot be used to directly infer the value of $\epsilon$.  Therefore, $C_q$ will be used as a free parameter which will be estimated by comparing the simulations to observations, as done for the level of anisotropy $\alpha$ (Equation~(\ref{eqn:2019_anisotropy_def})) and the source-polar angle $\theta_s$ (see Chapter~\ref{chap:scattering}).  Higher levels of density fluctuations (i.e. larger $C_q$ values) correspond to stronger scattering, which means that sources experience a higher degree of angular broadening, and photons spend more time propagating through the coronal medium.  As a result, the duration of the observed emission pulse increases with increasing $C_q$ values, leading to a delay in the observed peak-flux time, as well as longer decay times (see Chapter~\ref{chap:scattering}).

Similar to Chapter~\ref{chap:scattering}, $\omega_F = 1.1 \, \omega_{pe}$ and $\omega_H = 2 \, \omega_{pe}$ define the emission frequencies of the fundamental and harmonic sources, respectively (see Section~\ref{sec:plasma_emmission}).  According to the density model used (Equation~(\ref{eqn:2019_sim_dens_model})), the intrinsic location (denoted as $R_s$ in Figure~\ref{fig:prop_effects_cartoon}) of a fundamental source emitting at $\sim$32.5~MHz is found at a heliocentric distance $r_F \approx 1.8$~$\Rs$, whilst a harmonic source emitting at $\sim$32.5~MHz is found at $r_H \approx 2.2$~$\Rs$.  The simulations assume an intrinsic point source and an instantaneous injection of photons into the heliosphere (see Section~\ref{sec:anisotropic_simulations_2019}).  A total of $2 \times 10^5$ photons is used for every simulation run.  The simulated properties after the scattering screen alter by $\le 1\%$ (see Section~\ref{sec:anisotropic_simulations_2019}), allowing for the characterisation of the properties with a precision of $\le 1\%$.
Therefore, for a typical Type III burst source size of 20$\arcmin$ at $\sim$32~MHz \citep{2017NatCo...8.1515K}, the change in size past the scattering screen is $\le 0.2\arcmin$.  Consequently, the greatest source of uncertainty in the simulated properties stems from the finite number of photons used (i.e. the statistical error).  The FWHM size and position of the sources---and any associated uncertainties---are obtained by fitting the simulated radio images with a 2D elliptical Gaussian function, as done to the LOFAR images (see Section~\ref{sec:centroid_calc} and, e.g., \cite{2017NatCo...8.1515K}).
In analyses of radio observations, the temporal evolution of sources at a given frequency is obtained through measurements during the decay phase of the burst (see, e.g., \cite{2017NatCo...8.1515K}), which characterises any emissions occurring after the observed peak-flux time at a given frequency.
Therefore, the same approach is adopted in the work presented here.  The simulated time profiles are fit with an exponential function.  The decay time is then obtained as the HWHM of the fitted profile (i.e. the duration from the peak-flux time until the flux reaches half its maximum value), the one standard deviation of which is used as the uncertainty in the decay time.

\begin{figure}[ht!]
    \centering
	\includegraphics[width=1.0\textwidth, keepaspectratio=true]{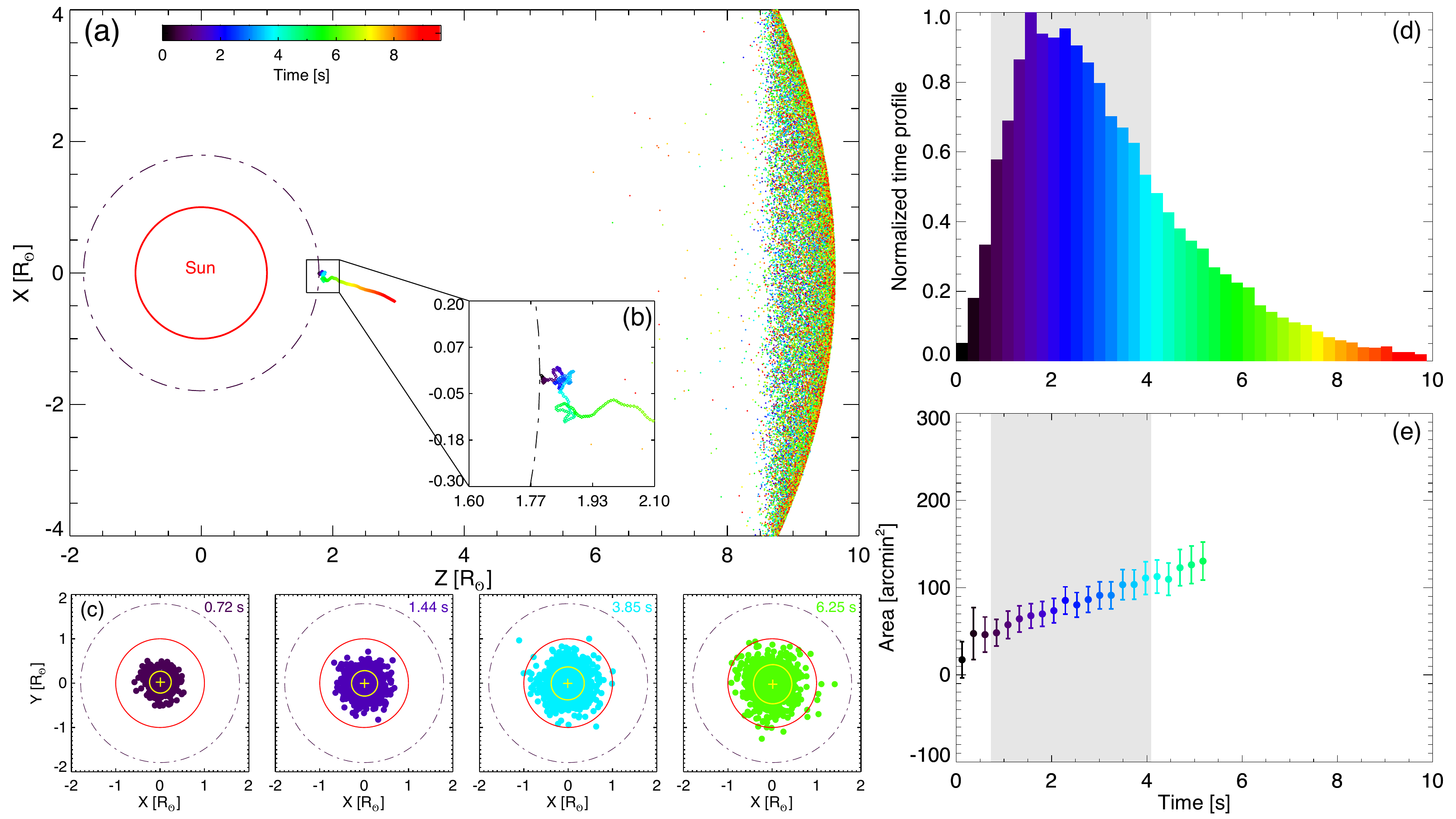}
    \caption[Illustration of the simulation outputs.]
    {Illustration of the ray-tracing simulation outputs assuming fundamental emissions at 32.5~MHz ($R_s = 1.8$~$\Rs$), isotropic scattering ($\alpha=1$), $C_q = 80 \, \mathrm{R_\sun^{-1}}$, and $\theta_s = 0\degr$, where a Sun-centred Cartesian coordinate system is used and the $z$-axis points towards the observer.  The colours represent the different arrival times of photons on the scattering screen, from which the free-space propagation time has been subtracted.
    (a)~Photon locations once they arrive at the scattering screen (found at 9.6~$\Rs$ for the given parameters).
    (b)~The path of a ray illustrating the strong scattering experienced near the source.  The black dashed line depicts the intrinsic emission location $R_s = 1.8$~$\Rs$.
    (c)~Snapshots demonstrating the evolution of the apparent source size with time, with respect to the solar limb (red curve).  Yellow circles and plus signs depict the source's FWHM size and observed centroid location, respectively.
    (d)~Photon flux at the scattering screen as a function of time (i.e. the source's time profile), normalised with respect to the peak flux.  The grey-shaded area indicates the FWHM duration of the observed pulse.
    (e)~Simulated FWHM area of the source and its associated one-standard-deviation uncertainty as a function of time.
    Figure taken from \cite{2020ApJ...905...43C}.
	}
    \label{fig:Chen_sim_outputs_demonstration}
\end{figure}

For the purposes of evaluating the need to consider anisotropic scattering, as concluded in Chapter~\ref{chap:scattering}, the simulations are run assuming both isotropic ($\alpha = 1$) and anisotropic ($\alpha \neq 1$) density fluctuations (as defined in Equation~(\ref{eqn:2019_anisotropy_def})).  Figure~\ref{fig:Chen_sim_outputs_demonstration} illustrates the simulation set-up and the obtained simulated source properties.  For this figure, it was assumed that a fundamental source located at $R_s = 1.8$~$\Rs$ (i.e. $\omega_F \approx 32.5$~MHz) and at a source-polar angle $\theta_s = 0\degr$ emits into a corona characterised by isotropic ($\alpha = 1$) density fluctuations of strength $C_q = 80 \, \mathrm{R_\sun^{-1}}$.
Panel \hyperref[fig:Chen_sim_outputs_demonstration]{(a)} depicts the location of the photons once they reach the scattering screen, which is found (in this case) at a heliocentric distance 9.6~$\Rs$.  The different colours represent the range of times that photons took to reach the scattering screen from the moment of emission, with photons shown in black being the fastest and those in red the slowest, as indicated by the colour bar.
The photon time-of-flight through free space was subtracted from the depicted times ($t$), such that
\begin{equation*}
	t = t_{screen} - \dfrac{(r_{screen} - R_s)}{c} \, ,
\end{equation*}
where $t_{screen}$ is the total time of travel from the intrinsic location $R_s$ until the location of the scattering screen $r_{screen} = 9.6$~$\Rs$, and $c$ is the speed of light.  Therefore, if photons were propagating through free-space, they would all be depicted as arriving at $t=0$~s.
  Panel \hyperref[fig:Chen_sim_outputs_demonstration]{(b)} is an inset depicting the path of a randomly-chosen photon ray, emphasising the strong scattering experienced near the emission source where the photon frequency $\omega$ is close to local plasma frequency $\omega_{pe}$.  At larger distances where $\omega \gg \omega_{pe}$, the scattering rate decreases and refraction off of large-scale density inhomogeneities becomes more significant, resulting in some ``focusing'' of the radio waves.  Panel \hyperref[fig:Chen_sim_outputs_demonstration]{(c)} shows snapshots of the source images obtained at different times during the arrival of the photons at the scattering screen, demonstrating the dynamics of the radio source.  The dots represent the photons (colour-coded to reflect their arrival times), whereas the yellow circles indicate the FWHM size of the source at the specific moment in time.  Given that isotropic scattering and a source-polar angle $\theta_s = 0\degr$ were assumed, the FWHM $x$-size of the source equals its y-size.  The yellow plus signs illustrate the apparent centroid positions, which overlap with the solar centre in the ($x$, $y$) plane of the sky when $\theta_s = 0\degr$.
Panel \hyperref[fig:Chen_sim_outputs_demonstration]{(d)} depicts the simulated time profile, and panel \hyperref[fig:Chen_sim_outputs_demonstration]{(e)} shows the FWHM area of the apparent source as a function of time, where the error bars represent a one-standard-deviation uncertainty.  The grey-shaded area in panels \hyperref[fig:Chen_sim_outputs_demonstration]{(d)} and \hyperref[fig:Chen_sim_outputs_demonstration]{(e)} represents the FWHM duration of the observed time profile.
Figure~\ref{fig:Chen_sim_outputs_demonstration} illustrates that for a $\sim$32.5~MHz source emitting in a medium characterised by $C_q = 80 \, \mathrm{R_\sun^{-1}}$ and isotropic density fluctuations, the peak of the time profile is delayed by 2.5~s, while the instantaneous injection of photons leads to a pulse with a FWHM duration of $\sim$3.5~s.

\subsection{Isotropic scattering simulations of fundamental Type IIIb emissions} \label{sec:typeIIIb_isotropic_F_sims}
Using the simulation set-up introduced in Section~\ref{sec:typeIIIb_sim_setup}, the time profile and temporal evolution of the area and position of a $\sim$32.5~MHz source is investigated.  In this section, the simulations are conducted assuming fundamental emissions, isotropic density fluctuations ($\alpha=1$), $C_q = 80 \, \mathrm{R_\sun^{-1}}$, and angles $\theta_s$ ranging from 0 to 8$\degr$.

LOFAR images of the $\sim$32.5~MHz sources suggest small source-polar angles, since the centroids are observed closer to the solar centre than the limb (cf. Figure~\ref{fig:kontar2017_source_img} and Section~\ref{sec:angular_dependence}).  The polar angle of the Type IIIb sources can also be roughly estimated from the LOFAR images using the analytical estimation for scattering-induced shifts derived by \cite{2018ApJ...868...79C} (detailed in Section~\ref{sec:TypeII_scattering}).  A $\sim$32~MHz source is expected to shift away from its true location by approximately 0.6~$\Rs$ due to scattering, meaning that the Type IIIb fundamental source emitted at $\sim$1.8~$\Rs$ is expected to shift to a heliocentric distance of $\sim$2.4~$\Rs$.  The centroid of the fundamental Type IIIb source was observed (in the plane of the sky) at coordinates (250, 370) with respect to the solar centre (given in arcseconds; see Figure~\ref{fig:kontar2017_source_img}), where $1 \, \Rs \simeq 960$~arcsec.  Therefore, the polar angle of the Type IIIb source can be approximated as $\theta_s = \sin^{-1} (250/(2.2 \times 960)) \approx 6.2\degr$, justifying the limited range of small angles used for the simulations.

\begin{figure}[htp!]
    \centering
    \vspace{-2em}
	\includegraphics[width=0.55\textwidth, keepaspectratio=true]{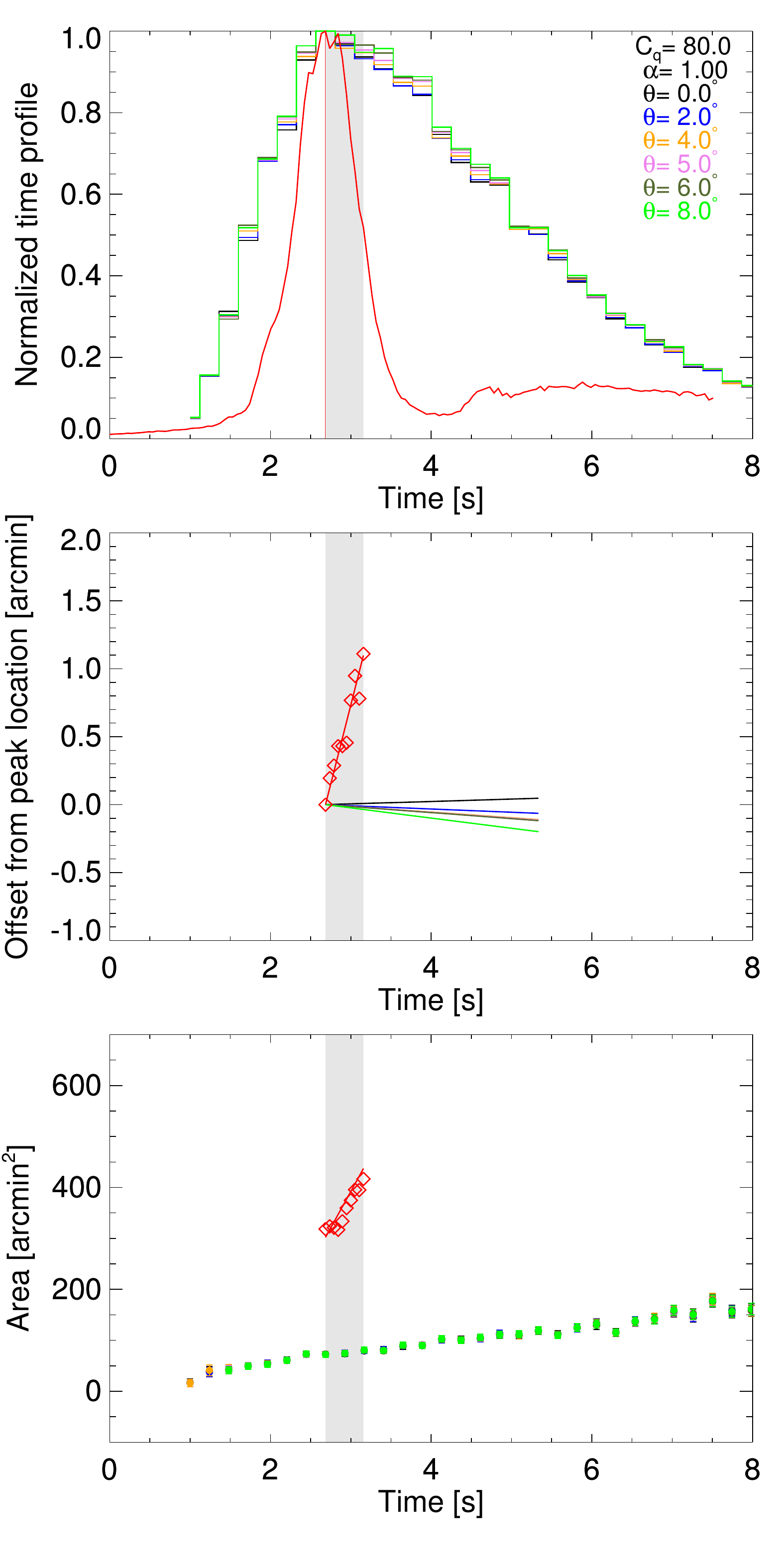}
	\vspace{-1.5em}
    \caption[Fundamental emission properties assuming isotropic scattering.]
    {
    Temporal evolution of the simulated source properties where fundamental emissions at 32.5~MHz, isotropic scattering ($\alpha=1$), and $C_q = 80 \, \mathrm{R_\sun^{-1}}$ were assumed, but the angle was varied from $\theta_s =$~0--8$\degr$.  Red data represents the observed source properties recorded by LOFAR, whereas simulated properties are colour-coded for the different angles as indicated by the legend in the top panel.  The top panel illustrates the (normalised) time profile, the middle panel illustrates the change in the source's heliocentric location from its position at the peak-flux time, and the bottom panel shows the source's (FWHM) area.  Grey-shaded areas indicate the decay time of the burst observed by LOFAR.
    Figure taken from \cite{2020ApJ...905...43C}.
	}
    \label{fig:Chen_iso_F_sim}
\end{figure}

Figure~\ref{fig:Chen_iso_F_sim} demonstrates the simulation outputs---given $C_q = 80 \, \mathrm{R_\sun^{-1}}$ and angles $\theta_s$ ranging from 0 to 8$\degr$---alongside the values obtained from the LOFAR observations (depicted in red).  The top panel shows the time profiles (normalised with respect to the peak-flux value), the middle panel shows the offset of the source location from its location during the peak-flux time, and the bottom panel shows the source area as a function of time.  The peak of the time profile observed by LOFAR was aligned with the peak of the simulated time profiles for comparison.  The grey-shaded regions in each panel illustrate the observed decay time of the $\sim$32.5~MHz Type IIIb stria, as obtained from the LOFAR observations \citep{2017NatCo...8.1515K}.  The simulated properties (in all panels) are colour-coded for the range of angles used, as indicated by the legend in the top panel.

It should be emphasised that both the observed and simulated source location and area demonstrate a change with time (as discussed throughout this chapter).  In other words, the source exhibits an areal expansion and a centroid location displacement.  This occurs despite the fact that a single source is simulated, which emits at a fixed frequency from a fixed intrinsic location (such that the location of the scattering screen for the specific source is also fixed).  The reason for the source motion and areal expansion (of both the LOFAR observation and simulations) is that the sub-second properties are probed, instead of a single ``snapshot''.  This implies that the observations and simulations represent the arrival of photons at the detector at sub-second intervals.  Due to geometric and radio-wave propagation effects, photons reach the detector at different times, meaning that the source's properties will evolve as photons gradually arrive at the detector.  Witnessing this temporal evolution of source properties would not be possible without the high temporal resolution of LOFAR's imaging observations, or the ability to produce images from the simulations.

The simulated pulse---produced by the instantaneously-emitting point source---has a broad FWHM duration of $\sim$3.5~s (for all considered angles), which is significantly longer than the FWHM of the observed pulse (found to be $\sim$1.0~s), as seen in the top panel of Figure~\ref{fig:Chen_iso_F_sim}.  Consequently, the simulated decay time is also longer than the observed one, being $\sim$2.5~s long instead of $\sim$0.5~s.

The simulated apparent source motion in the plane of the sky also disagrees with the observed motion, as illustrated in the middle panel of Figure~\ref{fig:Chen_iso_F_sim}.  During the decay time of the burst, the observed source was found to move by $\sim$65~arcsec in $\sim$0.5~s, whereas the simulated source moves by less than 5~arcsec during the same time period of $\sim$0.5~s, whilst only moving by $\sim$12~arcsec during the $\sim$2.5~s comprising the entirety of the simulated decay time.  In other words, the simulated source positions in the case of isotropic scattering do not change as much as the observed source positions (shown by the red data points).

Similar to the source positions, the simulated source areas also do not match the observed values or their rate of change.  The observed areas vary from $\sim$300--440~$\mathrm{arcmin^{2}}$ in the time period of $\sim$0.5~s defining the decay time of the observed fundamental emissions.  The simulated areas, however, range from $\sim$60--100~$\mathrm{arcmin^{2}}$ during the entire $\sim$2.5~s of the simulated decay time.

Even though the simulated decay time is considerably longer than that observed, the simulated apparent source sizes are smaller than the observed ones by a factor of $\sim$4.  It is worth noting that the observed source areas were deconvolved for the FWHM area of the LOFAR beams, which is $\sim$110~$\mathrm{arcmin^{2}}$ at $\sim$32.5~MHz (see Equations~(\ref{eqn:lofar_fov}) and (\ref{eqn:obs_source_contributions}); \cite{2017NatCo...8.1515K}).  Echoing the results presented in Section~\ref{sec:isotropic_vs_anisotropic_scatt_sim}, the simulated decay time is longer than the observed while the simulated sources are too small, meaning that no matter how weak or strong the scattering is set to be (i.e. what value of $C_q$ is chosen), the simulations will never simultaneously match the observations.  Stronger scattering will produce both larger sizes and larger decay times, and vice versa.  It is thus evident that the isotropic scattering assumption does not suffice in describing the observed source properties, as determined in Section~\ref{sec:isotropic_vs_anisotropic_scatt_sim}.  As such, it is necessary to consider anisotropic density fluctuations ($\alpha \neq 1$) when simulating radio-wave propagation effects in the solar corona.

\subsection{Anisotropic scattering simulations of fundamental Type IIIb emissions} \label{sec:typeIIIb_anisotropic_F_sims}
The need to consider anisotropic scattering in simulations of radio-wave propagation effects was re-evaluated using the temporal evolution of the observed properties of a single burst (unlike the method employed in Section~\ref{sec:isotropic_vs_anisotropic_scatt_sim}).  Consequently, in this section, the simulations are conducted assuming fundamental emissions from an instantaneously-emitting point source and anisotropic density fluctuations ($\alpha \neq 1$), enabling a direct comparison to the results presented in Section~\ref{sec:typeIIIb_isotropic_F_sims}, where isotropic scattering ($\alpha=1$) was invoked.  The considered variables are the level of anisotropy $\alpha$, level of density fluctuations $C_q$, and the source-polar angle $\theta_s$, which will be determined through a comparison with the LOFAR observations.

Single input parameters can be varied with every simulation run.  To identify the combination of parameters that results in the most accurate reproduction of the observed properties, the simulated properties are essentially gridded (as shown throughout this thesis).
For example, as was established in Chapter~\ref{chap:scattering}, the level of density fluctuations $\epsilon$ is the primary parameter affecting the observed source size, whereas the level of anisotropy $\alpha$ dominates the time-profile characteristics.  Therefore, by running simulations for a single value of $\epsilon$ and several anisotropies $\alpha$, a narrow range of anisotropies that can describe the observed time profile is identified.  The same can be done to identify a narrow range of $\epsilon$ values that can describe the observed sizes.  Finally, by simultaneously comparing multiple observed properties to the grid of simulated properties, the set of simulation inputs that best describes the observations is identified.

The time profile and the temporal evolution of the apparent size and position of a $\sim$32.5~MHz source is examined for anisotropy levels $\alpha=0.20$, 0.25, and 0.30, density fluctuation levels $C_q = 1200$, 2300, and 4300~$\mathrm{R_\sun^{-1}}$, and source-polar angles $\theta_s = 0$--8$\degr$.  Anisotropy values $\alpha < 1$ were chosen, in line with the results presented in Chapter~\ref{chap:scattering}, where the presence of stronger scattering in the perpendicular (to the radial) direction was demonstrated.

\begin{figure}[ht!]
    \centering
	\includegraphics[width=1.0\textwidth, keepaspectratio=true]{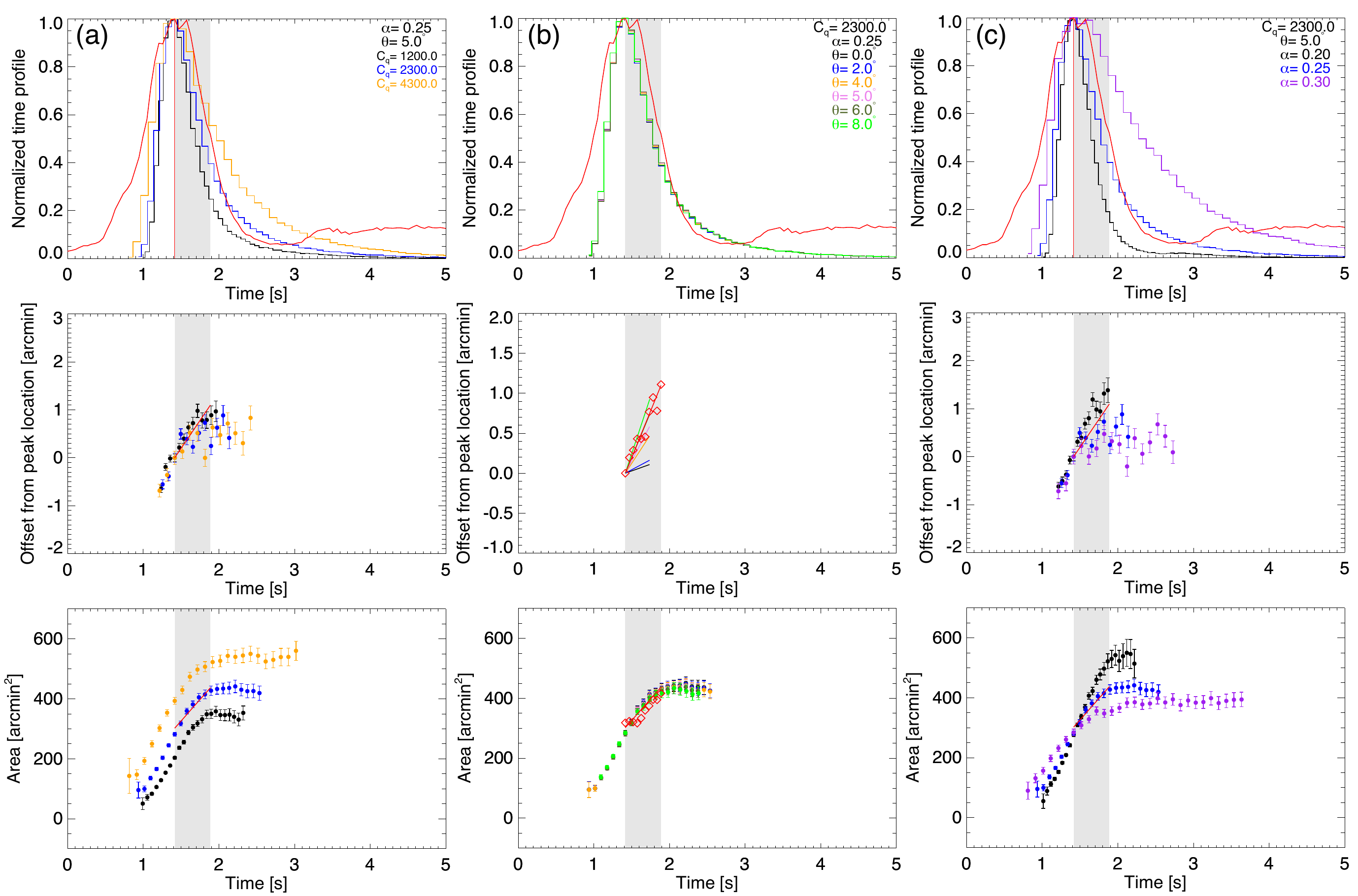}
    \caption[Fundamental emission properties assuming anisotropic scattering.]
    {Simulated fundamental properties of an instantaneously-emitting point source observed at $\sim$32.5~MHz, emitting into an anisotropic medium ($\alpha \neq 1$).
    Top panels show the time profile, middle panels show the source's shift from its peak-flux-time location, and bottom panels show the source's area.  Error bars represent the one-standard-deviation uncertainties.  Different input parameters were varied in each column, as indicated by the legends (in the top panels).
    Column (a) depicts the simulated properties for different density fluctuation levels ($C_q = 1200$, 2300, and 4300~$\mathrm{R_\sun^{-1}}$), where $\alpha=0.25$ and $\theta=5\degr$.
    Column (b) presents the results for $\alpha=0.25$, $C_q = 2300 \, \mathrm{R_\sun^{-1}}$, and angles $\theta_s = 0$, 2, 4, 5, 6, and 8$\degr$.
    Column (c) gives the simulation outputs for $C_q = 2300 \, \mathrm{R_\sun^{-1}}$, $\theta_s=5\degr$, and varying anisotropy levels $\alpha = 0.20$, 0.25, and 0.30.
    LOFAR data is shown in red, and the burst's observed decay time is illustrated by the grey-shaded areas.
    Figure taken from \cite{2020ApJ...905...43C}.
	}
    \label{fig:Chen_anis_F_sim}
\end{figure}

The simulation outputs are depicted in Figure~\ref{fig:Chen_anis_F_sim}, where---similar to Figure \ref{fig:Chen_iso_F_sim}---the top panels show the time profiles, the middle panels show the centroid's offset from its location at the peak-flux time, and the bottom panels show the source's area.  Column \hyperref[fig:Chen_anis_F_sim]{(a)} demonstrates the source properties obtained assuming $\alpha=0.25$, $\theta_s=5\degr$, but for a range of $C_q$ values: $C_q = 1200$, 2300, and 4300~$\mathrm{R_\sun^{-1}}$.  Column \hyperref[fig:Chen_anis_F_sim]{(b)}, on the other hand, presents the properties obtained assuming $\alpha=0.25$, $C_q = 2300 \, \mathrm{R_\sun^{-1}}$, and angles $\theta_s = 0$, 2, 4, 5, 6, and 8$\degr$.  For column \hyperref[fig:Chen_anis_F_sim]{(c)}, $C_q = 2300 \, \mathrm{R_\sun^{-1}}$, $\theta_s=5\degr$, and the anisotropy is varied between $\alpha = 0.20$, 0.25, and 0.30.  The data in red represent the observed properties recorded by LOFAR, whereas the remaining colours of each panel reflect the varying parameters, as indicated by the legends in the top panels.  The areas shaded in grey represent the decay time of the observed Type IIIb striae at $\sim$32.5~MHz.

The simulated decay times are $\sim$0.32, 0.50, and 0.72~s for $C_q = 1200$, 2300, and 4300~$\mathrm{R_\sun^{-1}}$, respectively (Figure~\ref{fig:Chen_anis_F_sim}\hyperref[fig:Chen_anis_F_sim]{(a)}).  As evident, the higher the level of density fluctuations (larger $C_q$ value), the longer the simulated decay time becomes.  The source sizes are also affected by the level of density fluctuations in a similar manner, as expected.  The largest source sizes are produced when the largest value of $C_q$ is assumed (4300~$\mathrm{R_\sun^{-1}}$).  For $C_q = 2300 \, \mathrm{R_\sun^{-1}}$, the source size changes from $\sim$280 to $\sim$430~$\mathrm{arcmin^{2}}$ during the $\sim$0.5~s of the observed decay time.  As can be seen from the bottom panel of Figure~\ref{fig:Chen_anis_F_sim}\hyperref[fig:Chen_anis_F_sim]{(a)}, these source sizes and their temporal evolution successfully reproduce the observed LOFAR sources and their motion (indicated by the red line; \cite{2017NatCo...8.1515K}).

The effects of the degree to which the source's position deviates from the observer's LoS (i.e. when $\theta_s > 0\degr$) are illustrated in Figure~\ref{fig:Chen_anis_F_sim}\hyperref[fig:Chen_anis_F_sim]{(b)}.  Unlike the case of isotropic scattering (Figure~\ref{fig:Chen_iso_F_sim}), anisotropic scattering generates an apparent motion of the source with time, where the apparent velocity of the source depends on the source-polar angle $\theta_s$.  The larger the source-polar angle, the larger the perceived displacement, as expected (see Section~\ref{sec:proj_effs_impact}).  It can also be seen that whilst varying the polar angle $\theta_s$ affects the apparent position of the source, it has an insignificant influence on its time profile and area.  It should be emphasised that no significant change is observed in the apparent source sizes due to the narrow range of angles $\theta_s$ probed (0--8$\degr$).  Otherwise, as illustrated in Section~\ref{sec:angular_dependence}, large polar angles can impact the perceived $x$-size (and thus area) of the sources to a considerable extent.

Figure~\ref{fig:Chen_anis_F_sim}\hyperref[fig:Chen_anis_F_sim]{(c)} suggests that all studied properties are impacted when the anisotropy level changes.  For $\alpha = 0.20$, 0.25, and 0.30, the obtained decay times are 0.24, 0.50, and 1.01~s, respectively.  The apparent source sizes range from $\sim$270--400~$\mathrm{arcmin^{2}}$ for $\alpha=0.20$, from $\sim$280--430~$\mathrm{arcmin^{2}}$ for $\alpha=0.25$, and from $\sim$280--380~$\mathrm{arcmin^{2}}$ for $\alpha=0.30$, during the $\sim$0.5~s decay time of the observed burst.

By varying the values of the input parameters of the simulations ($C_q$, $\theta_S$, and $\alpha$) and comparing the outputs to the observed properties, the values that best match the local coronal conditions for the specific event can be deduced.  The simulations demonstrate that the observed fundamental Type IIIb properties can be reproduced by assuming a point source located at a polar angle $\theta_s = 5\degr$ that simultaneously emits $\sim$32.5~MHz photons near the plasma frequency level, into a local coronal environment characterised by a level of density fluctuations $C_q = 2300 \, \mathrm{R_\sun^{-1}}$ and an anisotropy level $\alpha = 0.25$.

The simulated time profiles suggest that the intrinsic duration of the fundamental emission cannot be longer than $\sim$0.3~s (since the observed FWHM duration is $\sim$1.0~s and the simulated one is $\sim$0.7~s), otherwise the obtained profile will be too broad.  Given that the radio sources are approximated as Gaussian, the observed (deconvolved for the beam) source area is the sum of the intrinsic area and the expansion caused by scattering: $A_{obs} \simeq A_{true} + A_{scatt}$ (see Section~\ref{sec:centroid_calc}).  Therefore, the comparison of simulations to observations allows for the estimation of the intrinsic source size.
Simulations were also conducted assuming a finite source size for the fundamental emissions.  It was deduced that the intrinsic areas should be smaller than $\sim$50~$\mathrm{arcmin^{2}}$, otherwise, larger intrinsic sizes generate apparent areas that are too large and expansion rates that are smaller than that observed.

\subsection{Simulating harmonic Type IIIb emissions} \label{sec:typeIIIb_anisotropic_H_sims}

Following the successful reproduction of the sub-second temporal evolution of the fundamental emissions, the sub-second evolution of the harmonic emissions at the same frequency needs to be probed as well.  In other words, the aim is to successfully simulate the properties of the $\sim$32.5~MHz stria of the harmonic Type IIIb branch \citep{2017NatCo...8.1515K}.  The harmonic emissions are produced where the local plasma frequency $f_{pe} \approx 16$~MHz ($\omega_H \approx 2 \, \omega_{pe}$), which is found at a heliocentric distance $r_H \approx 2.2 \, \Rs$.  An intrinsic point source that emits instantaneously into an anisotropic turbulence medium is assumed.
Similar to Section~\ref{sec:typeIIIb_anisotropic_F_sims}, the simulations are run for anisotropies $\alpha=0.20$, 0.25, and 0.30, density fluctuation levels $C_q = 1200$, 2300, and 4300~$\mathrm{R_\sun^{-1}}$, and source-polar angles $\theta_s = 0$--8$\degr$.

\begin{figure}[t!]
    \centering
	\includegraphics[width=1.0\textwidth, keepaspectratio=true]{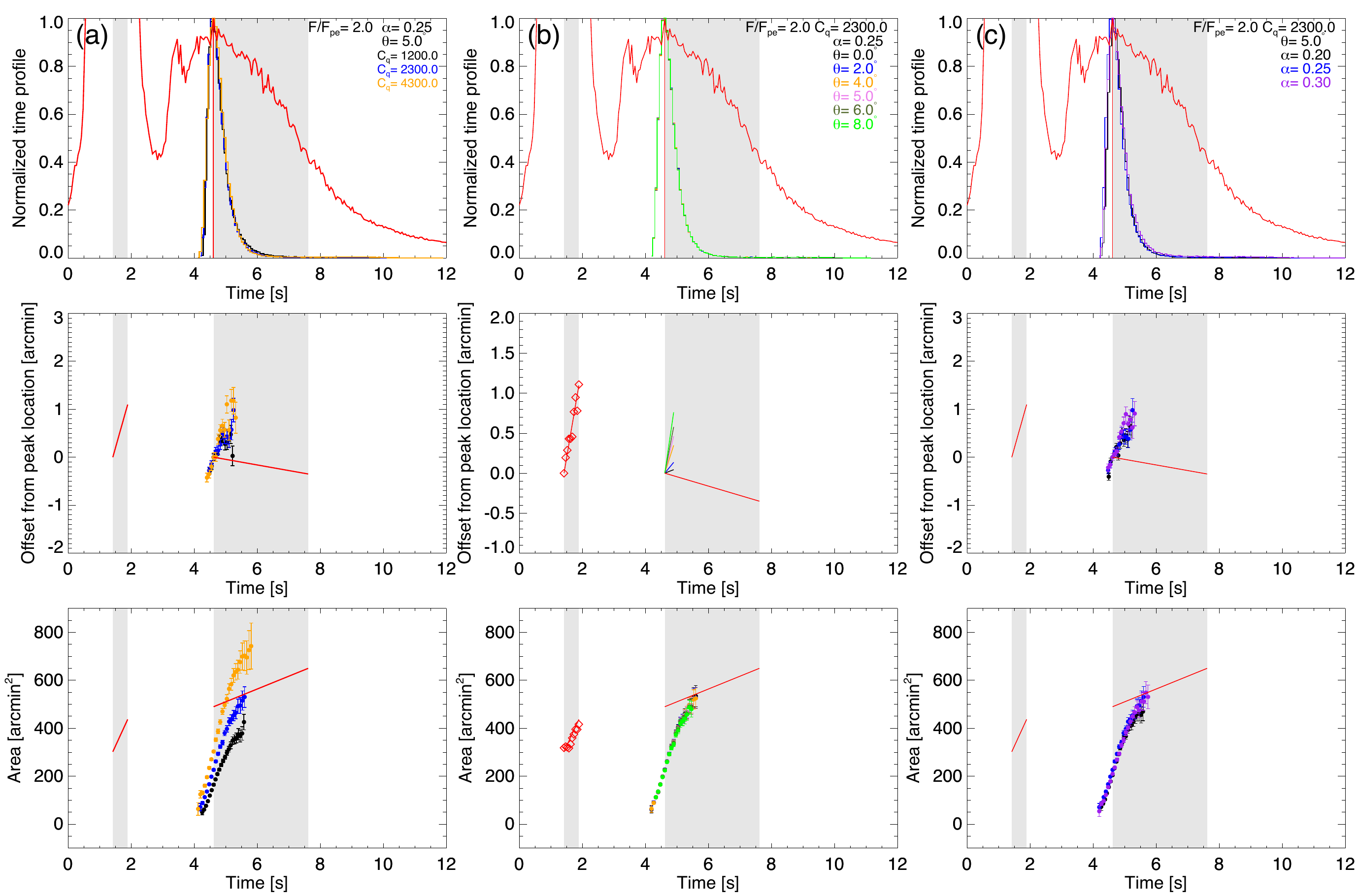}
    \caption[Harmonic emission properties assuming anisotropic scattering.]
    {Same as Figure~\ref{fig:Chen_anis_F_sim}, but for an instantaneously-emitting point source emitting at harmonic frequencies (near 32.5~MHz).  The observed fundamental source properties and decay time (thin grey-shaded area) are also depicted for direct comparison to the harmonic properties.
    Figure taken from \cite{2020ApJ...905...43C}.
	}
    \label{fig:Chen_anis_H_sim}
\end{figure}

Figure~\ref{fig:Chen_anis_H_sim} shows the simulated time profiles (top panels), the offset of the centroids with respect to the source position at the peak-flux time (middle panels), and the source sizes (bottom panels).  The level of density fluctuations $C_q$, level of anisotropy $\alpha$, and source-polar angle $\theta_s$ are varied in the same way as for Figure~\ref{fig:Chen_anis_F_sim}, as indicated by the legend and colour schemes used.  Similarly, the observed source properties (i.e. the LOFAR data) are indicated in red.  The thin grey-shaded areas between 1--2~s represent the observed decay time for the fundamental emissions, whereas the broader grey-shaded areas between 4--8~s represent the observed decay time of the harmonic Type IIIb stria.  The observed fundamental properties are included in Figure~\ref{fig:Chen_anis_H_sim} (indicated by the red data during the fundamental decay time) for a direct comparison with the observed harmonic properties.

It becomes immediately clear from the results depicted in Figure~\ref{fig:Chen_anis_H_sim} that none of the simulated harmonic properties agree with the observed properties of the harmonic emissions observed by LOFAR.  For example, the deduced parameters that matched the observed fundamental emissions ($C_q = 2300 \, \mathrm{R_\sun^{-1}}$, $\alpha = 0.25$, and $\theta_s = 5\degr$) generate a time profile, source positions, and areas similar to those for fundamental emissions ($\omega_F \approx \omega_{pe}$), but contradict the observed harmonic properties.  The obtained decay time for the harmonic source is only $\sim$0.4~s, compared to the observed $\sim$3~s.
The simulated harmonic source area is $\sim$300~$\mathrm{arcmin^{2}}$ near the peak-flux time, which is comparable to the obtained fundamental source area, but considerably smaller than the observed harmonic source of $\sim$500~$\mathrm{arcmin^{2}}$ (at the peak-flux time).  Additionally, the simulated centroid locations depict a rapid motion (similar to the fundamental source) and the source area changes at high rates, but the observed harmonic source is found to move significantly slower and expand far less rapidly with time.

As can be inferred, the simulated time profiles, source motions, and source sizes for an instantaneously-emitting point source at harmonic frequencies are inconsistent with the observed properties of the harmonic Type IIIb emissions.  This suggests that a harmonic source of finite size that emits photons over a finite time period needs to be considered.
A finite emission duration will result in a broader time profile, whereas a finite intrinsic source size will produce larger observed source sizes, as needed in order for the simulations to match the observations.

\subsection{Considering a harmonic source of finite size and finite emission duration} \label{sec:typeIIIb_corrected_H_sims}

The time that the electrons that form the beam (see Section~\ref{sec:plasma_emmission}) take to travel from the location of the fundamental emission ($r_F = 1.8 \, \Rs$) to the location from which the harmonic frequencies are emitted ($r_F = 2.2 \, \Rs$)---i.e. a distance of $\Delta r = 0.4 \, \Rs$ for the $\sim$32.5~MHz source---contributes to the observed time profile of the harmonic emissions.  The observed drift rate of Type IIIb solar radio bursts is used to estimate the speed of the electron beam exciting the Type IIIb emissions (see Equation~(\ref{eqn:v_exc})), found to be around $c/3$, where $c$ is the speed of light (e.g., \cite{2017NatCo...8.1515K}).  It can be assumed that the electron beam has a uniform spread of electron velocities between $c/6$ and $c/3$, such that the time-of-flight duration of electrons at the excitation location of harmonic emissions is
\begin{equation*}
	\Delta t = \dfrac{(6-3) \, \Delta r}{c} \, .
\end{equation*}
For the 32.5~MHz emissions, $\Delta r = 0.4 \, \Rs$ and thus $\Delta t \approx 3$~s.  Electron transport simulations by \cite{2018A&A...614A..69R} support such an expansion of the electron beam (based on the velocity distribution) and a corresponding increase of the emission duration.
Furthermore, a finite time is required for the production of harmonic emission in a given location, since the presence of Langmuir waves at the location does not imply an instantaneous conversion into radio waves (see Section~\ref{sec:plasma_emmission}; \cite{2014A&A...572A.111R}).
Therefore, when the estimated time-of-flight of the electrons is taken into account, the duration of the harmonic emission could be $\sim$3--4~s.

In order to simulate the effect of a harmonic source with a finite emission time, the harmonic emission is taken to be a Gaussian pulse ($\exp (-t^2 / 2 \sigma^2)$) with a standard deviation $\sigma = 2$~s (i.e. it has a FWHM duration of $\sim$4.7~s), which is chosen through comparison with the observed time profile.  This means that the observed profile is defined as the convolution of the intrinsic emission and the broadening caused by scattering.  Furthermore, the intrinsic harmonic source is taken to have a finite emission area, the size of which is determined through comparisons with the observations.

\begin{figure}[htp!]
    \centering
	\includegraphics[width=0.8\textwidth, keepaspectratio=true]{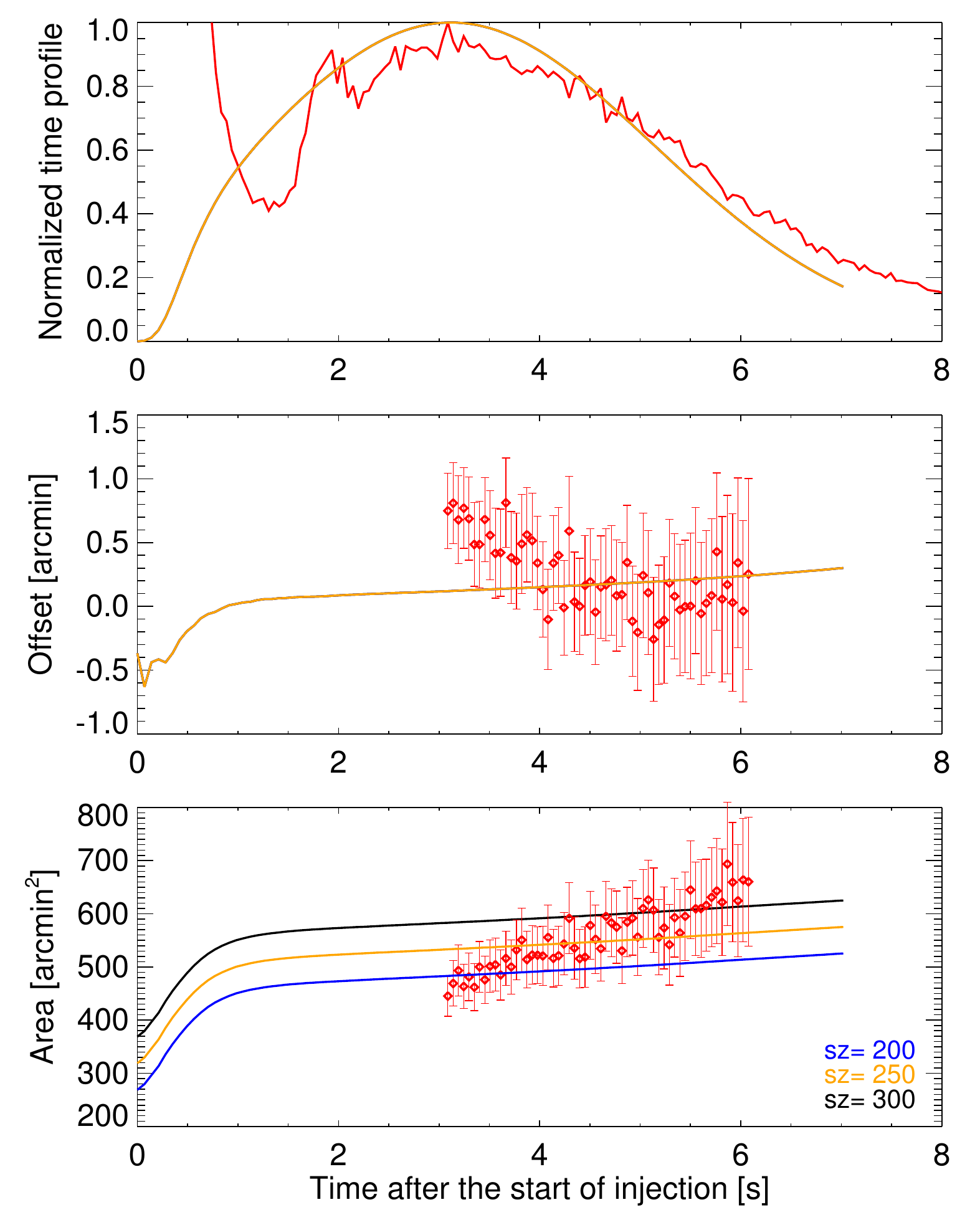}
    \caption[Harmonic emission properties assuming anisotropic scattering and a finite source size and emission duration.]
    {Simulated harmonic emission properties for a source of finite size and finite emission duration.  The simulations were conducted assuming $f_{pe} \approx 32.5$~MHz, $\alpha = 0.25$, $C_q = 2300 \, \mathrm{R_\sun^{-1}}$, $\theta_s = 5\degr$ (which successfully reproduced the fundamental emissions), an emission duration characterised by a Gaussian profile with a standard deviation of 2~s, and three different intrinsic source areas: $\sim$200~$\mathrm{arcmin^{2}}$ (blue), $\sim$250~$\mathrm{arcmin^{2}}$ (orange), and $\sim$300~$\mathrm{arcmin^{2}}$ (black).  The simulated time profile (top panel), source offset from its peak-flux-time location (middle panel), and the source area (bottom panel) are depicted along with the observed burst properties (red data).  The variation in intrinsic size does not affect the simulated time profile and source motion.
    Figure taken from \cite{2020ApJ...905...43C}.
	}
    \label{fig:Chen_anis_H_sim_corrected}
\end{figure}

The results for a harmonic source of finite size and finite emission duration (with a Gaussian profile) are presented in Figure~\ref{fig:Chen_anis_H_sim_corrected}.
The simulation's input parameters are defined as those that successfully reproduced the fundamental emissions ($C_q = 2300 \, \mathrm{R_\sun^{-1}}$, $\alpha = 0.25$, and $\theta_s = 5\degr$).
The time profile (top panel), the offset of the centroid position from its location at the peak-flux time (middle panel), and the source area (bottom panel) are depicted.  The illustrated results were obtained for a single emission duration (4.7~s) and three different intrinsic source sizes: $\sim$200~$\mathrm{arcmin^{2}}$ (blue curve), $\sim$250~$\mathrm{arcmin^{2}}$ (orange curve), and $\sim$300~$\mathrm{arcmin^{2}}$ (black curve).
The variation in intrinsic size does not affect the simulated time profile and centroid motion, hence the simulated data in the top two panels overlap.
The apparent time profile obtained has a slightly longer FWHM duration ($\sim$4.8~s) than the intrinsic pulse ($\sim$4.7~s).  A prolonged emission at the source generates a smaller centroid motion ($dr$) compared to the source that injects photons instantaneously (Figure~\ref{fig:Chen_anis_H_sim}).  Moreover, while the instantaneous harmonic emission results in a fast source motion ($dr/dt$; cf. Figure~\ref{fig:Chen_anis_H_sim}), the prolonged harmonic emission does not demonstrate a clear motion with time, making it consistent with the observed source properties.  To reproduce the slow centroid motion and areal expansion of the harmonic source, a continuous harmonic emission lasting for $\gtrsim 4$~s is required.
It can also be seen that a harmonic source with a physical (intrinsic) emission area of up to $\sim$200~$\mathrm{arcmin^{2}}$ produces scattered source areas that match the observed values---near the peak-flux time---more successfully that the other (larger) intrinsic sizes.  The $\sim$200~$\mathrm{arcmin^{2}}$ intrinsic source produces scattered source areas of $\sim$490~$\mathrm{arcmin^{2}}$ (at the peak-flux time), i.e. in good agreement with those observed ($\sim$500~$\mathrm{arcmin^{2}}$).  The peak-flux time is used to compare the simulated outputs to the observations (instead of the entire decay phase), as there are smaller uncertainties in the observed-value estimations near the peak time (see Section~\ref{sec:centroid_calc}).  In other words, an intrinsic source area of $\sim$200~$\mathrm{arcmin^{2}}$ is inferred (over the other values) because it produces observed areas that best match the most reliable part of the observations.

\subsection{Discussion and final remarks} \label{sec:typeIIIb_discussion}
In this chapter, observations with high temporal and spatial resolutions provided by LOFAR are utilised to take advantage of the ability to trace the temporal evolution of source properties at a single frequency.  The sub-second evolution of fundamental and harmonic Type IIIb radio sources emitted at $\sim$32.5~MHz was investigated in the context of anisotropic scattering simulations.  The simulations were set to reproduce the time profile and temporal evolution of the source position and size of a point source from which all photons are injected into the solar corona simultaneously.

Simulation results for the fundamental source were presented for both isotropic and anisotropic density fluctuations.  It was demonstrated that the isotropic scattering assumption cannot sufficiently explain the observed source properties, confirming the conclusions of Chapter~\ref{chap:scattering}.  The anisotropic scattering assumption, on the other hand, resulted in source properties that agreed with the observed ones.  Through this comparison of simulated to observed properties, parameters describing the local coronal conditions have been inferred.  The level of density fluctuations $C_q$ was found to be 2300~$\mathrm{R_\sun^{-1}}$ (for the adopted model; see Section~\ref{sec:dens_fluct_spec}), the level of anisotropy $\alpha = 0.25$, and the source-polar angle $\theta_s = 5\degr$.  The combination of these parameters successfully reproduced the observed properties of the $\sim$32.5~MHz fundamental emissions of the Type IIIb burst observed with LOFAR, namely, the time profile and the absolute values and temporal evolution of the source positions and source areas.

\begin{figure}[t!]
\centering

\captionsetup[subfigure]{position=top, aboveskip=0em, belowskip=0em}

\begin{subfigure}[t]{0.5\textwidth}
    \centering
    \includegraphics[height=0.93\textwidth, keepaspectratio=true]{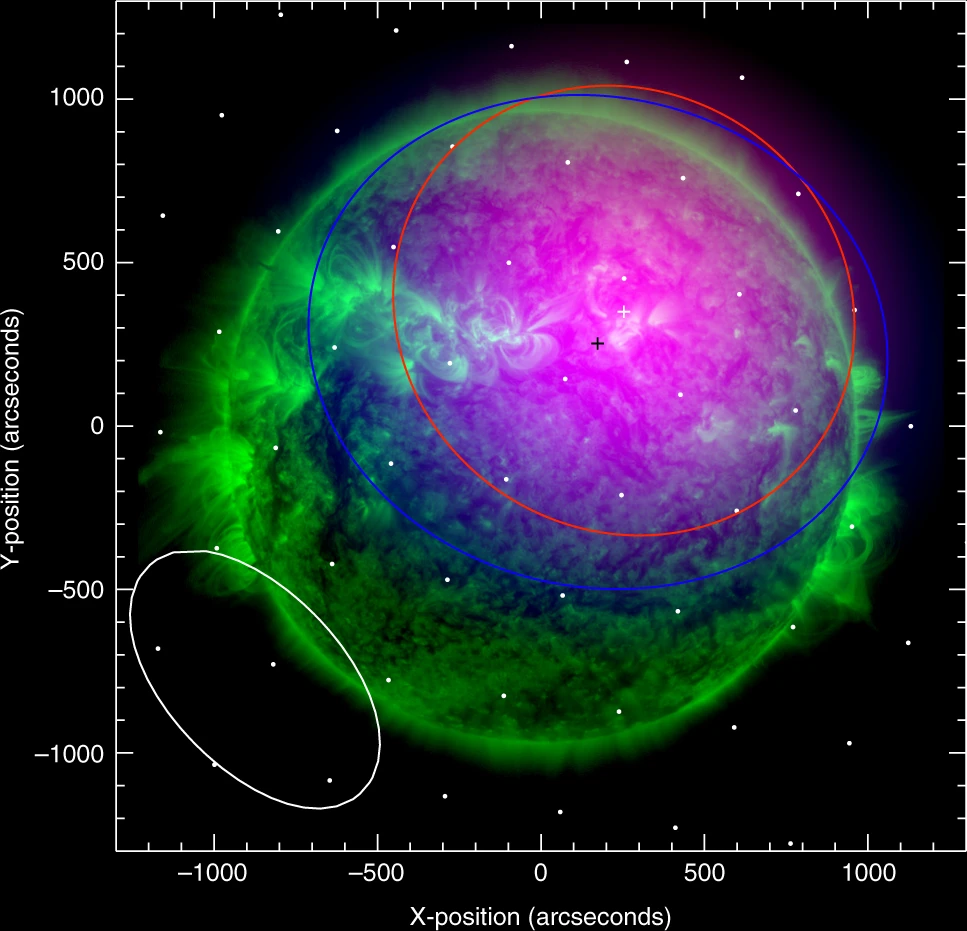}
    \caption{}
    \label{fig:kontar2017_source_img}
\end{subfigure}%
\begin{subfigure}[t]{0.5\textwidth}
    \centering
    \includegraphics[height=0.94\textwidth, keepaspectratio=true]{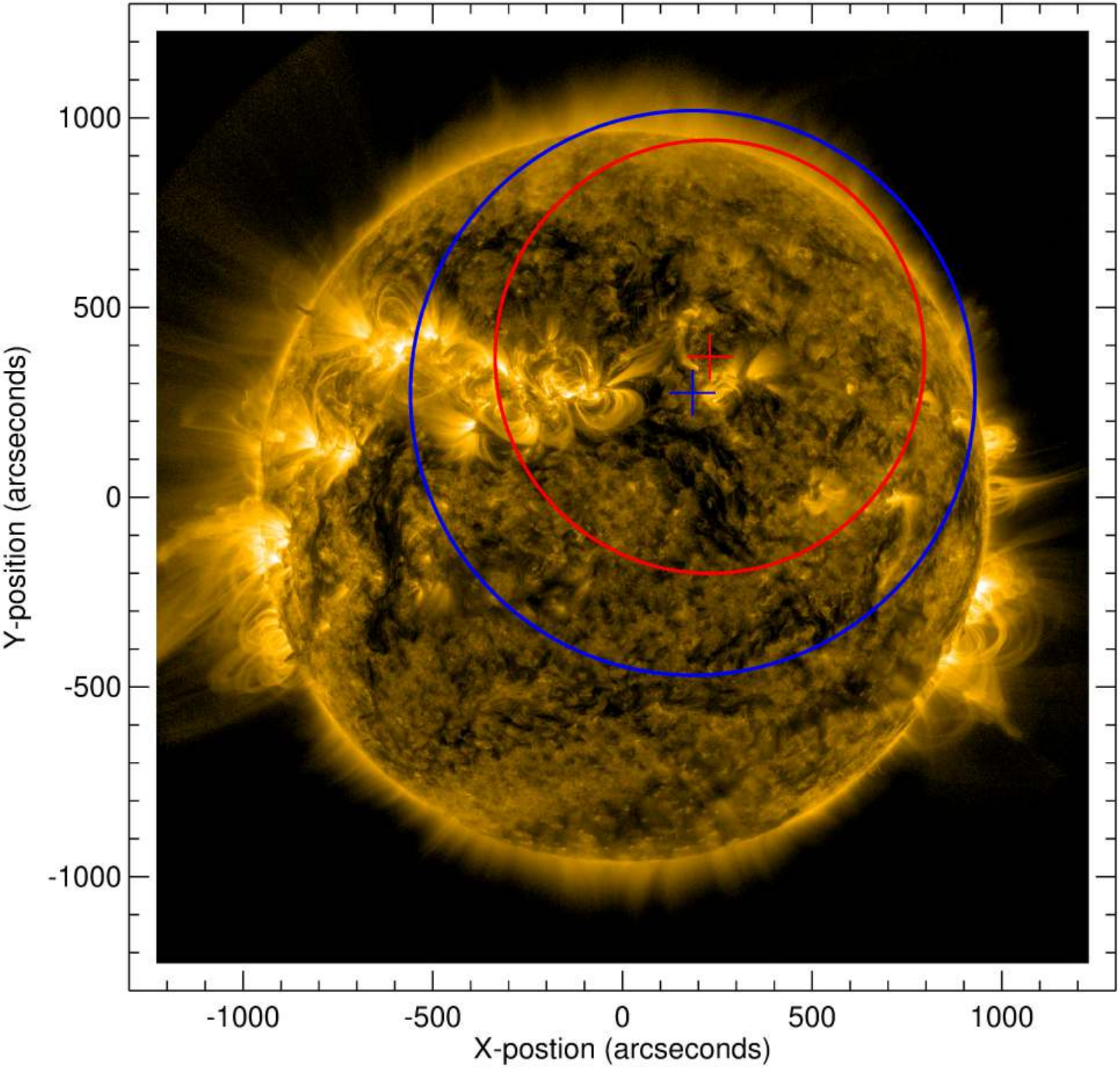}
    \caption{}
    \label{fig:chen_sim_obs_sources}
\end{subfigure}

\caption[Observed vs simulated fundamental and harmonic sources.]
{Observed and simulated (expected) sources at 32.5~MHz, shown with respect to the Sun (imaged using AIA 171~\AA\ data).
(a)~Observed fundamental and harmonic FWHM source sizes and associated centroids for the Type IIIb burst recorded by LOFAR.  The red ellipse and white plus sign depict the fundamental source and its centroid position, respectively.  The blue ellipse and black plus sign depict the harmonic source and its centroid position, respectively.  The fundamental and harmonic sizes are shaded for emphasis in magenta and blue, respectively.  The FWHM beam size of LOFAR at 32.5~MHz is also shown (white ellipse) along with the central locations of the tied-array beams (white dots).
(b)~Simulated fundamental (red) and harmonic (blue) FWHM source sizes and the associated centroid positions (shown in red and blue plus signs, respectively).  The sources were simulated using $\alpha = 0.25$, $C_q = 2300 \, \mathrm{R_\sun^{-1}}$, and $\theta_s = 5\degr$.  For the harmonic emissions, a source of finite size and emission duration was assumed.
Panel (a) was taken from \cite{2017NatCo...8.1515K} and reproduced under \href{https://creativecommons.org/licenses/by/4.0/}{\color{black}{\textsl{CC BY 4.0}}}, and panel (b) was taken from \cite{2020ApJ...905...43C}.
}
\label{fig:TypeIIIb_apparent_FH_sources}
\end{figure}

An attempt to reproduce the $\sim$32.5~MHz harmonic emissions of the Type IIIb burst was made using the same parameters inferred from the fundamental emissions.  While fundamental emissions could be successfully reproduced assuming an instantaneous emission from a point source, the harmonic emissions could not.  Significant discrepancies were identified between the observed and simulated harmonic time profiles, source areas, and sizes, as well as their temporal evolution.  These discrepancies could not be redeemed by varying the values of the input parameters, as changing a parameter to improve one simulated property would worsen another.  It was concluded that the harmonic emissions cannot be described by a point source that emits all photons into the corona instantaneously.  Instead, it was found that a source of a finite size and finite emission duration is required.  The harmonic source properties were successfully described when a $\sim$200~$\mathrm{arcmin^{2}}$ intrinsic source that emits continuously for $\sim$4.7~s was assumed.  The intrinsic duration of harmonic sources was related to electron transport effects, a dominant contribution to which came from the time taken ($\sim$3~s) by the electron beam to travel between the location of initial excitation at 1.8~$\Rs$ (fundamental emissions) until the region where wave-wave interactions excite harmonic emissions at 2.2~$\Rs$ (see Section~\ref{sec:plasma_emmission}).
It should be highlighted, though, that while the harmonic emissions required a finite intrinsic source size in order to be reproduced, the fundamental emissions were not restricted to the assumption of an intrinsic point source.  As discussed in Section~\ref{sec:typeIIIb_anisotropic_F_sims}, it was estimated that fundamental emissions originating from a source that is up to $\sim$50~$\mathrm{arcmin^{2}}$ in size would also maintain agreement between the simulated and observed properties.

Figure~\ref{fig:TypeIIIb_apparent_FH_sources} shows a side-by-side comparison of the observed fundamental and harmonic sources (left panel) and the simulated ones (right panel), with respect to the Sun.  It is clear that the two panels are alike, with respect to both the areal expansion of the sources and the heliocentric centroid locations.  As described in Section~\ref{sec:anisotropic_simulations_2019}, the applied simulations assume a radial density model.  As such, the source centroids and their scattering-induced shifts are defined along the $x$-direction (as portrayed throughout Chapter~\ref{chap:scattering}).  Given that the simulations consider a spherically-symmetric corona, the azimuthal angle (i.e. the angle from the $x$-axis to the source) can be given any arbitrary value without impacting the interpretation of the simulations.  As such, the sources in Figure~\ref{fig:chen_sim_obs_sources} were (azimuthally) rotated to the degree required to match the sources depicted in Figure~\ref{fig:kontar2017_source_img}.

It has been demonstrated that the observed fundamental and harmonic source properties (including their temporal characteristics) can be successfully described within the framework of radio-wave propagation effects where anisotropic scattering dominates.  Although the analysed observation was of a Type IIIb radio burst, the arguments presented in this chapter could be applicable to all radio emissions resulting from the plasma emission mechanism (see Sections~\ref{sec:all_bursts} and \ref{sec:plasma_emmission}).  It must, however, be acknowledged that the parameters describing the turbulence in the vicinity of an emitting radio source can vary from one heliocentric distance to another, and from event to event.

\section{Simulating the Observed Properties of Drift-Pair Solar Radio Bursts} \label{sec:drift_pairs}

\subsection{Typical characteristics of Drift-pair bursts} \label{sec:driftpairs_properties}
Drift-pair solar radio bursts (Figure~\ref{fig:Kuznetsov2020_lofar_obs}\hyperref[fig:Kuznetsov2020_lofar_obs]{a}) are a rare and non-classical type of solar radio emissions (cf. Section~\ref{sec:all_bursts}) that have been observed in the low-frequency domain, between $\sim$10--100~MHz.  First identified spectrally by \cite{1958AuJPh..11..215R}, they are fine structures with a very characteristic narrowband morphology: two almost-identical parallel stripes that repeat each other in time (instead of frequency), typically separated by $\sim$1--2~s.  Although this range of temporal separations between the two components is true for all Drift-pair bursts---irrespective of the frequency probed---\cite{2005SoPh..231..143M} found that there is a slight decrease of the observed delay between the components with decreasing frequency.  It is worth mentioning, though, that the relation between the delay and frequency remains unclear, given previous observations by \cite{1978A&A....70..801M} that suggested a constant delay with frequency.  Notably, both studies (\cite{1978A&A....70..801M} and \cite{2005SoPh..231..143M}) can be regarded as ambiguous since neither presented the uncertainties in their measurements.  Therefore, the dependence of the delay on frequency is still to be confirmed.

Both positive and negative frequency-drift rates are observed, where the negative frequency-drift bursts are sometimes referred to as ``forward'' and those with positive drift values are referred to as ``reverse'' \citep{1971A&A....12..371D, 1984A&A...130...39D}.  Drift-pair bursts of positive frequency drifts are more commonly observed.  The frequency-drift rates tend to increase with the emission frequency, having an absolute value ($\abs{df/dt}$) of $\sim$1--2~$\MHzs$ at around 30~MHz.  It has been noted that their frequency-drift rates are between those of Type II and Type III bursts (see Sections~\ref{sec:typeIIs} and \ref{sec:typeIIIs}), specifically, around 10 times higher than typical Type II drift rates and $\sim$3 times lower than typical Type III drifts, at the same frequencies \citep{1985srph.book.....M}.
The inferred exciter speed from these drift rates ($\sim$20,000~$\kms$ at $\sim$30~MHz; see Section~\ref{sec:f_vs_R_relation}) suggests that whistler waves are a likely exciting agent of Drift-pair bursts, since they are capable of propagating both towards and away from the Sun, accounting for both the forward and reverse bursts \citep{2020ApJ...898...94K}.
Both Drift-pair components are characterised by the same frequency drift (hence parallel) and they both appear to start and end in dynamic spectra at the same frequencies, whereas the intensity of the two components can differ.  The duration of each component at a fixed frequency is $\sim$1~s, although bursts with negative drifts are found to be somewhat shorter in duration than those with positive drifts \citep{2005SoPh..231..143M}.
The intriguing similarity between the first (in time; ``leading'') and second (``trailing'') components prompted---from the very beginning---the proposition that the trailing component of Drift-pair bursts is the mere reflection of the leading one (see Section~\ref{sec:driftpairs_radio_echo_theory} and \cite{1958AuJPh..11..215R}).

Following imaging observations of Drift-pair bursts, \cite{1979PASAu...3..379S} found that the emission sources of the leading and trailing components of the bursts are virtually co-spatial when imaged at the same frequency (in that case, 43~MHz).  Even though these observations were conducted with a high angular resolution of $\sim$4$\arcmin$, the temporal resolution was $\sim$3~s, therefore, insufficient for resolving the dynamics of the two components which tend to be separate by $\lesssim 2$~s.  The only other published study of Drift-pair source sizes and positions (prior to the results presented in this chapter) was that of \cite{2019A&A...631L...7K}, who conducted high-resolution multi-frequency imaging observations of a Drift-pair burst observed by LOFAR, examining the evolution of the radio sources at both a fixed frequency and along the components, with a cadence of $\sim$0.01~s.  Their spatially-resolved observations showed that the sources of both the leading and trailing components propagate in the same direction and along the same trajectory, separated from each other by a certain amount of time.

\subsection{LOFAR observation of a Drift-pair burst} \label{sec:driftpairs_LOFAR_properties}
\begin{figure}[ht!]
\centering

\centerline{\includegraphics[width=0.495\linewidth]{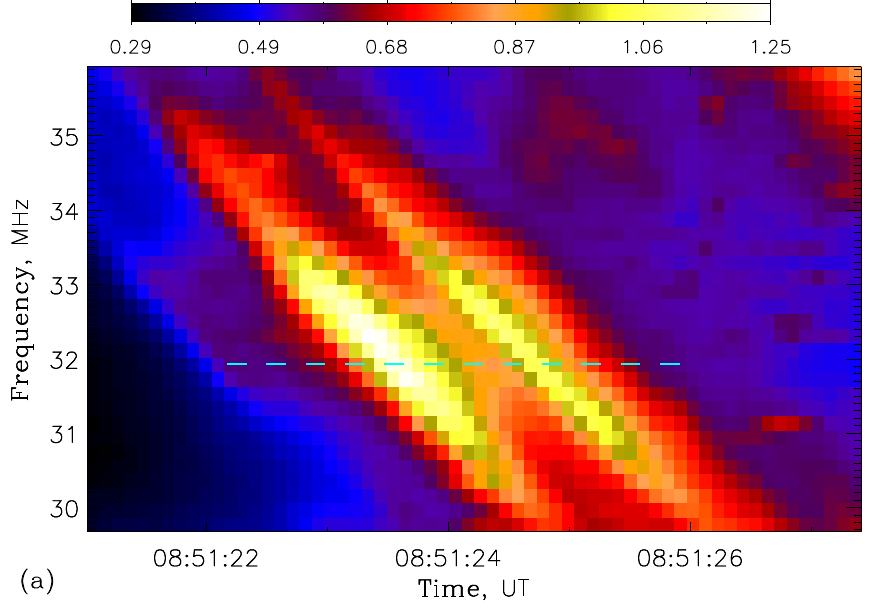}~
\includegraphics[width=0.495\linewidth]{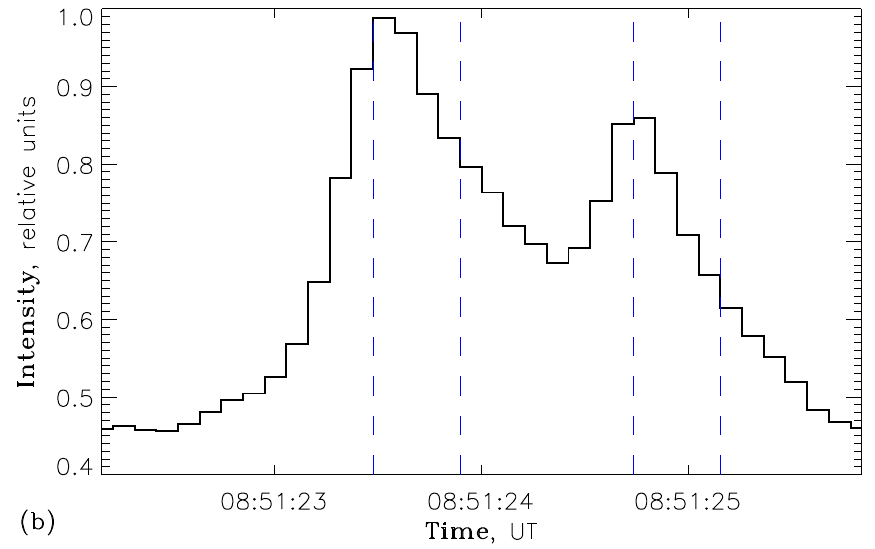}}
\centerline{\includegraphics[width=0.495\linewidth]{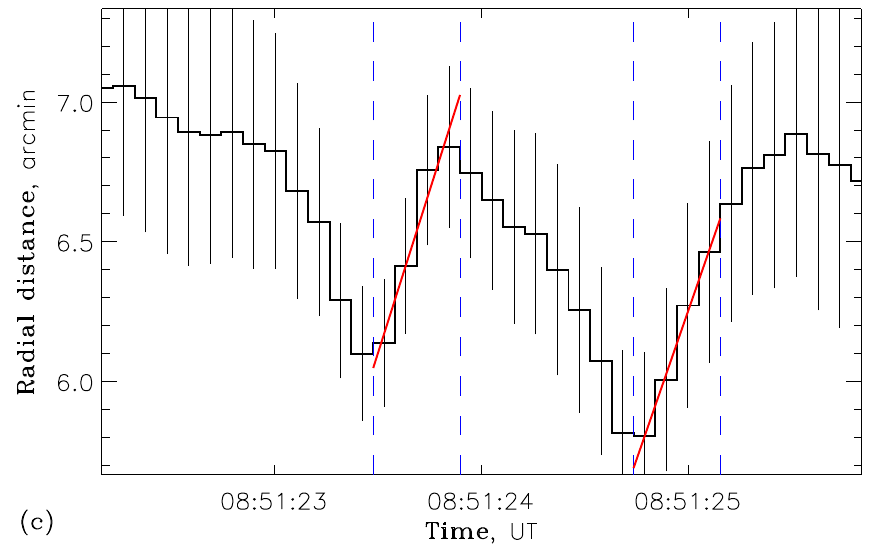}~
\includegraphics[width=0.495\linewidth]{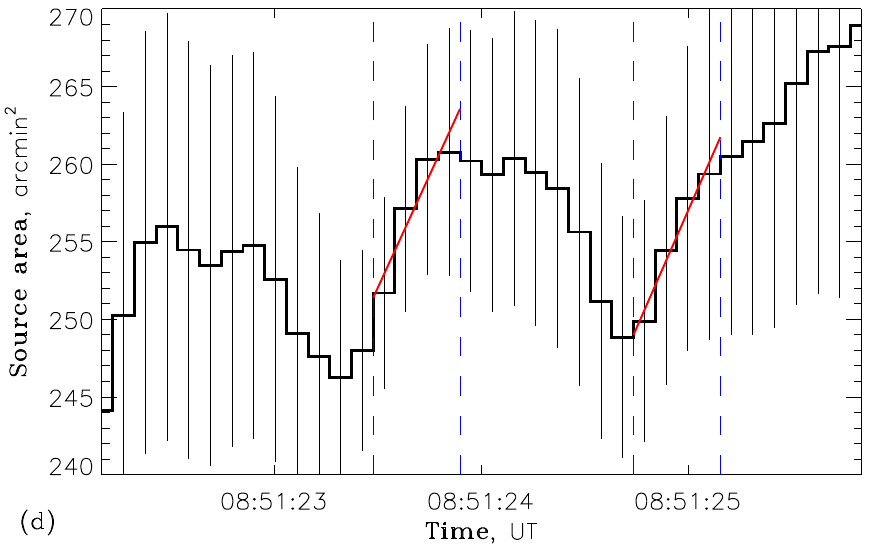}}
\caption[LOFAR observations of a Drift-pair burst and its source properties.]
{(a)~Dynamic spectrum of a Drift-pair burst recorded by LOFAR on 12 July 2017 (in relative intensity units).  The blue dashed line indicates the imaged time range and frequency ($\sim$32~MHz).
(b)~Observed time profile depicting the normalised (with respect to the maximum value) intensity of the emissions at $\sim$32~MHz.  Blue dashed lines indicate the peak time and estimated HWHM decay times of each component.
(c)~Projected (in the plane of the sky) heliocentric source position as a function of time.  Red lines depict the fits used to estimate the source's radial velocity for each component.  Blue dashed lines correspond to the times annotated in panel (a), i.e. the peak-intensity time and HWHM decay time for each component.  Error bars represent a one-standard deviation uncertainty.
(d)~Source area (and associated one-standard-deviation errors) as a function of time.  Red lines depict the fits used to estimate the source's areal expansion for each component in the time intervals indicated by the blue dashed lines.
Figure taken from \cite{2020ApJ...898...94K}.
}

\label{fig:Kuznetsov2020_lofar_obs}
\end{figure}

Figure~\ref{fig:Kuznetsov2020_lofar_obs}\hyperref[fig:Kuznetsov2020_lofar_obs]{a} depicts a Drift-pair burst observed by LOFAR on 12 July 2017 between 30--70~MHz.  The specific Drift-pair burst was first reported and analysed by \cite{2019A&A...631L...7K}.  It is worth emphasising that the spectral and temporal resolutions with which the presented event was recorded are $\sim$12.2~kHz and $\sim$0.01~s, respectively, i.e. much higher than those of \cite{1979PASAu...3..379S} which were limited by a temporal resolution of $\sim$3~s.  The emission sources were well-resolved too, given a synthesised FWHM beam size (Equation~(\ref{eqn:lofar_res})) of $\sim$10$\arcmin$ and a beam separation of $\sim$6$\arcmin$ at 32~MHz.  For the presented analysis of this Drift-pair burst, the temporal resolution was rebinned and decreased to $\sim$0.1~s.

As illustrated on the dynamic spectrum (Figure~
\ref{fig:Kuznetsov2020_lofar_obs}\hyperref[fig:Kuznetsov2020_lofar_obs]{a}), both components of the burst are imaged at $\sim$32~MHz.  The trailing component is temporally separated from the leading component by $\sim$1.2~s; a typical (peak-to-peak) delay.  The time profile of the Drift-pair components is illustrated in Figure~\ref{fig:Kuznetsov2020_lofar_obs}\hyperref[fig:Kuznetsov2020_lofar_obs]{b}, where the normalised intensity (with respect to the peak intensity value) is given.  The temporal evolution of the radial source position and the source size (for the same time interval) are illustrated in Figures~\ref{fig:Kuznetsov2020_lofar_obs}\hyperref[fig:Kuznetsov2020_lofar_obs]{c} and \ref{fig:Kuznetsov2020_lofar_obs}\hyperref[fig:Kuznetsov2020_lofar_obs]{d}, respectively.  The source parameters and the associated one-standard-deviation errors were obtained by fitting LOFAR's emission images with a 2D elliptical Gaussian function, as demonstrated in Section~\ref{sec:centroid_calc}.  During the decay time of both components (indicated by the blue dashed lines in panels \hyperref[fig:Kuznetsov2020_lofar_obs]{(b)}--\hyperref[fig:Kuznetsov2020_lofar_obs]{(d)}, for each component), the emission source demonstrates a clear radial motion away from the Sun, as well as an increase in its FWHM area.  On average, the source's radial position increases by $\sim$2.2$\arcmin$ per second (in the plane of the sky), corresponding to a speed $dr/dt \simeq c/3$.  The source area expands at a rate $dA/dt \simeq 30$~$\mathrm{arcmin^{2} \, s^{-1}}$.
Using the observed source area, the anisotropy of the scattering process can be estimated \citep{2017NatCo...8.1515K, 2018SoPh..293..115S}.  As described in Section~\ref{sec:centroid_calc}, the observed area $A_{obs}$ is the convolution of the instrument's beam area $A_{beam}$ and the real source area $A_{real}$, such that $A_{obs} = A_{real} + A_{beam}$ (where $A_{real}$ includes the scattering-induced broadening: $A_{real} = A_{true} + A_{scatt}$).
Given that the observed plane-of-sky source area $A_{obs} \simeq 250$~$\mathrm{arcmin^{2}}$ and LOFAR's beam area $A_{beam} \approx 100$~$\mathrm{arcmin^{2}}$, the real area $A_{real} \simeq 150$~$\mathrm{arcmin^{2}}$.  Assuming the source is nearly circular, this corresponds to a linear size of $\sim$14$\arcmin$ across the observer's LoS ($\Delta r_{\perp} \approx 14\arcmin$) which, as mentioned, includes the effects of scattering that enlarge the intrinsic source size.  On the other hand, the scattered source size along the LoS, $\Delta r_{\parallel}$---which defines the width of the light curve---cannot exceed the value $c \Delta t$, where $\Delta t$ is the duration of a single Drift-pair component.  It can be seen from Figure~\ref{fig:Kuznetsov2020_lofar_obs}\hyperref[fig:Kuznetsov2020_lofar_obs]{b} that the duration of each component of the studied burst is $\sim$0.6~s, meaning that $\Delta r_{\parallel} \lesssim 4\arcmin$.  It can therefore be deduced that $\Delta r_{\perp} \gg \Delta r_{\parallel}$.  This suggests that scattering is highly anisotropic, specifically, it is stronger in the perpendicular direction (to the LoS) compared to the parallel one (as deduced in Chapter~\ref{chap:scattering} and Section~\ref{sec:typeIIIb_sim_setup}).

\subsection{Probing the radio echo hypothesis} \label{sec:driftpairs_radio_echo_theory}
As mentioned in Section~\ref{sec:driftpairs_properties}, the similarities between the leading and trailing components of Drift-pair bursts led to the hypothesis that the trailing component is a reflection of the leading component, an effect termed as the radio ``echo'' \citep{1958AuJPh..11..215R}.  In other words, both components originate from the same emission source, but some of the radiation propagates directly to the observer and some does not, thus following different paths and reaching the observer at two distinct times.  It was proposed that the reflection occurs in regions of the solar corona which are closer to the Sun than the emission source, and are therefore denser.  The justification was that these denser regions force the emitted radiation to reflect and propagate back towards the observer, since it cannot propagate through plasma levels at or below the cut-off frequency ($f_{pe}$; see Section~\ref{sec:plasma_emmission}).  As discussed in Section~\ref{sec:plasma_emmission}, in order for radio-wave propagation to occur, the ratio of the emission frequency to the local plasma frequency $f/f_{pe}$ must be $\gtrsim 1$.  The larger this ratio is, the longer the delay between the direct and reflected rays (and thus Drift-pair components) is expected to be, given that photons emitted from the source need to travel a longer distance before they encounter a region where $f = f_{pe}$, which reflects them.

The observed time delay between the two components ($\sim$1--2~s) was thought to be too large to result from fundamental emissions, so it was argued that in order to reproduce such long delays, the point of emission needed to be farther away from the region of reflection (where $f \rightarrow f_{pe}$) than fundamental emissions ($f_F \approx f_{pe}$), suggesting that Drift-pair bursts were the result of harmonic emissions ($f_H \approx 2 \, f_{pe}$; \cite{1958AuJPh..11..215R}).

Moreover, the echo hypothesis was questioned altogether as it was believed to be unable to explain the observed properties of Drift-pair bursts.  Specifically, it was predicted that: (i) the reflected component should be less intense and more diffuse than the direct one (due to scattering effects; \cite{1974SoPh...35..153R}), (ii) the reflected rays should correspond to different source positions than those produced by the direct rays, especially for sources located farther from the solar centre (see \cite{1982srs..work..182M}), and (iii) the delay between the two components should increase with the emission frequency \citep{1982srs..work..182M}.  These predictions, however, did not agree with the observed properties of Drift-pair bursts.  The degree of circular polarisation of the two components favoured fundamental emissions \citep{1979PASAu...3..379S, 1984A&A...130...39D}, the time-profiles of the two components appeared to be nearly-identical \citep{1958AuJPh..11..215R}, the source positions of the two components were found to spatially coincide \citep{1979PASAu...3..379S}, and the time delay between the two components appeared (at that time) to be constant at the emission frequencies observed \citep{1978A&A....70..801M, 1982srs..work..182M}.  It should be reiterated, though, that more recent studies of Drift-pair bursts \citep{2005SoPh..231..143M} suggested that the time delay has an inverse dependence on the emission frequency (as described in Section~\ref{sec:driftpairs_properties}).

These predictions and the associated criticism were made under the assumption that scattering in the solar corona is \textit{isotropic}.  However, it is now known that only \textit{anisotropic} scattering can successfully account for the observed properties of radio bursts (see Chapter~\ref{chap:scattering} and Section~\ref{sec:typeIIIb_sim_setup}).  For example, as was established in Chapter~\ref{chap:scattering}, scattering resulting from anisotropic density fluctuations can be very strong and still produce highly-directional emissions.  As such, the radio echo hypothesis ought to be probed within the framework of anisotropic radio-wave scattering, studying the behaviour of both the direct and reflected rays as they propagate through the turbulent coronal medium.  Hence, in the upcoming sections, the properties and temporal evolution of the Drift-pair burst observed by LOFAR (Figure~\ref{fig:Kuznetsov2020_lofar_obs}) are compared to the radio-wave propagation simulations described in Chapter~\ref{chap:scattering}.

\subsubsection{Ray-tracing simulation set-up}
The simulations were set up as described in Sections~\ref{sec:anisotropic_simulations_2019} and \ref{sec:typeIIIb_sim_setup}, where a stationary point source was taken to instantaneously and isotropically inject $\sim$\vphantom{}$10^4$ photons of frequency $f \simeq 35.2$~MHz into the heliosphere.  Similar to Chapter~\ref{chap:scattering}, the level of density fluctuations $\epsilon$, the level of anisotropy $\alpha$, and the source-polar angle $\theta_s$ are varied in this analysis.  Additionally, the ratio of the source's emission frequency to the local plasma frequency $f/f_{pe}$ is also varied in this study, examining its impact on the delay between the Drift-pair components.

Due to the finite number of photons used, the simulated parameters will have an associated statistical error.  Particularly, the uncertainty on the simulated source size and area is the lowest at the simulated peak-flux time, where the number of ``arriving'' photons is the largest.  Moreover, the simulations do not consider contributions from continuous background emissions or randomly-varying radio noise, which complicate real observations (cf. Figure~\ref{fig:Kuznetsov2020_lofar_obs}).
Similar to Chapter~\ref{chap:scattering} and Section~\ref{sec:typeIIIb_sim_setup}, the time-of-flight of photons is subtracted from the depicted simulation outputs, such that the simulated delays represent those caused by radio-wave propagation effects.  Namely, if photons were to propagate through free space, they would be depicted as arriving at time $t=0$.

\subsection{The impact of eliminating scattering effects} \label{sec:driftpairs_no_scattering}

\begin{figure}[t!]

\centerline{\includegraphics[width=0.49\linewidth]{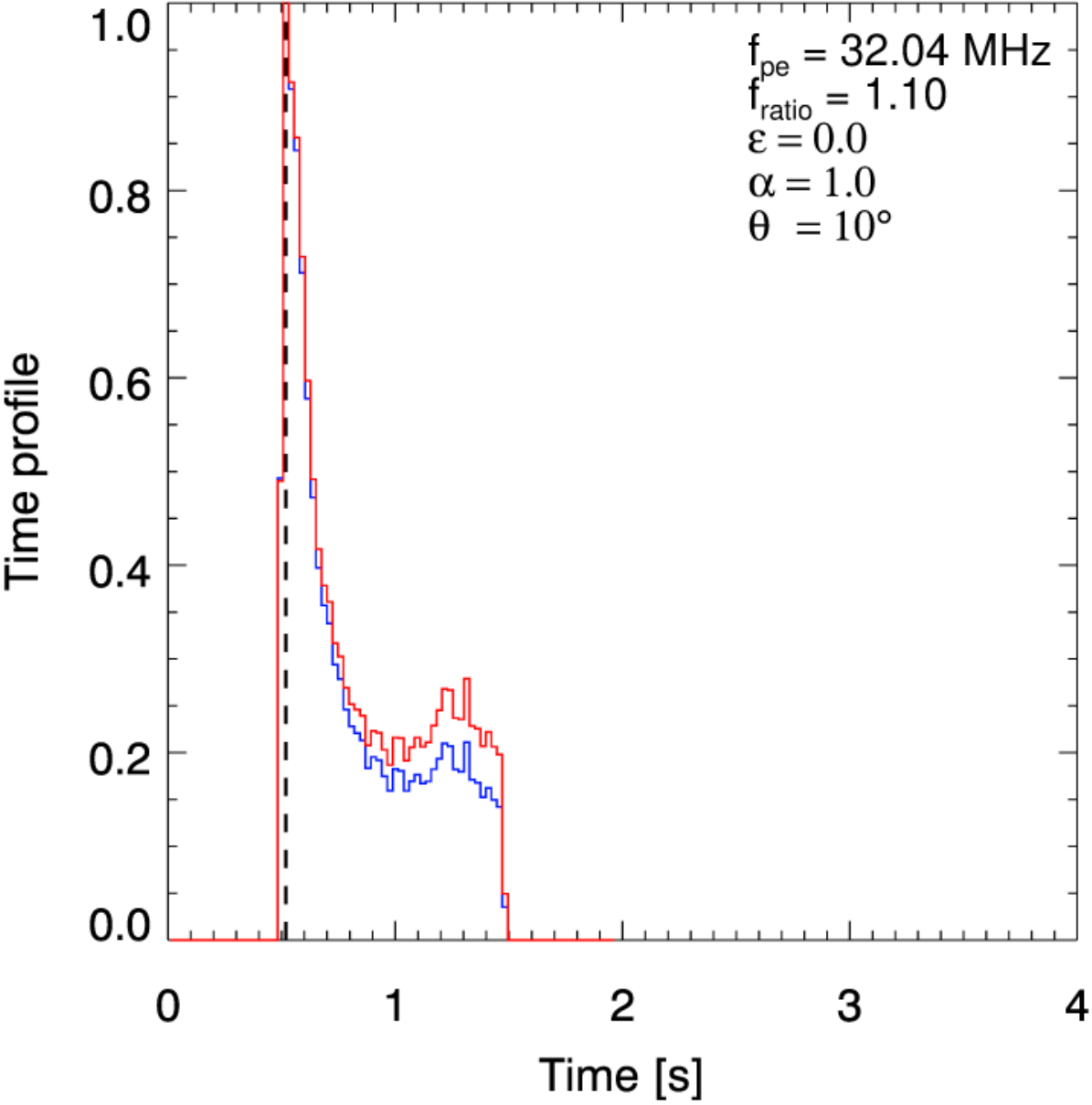}~
\includegraphics[width=0.49\linewidth]{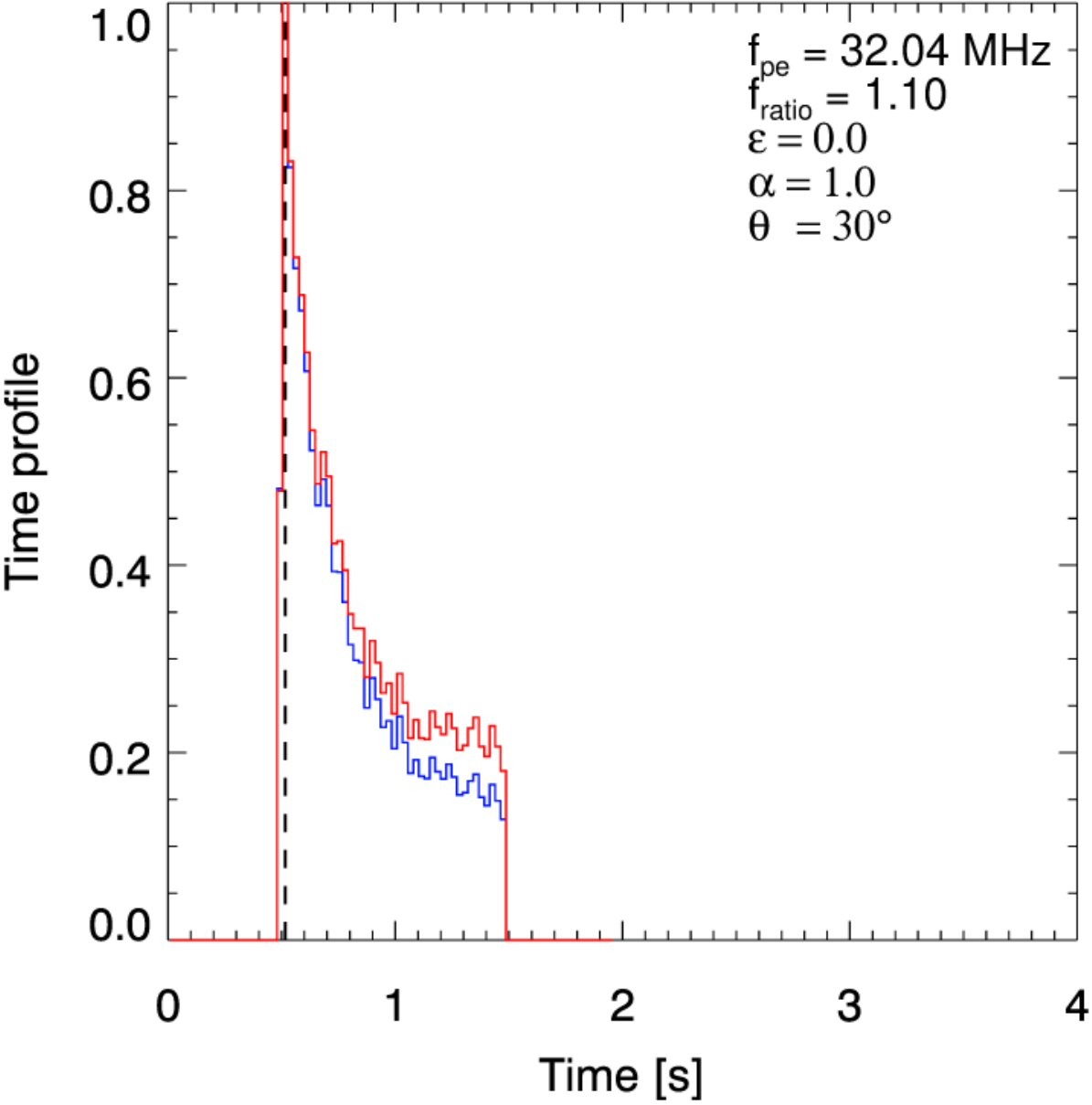}}
\centerline{\includegraphics[width=0.49\linewidth]{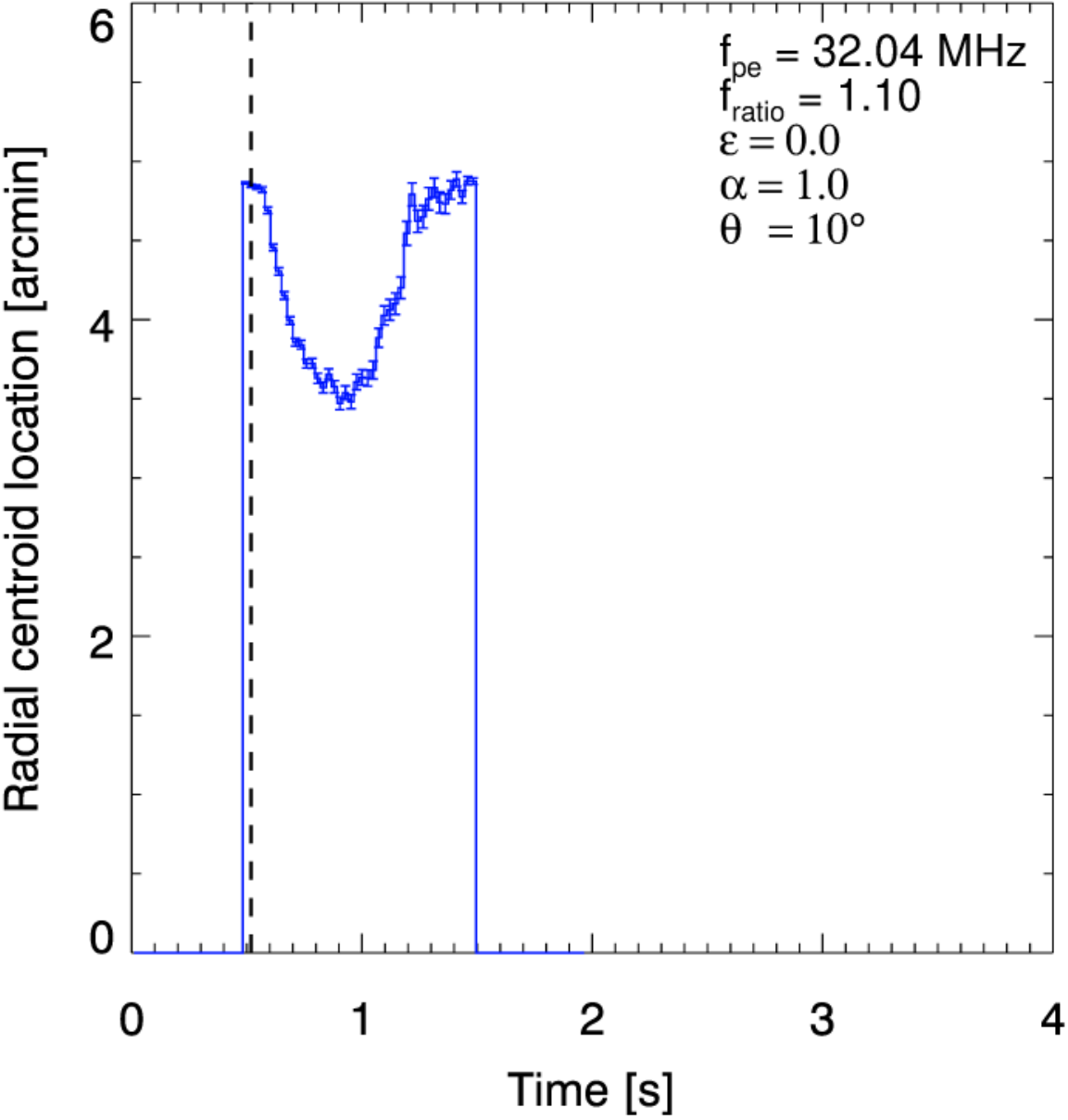}~
\includegraphics[width=0.49\linewidth]{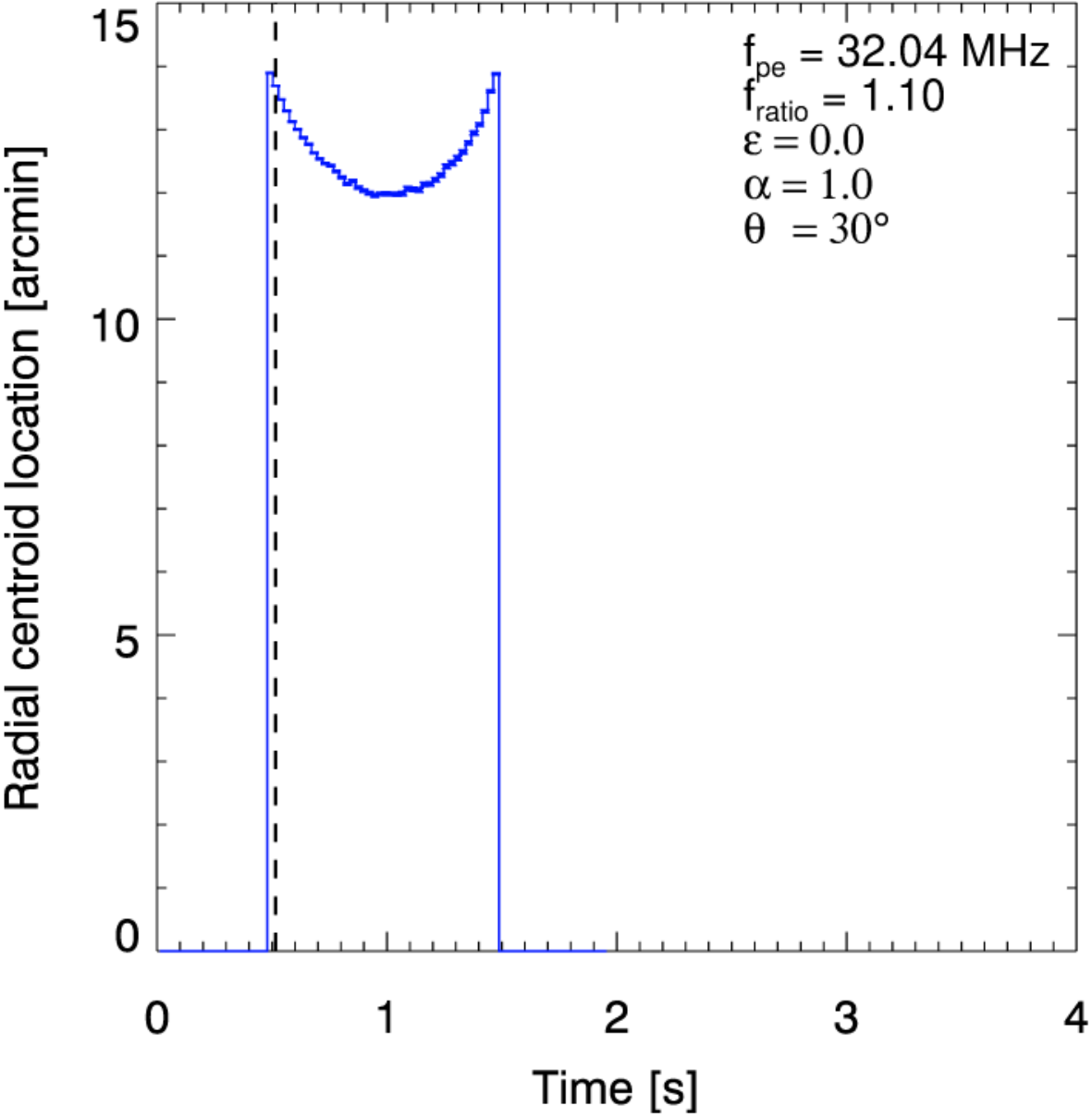}}
\caption[Simulation outputs assuming no scattering.]
{Simulated time profiles (top panels) and heliocentric source locations (bottom panels) for a point source emitting at 35.2~MHz (where $f/f_{pe}=1.10$) into a corona without small-scale density fluctuations ($\epsilon = 0$).  The left and right columns show results for $\theta_s=10\degr$ and 30$\degr$, respectively.  The time profiles are normalised with respect to the peak flux.  Red curves depict the apparent flux curve where no absorption is considered, whereas blue curves show the results that include the effects of collisional absorption.  Black dashed lines indicate the peak-flux time of the light curve and error bars represent the one-standard-deviation uncertainties.
Figure taken from \cite{2020ApJ...898...94K} and then adapted.
}

\label{fig:Kuznetsov2020_no_scattering}
\end{figure}

To visualise the impact of a scattering-free corona on the simulated Drift-pair properties, the simulations are first run assuming no density fluctuations, i.e. $\epsilon = 0$.  Hence, no anisotropy level can be considered (the lack of anisotropy is annotated as $\alpha = 1$ in Figure~\ref{fig:Kuznetsov2020_no_scattering}).  This implies that in the absence of scattering, the radio source properties are entirely determined by (large-scale) refraction and reflection.  Fundamental emissions are assumed, where the ratio between the emission and local plasma frequency $f/f_{pe} = 1.10$.  The simulated time profiles and source locations are obtained for angles $\theta_s = 10\degr$ and 30$\degr$, as illustrated by the left and right panels of Figure~\ref{fig:Kuznetsov2020_no_scattering}, respectively.

When the source is located closer to the solar centre ($\theta_s = 10\degr$), the intensity of the burst decays rapidly and a secondary peak---resulting due to the reflection of radio waves (i.e. the reflected component)---appears $\sim$0.7~s after the first.  When the source is located farther from the solar centre ($\theta_s = 30\degr$), the intensity decays more gradually, but no double-peak structure is present as the secondary peak is lost in the tail of the first component and cannot be distinguished.
It is also found that the absolute intensity of the peak decreases when the source is located at $\theta_s = 30\degr$, compared to its value when at $\theta_s = 10\degr$.
As evident, none of these time profiles correspond to the observed one (Figure~\ref{fig:Kuznetsov2020_lofar_obs}\hyperref[fig:Kuznetsov2020_lofar_obs]{b}), where the two components are clearly visible and are separated in time by $\sim$1.2~s.  The simulated radial source positions, however, show even greater discrepancies from the observed ones (Figure~\ref{fig:Kuznetsov2020_lofar_obs}\hyperref[fig:Kuznetsov2020_lofar_obs]{c}).  Neither $\theta_s = 10\degr$ nor $\theta_s = 30\degr$ produce heliocentric distances that are comparable to the observed, but also fail to reproduce the observed temporal evolution of the apparent sources.  However, the centroid locations of the direct and reflected components for $\theta_s = 10\degr$ virtually coincide (at the peak-flux time), which agrees with the observed behaviour.
This is not surprising given that for sources emitting near $\sim35$~MHz where $f/f_{pe} = 1.10$ (fundamental), the projected distance (in arcminutes) between the location of radio-wave excitation and the nearest reflection region is $\sim$\vphantom{}$0.75 \, \sin \theta_s$ (i.e. $< 1\arcmin$) and decreases with decreasing angle $\theta_s$.  Moreover, the simulated source areas (for $\epsilon=0$) were found to be much smaller than those observed, as they did not exceed 5~$\mathrm{arcmin^{2}}$ (cf. Figure~\ref{fig:Kuznetsov2020_lofar_obs}).
It should be reiterated that the simulations suggest a displacement of the centroid location with time, despite that a fixed intrinsic source position and a fixed emission frequency are assumed.  As explained in Section~\ref{sec:typeIIIb_isotropic_F_sims}, this occurs due to the fact that the sub-second evolution of the radio sources is probed, and as photons arrive at the detector at different times (due to radio-wave propagation effects), the estimated source properties---like the position and area---change.  This effect is also visible in sub-second observations, like those of LOFAR presented in this chapter.

Overall, it is evident that a medium in which small-scale density fluctuations (and thus scattering) are absent---and only refraction and reflection effects are present---cannot account for the observed properties of Drift-pair bursts.

\subsection{Quantitative generation of radio echoes and the need for anisotropic scattering} \label{sec:driftpairs_radio_echo}

\begin{figure}[htp!]

\centerline{\includegraphics[width=0.322\textwidth]{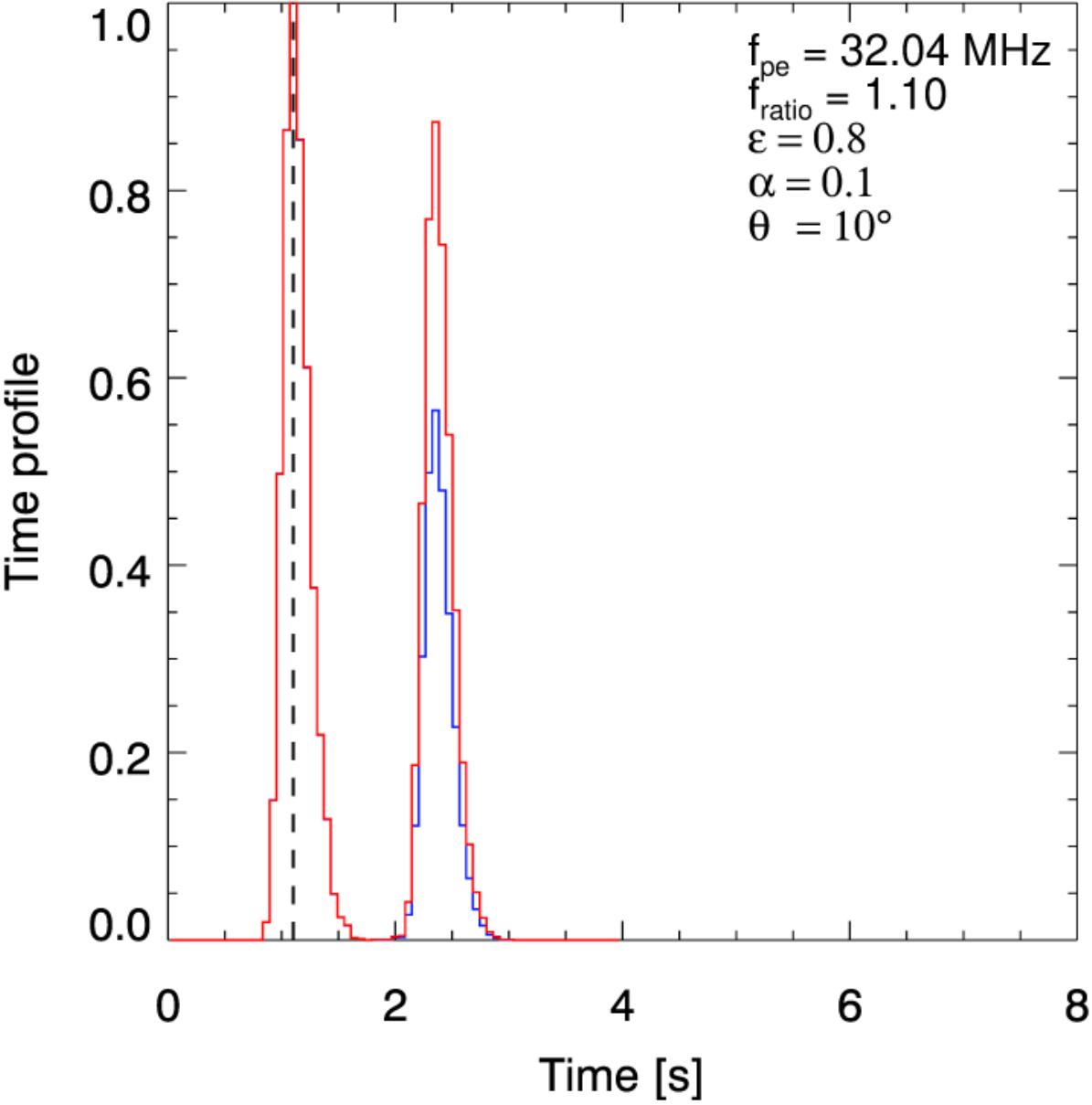}~
\includegraphics[width=0.322\textwidth]{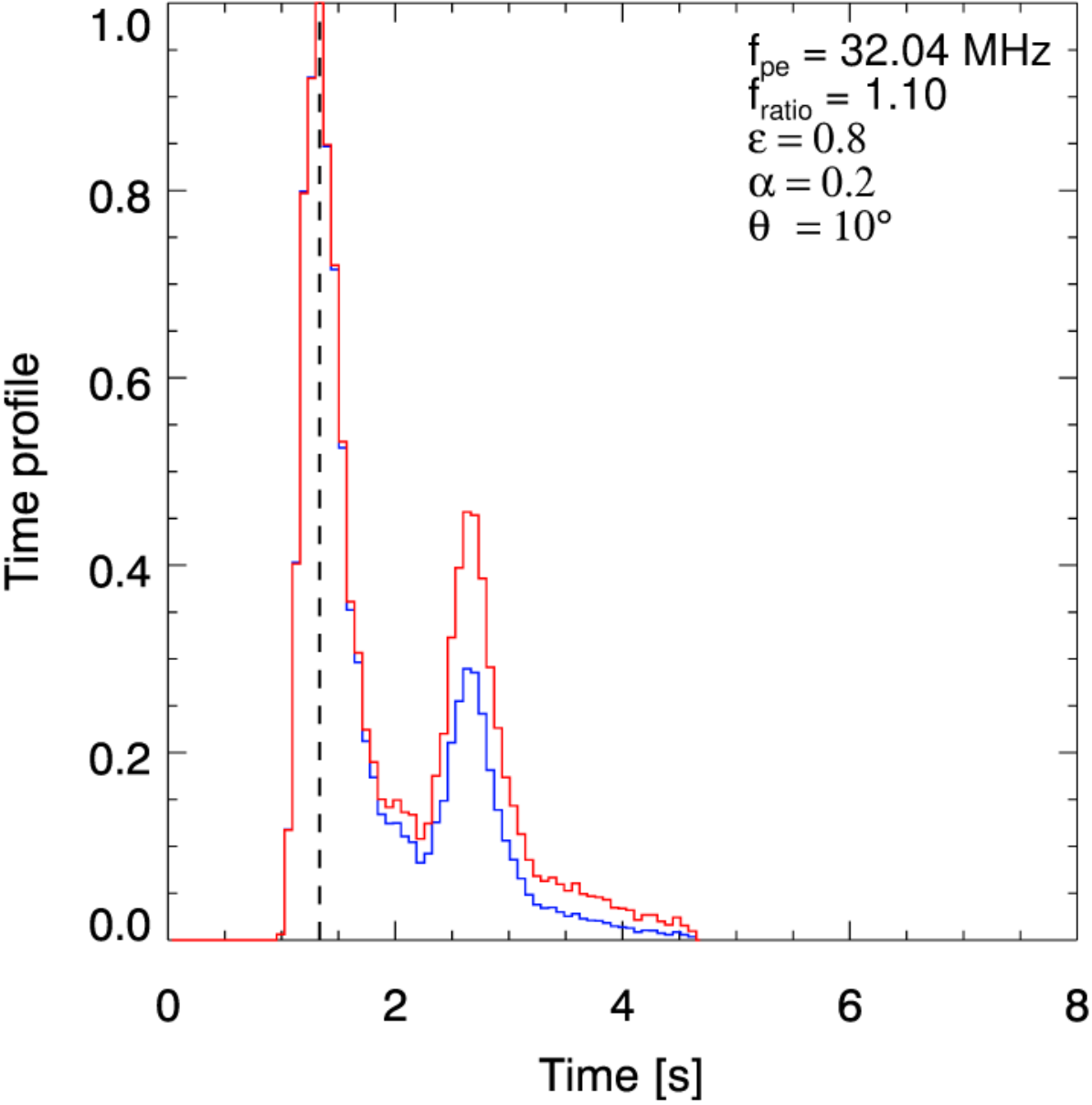}~
\includegraphics[width=0.322\textwidth]{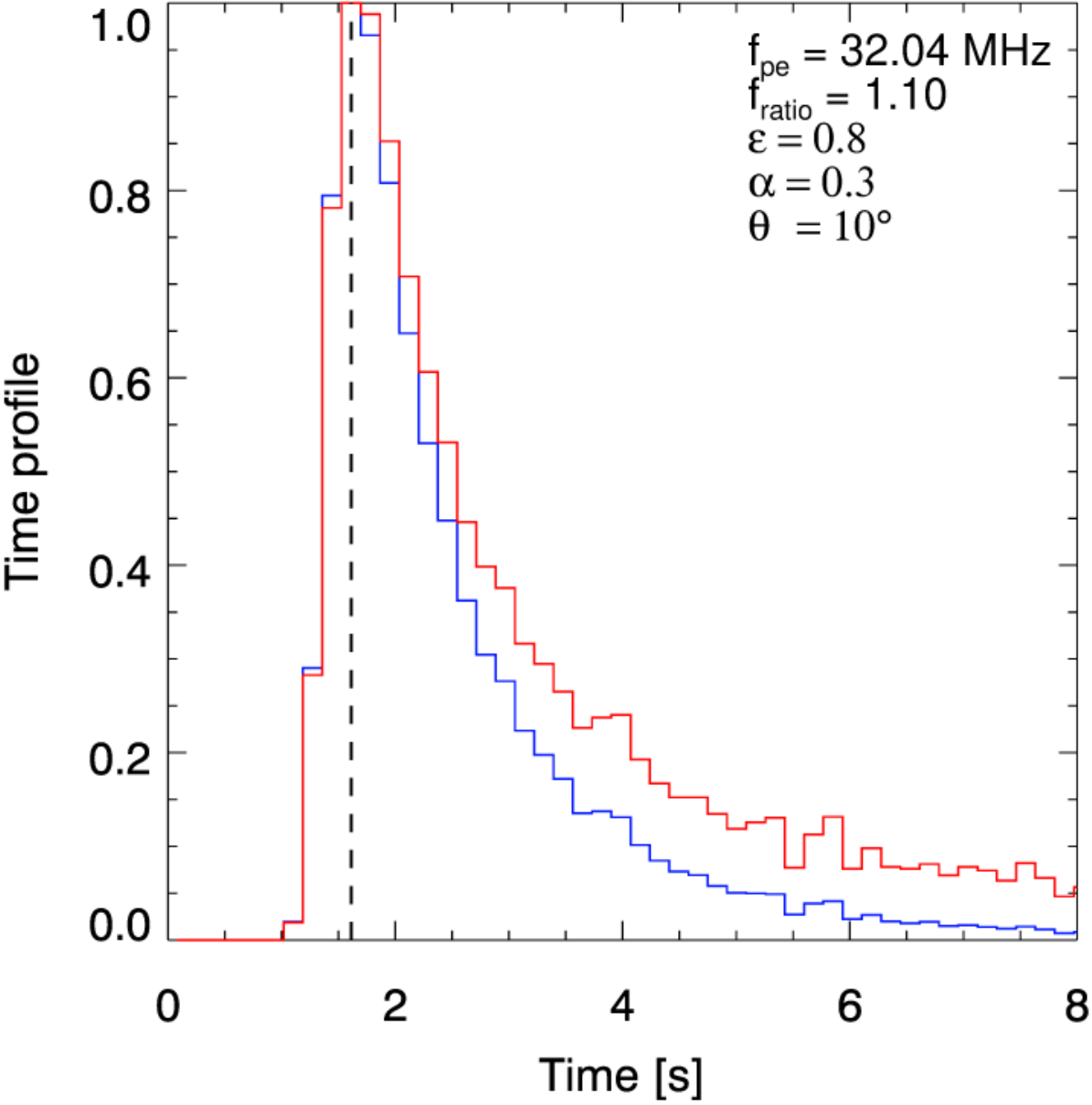}}
\centerline{\includegraphics[width=0.322\textwidth]{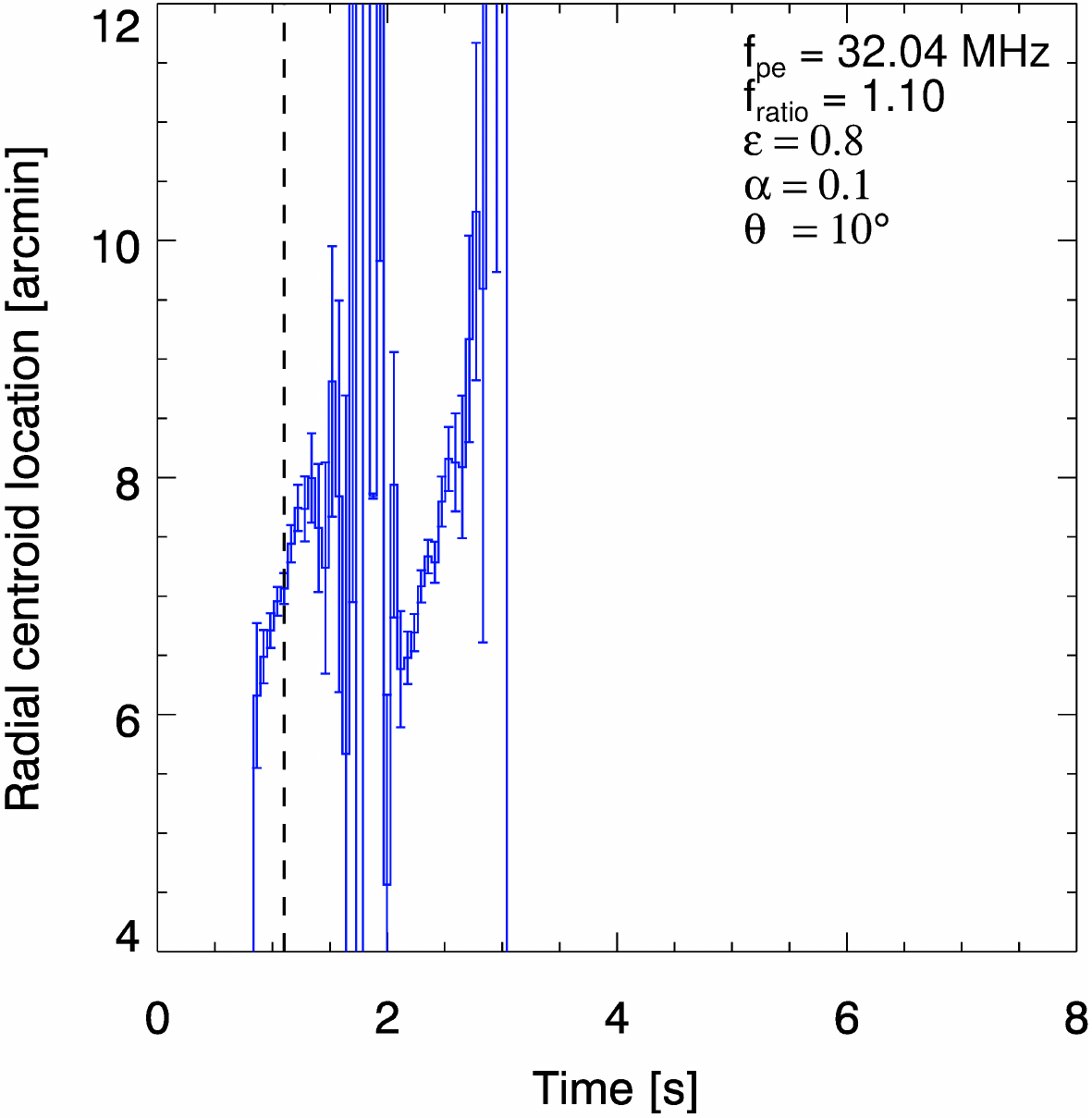}~
\includegraphics[width=0.322\textwidth]{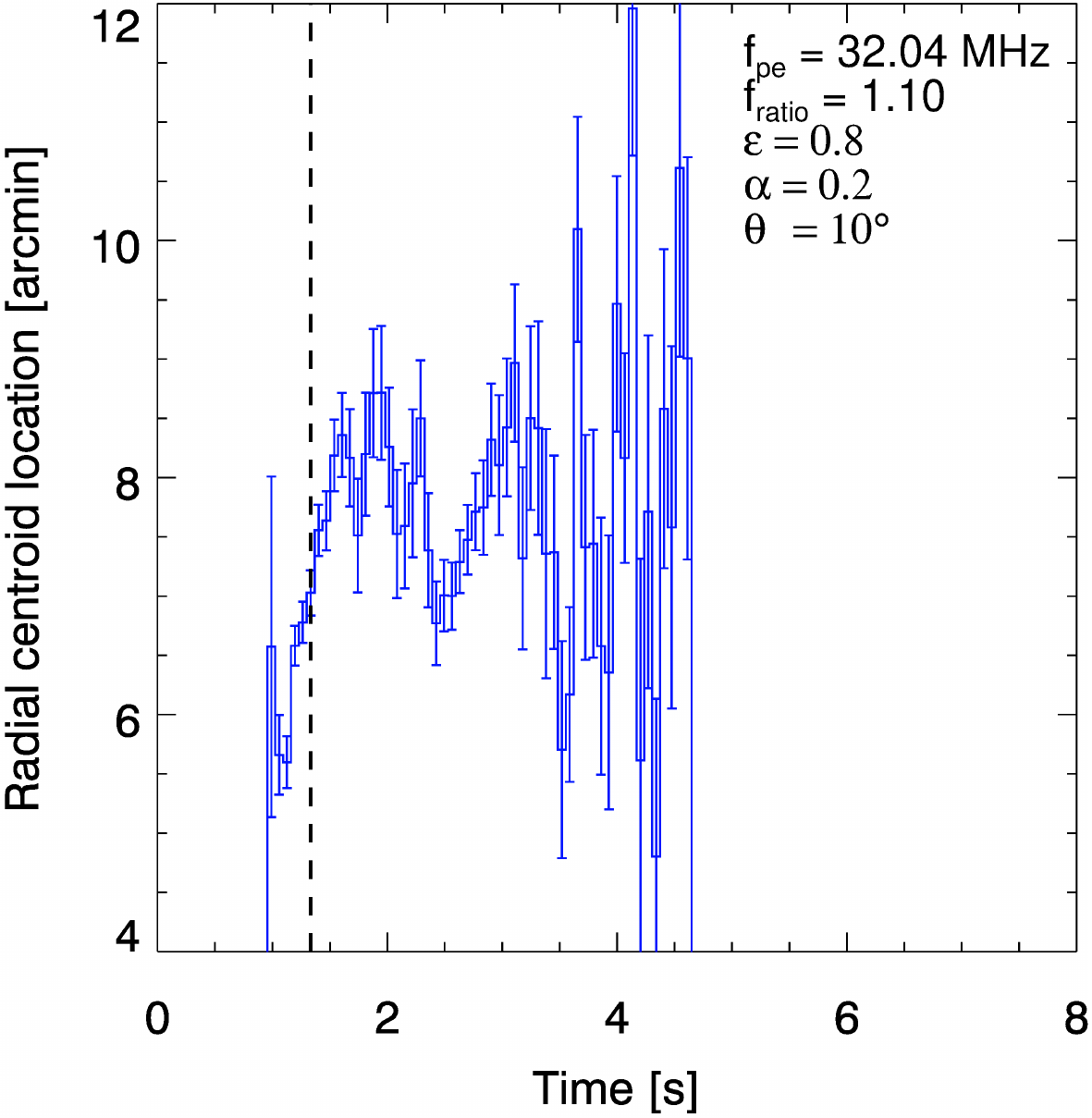}~
\includegraphics[width=0.322\textwidth]{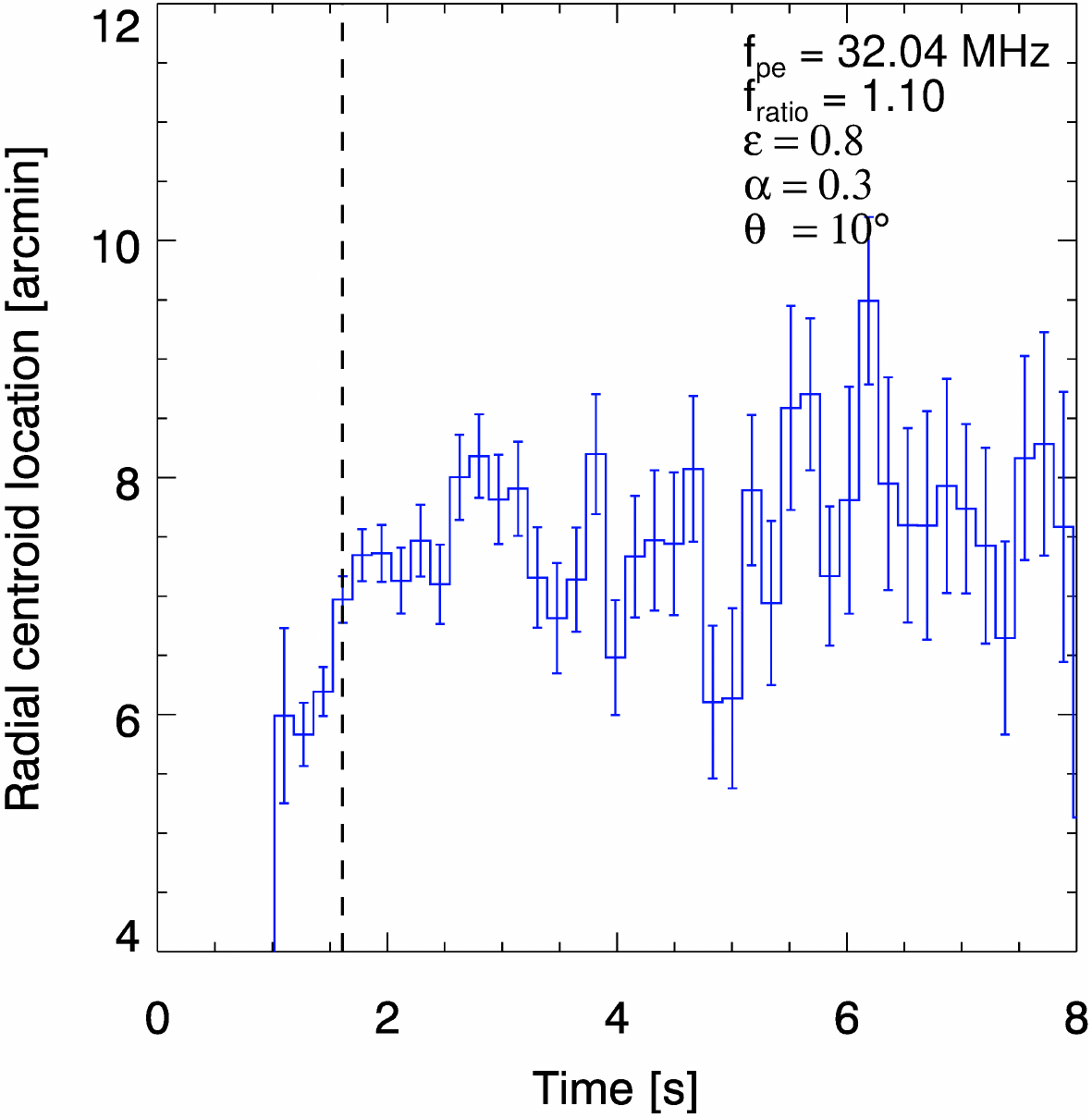}}
\centerline{\includegraphics[width=0.322\textwidth]{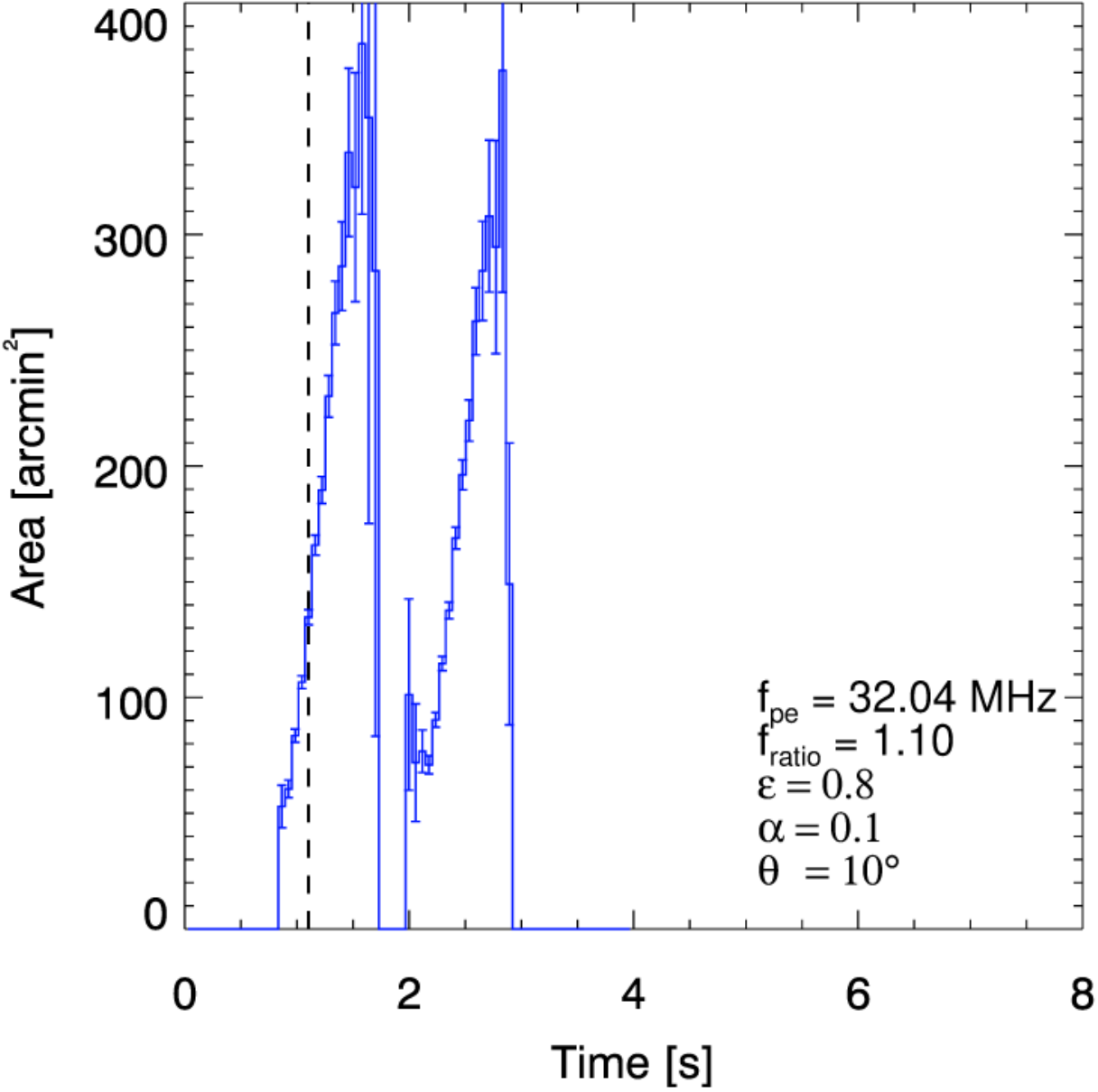}~
\includegraphics[width=0.322\textwidth]{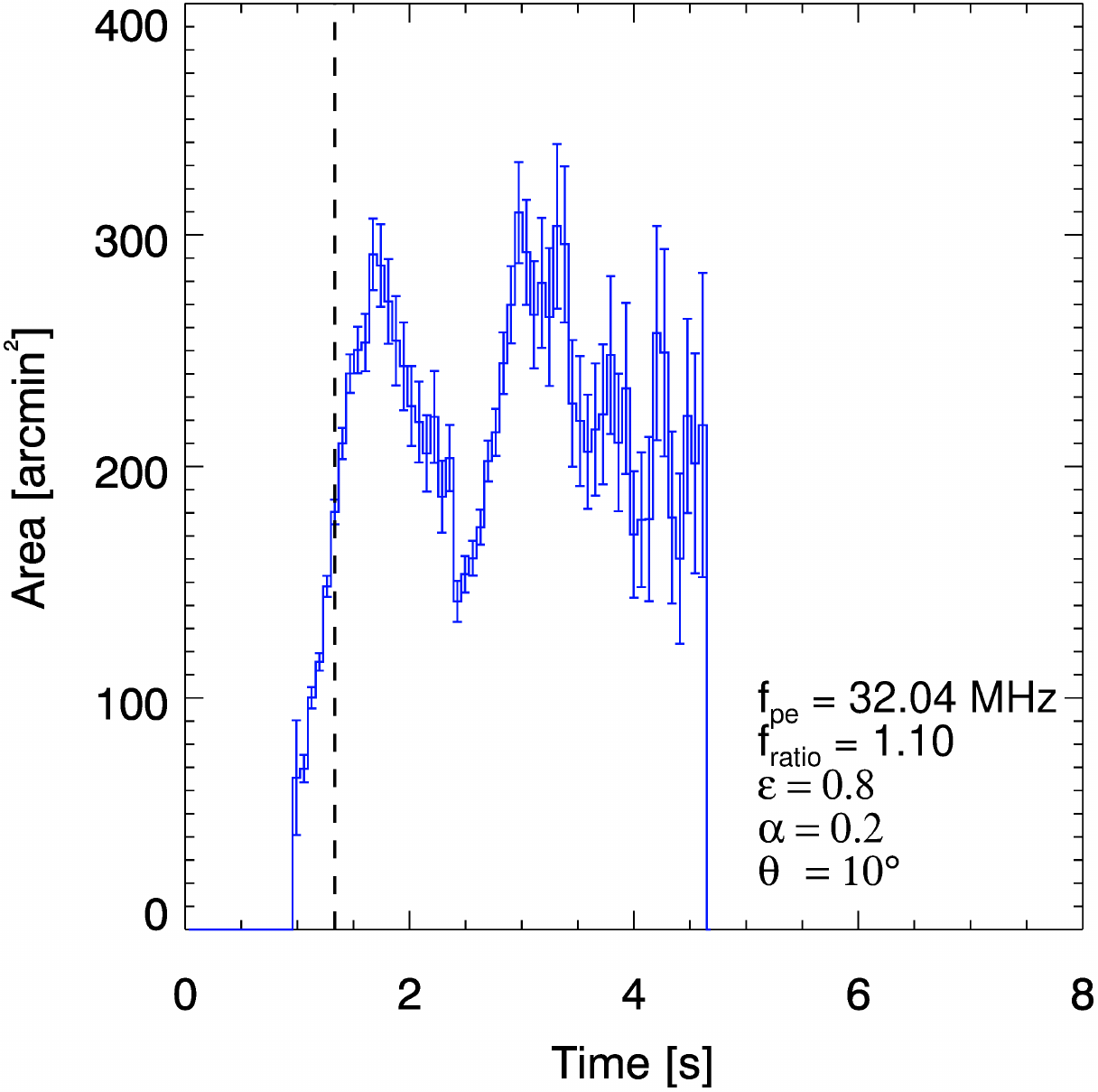}~
\includegraphics[width=0.322\textwidth]{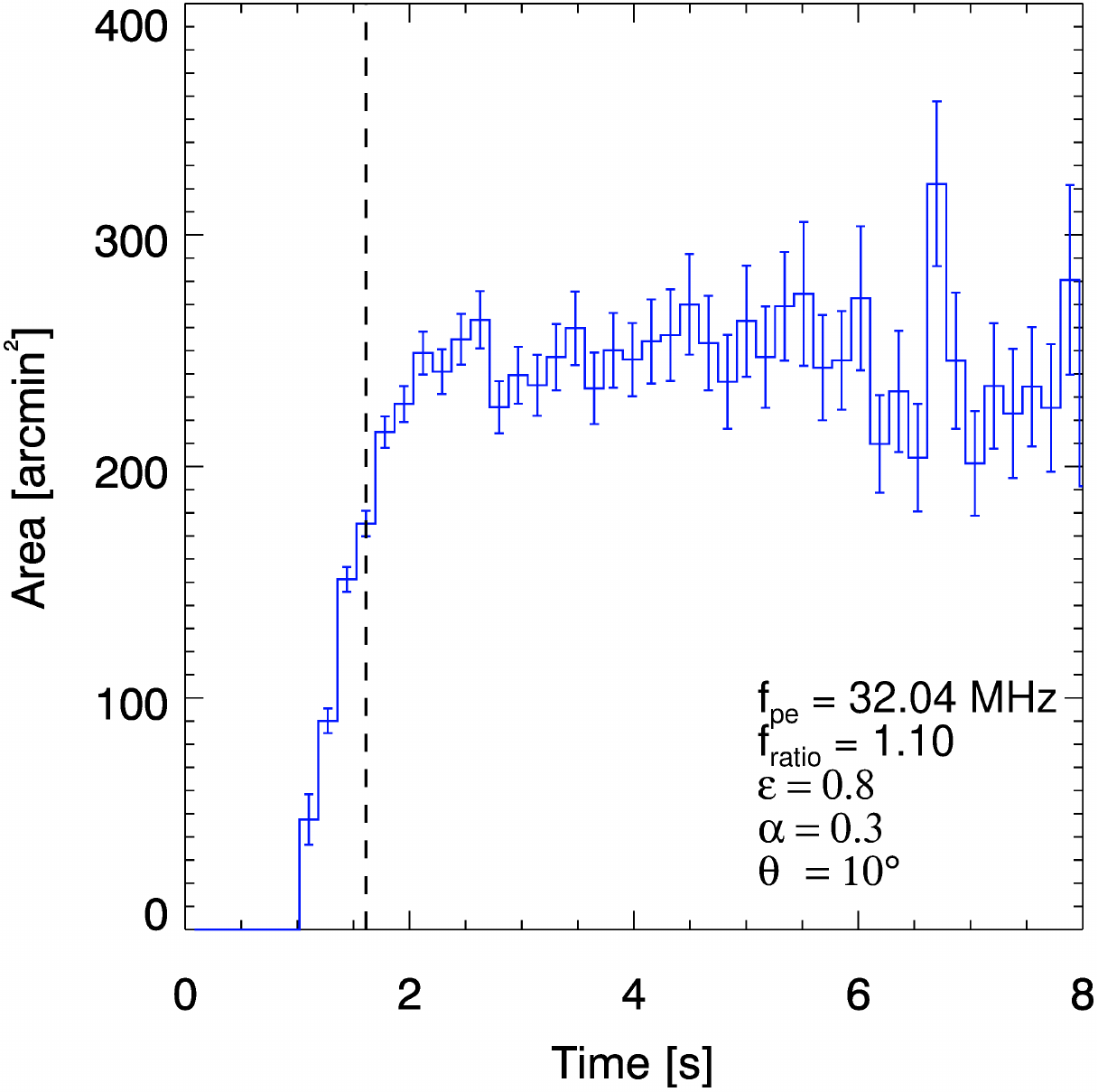}}
\caption[Simulation outputs for varying anisotropy levels.]
{Simulated properties for a source emitting at 35.2~MHz (where $f/f_{pe}=1.10$), assuming $\epsilon=0.8$, $\theta_s = 10\degr$, and levels of anisotropy $\alpha=0.1$ (left column), $0.2$ (middle column), and \textbf{$0.3$} (right column).  Top row: normalised time profiles (with respect to the peak-flux value), where the blue and red lines represent the apparent emission with and without collisional absorption, respectively.  Middle row: projected (in the plane of the sky) heliocentric source locations.  Bottom row: apparent (FWHM) source areas.  Error bars represent the one-standard-deviation uncertainty and black dashed lines indicate the peak-flux time.
Figure taken from \cite{2020ApJ...898...94K} and then adapted.
}

\label{fig:Kuznetsov2020_anisotropy_effects}
\end{figure}

Given the inability of the scattering-free coronal medium ($\epsilon = 0$) to reproduce the observed Drift-pair properties, a level of density fluctuations is introduced in order to compare simulations accounting for scattering to the observations.  The level of density fluctuations is defined as $\epsilon = 0.8$, fundamental emissions at $f = 35.2$~MHz where $f = 1.10 \, f_{pe}$ are taken, the source-polar angle is set as $\theta_s = 10\degr$, and the level of anisotropy is varied between $\alpha = 0.1$, 0.2, and 0.3.  Figure~\ref{fig:Kuznetsov2020_anisotropy_effects} illustrates the simulation outputs where the left column depicts the source properties for $\alpha=0.1$, the middle column for $\alpha=0.2$, and the right column for $\alpha=0.3$.  From top to bottom, the rows depict the obtained time profiles, the source's heliocentric motion, and the source's areal evolution.  It can be seen that the higher the anisotropy level (with $\alpha=0.1$ being the highest shown), the more pronounced the double-peak structure becomes---for all source properties.

For $\alpha=0.1$, the time profile indicates that the reflected component is delayed by $\sim$1.25~s with respect to the direct one (from peak to peak), and that both components have a short FWHM duration, estimated to be $\sim$0.5~s.  The amplitude of the reflected component is lower than that of the direct one, even when absorption is ignored in the simulations (red curve).  The overall shape of the simulated time profile (considering absorption; blue line) agrees with the observed time profile (Figure~\ref{fig:Kuznetsov2020_lofar_obs}\hyperref[fig:Kuznetsov2020_lofar_obs]{b}).  The source locations of both components coincide, with both being found at a heliocentric distance of $\sim$7.0$\arcmin$ at the peak-flux time.  The sources also move away from the Sun at a rate of $\sim$4.0~$\mathrm{arcmin \, s^{-1}}$, which is higher than the observed speed of $\sim$2.2~$\mathrm{arcmin \, s^{-1}}$.  Both components are found to have nearly-identical source areas, being $\sim$140~$\mathrm{arcmin^{2}}$ at the peak-flux time, while they expand at a rate of $\sim$520~$\mathrm{arcmin^{2} \, s^{-1}}$.  The simulated source size agrees well with the observed size of $\sim$150~$\mathrm{arcmin^{2}}$, once the observations are deconvolved from the LOFAR beam which has an area $A_{beam} = 100 \, \mathrm{arcmin^{2}}$ (i.e. the simulations are compared to $A_{real}$; see Section~\ref{sec:driftpairs_LOFAR_properties}).
However, similar to the simulated source speed, the source's simulated expansion rate of 520~$\mathrm{arcmin^{2} \, s^{-1}}$ is considerably higher than the value of 30~$\mathrm{arcmin^{2} \, s^{-1}}$ inferred from the observations (Section~\ref{sec:driftpairs_LOFAR_properties}).  The discrepancy between the observed and simulated speeds and expansion rates can be attributed to the fact that the values obtained from the LOFAR observations represent a combined source that includes the contributions of a variable bursty signal and a background continuum.  In other words, the observations are the weighted average of the locations and sizes of the corresponding sources, leading to reduced variation rates of the source parameters compared to what the simulations suggest.

When the level of anisotropy is decreased to $\alpha=0.2$ (middle column of Figure~\ref{fig:Kuznetsov2020_anisotropy_effects}), the peaks of both components become broader---having a FWHM duration of $\sim$0.7~s (as opposed to $\sim$0.5~s for $\alpha=0.1$)---and the delay between the two components becomes sightly longer ($\sim$1.30~s instead of $\sim$1.25~s).  The most striking impact on the time profile, however, is the considerable decrease in relative amplitude of the reflected component.  The apparent source positions, on the other hand, are unaffected.  The centroid locations of both components remain at $\sim$7.0$\arcmin$ from the solar centre (at the peak-flux time).  The simulated source areas demonstrate a small increase from $\sim$140~$\mathrm{arcmin^{2}}$ to $\sim$150~$\mathrm{arcmin^{2}}$ (for $\alpha=0.1$ and 0.2, respectively), whereas the expansion rate decreases considerably to a value of $\sim$370~$\mathrm{arcmin^{2} \, s^{-1}}$ (from $\sim$520~$\mathrm{arcmin^{2} \, s^{-1}}$).

When the anisotropy is decreased even further to $\alpha=0.3$ (right column of Figure~\ref{fig:Kuznetsov2020_anisotropy_effects}), the differences become even more prominent.  The time profile of each component becomes so broad that the contribution of the reflected emissions is completely engulfed in the tail of the direct emissions and cannot be distinguished.  Moreover, the simulated source area at the peak-flux time increases to $\sim$180~$\mathrm{arcmin^{2}}$, which is too large compared to the observed size of $\sim$150~$\mathrm{arcmin^{2}}$.

It is also found that, as the level of anisotropy becomes weaker, the peak intensity of the primary component is observed at a later time, and the absolute value of the primary component's peak intensity decreases.  The simulated attenuation of the signal (both for the primary and reflected components), the delay in the peak intensity's arrival time, the time-broadening of the components, and the subsequent increase in the delay between the two components, are all consequences of the weaker anisotropy.  Specifically, weaker anisotropy corresponds to less directional emissions, implying that the photons' path is less restricted and so photons spend more time in the corona before they reach the observer, which also contributes to the attenuation (through free-free absorption) of the signal.  The relative intensity of the reflected component is affected to a large degree, since the reflected photons travel an additional distance (compared to the direct emissions) before they reach the observer, thus spending more time in the collisional coronal medium.

Overall, the simulated properties suggest that the formation of Drift-pair bursts requires density fluctuations that are characterised by significantly-strong anisotropy levels, favouring values of $\alpha \lesssim 0.1$--0.2.

\subsection{Dependence of properties on centre-to-limb variations} \label{sec:driftpairs_centre_to_limb_dependence}
The dependence of source properties on the polar angle is also probed.  Simulations are run assuming $f = 35.2$~MHz, $f/f_{pe} = 1.10$, $\epsilon = 0.8$, $\alpha = 0.1$, and a source-polar angle that varies between $\theta_s = 10\degr$, 30$\degr$, and 50$\degr$.

\begin{figure}[ht!]

\centerline{\includegraphics[width=0.322\textwidth]{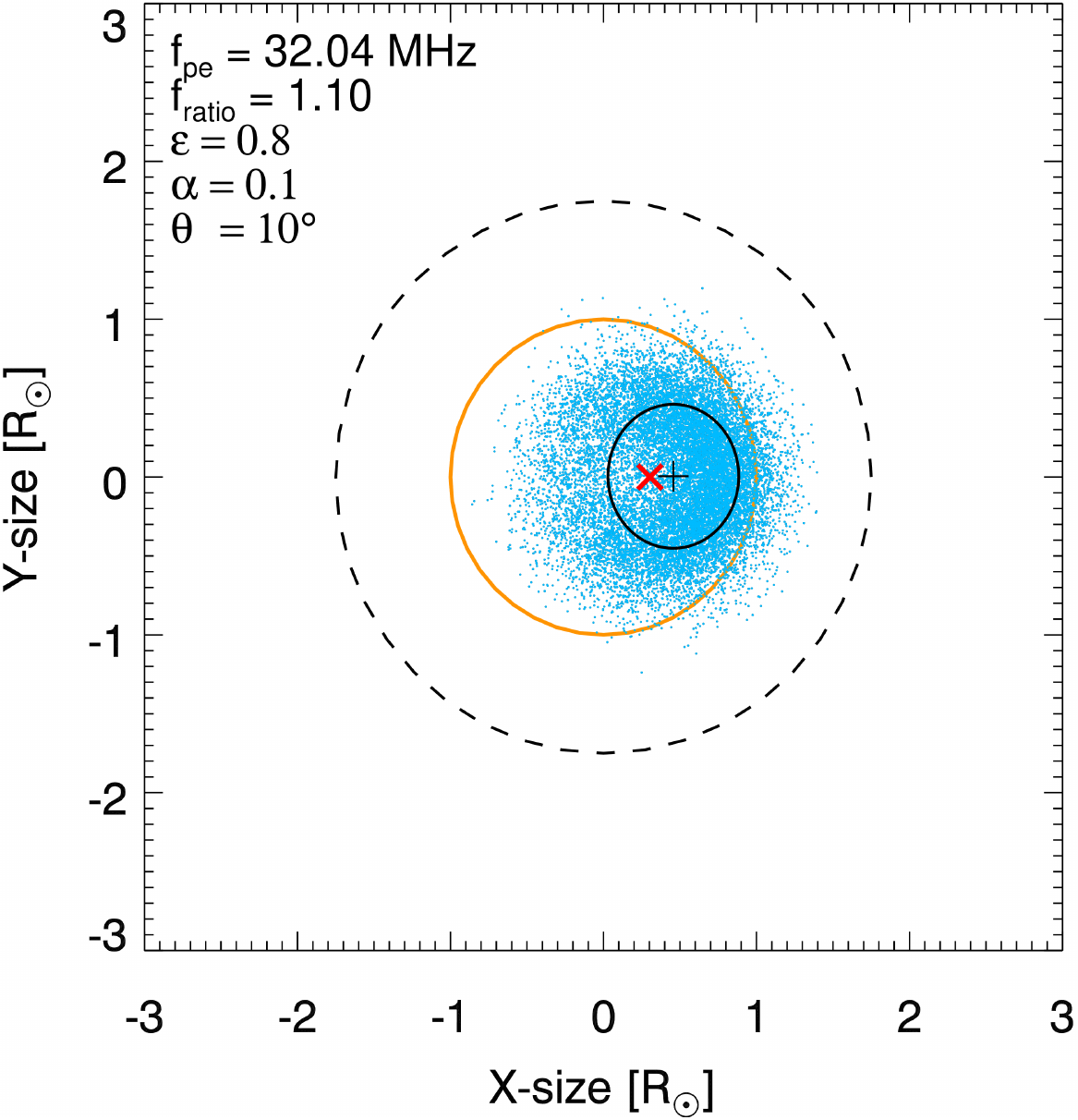}~
\includegraphics[width=0.322\textwidth]{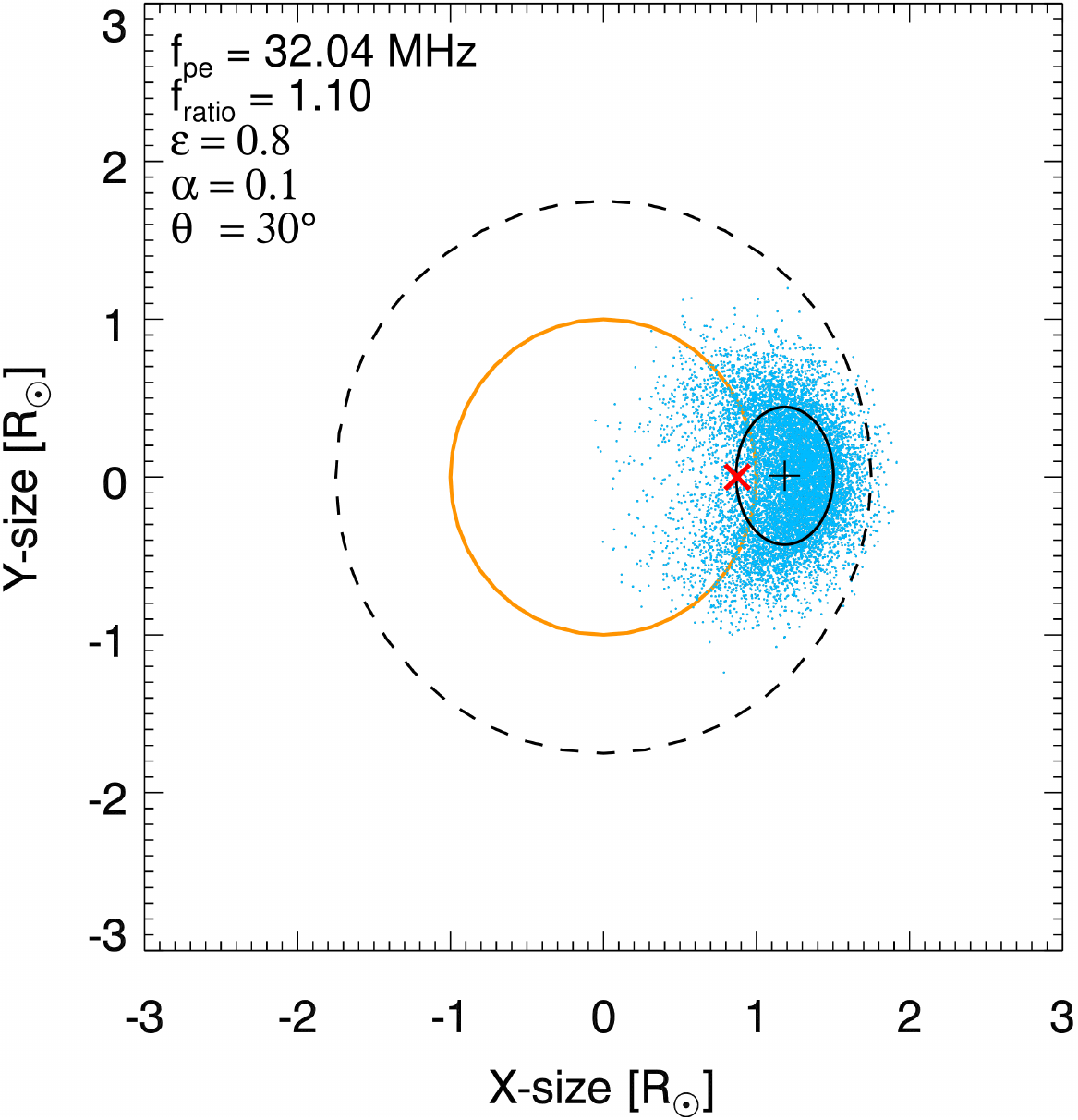}~
\includegraphics[width=0.322\textwidth]{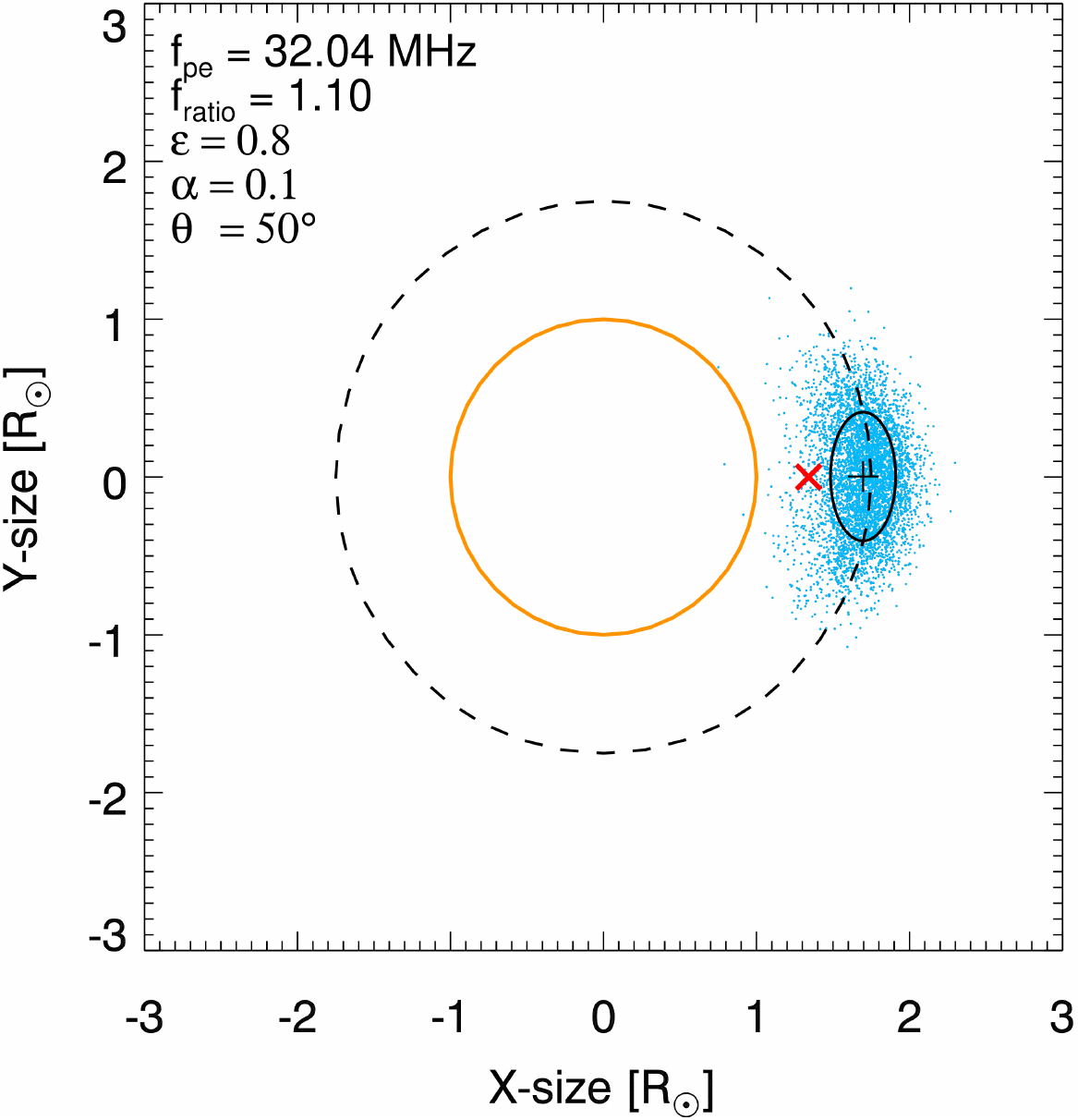}}
\caption[Dependence of the simulated emission images on the source-polar angle.]
{Simulated radio images of a source emitting at 35.2~MHz (where $f/f_{pe}=1.10$), assuming $\epsilon = 0.8$, $\alpha=0.1$, and a source-polar angle $\theta_s = 10\degr$ (left), 30$\degr$ (middle), and 50$\degr$ (right).  Each blue dot represents a photon, the black dashed circle indicates the projected (in the plane of the sky) heliocentric location of the intrinsic source, and the orange circle indicates the solar limb.  The source's FWHM area is illustrated by the black ellipse, whereas the (projected) intrinsic and apparent centroid locations are indicated by the red cross and black plus sign, respectively.
Figure taken from \cite{2020ApJ...898...94K} and then adapted.
}

\label{fig:Kuznetsov2020_source_vs_angle}
\end{figure}

The effect of the source-polar angle on the simulated radio images is illustrated in Figure~\ref{fig:Kuznetsov2020_source_vs_angle}.  Images for angles $\theta_s = 10\degr$, 30$\degr$, and 50$\degr$ are shown from left to right, respectively.  The images represent the time-integrated brightness distribution ($I(x, y)$; see Section~\ref{sec:anisotropic_simulations_2019}), which includes both the direct and reflected components.  As anticipated, the observed source centroid (black plus sign) is found farther away from the intrinsic source location (red cross) and the solar centre, due to the impact of radio-wave propagation effects (predominantly scattering) on the emitted photons.

\begin{figure}[t!]

\centerline{\includegraphics[width=0.322\textwidth]{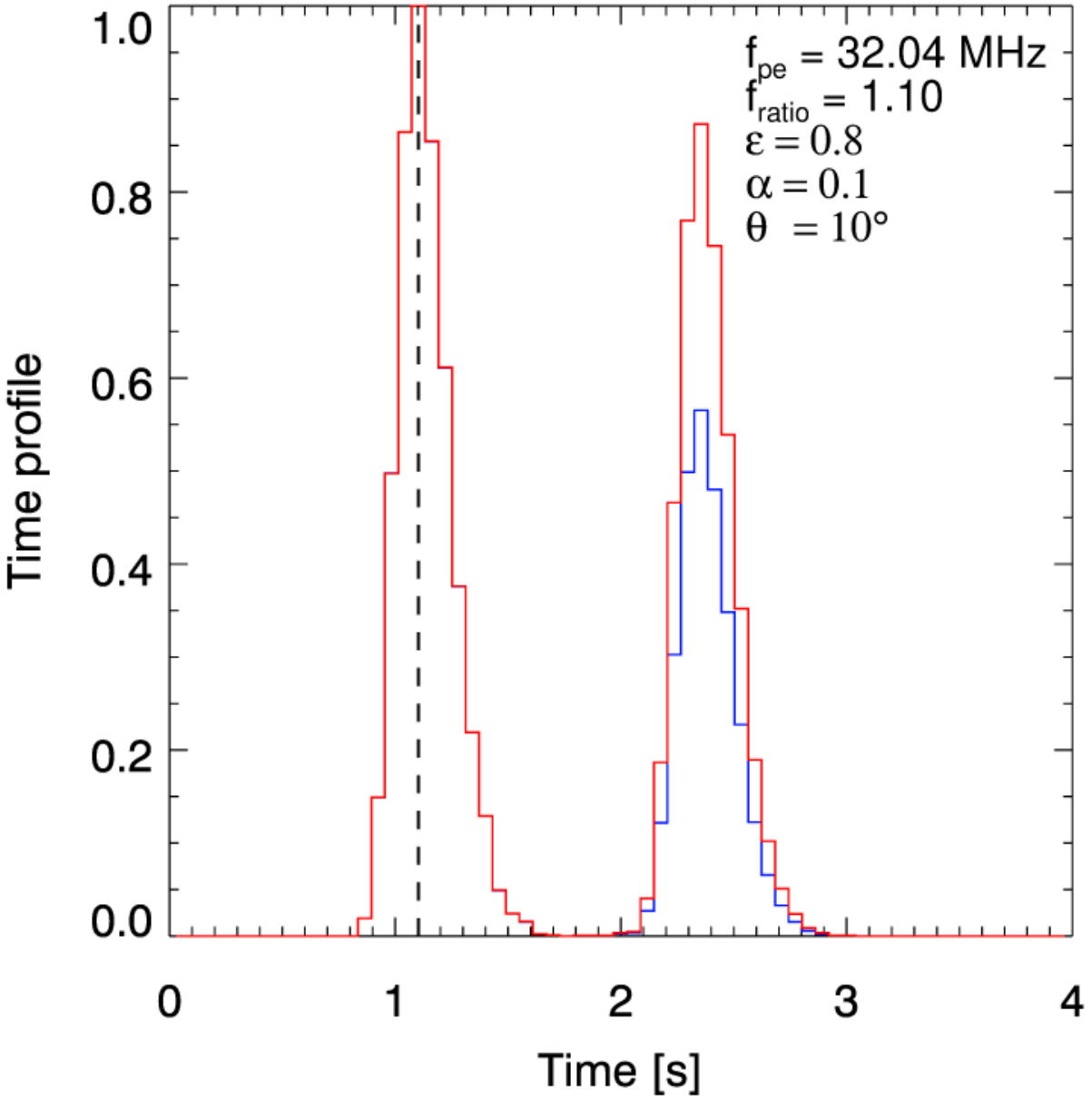}~
\includegraphics[width=0.322\linewidth]{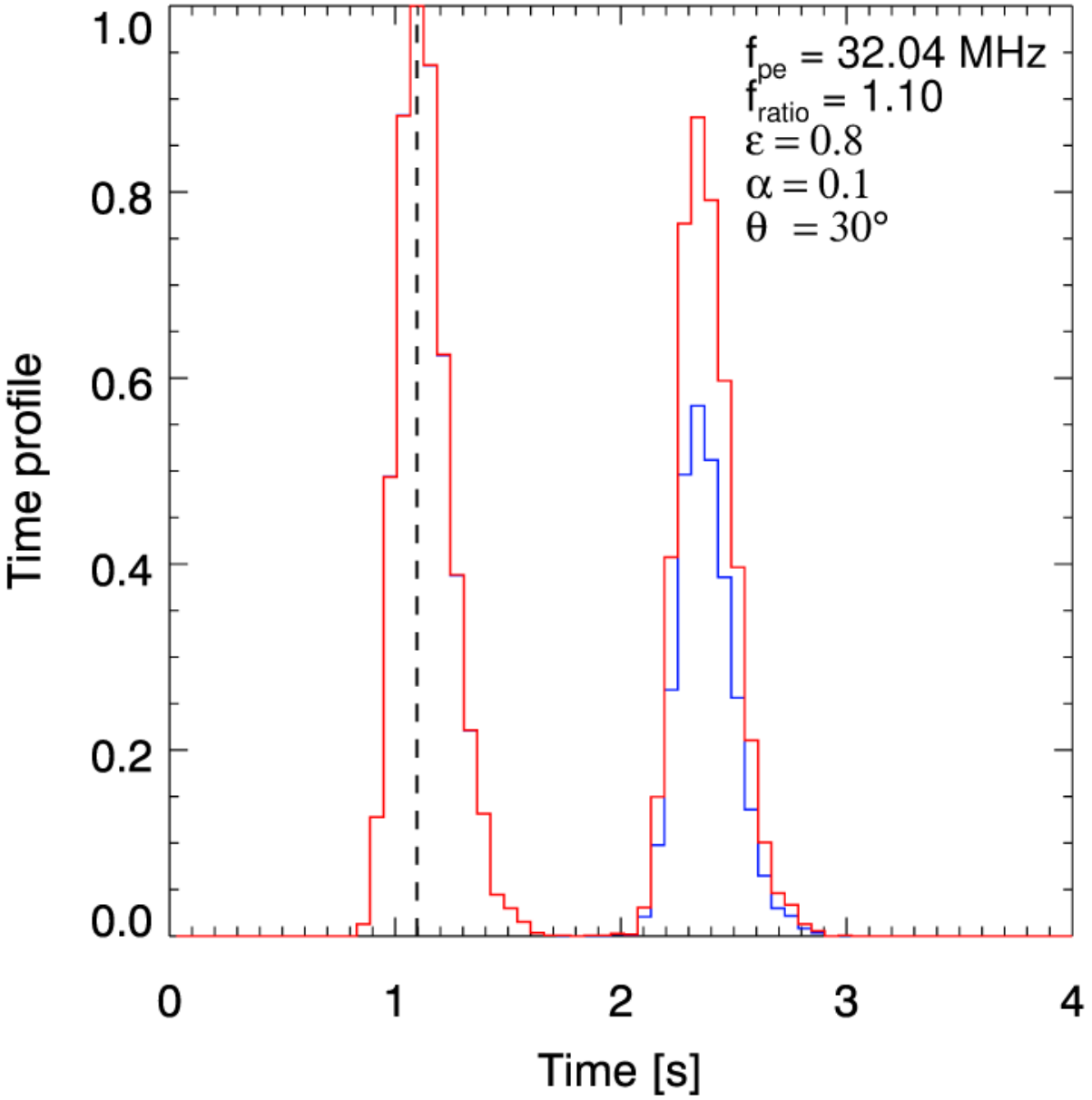}
\includegraphics[width=0.322\linewidth]{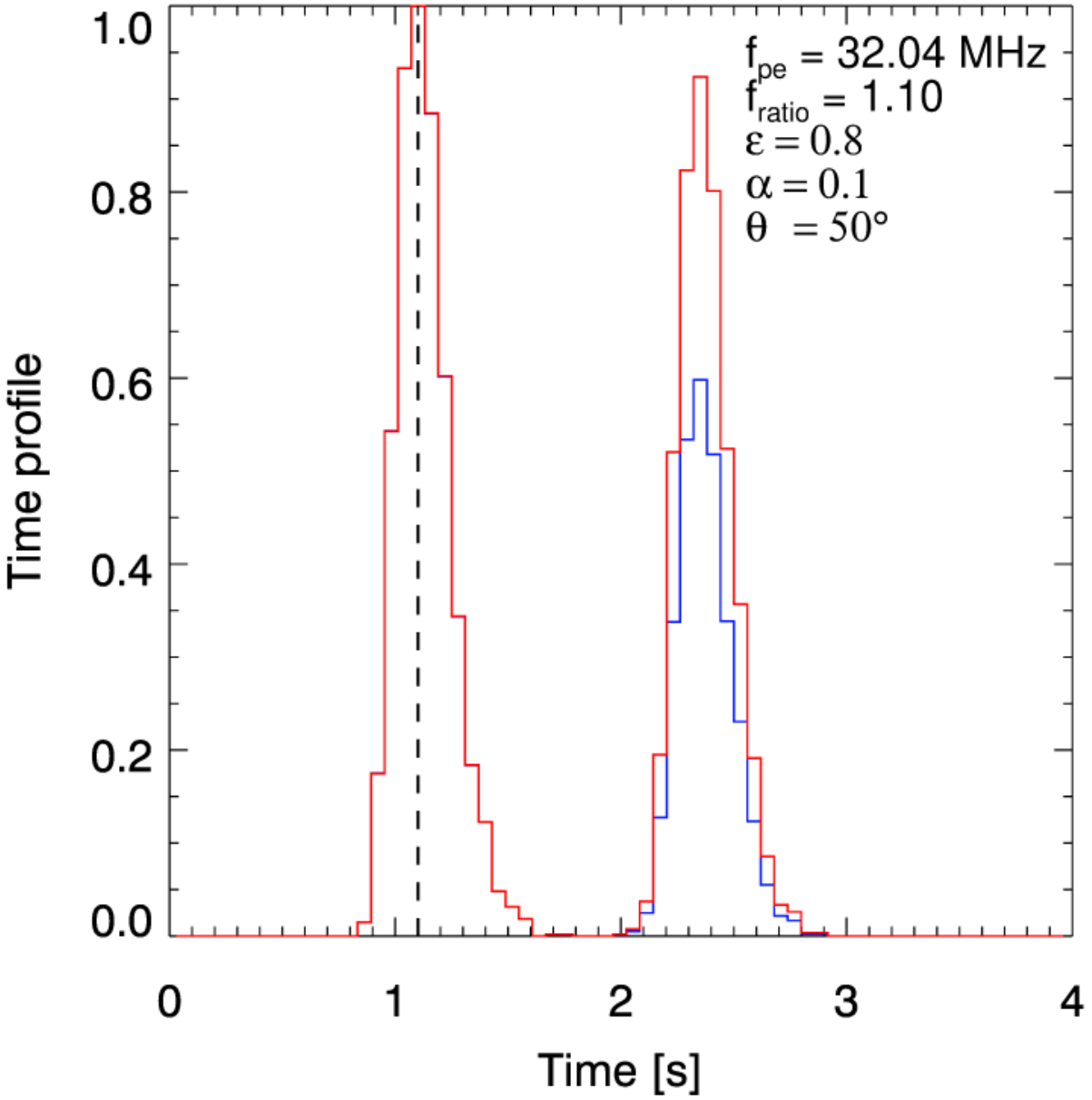}}
\centerline{\includegraphics[width=0.322\textwidth]{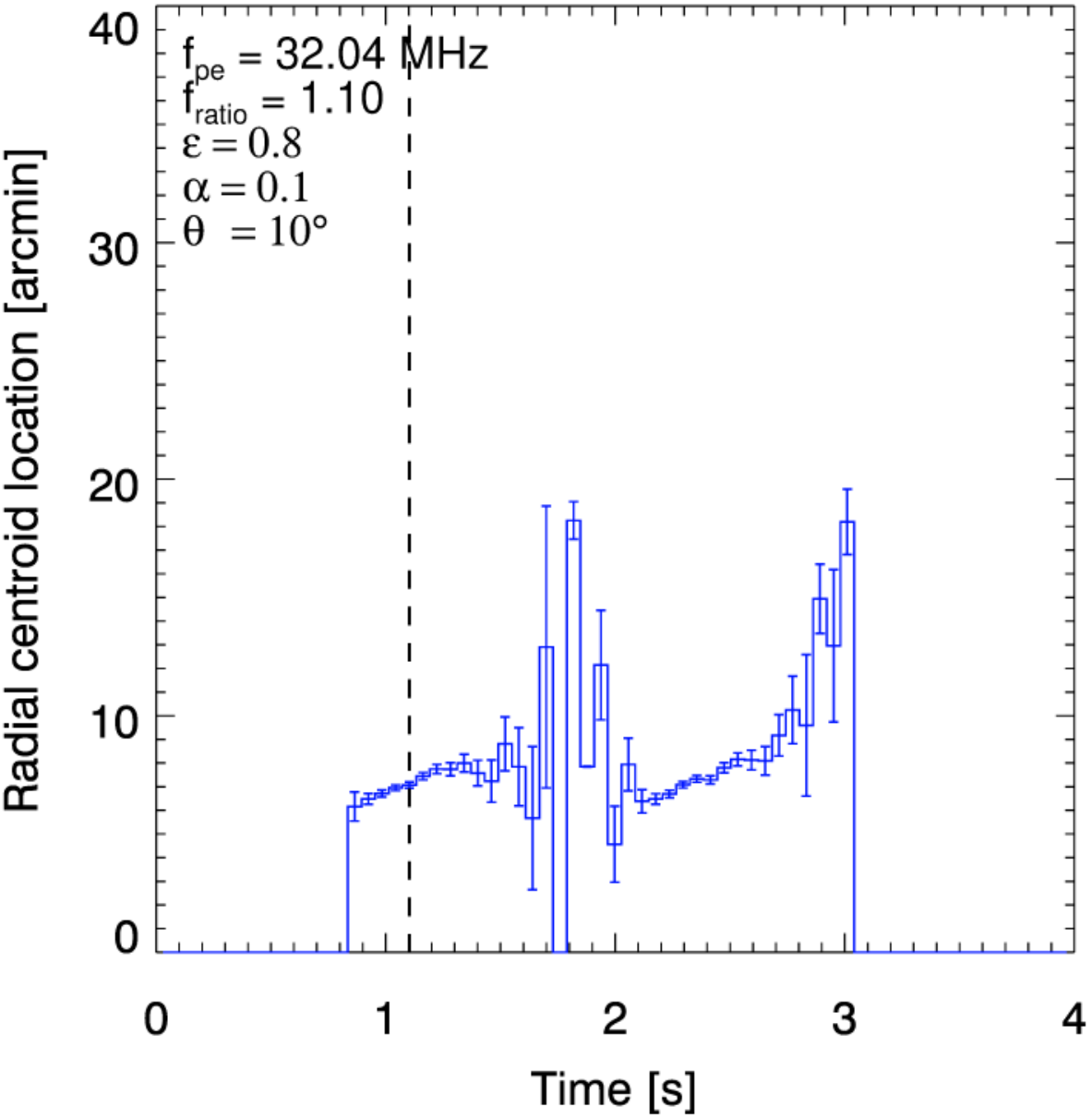}~
\includegraphics[width=0.322\linewidth]{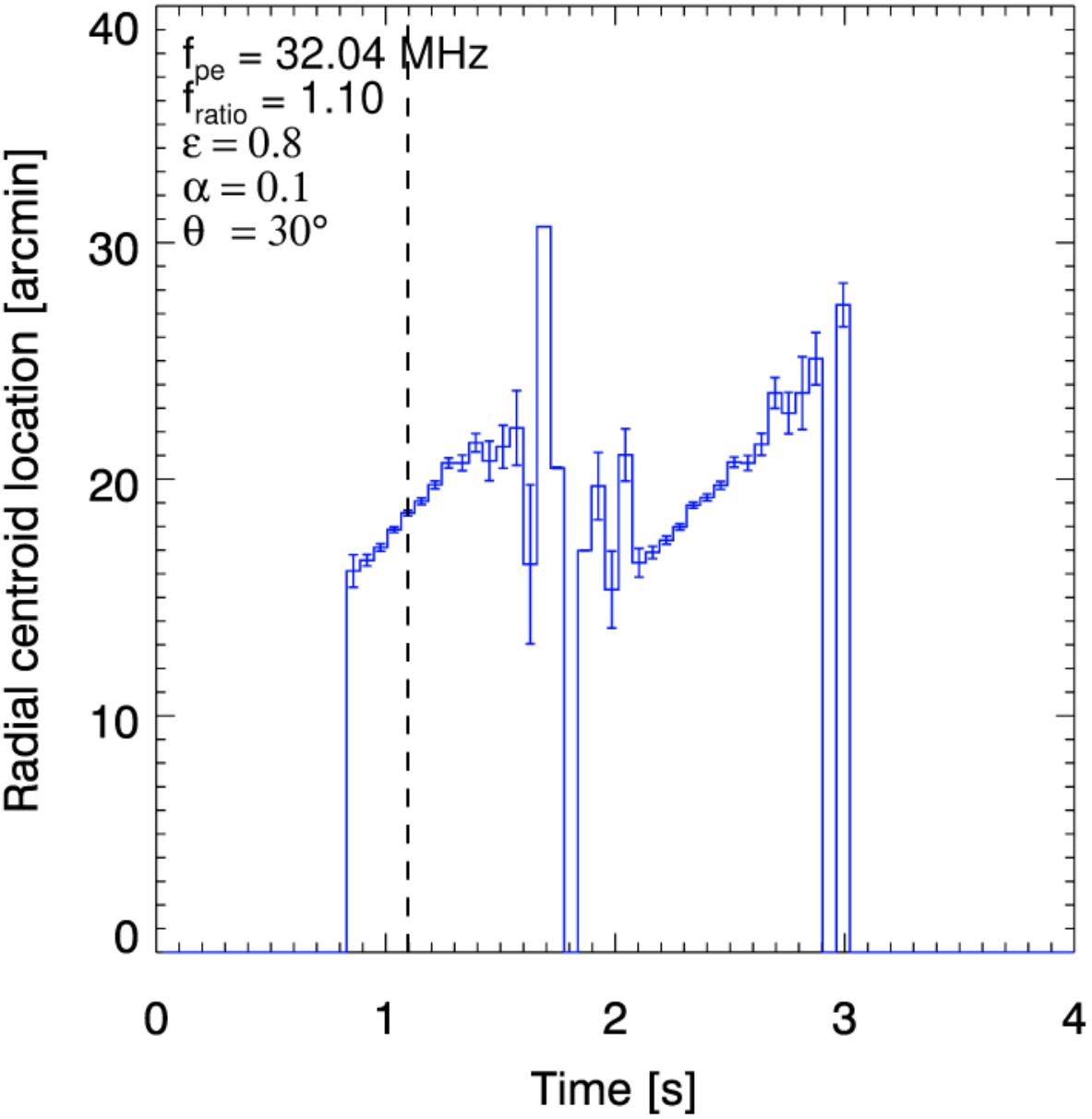}
\includegraphics[width=0.322\linewidth]{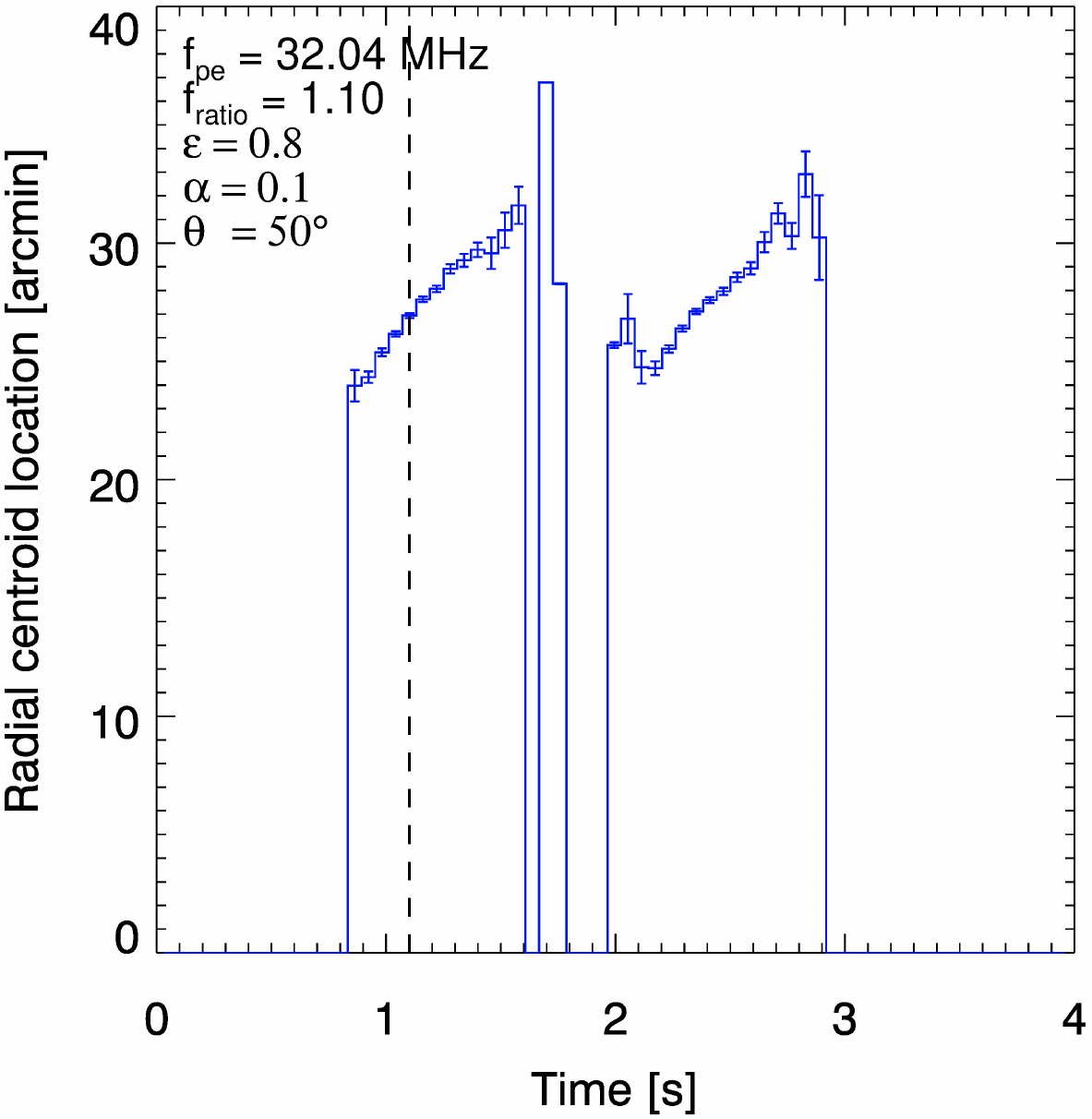}}
\centerline{\includegraphics[width=0.322\textwidth]{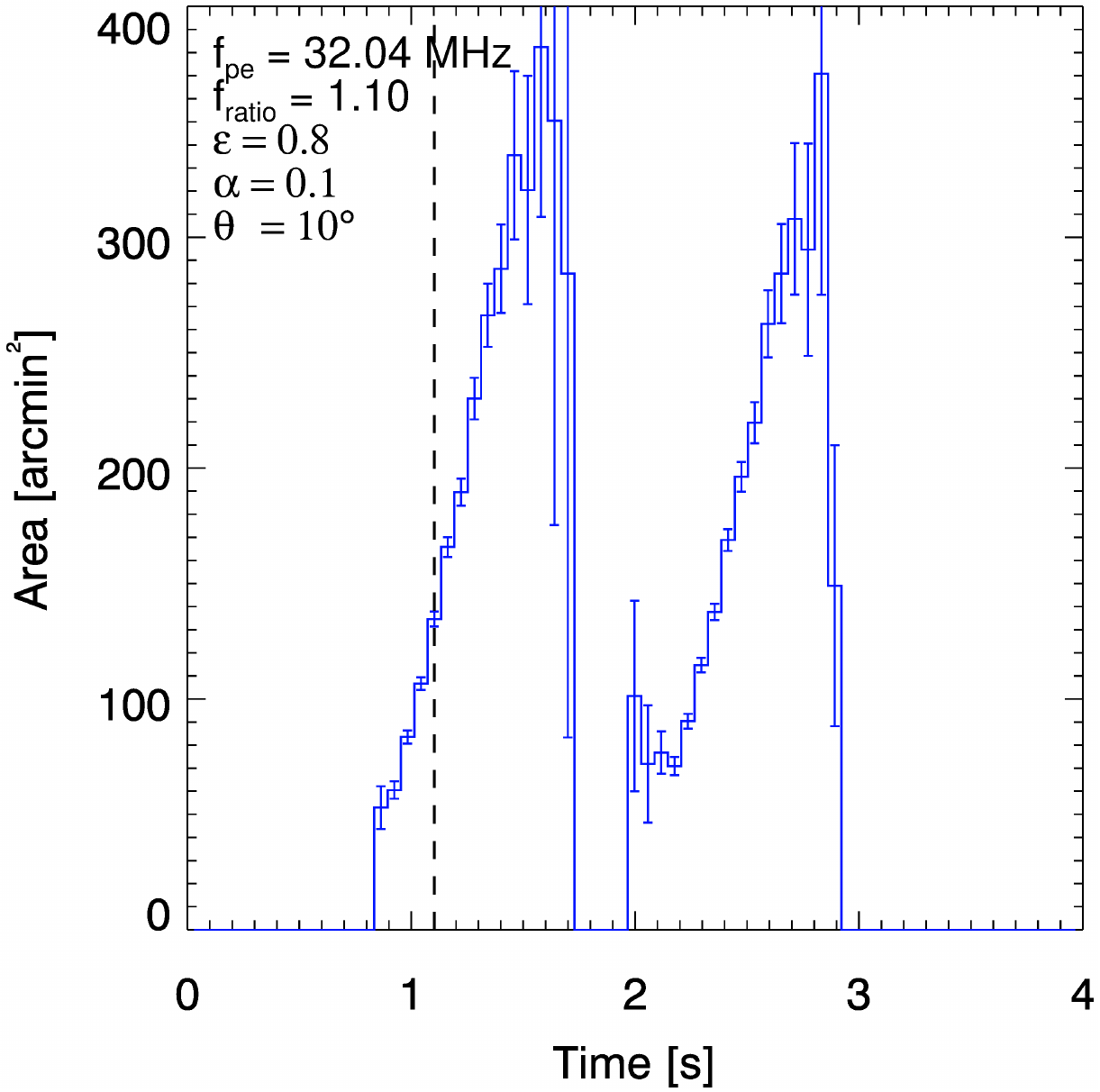}~
\includegraphics[width=0.32\linewidth]{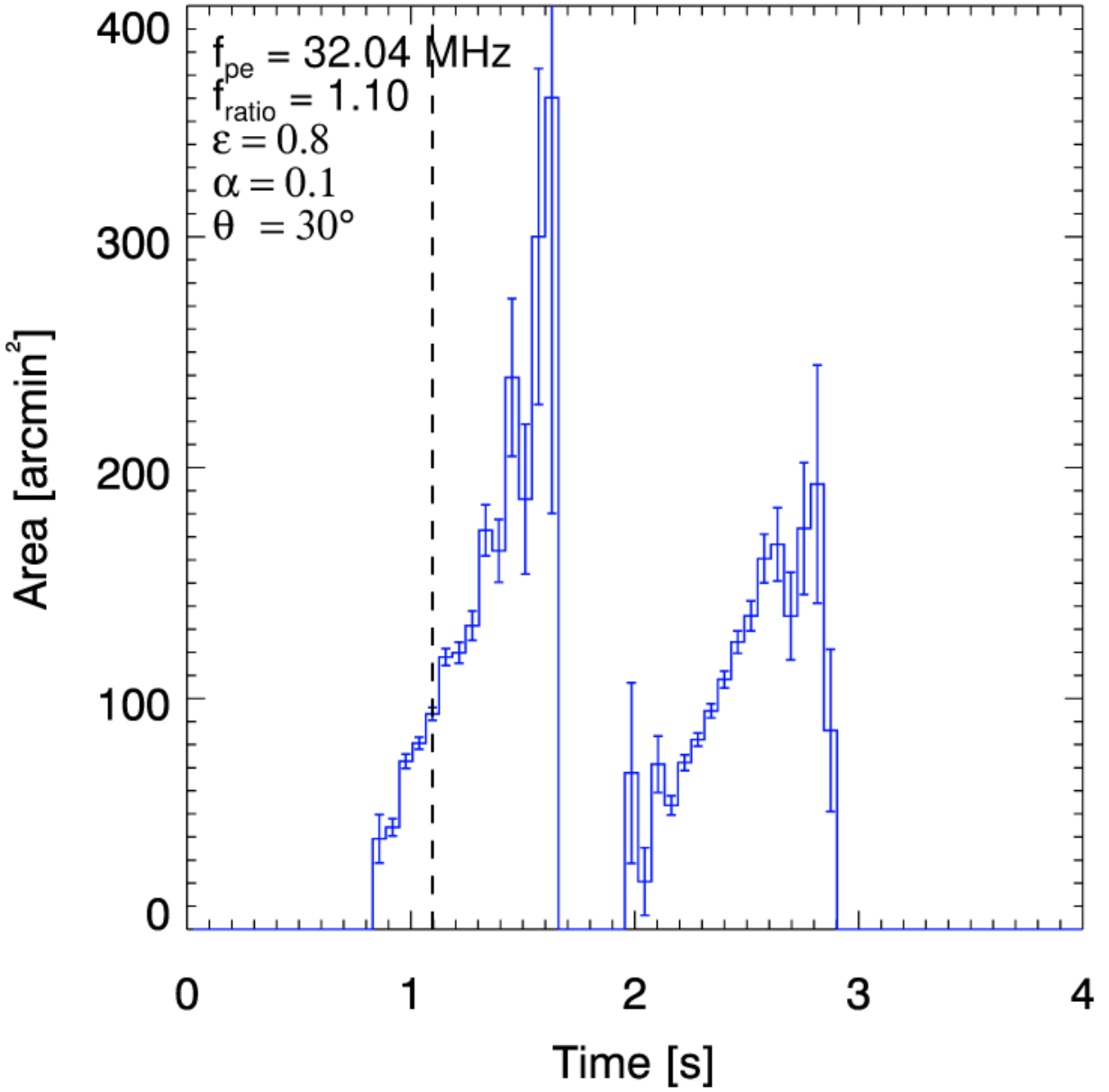}
\includegraphics[width=0.322\linewidth]{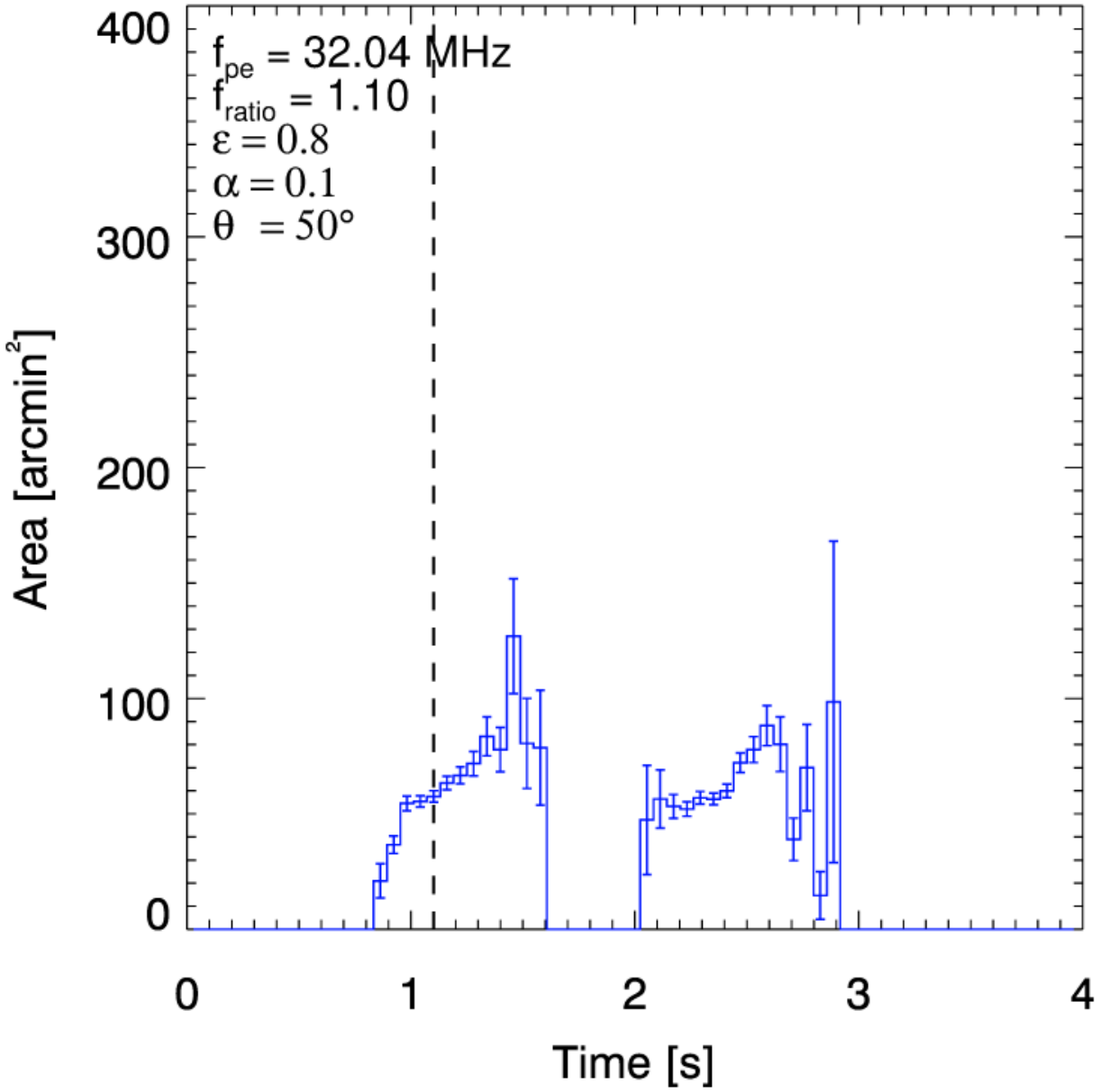}}
\caption[Simulation outputs for varying source-polar angles.]
{Simulated properties for a source emitting at 35.2~MHz (where $f/f_{pe}=1.10$), assuming $\epsilon=0.8$, $\alpha=0.1$, and source-polar angles $\theta_s = 10\degr$ (left column), $30\degr$ (middle column), and \textbf{$50\degr$} (right column).  Top row: normalised time profiles (with respect to the peak-flux value), where the blue and red lines represent the apparent emission with and without collisional absorption, respectively.  Middle row: projected (in the plane of the sky) heliocentric source locations.  Bottom row: apparent (FWHM) source areas.  Error bars represent the one-standard-deviation uncertainty.  Black dashed lines indicate the peak-flux time.
The data shown in this figure was also published in \cite{2020ApJ...898...94K}.
}

\label{fig:Kuznetsov2020_obs_angle_effects}
\end{figure}

Figure~\ref{fig:Kuznetsov2020_obs_angle_effects} presents the impact of varying the source-polar angle on the observed time profile, the source locations, and source sizes.  The left, middle, and right columns show results for $\theta_s = 30\degr$, 30$\degr$, and 50$\degr$, respectively, and time profiles are depicted in the top row, radial source locations in the middle row, and source areas in the bottom row.

All simulated time profiles look similar, irrespective of the source-polar angle assumed.  The double-peak structure is clearly visible for all cases, the relative intensity is fairly constant (although a small increase with increasing angle $\theta_s$ was estimated), the delay in photon arrival remains the same, and the delay between the two components is $\sim$1.25~s (for all angles $\theta_s$), consistent with the observed value ($\sim$1.2~s).
However, the absolute value of the primary component's peak intensity is also found to decrease with increasing angle $\theta_s$.
It can also be seen that for all source-polar angles, the apparent heliocentric location of the reflected component coincides with the location of the direct component.  The sources of the two components coincide both because fundamental emissions are considered, and because scattering has a significant effect on the propagating photons.  Specifically, the projected distance between the intrinsic source location and the nearest point of reflection for a fundamental source emitting at $\sim$35~MHz is short, less than 1$\arcmin$ ($\sim$\vphantom{}$0.75 \, \sin \theta_s$, as described in Section~\ref{sec:driftpairs_no_scattering}).  This distance, however, is larger for harmonic emissions as the photons need to travel farther until they encounter frequencies equal to $f_{pe}$ (i.e. $f_F \approx f_{pe}$ but $f_H \approx 2 f_{pe}$), so the sources of the direct and reflected harmonic components are not expected to coincide.  Moreover, anisotropic scattering results in a narrow directivity pattern (as illustrated in Section~\ref{sec:sim_TypeIIIb_properties}), meaning that the possible range of trajectories that photons can follow before they reach the observer are restricted, disabling the reflected emissions from appearing at a different heliocentric location than the direct ones.
In other words, the level of anisotropy dictates the relative intensity, the broadening, and delay between the two components (irrespective of the angle $\theta_s$ probed), and accounts for the coincidence in the source locations of the direct and reflected emissions.  However, the value of the angle $\theta_s$ affects the absolute intensity of the emissions.

The simulated source position and source speed (in the plane of the sky) increase with polar angle $\theta_s$, as expected (cf. Section~\ref{sec:angular_dependence} and \ref{sec:proj_effs_impact}).  The apparent source speed reaches a maximum of $\sim$10~$\mathrm{arcmin \, s^{-1}}$ at $\theta_s = 30\degr$--50$\degr$ (whereas it is $\sim$4.0~$\mathrm{arcmin \, s^{-1}}$ at $\theta_s = 10\degr$).  On the other hand, as the source-polar angle increases and the source moves away from the solar centre, the source area decreases (consistent with the results obtained in Section~\ref{sec:angular_dependence}), reducing from $\sim$180 to $\sim$70~$\mathrm{arcmin^{2}}$ (for $\theta_s$ from 10$\degr$ to 50$\degr$).  The areal expansion rate also decreases significantly with increasing values of angle $\theta_s$, reducing from $\sim$520 to $\sim$50~$\mathrm{arcmin^{2} \, s^{-1}}$.

It is found that both the observed source size and emission intensity (i.e. number of photons reaching the observer) decrease with increasing source-polar angle $\theta_s$.
Even though anisotropic scattering and refraction can produce the characteristic double-peak time profiles of Drift-pair bursts---as well as co-spatial source locations---for all source-polar angles (Figure~\ref{fig:Kuznetsov2020_obs_angle_effects}), sources that are located closer to the solar centre ($\theta_s \rightarrow 0\degr$) are more likely to produce Drift-pair bursts, as they correspond to higher (absolute) radio fluxes (including higher signal-to-noise and signal-to-background ratios), meaning that the probability of detecting the reflected emissions is higher.

\begin{SCfigure}[2][!ht]

\includegraphics[width=0.35\textwidth, keepaspectratio=true]{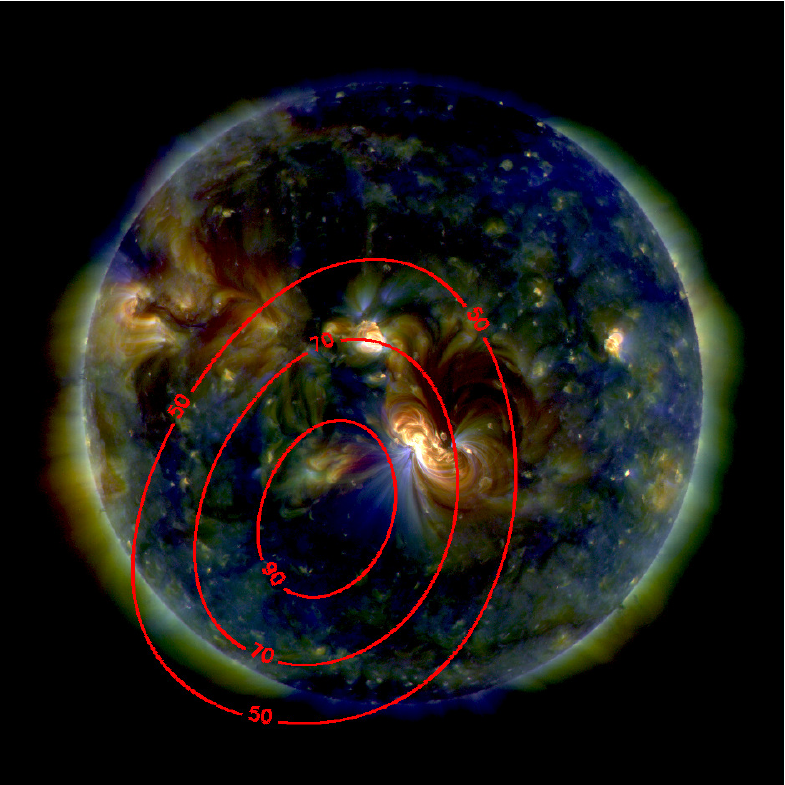}
\caption[LOFAR image of a Drift-pair burst source.]
{Leading Drift-pair burst component imaged by LOFAR (at the peak-flux time and $\sim$32~MHz).  The contours represent the 50\%, 70\%, and 90\% maximum-intensity levels.  The Sun is illustrated using a composite AIA EUV image.
Figure taken from \cite{2020ApJ...898...94K}.
}

\label{fig:Kuznetsov2020_lofar_sdo_img}
\end{SCfigure}

A comparison of the simulated source properties (Figure~\ref{fig:Kuznetsov2020_obs_angle_effects}) to the observed ones (Figure~\ref{fig:Kuznetsov2020_lofar_obs}) suggests that the studied Drift-pair burst was emitted at a source-polar angle $\theta_s \lesssim 10\degr$ (in a medium with strongly-anisotropic density fluctuations; $\alpha \approx 0.1$).  LOFAR emission images of this burst (Figure~\ref{fig:Kuznetsov2020_lofar_sdo_img}) support this conclusion, as the observed source appears on the solar disk and relatively close to the solar centre \citep{2020ApJ...898...94K}.

\subsection{Dependence of properties on the emission-to-plasma frequency ratio} \label{sec:driftpairs_frequency_dependence}

\begin{figure}[ht!]

\centerline{\includegraphics[width=0.7\linewidth]{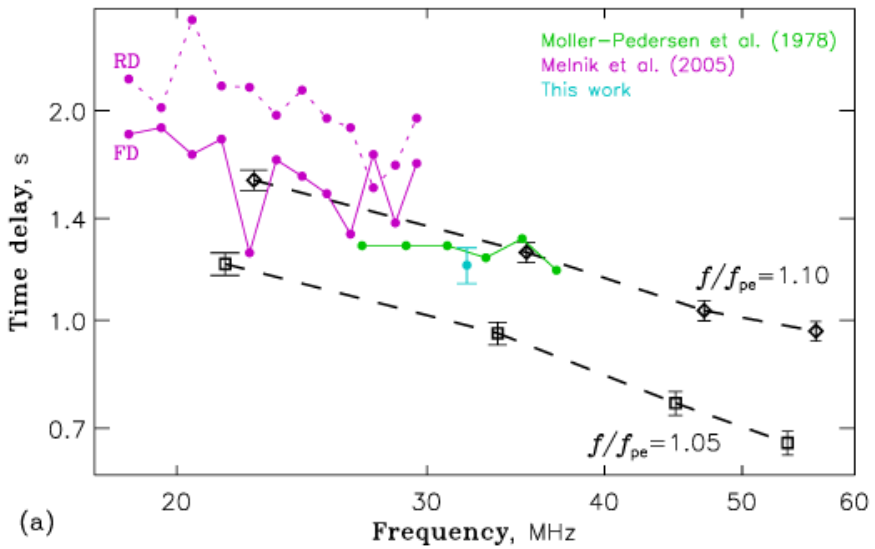}}
\centerline{\includegraphics[width=0.7\linewidth]{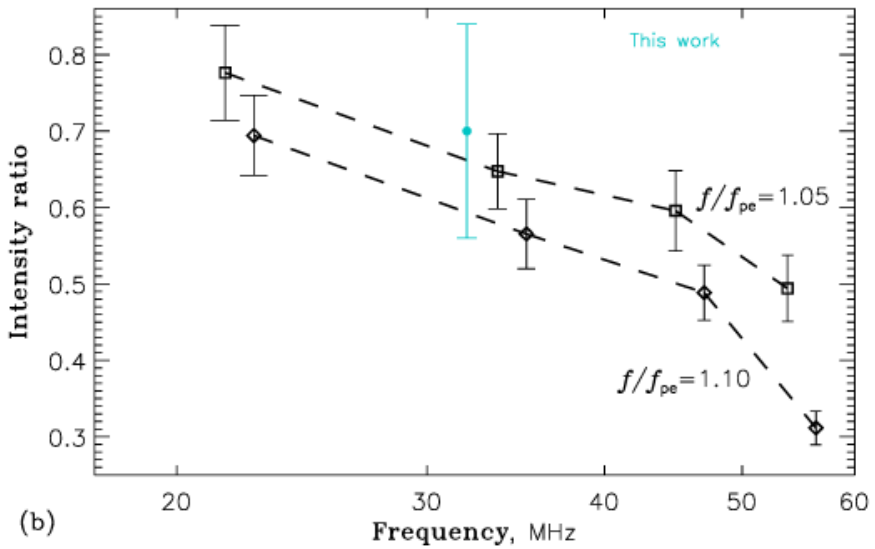}}
\caption[Dependence of Drift-pair burst properties on the emission frequency.]
{Drift-pair burst properties as a function of the emission frequency (in MHz).
(a)~Time delay (in seconds) observed between the direct and reflected components.
(b)~Intensity ratio between the two components (i.e. the relative intensity of the second component).
Black dashed lines show the simulation results for a source emitting in a medium characterised by $\epsilon=0.8$ and $\alpha=0.1$, and located at a polar angle $\theta_s=10\degr$.  Results for two emission-to-plasma frequency ratios are shown: $f/f_{pe}=1.05$ (annotated with a square) and $f/f_{pe}=1.10$ (annotated with a diamond).  Data from observational studies (including from the work presented in this chapter) is also illustrated, as indicated by the legends.  Error bars (where provided) represent the one-standard-deviation uncertainty.
Figure taken from \cite{2020ApJ...898...94K} and then adapted.
}

\label{fig:Kuznetsov2020_delay_intens_vs_freq}
\end{figure}

By definition, a radio source emits at frequencies $f$ that are above the local plasma frequency $f_{pe}$ (see Section~\ref{sec:plasma_emmission}), although the exact ratio between the two frequencies is unknown.  So far in Chapters~\ref{chap:scattering} and \ref{chap:observation_simulations}, this ratio was taken to be $f/f_{pe} = 1.10$ for the purposes of simulating radio properties.  In order to investigate the impact of this value on the simulated properties, the results presented so far throughout Section~\ref{sec:drift_pairs} are repeated for both $f/f_{pe} = 1.05$ and 1.10, but also for several frequencies between 20 and 60~MHz, assuming $\epsilon=0.8$, $\alpha=0.1$, and $\theta_s=10\degr$.  The output of these simulations is summarised in Figure~\ref{fig:Kuznetsov2020_delay_intens_vs_freq}, where panel \hyperref[fig:Kuznetsov2020_delay_intens_vs_freq]{(a)} depicts the time delay between the burst components against the emission frequency, and panel \hyperref[fig:Kuznetsov2020_delay_intens_vs_freq]{(b)} depicts the intensity ratio between the two components (against frequency).  Simulation outputs are indicated in black squares for $f/f_{pe} = 1.05$ and in black diamonds for $f/f_{pe} = 1.10$.  Simulated values are depicted along with values obtained from Drift-pair observations, both from this study (light blue; see Figure~\ref{sec:driftpairs_LOFAR_properties}) and from previous studies (\cite{1978A&A....70..801M}, shown in green, and \cite{2005SoPh..231..143M}, shown in magenta), as indicated by the legend.  The values from \cite{2005SoPh..231..143M} distinguish between Drift-pair bursts with negative frequency-drift rates (i.e. ``forward''; annotated with an ``FD'' and a solid magenta line) and those with positive frequency-drift rates (i.e. ``reverse''; annotated with an ``RD'' and a dashed magenta line).  It should also be noted that \cite{1978A&A....70..801M} and \cite{2005SoPh..231..143M}---who measured the time delay between the components, but not the intensity ratio---do not report the uncertainties in the values obtained from the observations.

Figure~\ref{fig:Kuznetsov2020_delay_intens_vs_freq}\hyperref[fig:Kuznetsov2020_delay_intens_vs_freq]{a} indicates that the time delay between the direct (leading) and reflected (trailing) Drift-pair components decreases with increasing frequency, where the delay can be (approximately) characterised as a function of $f^{-1/2}$.  Therefore, the simulations predict that there is indeed an inverse relation between the observed time delay and frequency (as hinted by \cite{2005SoPh..231..143M}), a dependence that could not be confidently inferred from the currently-available observations (due to ambiguities discussed in Section~\ref{sec:driftpairs_properties}).
A reason for this dependency is the non-linear relation of the coronal density to the radial distance from the Sun.  The coronal density decreases faster at distances closer to the Sun compared to larger distances, which means that higher-frequency sources are closer to their reflection point (where $f \rightarrow f_{pe}$) than lower-frequency sources, and thus travel for shorter times before they reach the observer.  Furthermore, the delay between the components is shorter when the emission ratio is lower, i.e. $f/f_{pe} = 1.05$ (instead of $f/f_{pe} = 1.10$).  This is also expected given that---for a given plasma frequency $f_{pe}$---a photon emitted at $f = 1.05 \, f_{pe}$ is physically closer to the region which can reflect it (i.e. where $f \simeq f_{pe}$) compared to a photon emitted at $f = 1.10 \, f_{pe}$, meaning that it travels a shorter distance (and thus for less time) before it reaches the observer (as stated in Section~\ref{sec:driftpairs_radio_echo_theory}).
The dependence of the time delay between components on the frequency ratio $f/f_{pe}$ can be used as a diagnostic tool for estimating the characteristic wavenumber $k_L$ of Langmuir waves (see Equation~(\ref{eqn:Langmuir_wavenum})), which are responsible for plasma emissions (see Section~\ref{sec:plasma_emmission}).
The observational data appears to agree to a higher degree with the simulated values for $f/f_{pe} = 1.10$ than for $f/f_{pe} = 1.05$.

Figure~\ref{fig:Kuznetsov2020_delay_intens_vs_freq}\hyperref[fig:Kuznetsov2020_delay_intens_vs_freq]{b} indicates that the intensity ratio between the direct and reflected Drift-pair burst components also decreases with increasing frequency.
This trend is caused by the fact that higher-frequency sources emit in denser regions of the corona (i.e. closer to the Sun), implying that higher-frequency photons undergo stronger collisional (free-free) absorption.  This can explain why Drift-pair bursts are predominantly observed at low frequencies ($\lesssim 100$~MHz), since at higher frequencies the collisional absorption becomes so strong that the reflected component cannot be resolved.  Additionally, for a fixed plasma frequency $f_{pe}$, reflected photons with frequency $f = 1.05 \, f_{pe}$ are absorbed less than reflected photons with frequency $f = 1.10 \, f_{pe}$, given that they do not need to travel as far in denser plasmas in order to reach the reflection point, so they spend relatively less time in the turbulent coronal medium and are affected less by collisional damping.  As such, the intensity ratio between the direct and reflected components is higher (at a given frequency) for photons emitted at $f = 1.05 \, f_{pe}$ compared to the ratio for photons emitted at $f = 1.10 \, f_{pe}$.  Given the limited number of observational data (and large uncertainties) for the relative intensity of the two components, no statement can be made as to which frequency ratio produces intensities that best match the observations.  It is, however, clear that the variation of the emission-to-plasma frequency ratio does not affect the relative intensity of the reflected component to the extent it affects the time delay between the two components.

\subsection{Discussion and final remarks} \label{sec:driftpairs_discussion}
The puzzling characteristic of Drift-pair bursts is that---unlike other radio bursts---their two components repeat in time and not in frequency.  This morphology was attributed to the reflection of the emitted radio waves off of denser coronal regions, referred to as the radio echo \citep{1958AuJPh..11..215R}.

As described in Section~\ref{sec:driftpairs_radio_echo_theory}, the radio echo theory received some criticism which included the arguments that (i) the reflected rays should be broadened by scattering (and thus have a broader time profile), and (ii) that the source positions of the two components should be considerably different \citep{1982srs..work..182M}.
However, the radio echo hypothesis and its relevant predictions were criticised at the time under the assumption that density inhomogeneities in the corona (and scattering) were \textit{isotropic} \citep{1974SoPh...35..153R, 1978A&A....70..801M}.  Given the recently-improved understanding of radio-wave scattering (see Chapter~\ref{chap:scattering}; \cite{2019ApJ...884..122K}) and the (rough) estimations enabled by spatially-resolved emission images (see the estimation in Section~\ref{sec:driftpairs_LOFAR_properties}; \cite{2018SoPh..293..115S}), the radio echo hypothesis needed to be tested under the assumption of \textit{anisotropic} density fluctuations.  Therefore, the ray-tracing simulations presented in Chapter~\ref{chap:scattering}---which account for anisotropic scattering---have been utilised.

It has been demonstrated that the features and properties of Drift-pair bursts can be quantitatively reproduced assuming \textit{fundamental} emissions, if and when the anisotropy of density fluctuations is sufficiently strong.  Specifically, the signature double structure of Drift-pair bursts is formed when the anisotropy $\alpha \lesssim 0.2$, but $\alpha \approx 0.1$ is required in order to produce components which are temporally separated by the required amount ($\sim$1.25~s, assuming fundamental emissions) and whose intensities are as similar as observed.  The main contributor to the attenuation of the reflected component is the collisional damping experienced during the propagation of photons in regions of denser plasma, before they are reflected back towards the observed.
The strong anisotropy levels also result in highly-directional emissions for both the direct and reflected rays, meaning that both components have a time profile with similar (and sufficiently short) FWHM durations, as the photons' path---and hence the time spent in the corona---is restricted.
The duration of each component needs to be shorter than the time delay between the components in order for the characteristic double structure to be observed, and thus identified as a Drift-pair burst.  The simulations have also reproduced apparent sources for the direct and reflected emissions which spatially coincide and demonstrate the same radial motion (given a certain temporal delay between them).  In addition to that, it was shown that emissions from sources which are located closer to the solar centre ($\theta_s \rightarrow 0\degr$)---in the plane of the sky---are more likely to be observed as Drift-pair bursts.
Contrary to previous suggestions \citep{1958AuJPh..11..215R}, time profiles with similar relative intensities, FHWM durations, and delays between the two components were produced for a wide range of projected source positions on the solar disk (i.e. no significant variation with the source-polar angle $\theta_s$ was identified), as the anisotropy appears to dictate these characteristics.
However, thanks to the anisotropy resulting in directivity patterns that are predominantly in the radial direction (see Section~\ref{sec:sim_TypeIIIb_properties}), emissions from sources located at larger source-polar angles will appear to be fainter, as fewer photons reach the observer (i.e. the absolute intensity value is affected).  This does not only impact the apparent intensity of the radiation, but also the brightness of the source emissions relative to that of the background continuum (resulting in a poorer signal-to-noise ratio).  Therefore, Drift-pair bursts are more likely to be observed when the source is located closer to the solar centre, consistent with the observed centre-to-limb variation statistics \citep{1974A&A....37..163M}.  The frequency relation of the time delay and the intensity ratio between the two components was also investigated, leading to the conclusion that both of these properties decrease with increasing emission frequency.  Furthermore, the dependence of the time delay and intensity ratio on the emission-to-plasma frequency ratio $f/f_{pe}$ was examined.  It was found that the time delay increases with increasing $f/f_{pe}$ values, whereas the intensity ratio decreases.  The observed time delay is predominantly affected by the $f/f_{pe}$ ratio as it determines the path difference between the direct and reflected components, implying that Drift-pair bursts can be used to infer the local plasma frequency, and by extent, diagnose the plasma emission mechanism.

The fact that strong anisotropies generate direct and reflected components which have a similar duration to each other, can explain the lack of radio echo observations in other types of radio bursts observed at similar frequencies (a concern first raised by \cite{1958AuJPh..11..215R}).  For example, Type III bursts (see Section~\ref{sec:typeIIIs}) tend to last for over 1~s, meaning that if a reflected component is present it will be masked by the contribution (i.e. lost in the tail) of the direct component.  In addition to that, the anisotropy needs to be sufficiently high ($\alpha \lesssim 0.2$) for the reflected component to be distinguishable, which does not appear to be the case for observations of Type IIIb bursts reproduced using the same mathematical model applied in this chapter ($\alpha > 0.2$ was required; see Chapter~\ref{chap:scattering} and Section~\ref{sec:typeIIIb_sim_setup}).

An understanding of why Drift-pair bursts tend to be observed within a very limited range of frequencies ($\sim$10--100~MHz) is also obtained.  The higher-frequency boundary is likely due to the collisional (free-free) absorption, which is stronger at denser plasma regions and thus impacts higher frequencies more than lower frequencies.  Collisional damping affects the reflected rays the most, since they propagate to---and spend more time in---denser regions before they are reflected towards the observer.  Therefore, the relative amplitude of the reflected component (with respect to the direct) decreases at higher frequencies, until $\gtrsim 100$~MHz, after which it becomes too faint to be distinguished.  On the other hand, the lower-frequency boundary likely arises due to instrumental limitations.  The ionospheric cut-off at $\sim$10~MHz (see Section~\ref{sec:radio_obs}) prevents ground-based observations from being conducted at lower frequencies, whereas space-based radio instruments which observe at lower frequencies do not (as of yet) have a sufficient temporal and spectral resolution---nor the sensitivity---to resolve the double structure of Drift-pair bursts.

\section{Conclusions} \label{sec:obs_vs_sim_chapter_conclusions}
In this chapter, observations of high temporal and spectral resolutions provided by LOFAR are utilised to take advantage of the ability to trace the temporal evolution of source properties at a single frequency.  LOFAR enables the simultaneous observation (and imaging) of several source properties as a function of time, with very short time intervals of $\sim$0.01s.

The properties of a Type IIIb and Drift-pair solar radio burst, observed by LOFAR near 32~MHz, were analysed at sub-second scales and quantitatively reproduced using ray-tracing simulations that allow for an anisotropic scattering description.  It was demonstrated---in both cases---that isotropic scattering ($\alpha=1$) cannot account for all of the observed source properties simultaneously.  Therefore, the necessity to describe plasma emissions within the framework of an anisotropic turbulent medium (where scattering is stronger in the perpendicular direction, i.e. $\alpha < 1$) was reaffirmed (see Chapter~\ref{chap:scattering}).  Besides the level of anisotropy $\alpha$, the influence of the level of density fluctuations $\epsilon$ and the source-polar angle $\theta_s$ was also probed.

The sub-second temporal evolution of both the fundamental and harmonic emissions of a Type IIIb burst was studied for the first time.  It was found that whilst the fundamental emissions can be successfully described assuming an instantaneously-emitting point source, the harmonic emissions cannot.  Instead, a finite size and finite emission duration was necessary to describe the harmonic properties using the same parameters that described the fundamental emissions.  It is worth mentioning, however, that the observed fundamental emissions could also be reproduced when a finite source size was assumed, albeit much smaller than that required to reproduce the harmonic emissions.

Regarding the Drift-pair burst, the simulations demonstrated that the observed properties and their evolution for both the leading and trailing components can be described using the radio echo hypothesis when fundamental emissions are considered.  In other words, it was shown that thanks to reflection at denser plasma regions, radio waves emitted from a single source can form a reflected component that is almost identical to that of the direct radio waves (but is observed with a certain delay), such that the simulated characteristics are consistent with the observed ones.  It was, however, illustrated that Drift-pair bursts can only form under certain conditions, specifically, when the anisotropy level is very high ($\alpha \approx 0.1$) and (preferably) when the source-polar angle $\theta_s$ is small.  The dependence of the observed time delay and relative intensity between the direct and reflected components on both the emission frequency ($f$) and the emission-to-plasma frequency ratio ($f/f_{pe}$) was also investigated.  It was indicated that $f/f_{pe}$ ratio can influence the observed delay between the components to a considerable extent, with larger values resulting in larger delays.

\cleardoublepage
\chapter{Split-Band Type II Bursts} \label{chap:split-band_typeII}

\textit{These results were published in \cite{2018ApJ...868...79C} and \cite{2020ApJ...893..115C}.}

\section{Debated Interpretations of Split-Band Type II Burst Images} \label{sec:previous_split-band_TypeII_imaging}
Type II radio bursts (introduced in Section~\ref{sec:typeIIs}) that demonstrate band splitting have often been observed.  As of this day, the most widely-accepted interpretations of band splitting are the \cite{1974IAUS...57..389S, 1975ApL....16R..23S} model and the \cite{1983ApJ...267..837H} model, each of which makes opposing predictions regarding the location of the subband sources.  As detailed in Section~\ref{sec:bandsplitting_models}, if band splitting results from the mechanism described by \cite{1974IAUS...57..389S, 1975ApL....16R..23S}, the upper- and lower-frequency subband sources are expected to be virtually co-spatial.  If, however, the \cite{1983ApJ...267..837H} mechanism is at play, the subband sources are expected to be physically separated.

Over the decades, imaging observations of split-band Type II bursts have revealed significant observed separations between the two sources.  \cite{1974IAUS...57..389S, 1975ApL....16R..23S} were aware of the large separations observed (from $1\arcmin$--$4\arcmin$, corresponding to $\sim$0.06--0.25~$\mathrm{R_\sol}$), and presented several arguments as to why the apparent separation did not represent the intrinsic nature of the split-band Type II sources.  One of the explanations was based on the inability to observe both Type II subbands at the same time, due to the limitations of the available instruments, thus introducing time-delay ambiguities in the observations.  Namely, only a few, fixed frequencies could be imaged (no more than two frequencies, at that time), as described in Section~\ref{sec:radio_obs}.
Specifically, \cite{1974IAUS...57..389S, 1975ApL....16R..23S} examined the source locations when both subbands were imaged at 80~MHz (i.e. at the same frequency but at different times), implying that the shock travelled away from the Sun between the times at which the upper- and lower-frequency subband sources were probed.
Thus, it was argued that the sources could be co-spatial and that the observed separation was artificial.  The other explanation regarded the enhanced scattering and refraction that the lower-frequency photons emitted upstream of the shock would experience compared to the higher-frequency photons.  While it was acknowledged that the observed source location would not represent the true location, the need for a quantitative estimation of the source's displacement was highlighted.

\section{Imaging Spectroscopy of a Split-Band Type II Burst with LOFAR}
\subsection{Overview of the observations} \label{sec:2018typeII_overview}
\begin{figure}[ht!]
    \centering
	\includegraphics[width=1.0\textwidth, keepaspectratio=true]{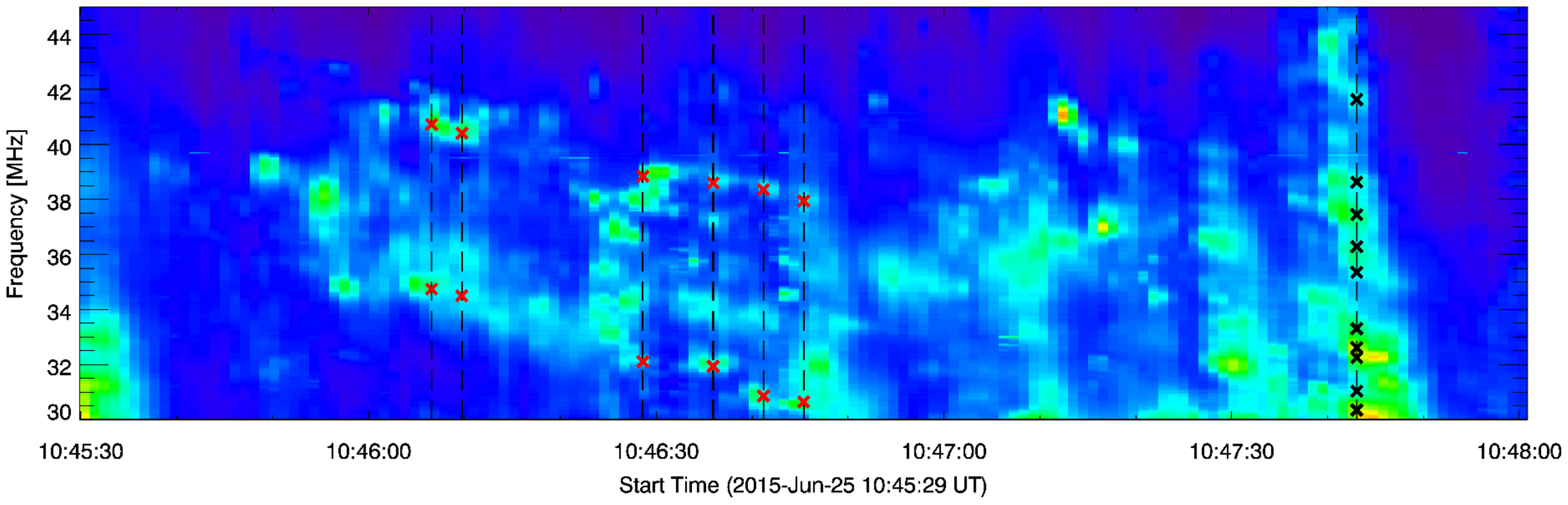}
    \caption[Split-band Type II burst observed with LOFAR on 25 June 2015.]
    {Dynamic spectrum illustrating part of the radio emissions observed by the LOFAR LBA antenna between 10:45:30 and 10:48:00~UT on 25 June 2015.  The black dashed lines illustrate single-time moments, whereas crosses represent the locations at which emission images were fitted so that source centroid estimations could be obtained.  Both subbands of the split-band Type II burst were imaged (red crosses), with points from each of the upper- and lower-frequency subbands imaged at the exact same time, at six different moments covering the duration of the observed band splitting.  Points along the observed Type III burst were also selected for imaging (black crosses).
    Figure taken from \cite{2018ApJ...868...79C}.
	}
    \label{fig:2018_dynspec}
\end{figure}

A Type II burst that experiences band splitting was observed on 25 June 2015 by LOFAR between $\sim$10:46 and 10:48~UT, as shown in Figure~\ref{fig:2018_dynspec} \citep{2018ApJ...868...79C}.  The fluctuations of intensity along the two subbands are similar and both subbands evolve in frequency-time in a synchronised manner (suggesting simultaneous propagation through the same density region), producing parallel-like lanes, both defining features of split-band Type II bursts (see Section~\ref{sec:typeIIs}).  The frequency drift rate $df/dt$ was estimated from the dynamic spectrum in Figure~\ref{fig:2018_dynspec} (see Section~\ref{sec:all_bursts}) to be $\sim$\nolinebreak$-$0.1~$\MHzs$ and the relative frequency split $\Delta f_s/f \approx 0.21$ (cf. Figure~\ref{fig:bandwidths_cartoon}), both characteristic of Type II bursts.  Type III bursts---indicators of open magnetic fields---were observed to intersect parts of the Type II emissions, as depicted in Figure~\ref{fig:2018_dynspec}.

The LOFAR observation was conducted between 30--80~MHz using 24 core stations in the LBA outer configuration, and utilising the coherent Stokes beam-formed mode recording only the Stokes I information (see Section~\ref{sec:radio_obs}).  A mosaic of 169 individual beams formed a tied-array beam and covered a hexagonal area that extended up to $\sim$2.5~$\Rs$ from the Sun (see Section~\ref{sec:tied-array}).  Two additional beams were used for flux calibration purposes throughout the duration of the observation; one pointed at Tau A (a well-described point source) and one pointed at the ``empty sky'' (a part of the sky lacking bright radio sources at the frequencies of interest; see Section~\ref{sec:flux_cal}).
The configuration resulted in a temporal resolution of $\sim$0.01~s, a spectral resolution of $\sim$12.2~kHz, and a sensitivity of $\lesssim$~0.03~sfu per beam.  The average separation between the central locations of the beams was approximately 6.1$\arcmin$.  The synthesised beams had a FWHM of $\sim$10$\arcmin$ at 30~MHz.  For the analysis and presentation of this observation, the spectral and temporal resolution were reduced to $\sim$24.4~kHz and $\sim$1~s, respectively, achieved by rebinning the data which improves the processing time as well as the signal to noise ratio.

A CME eruption was observed in white-light coronagraphic images obtained by the LASCO instrument onboard \textit{SOHO} (see Section~\ref{sec:solar_activities}).  The CME appeared at $\sim$10:57~UT in LASCO's C2 coronagraph which images distances from $\sim$2.2--6~$\Rs$ with a temporal resolution of approximately 12 minutes.  The origin of the CME on the solar surface was probed using EUV data obtained by the AIA instrument onboard \textit{SDO}, which observes with a 12~s cadence.

\subsection{Probing the CME-Type II relation} \label{sec:TypeII-CME_relation}
\begin{figure}[ht!]
    \centering
	\includegraphics[width=1.0\textwidth, keepaspectratio=true]{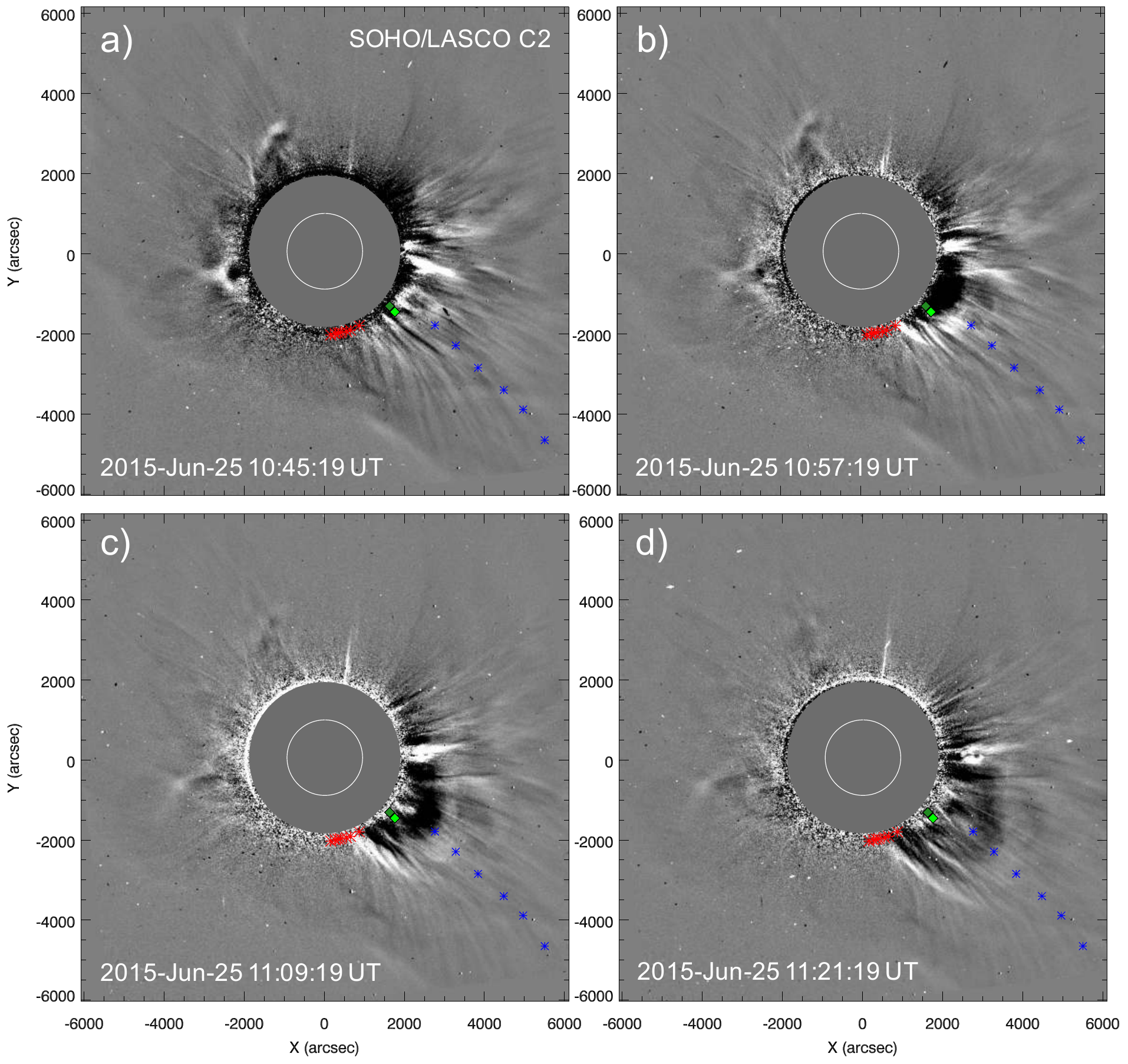}
    \caption[Consecutive white-light coronagraphic images and CME tracking.]
    {Panels (a)--(d) illustrate consecutive LASCO C2 running-difference images taken $\sim$12 minutes apart, used to track CME features.  Brighter structures in the LASCO FoV reflect relative increases in intensity, and vice versa.  Blue asterisks illustrate the tracking of the CME front, whereas red asterisks indicate the tracking of the CME's lateral expansion at a constant height of $\sim$2.2~$\Rs$.  The dark green and light green diamonds indicate the observed positions of the upper- and lower-frequency Type II subbands, respectively, at 10:46:29~UT.
    Figure taken from \cite{2018ApJ...868...79C}.
	}
    \label{fig:2018_cme_tracking}
\end{figure}

Given that Type II bursts are often excited by CME-driven shocks, the spatial and temporal relation of the CME to the Type II burst is investigated.  The CME appears to emerge from the south-west part of the Sun.  As it approaches the boundaries of the C2 FoV, it begins to dissolve into the coronal background which has been strongly disturbed by the residual structures of an earlier eruption recorded by C2 at $\sim$8:36~UT.  Figure~\ref{fig:2018_cme_tracking} shows the spatial evolution of the CME over time, with its first appearance at $\sim$10:57~UT.  Panels (\hyperref[fig:2018_cme_tracking]{a})--(\hyperref[fig:2018_cme_tracking]{d}) show consecutive running-difference images obtained at 12-minute time intervals, used to highlight the CME features and enable a more reliable tracking of the CME.  Structures that are brighter indicate relative increases in intensity, whereas structures that are darker indicate relative decreases in intensity.  The CME's features were tracked throughout the event's appearance in the C2 FoV.  The expansion of the CME's front was estimated using the locations indicated by the blue asterisks.  The lateral expansion of the CME's flank was tracked at a constant height of 2.2~$\Rs$, as shown by the red asterisks.  The dark green and light green diamonds indicate the imaged centroid locations of the Type II upper- and lower-frequency subbands, respectively, at 10:46:29~UT.

\begin{figure}[ht!]
    \centering
	\includegraphics[width=1.0\textwidth, keepaspectratio=true]{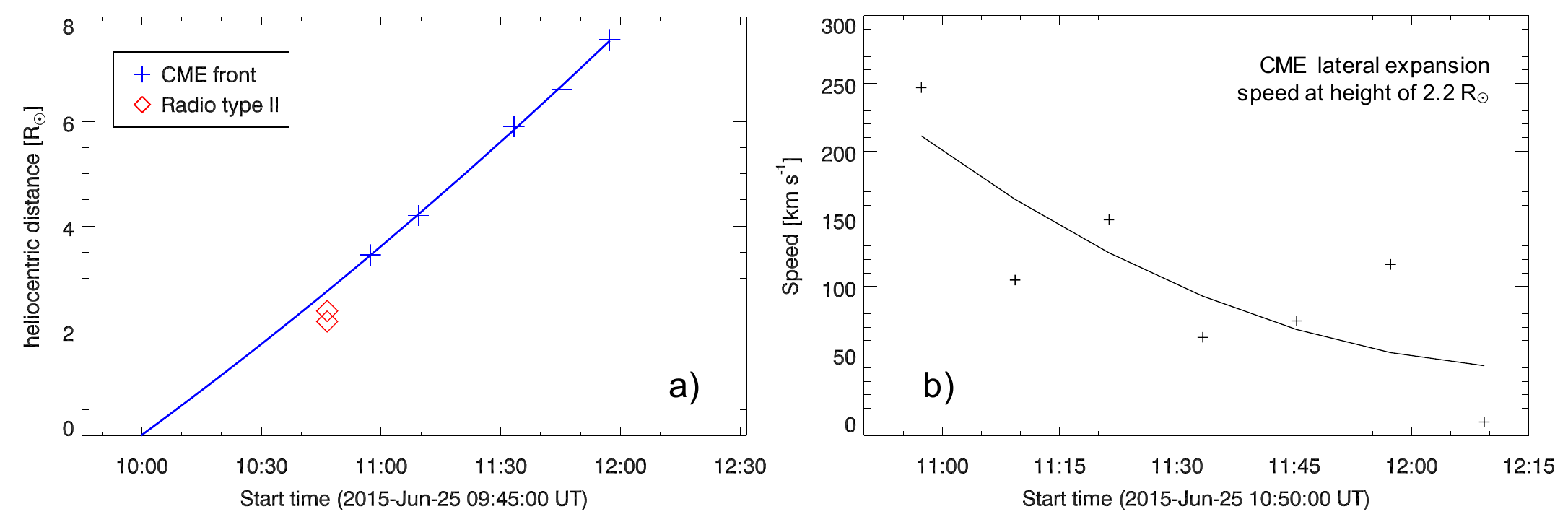}
    \caption[Heliocentric distance and lateral expansion speed of the CME as a function of time.]
    {(a)~Heliocentric height of the CME's front as a function of time (blue plus signs) obtained within the C2 FoV (see Figure~\ref{fig:2018_cme_tracking}).  The red diamonds represent the heliocentric height of the Type II subband sources as imaged at 10:46:29~UT.  A back-extrapolation of the fit through the CME's tracked trajectory (blue line) provides the CME's approximated height at the time of the Type II emissions, as well as an estimation of the CME's eruption time (i.e. $\sim$10:15~UT at 1~$\Rs$).  The mean plane-of-sky CME speed of was also derived from the fit.
    (b)~Lateral expansion speed of the CME measured at a constant height of $\sim$2.2~$\Rs$, obtained by tracking the CME's flank features in the C2 FoV, as shown in Figure~\ref{fig:2018_cme_tracking}.
    Figure taken from \cite{2018ApJ...868...79C}.
	}
    \label{fig:2018_cme_fits}
\end{figure}

Figure~\ref{fig:2018_cme_fits}\hyperref[fig:2018_cme_fits]{a} shows the heliocentric distance of the tracked CME front (blue plus signs) as a function of time.  The mean plane-of-sky speed of the CME front was estimated to be $\sim$740~$\kms$ in the C2 FoV, by applying a non-linear (second order polynomial) fit through the heliocentric distances.  A back-extrapolation of the fit is used to infer that the CME was at a height of $\gtrsim$~2.5~$\Rs$ above the solar centre during the Type II emissions (represented by the red diamonds), and that the CME erupted at approximately 10:15~UT.  The time of the CME's onset could not be estimated from X-ray data since a strong flare (of magnitude M7.9) which occurred at $\sim$8:00~UT masked the contribution of the CME of interest.  Figure~\ref{fig:2018_cme_fits}\hyperref[fig:2018_cme_fits]{b} illustrates the speed with which the CME's flank expanded in the lateral direction (at 2.2~$\Rs$) as a function of time.  An overall deceleration with progressing time can be observed, as indicated by the non-linear fit.

Given the lack of a clear impulsive phase related to the CME associated with the Type II emissions (first imaged at $\sim$10:57~UT) and inability to distinguish the CME's onset time in X-ray data, \textit{SDO}/AIA data was used to examine the surface of the Sun near the time of appearance of the CME in order to estimate its launch time, as well as identify its region of origin.  Significant coronal dimming was observed at $\sim$9:50~UT on the southern part of the active region from which the stronger CME at $\sim$8:36~UT emerged (see Figure~\ref{subfig:2018_lasco_fit}).  Coronal dimming signals density depletion and mass loss, consistent with CME eruptions (see Section~\ref{sec:solar_activities}).  It is thought to be a powerful diagnostic of the early phases of CMEs as it relates to the outward flow of solar material \citep{2004psci.book.....A}.  The time of the observed dimming also agrees with the estimated CME eruption time ($\sim$10:15~UT), providing additional evidence that the CME originated from the specific part of the active region.

To summarise the sequence of events, following a strong flare observed at $\sim$8:00~UT and originating from an active region on the west side of the Sun, a strong CME appears in the C2 FoV at $\sim$8:36~UT.  This CME strongly disturbs the coronal environment for hours to come.  At $\sim$9:50~UT, coronal dimming is observed on the edge of the same active region.  The dimming is associated with a second, weaker CME, estimated to have erupted at $\sim$10:15~UT.  The Type II emissions are observed at $\sim$10:46~UT, and are thus related to the second, weaker CME.  This weaker CME eventually appears in the C2 FoV at $\sim$10:57~UT.

\subsection{LOFAR imaging of the split-band Type II burst} \label{sec:imaged_bandsplitting}

The locations of the Type II emission sources are represented by centroids which are calculated by fitting a 2D elliptical Gaussian on the LOFAR emission images (see Section~\ref{sec:centroid_calc}), applied---in this case---on the 70\% maximum intensity level in order to eliminate background noise contributions.  The uncertainties on the centroid estimations (utilised throughout this chapter) were also obtained from the 2D elliptical Gaussian fit using the expressions presented in Equation~\ref{eqn:gauss_centroid_error}.  The crosses in Figure~\ref{fig:2018_dynspec} indicate the time-frequency points at which the centroid locations of the radio emission sources were calculated.  The moments when Type II emissions were imaged are indicated by red crosses, whereas black crosses are used to indicate the imaged Type III emissions.  All Type III sources were imaged at a single moment in time, depicting the spectral evolution of the sources without contributions from their temporal motion.  Both subbands of the Type II burst were imaged, with points from each of the upper- and lower-frequency subband selected at the exact same time.  This ensures that when imaged, no time-delay ambiguities will affect the relative positions of the subband sources, addressing the first concern of \cite{1974IAUS...57..389S, 1975ApL....16R..23S}.

\begin{figure}[ht!]
    \centering
    
	\captionsetup[subfigure]{aboveskip=+0.5em, belowskip=-0em}
	
\begin{subfigure}[t]{0.496\linewidth}
    \centering
    \includegraphics[width=1\textwidth, keepaspectratio=true]{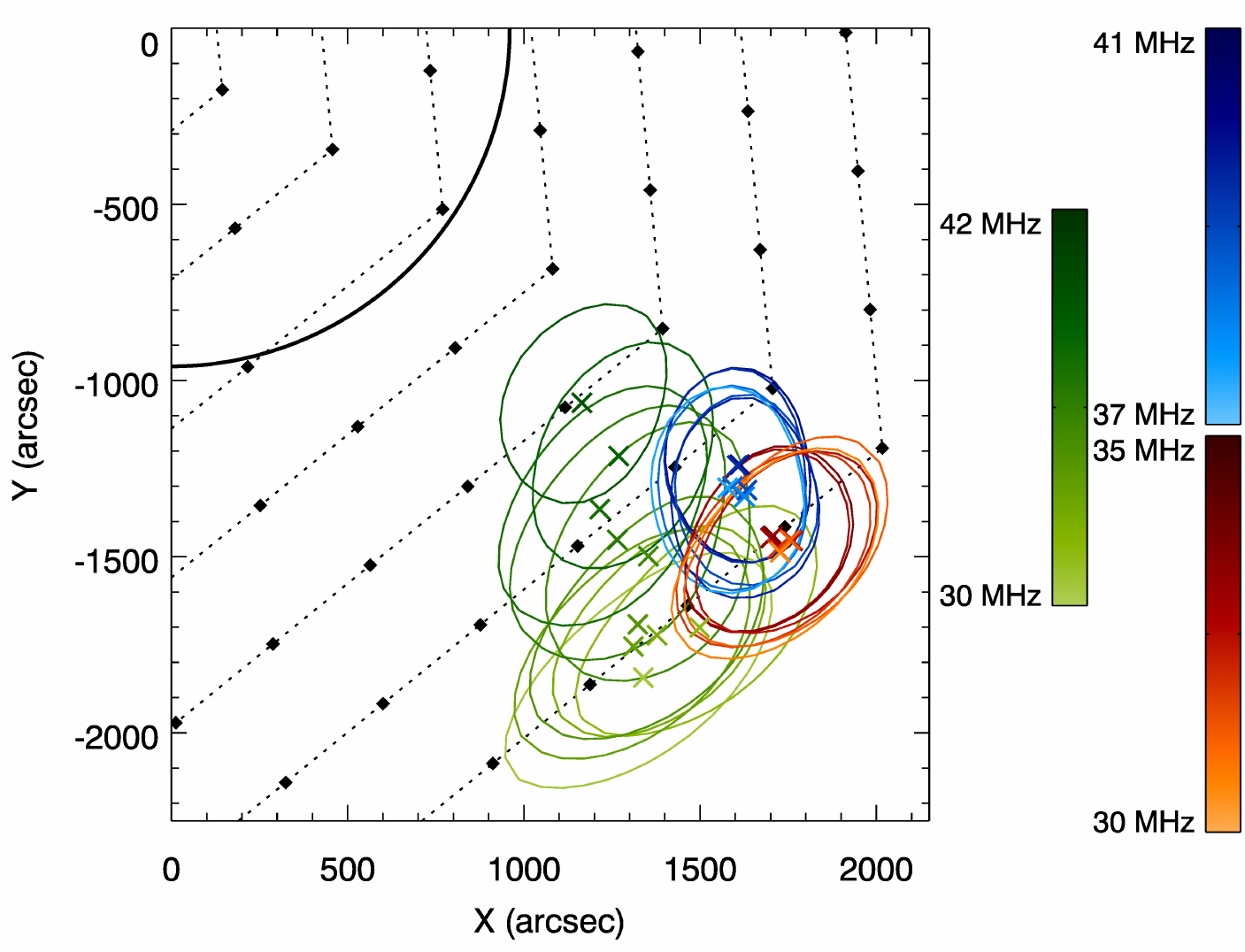}
    \caption{}
    \label{subfig:2018_lofar_img}
\end{subfigure}
\begin{subfigure}[t]{0.496\linewidth}
    \centering
    \includegraphics[width=1\textwidth, keepaspectratio=true]{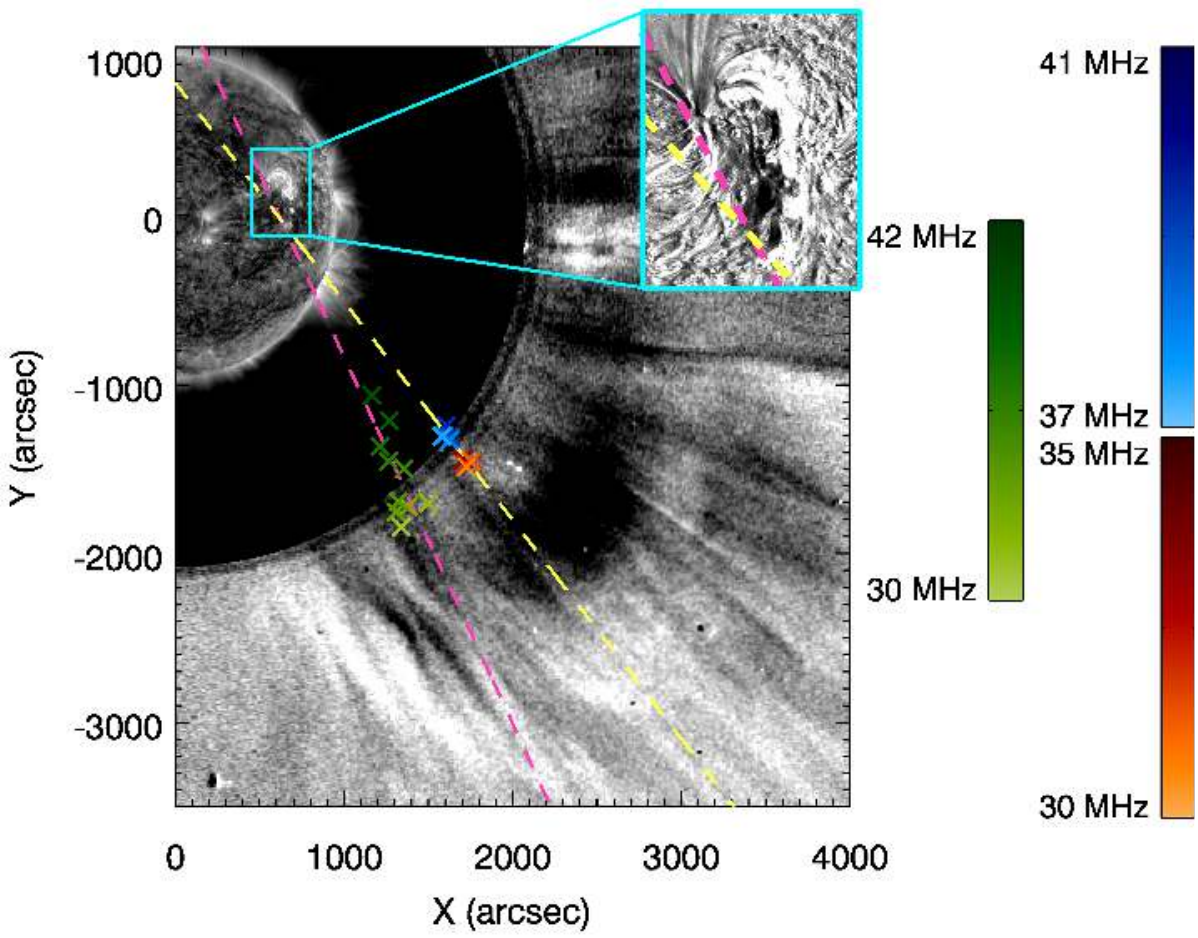}
    \caption{}
    \label{subfig:2018_lasco_fit}
\end{subfigure}

    \caption[LOFAR imaging of radio emissions and combined multi-wavelength observations.]
    {(a)~Centroid locations with respect to the solar limb (solid black curve) and respective 90\% maximum intensity contours, as obtained through LOFAR's emission images at the locations indicated in Figure~\ref{fig:2018_dynspec}.  The blue centroids represent the upper-frequency Type II subband sources, the red centroids represent the lower-frequency subband sources, and the green centroids represent the Type III centroids.  The colour schemes reflect the progression from high frequencies (dark colours) to low frequencies (bright colours), as indicated by the colour bars.  The filled black diamonds indicate the central locations of the LOFAR beams, which in collective form the tied-array beam.
    (b)~The same radio sources obtained from the LOFAR emission images as in panel (a), shown along with EUV 171~\AA\ data from \textit{SDO}/AIA which depicts activities on the solar surface, as well as a running-difference image of white-light data from the \textit{SOHO}/LASCO/C2 coronagraph highlighting the CME eruption and the coronal streamer.  A linear fit was applied through both the Type II centroids (yellow line) and Type III centroids (magenta line).  The fits appear to point back towards the active region and intersect above the area of the observed dimming.  The inset is a running-ratio image of \textit{SDO}/AIA data at 193~\AA\ indicating the area experiencing dimming, from which the CME is thought to have originated, while emphasising the point of intersection of the linear fits.
    Figure taken from \cite{2018ApJ...868...79C} and then adapted.
    }  
    \label{fig:2018_centroids}
\end{figure}

The resulting source locations are indicated in Figure~\ref{fig:2018_centroids}.  The upper- and lower-frequency subband sources of the Type II burst are presented in blue and red colour schemes, respectively, whereas the Type III sources are presented in a green colour scheme.  
Figure~\ref{subfig:2018_lofar_img} illustrates the calculated centroid locations (in crosses) and their respective 90\% maximum intensity contours, with respect to the solar limb (solid black curve) and the central locations of the LOFAR beams (black diamonds).  A combination of LASCO, AIA, and LOFAR data is indicated in Figure~\ref{subfig:2018_lasco_fit}, where the spatial relation of the radio emissions to the solar activities is illustrated.  Similar to Figure~\ref{fig:2018_cme_tracking}, brighter structures in LASCO's FoV reflect relative increases in intensity, whereas darker structures reflect relative decreases in intensity.  The Type III sources appear to trace the streamer that is located south of the CME, which arose during the eruption of the preceding CME event at $\sim$8:36~UT.  The Type II sources appear to be located at the southern flank of the CME, where compression between the CME and the streamer is likely to occur.  A linear fit was applied through the Type II and Type III sources (see the respective yellow and magenta lines).  Both lines appear to point towards the region of the dimming where they also intersect one another, as emphasised by the inset of Figure~\ref{subfig:2018_lasco_fit}, indicating that the exciters of the radio sources have potentially originated from that active region.

Other Type III (or Type III-like) bursts were also observed both before and after the Type II emissions shown in Figure~\ref{fig:2018_dynspec}.  Imaging of those Type III bursts illustrated a similar behaviour to the Type III sources presented in this section, as they were found to propagate away from the Sun with decreasing frequency, and were observed south of the Type II centroids.

\begin{figure}[ht!]
    \centering
	\includegraphics[width=0.5\textwidth, keepaspectratio=true]{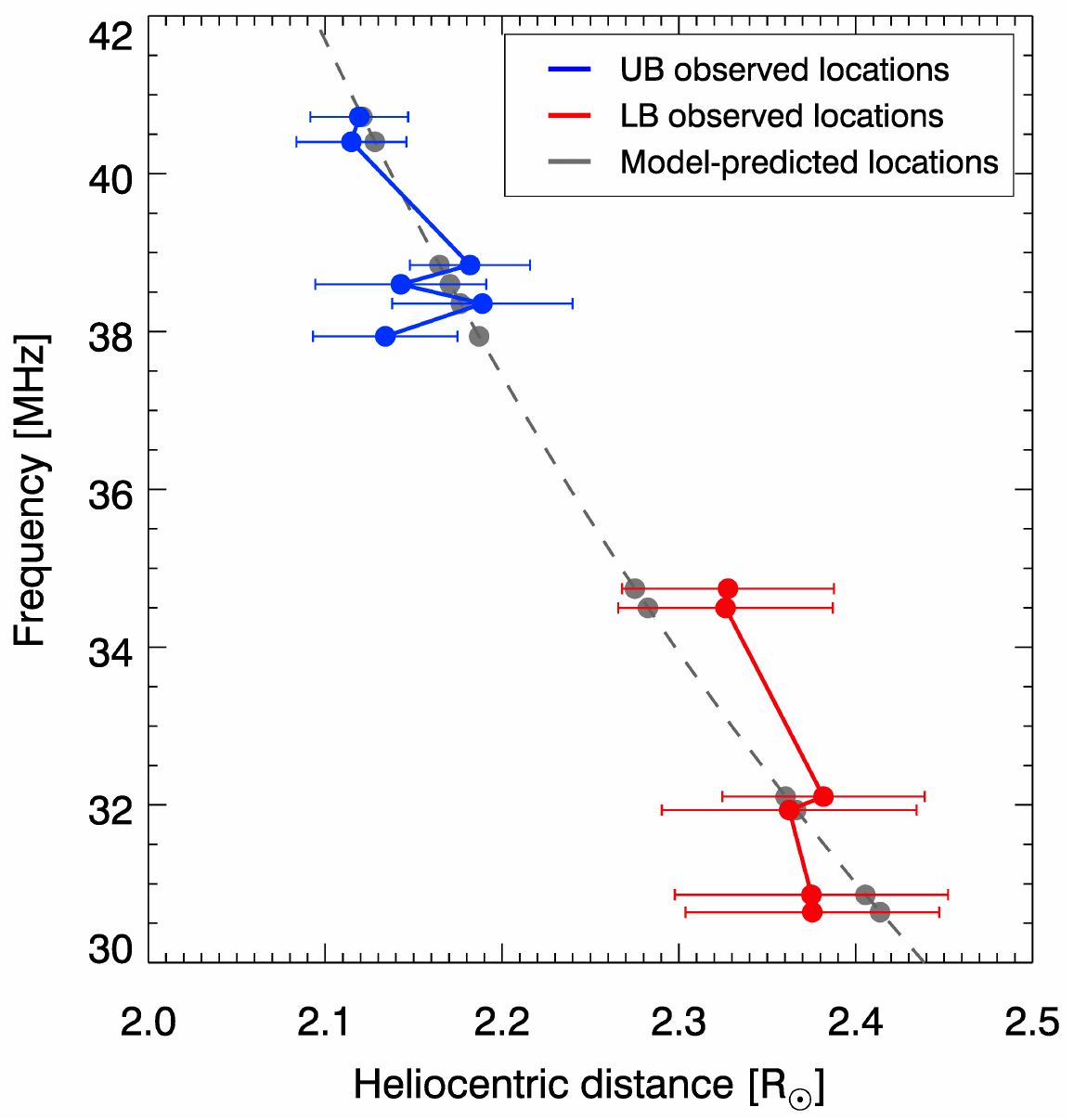}
    \caption[Comparison of observed and model-predicted source locations.]
    {
    Comparison of the observed heliocentric source locations with the heliocentric locations estimated using the 4.5$\times$Newkirk coronal density model (grey dashed line).  The upper-frequency subband sources (UB) of the Type II burst are illustrated by blue circles, whereas the lower-frequency sources (LB) are illustrated by red circles.  The model-predicted heliocentric locations of sources at equivalent frequencies were calculated and indicated by grey circles.  The error bars represent the uncertainties on the obtained centroid estimations.
	Figure taken from \cite{2018ApJ...868...79C}.
    }    
    \label{fig:2018_locations_vs_model}
\end{figure}

Despite LOFAR's capabilities allowing for time-delay ambiguities in imaging observations to be eliminated, a significant average separation of $\sim$\nolinebreak$0.2 \pm 0.05~\Rs$ was observed between the upper- and lower-frequency subband sources, as depicted in Figures~\ref{fig:2018_centroids} and \ref{fig:2018_locations_vs_model}.  This result weakens the arguments made by \cite{1974IAUS...57..389S, 1975ApL....16R..23S} and instead agrees with the \cite{1983ApJ...267..837H} prediction for split-band Type II bursts, i.e. that the intrinsic emission sources are spatially separated.

As can be seen in Figure~\ref{subfig:2018_lofar_img}, sources which are observed at the lowest frequencies (near 30~MHz) appear at the boundaries of LOFAR's FoV, and so it is possible that these emissions were not fully imaged, therefore impacting the estimation of the centroid locations.  Given this observational limitation, the imaged separation between the upper- and lower-frequency subband sources can be considered as a lower-limit of their true plane-of-sky separation.

The heliocentric distance of each Type II source was estimated using the obtained centroid locations, and was then compared to the Newkirk density model, as shown in Figure~\ref{fig:2018_locations_vs_model}.  It was found that the heliocentric locations of both the upper- (blue) and lower-frequency (red) subbands are best matched by the 4.5$\times$Newkirk model (shown by the grey dashed line), described in Equation~(\ref{eqn:r_Newkrik}) (i.e. $N$=4.5; see Section~\ref{sec:f_vs_R_relation}).  This result also agrees with the \cite{1983ApJ...267..837H} model expectation, as both subbands are found within the same atmosphere, i.e. no density jump is observed between them and so they are described by the same density model.  It is worth noting, however, that the deduced coronal density (4.5$\times$Newkirk) corresponds to relatively high electron densities.  An explanation for the high densities inferred is presented in Section~\ref{sec:scattering_TypeII_consequences} (and schematically illustrated in Figure~\ref{fig:2018_cartoon}).

\section{Considering Projection Effects in Split-Band Type II Burst Observations} \label{sec:proj_effs_in_split-band_typeIIs}
\subsection{The impact of projection effects} \label{sec:proj_effs_impact}
Several factors, like radio-wave propagation effects in the solar corona (see Section~\ref{sec:prop_effs}), can impair one's ability to observe the true 3D nature and evolution of radio sources.  Another such example is projection effects; the process where 3D information is lost during the translation to the 2D plane-of-sky depiction.

The observed heliocentric distance $R_{obs}$ of a source will vary according to the angle of observation $\theta_s$ (the angle from the observer's LoS to the source, illustrated in Figure~\ref{fig:prop_effects_cartoon}), so that $R_{obs} = R_{true} \, \sin(\theta_s)$.  Similarly, the observed separation between two sources also depends on angle $\theta_s$.  The true heliocentric distance and separation is observed when $\theta_s = 90\degr$, but as $\theta_s \rightarrow 0\degr$ the heliocentric distance and separation become increasingly underestimated.

\begin{figure}[b!]
    \centering
	\includegraphics[width=0.9\textwidth, keepaspectratio=true]{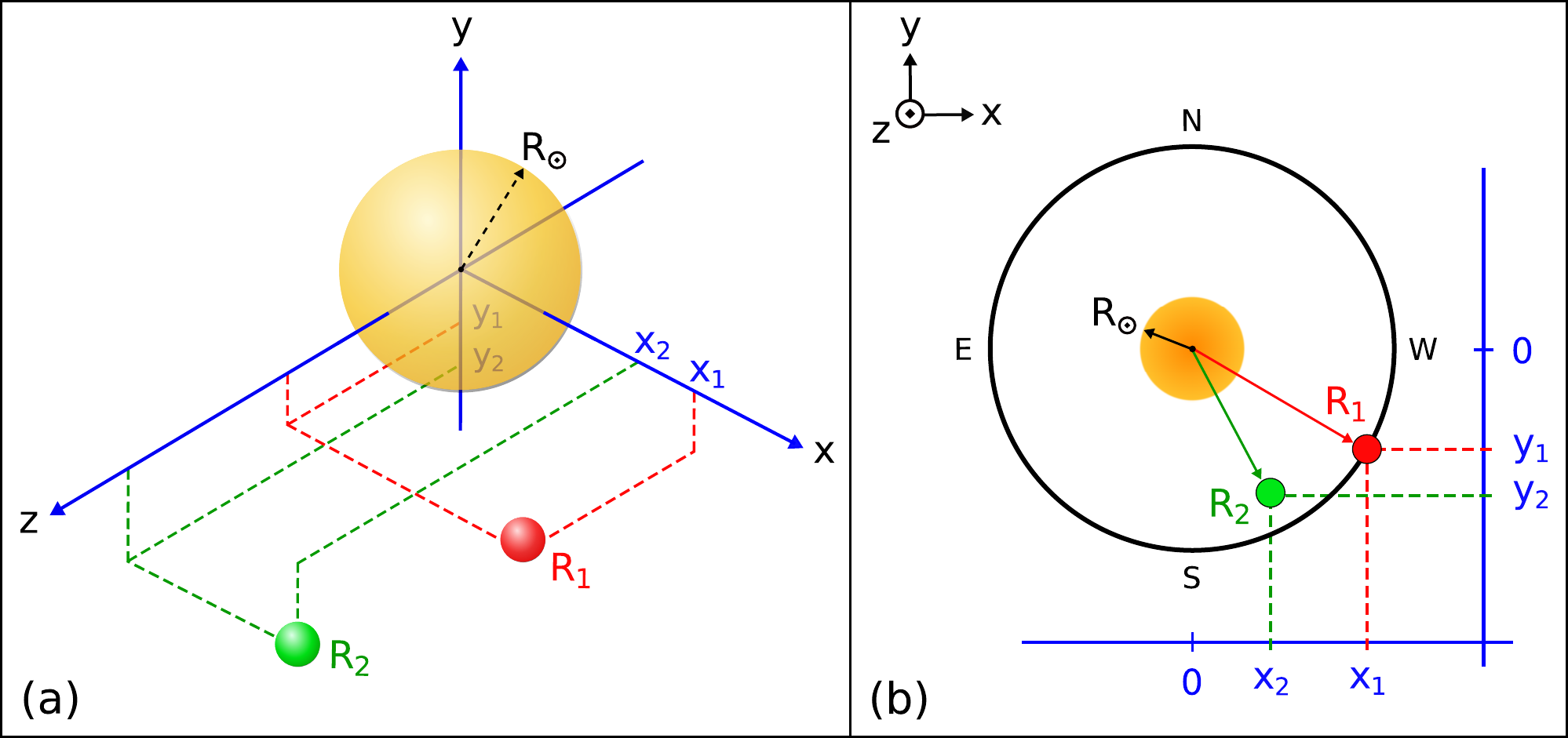}
    \caption[Schematic illustration of the impact of projection effects on observations.]
    {Schematic illustration of the impact of projection effects on 2D plane-of-sky observations.  The observer's LoS is parallel to the $z$-axis, defining the $xy$-plane as the plane of the sky.
    (a)~A 3D depiction of two sources shown in red and green, emitted at heliocentric distances $\vec{R}_1$ and $\vec{R}_2$, respectively, where $\vert \vec{R}_2 \vert > \vert \vec{R}_1 \vert$.
    (b)~The corresponding plane-of-sky projection where the green source is perceived as being closer to the solar centre than the red source, due to projection effects.
    Figure taken from \cite{2020ApJ...893..115C}.
    }  
    \label{fig:2020_proj_effs_cartoon}
\end{figure}

The relative positions of multiple sources can also be misrepresented when observations are limited to 2D information.  Figure~\ref{fig:2020_proj_effs_cartoon} illustrates the impact that projection effects can have on the interpretation of radio emissions.  Here (and throughout this thesis), the observer's LoS is taken to be parallel to the $z$-axis, and thus the plane of the sky is the $xy$-plane.  Figure~\ref{fig:2020_proj_effs_cartoon}\hyperref[fig:2020_proj_effs_cartoon]{a} illustrates the true 3D nature of two sources, where the red source is located at heliocentric distance $\vec{R}_1 = (x_1, y_1, z_1)$ and the green source is located at $\vec{R}_2 = (x_2, y_2, z_2)$, such that $\vert \vec{R}_2 \vert > \vert \vec{R}_1 \vert$.  In other words, the true location of green source is found farther away from the Sun than that of the red source.

Figure~\ref{fig:2020_proj_effs_cartoon}\hyperref[fig:2020_proj_effs_cartoon]{b}, however, represents the scenario where only the 2D plane-of-sky information is available to the observer and the out-of-plane distance ($z_1$ and $z_2$) of the sources is unknown.  From the perspective of the observer in the plane-of-sky depiction, the green source appears closer to the solar centre than the red source, so that $ R_2^2 > R_1^2$ and $x_2^2 + y_2^2 < x_1^2 + y_1^2$.

\subsection{Estimating projection effects using images of split-band Type II bursts} \label{sec:bandsplitting_proj_effs_estimation}
\begin{figure}[t!]
    \centering
	\includegraphics[width=1.0\textwidth, keepaspectratio=true]{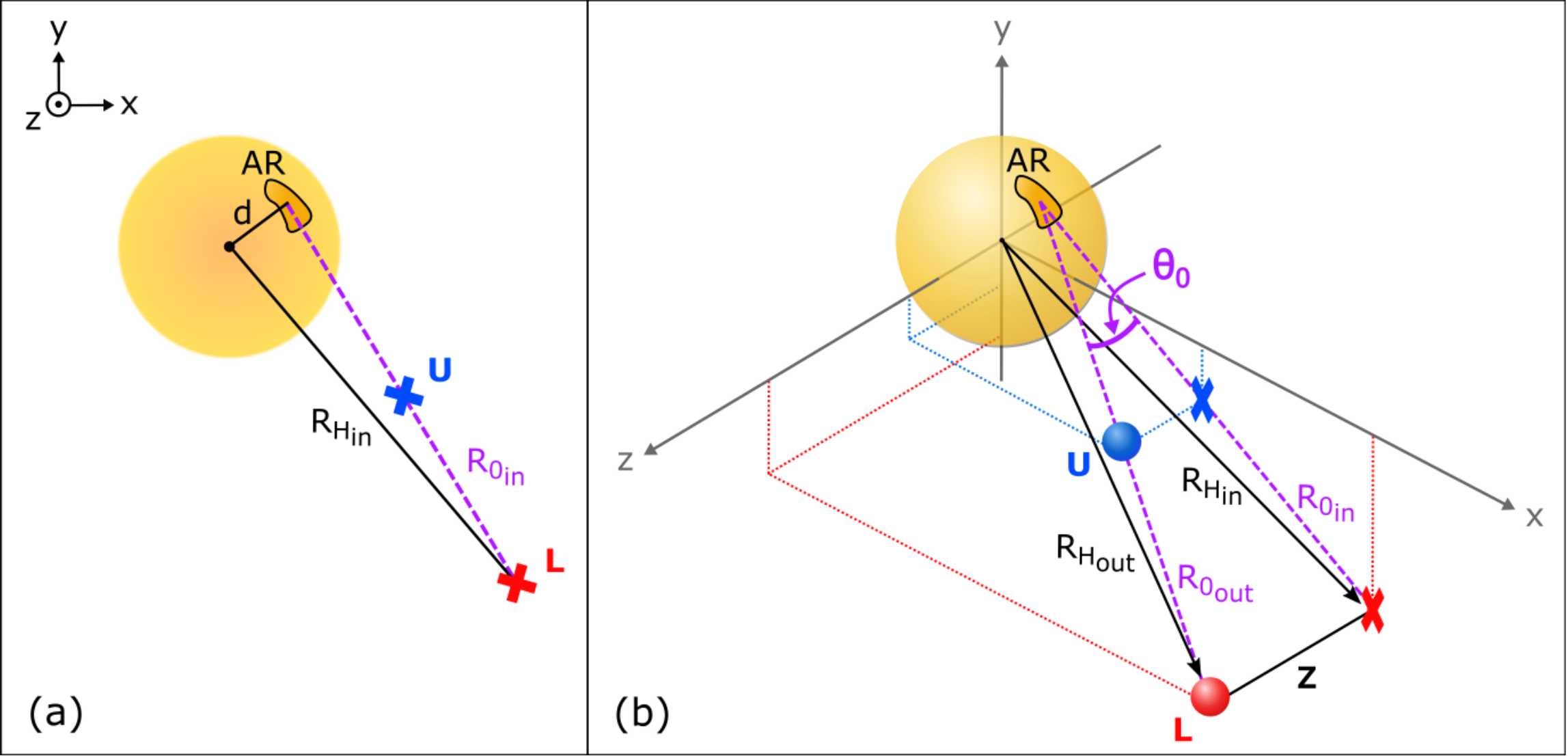}
    \caption[Schematic illustration of how the out-of-plane locations of split-band Type II sources are estimated.]
    {
    Schematic representation of the geometry that allows for the estimation of the out-of-plane locations of split-band Type II sources, by utilising simultaneous imaging of both subbands.  It is necessary that both sources follow the same trajectory away from the Sun, as indicated by the purple lines.  The upper-frequency source (U) is depicted in blue, whereas the lower-frequency source (L) is depicted in red.  The observed (in-plane) source locations are indicated with crosses, whereas the true source locations are indicated by spheres.  For the sake of clarity, descriptive labels were added only for one of the two subband sources displayed, specifically for the lower-frequency source.
    (a)~The quantities that can be obtained directly from the 2D emission images are: the observed source's in-plane heliocentric distance ($R_{H_{in}}$), its in-plane distance from a specific region of the Sun ($R_{0_{in}}$), and the heliocentric distance of that region of the Sun ($d$), shown here as a location within an active region (AR).
    (b)~A 3D depiction of the geometry allowing for the estimation of the out-of-plane distance $z$ of the source, where the angle between its in-plane and out-of-plane trajectories can be obtained (indicated as angle $\theta_0$).  The source's out-of-plane heliocentric distance ($R_{H_{out}}$) and out-of-plane distance from the region of the Sun ($R_{0_{out}}$) are also indicated.
	}
    \label{fig:typeII_out_of_plane_cartoon}
\end{figure}

Imaging observations of split-band Type II radio bursts can be used to estimate the out-of-plane location of the radio sources \citep{2018ApJ...868...79C} by applying either the \cite{1974IAUS...57..389S, 1975ApL....16R..23S} or the \cite{1983ApJ...267..837H} interpretation of band splitting.  Such estimation requires that the sources of the upper- and lower-frequency bands which are to be compared are imaged at the exact same moment in time.  Furthermore, the observation should support the assumption that both sources follow the same straight trajectory away from the surface of the Sun, i.e. that they can both be characterised by the same angle $\theta_{0}$.  Angle $\theta_0$ is defined as the angle between the vector describing the in-plane trajectory of the sources and the vector describing their out-of-plane trajectory, from a specific region of the Sun (see Figure~\ref{fig:typeII_out_of_plane_cartoon}).  The fit through the Type II sources in Figure~\ref{subfig:2018_lasco_fit} indicates that both subbands propagate along the same path away from the active region of origin, meaning that this assumption is valid for the observation presented in this chapter.

Assuming that the upper-frequency source is emitted at a location with density $n_U$ and the lower-frequency source is emitted at a location with density $n_L$, the Newkirk density model (see Equation~(\ref{eqn:n_Newkirk})) can be invoked and the density ratio of the two locations can be expressed as
\begin{equation} \label{eqn:dens_ratio_U&L}
	\dfrac{n_L}{n_U} = \dfrac{N_L \cdot n_0 \cdot 10^{4.32 R_\sol/R_L}}{N_U \cdot n_0 \cdot 10^{4.32 R_\sol/R_U}} \, ,
\end{equation} 
where $R_U$ and $R_L$ denote the true out-of-plane heliocentric distances ($R_{H_{out}}$; see Equation~(\ref{eqn:R_Hout})) of the upper- and lower-frequency sources, respectively, and $N_U$ and $N_L$ are the corresponding density multiplicative factors, as described in Equation~(\ref{eqn:n_Newkirk}).
Taking the logarithm, substituting for $n_U = (f_U/\kappa)^2$ and $n_L = (f_L/\kappa)^2$ (see Section~\ref{sec:f_vs_R_relation} and Equation~(\ref{eqn:fpe_vs_ne}))---where $f_U$ and $f_L$ represent the emission frequencies of the two sources---and re-arranging Equation~(\ref{eqn:dens_ratio_U&L}), leads to:
\begin{equation} \label{eqn:r_ratio_U&L}
	\dfrac{R_L}{R_U} = 1 - \dfrac{2}{4.32 R_\sol} \cdot R_L \left[\log_{10} \left( \dfrac{f_L}{f_U} \right) + \log_{10} \left( \dfrac{N_U}{N_L} \right) \right] \, .
\end{equation}
This relation can be applied on all split-band Type II observations, irrespective of the band-splitting mechanism believed to be at play.  In other words, it can be applied whether the observations suggest that both the upper- and lower-frequency subband sources are found upstream of the shock front, or the observations suggest that the subband sources are found on opposite sides of the shock front (see Section~\ref{sec:bandsplitting_models}).

If the locations of both sources are emitted upstream of the shock where no density jump occurs between them---like in the \cite{1983ApJ...267..837H} model---the same density model determines the locations of both sources.  In this case \citep{2018ApJ...868...79C}, the ratio of the multiplicative factors
\begin{equation} \label{eqn:N_ratio_Holman}
	\dfrac{N_U}{N_L} = 1 \, .
\end{equation}
If, on the other hand, the sources are thought to emit from opposite sides of the shock front---like in the \cite{1974IAUS...57..389S, 1975ApL....16R..23S} model---a density jump occurs between the two sources.  The Rankine-Hugoniot conditions can be invoked to infer the density jump across the shock front using the observed (average) frequency split $\Delta f_s$ of the two subbands (see Equation~(\ref{eqn:shock_density_jump_n2/n1}) and Section~\ref{sec:bandsplitting_models}), estimating the ratio of the 
multiplicative factors as:
\begin{equation} \label{eqn:N_ratio_Smerd}
	\dfrac{N_U}{N_L} = \left( \dfrac{f_U - f_L}{f_L} + 1 \right)^2 \, .
\end{equation}

An expression for the out-of-plane heliocentric distance ($R_{H_{out}}$) of each source can be obtained using geometric relations which include parameters than can be directly acquired from the 2D images \citep{2018ApJ...868...79C}.  The out-of-plane distance $R_{H_{out}}$ of a source is a function of the source's in-plane heliocentric distance $R_{H_{in}}$ (which is obtained from the images) and a certain distance $z$ out of the plane (see Figures~\ref{fig:2020_proj_effs_cartoon} and \ref{fig:typeII_out_of_plane_cartoon}):
\begin{equation} \label{eqn:R_Hout}
R_{H_{out}} = \sqrt{z^2 + R_{H_{in}}^2} \,.
\end{equation}

The out-of-plane distance $z$ of the source is an unknown parameter which can be calculated thanks to the the assumption that both sources propagate away from the Sun along the same trajectory.  This allows for the out-of-plane distance $z$ to be expressed as:
\begin{equation} \label{eqn:out_of_plane_dist}
z = R_{0_{in}} \cdot \tan(\theta_0) \,.
\end{equation}
Here, $R_{0_{in}}$ is the in-plane distance of the source from a specific region on the solar surface.  For the Type II observation presented in this chapter, this region was considered to be a point along the linear fit through the Type II centroids and within the area of the active region in which dimming was observed, believed to be the origin of the CME exciting the radio emission (see Figure~\ref{subfig:2018_lasco_fit}).  The (heliocentric) location of this region of origin can also be calculated from the 2D images (see distance $d$ in Figure~\ref{fig:typeII_out_of_plane_cartoon}\hyperref[fig:typeII_out_of_plane_cartoon]{a}).  Similar to $R_{H_{in}}$, distance $R_{0_{in}}$ can also be calculated from the observed emission images.  Angle $\theta_0$ is an unknown parameter that can, however, be computed using the mathematical relation given in Equation~(\ref{eqn:r_ratio_U&L}).  The solution is the value of angle $\theta_0$ that satisfies the equality in Equation~\ref{eqn:r_ratio_U&L}, given that angle $\theta_0$ must have the same value for both the upper- and lower-frequency sources (see Figure~\ref{fig:typeII_out_of_plane_cartoon}\hyperref[fig:typeII_out_of_plane_cartoon]{b}), and the resulting ratio of the out-of-plane heliocentric distances ($R_L/R_U$) must agree with the right-hand side of the equation, satisfying the equality.

An advantage of estimating the out-of-plane heliocentric distances of the subband sources (instead of the in-plane distances), is that they can be compared to the distances predicted by the assumed density model (like in Figure~\ref{fig:2018_locations_vs_model}), which allows for a more realistic multiplicative factor $N$ to be obtained (see Equation~(\ref{eqn:n_Newkirk})).  This will in turn enable a more realistic estimation of the local coronal density from the studied radio observations (and thus inferred exciter speed; see Section~\ref{sec:f_vs_R_relation}), potentially influencing other further-inferred parameters describing the local coronal conditions (like those described in Section~\ref{sec:bandsplitting_models}).

The observational limitations (discussed in Section~\ref{sec:imaged_bandsplitting}) present in this observation, however, did not allow for the computation of the out-of-plane distances with confidence.
Specifically, the lower frequency sources are observed at the edge of LOFAR's FoV, implying that their centroid locations may be underestimated.  This could explain why a single value (or a narrow range) of the angle $\theta_0$ could not be obtained from the current observations.  However, this model could prove to be beneficial for analyses of observations limited to 2D plane-of-sky information.  Moreover, this method can be applied to, for example, Type III bursts (where Equation~(\ref{eqn:N_ratio_Holman}) should be used, as no density jump is present), assuming that different frequencies can be imaged at the exact same time and that the sources follow the same straight trajectory away from the Sun.
Future observations conducted with instruments that possess the necessary imaging capabilities (i.e. image multiple frequencies simultaneously and have a sufficiently large FoV), like LOFAR, can be used to test the mathematical model presented in this section.

\section{Accounting for Scattering Effects in Split-Band Type II Bursts} \label{sec:TypeII_scattering}
The subband sources of a split-band Type II burst were imaged at the same time, eliminating time-delay ambiguities, but a significant average separation between the upper- and lower-frequency subband sources was still observed ($\sim$\nolinebreak$0.2 \pm 0.05~\Rs$; see Figure~\ref{subfig:2018_lofar_img}).  As mentioned in Section~\ref{sec:previous_split-band_TypeII_imaging}, besides the argument of time-delay ambiguities in observations, \cite{1974IAUS...57..389S, 1975ApL....16R..23S} argued that scattering effects can also distort the intrinsic separation between sources, as lower-frequency sources are shifted away from their true location more than higher-frequency sources, in a way that considerable separations are perceived.  Later studies highlighted the dominance and significance of scattering effects in the solar corona \citep{2017NatCo...8.1515K}.  The impact, however, of scattering effects on split-band Type II sources was never quantitatively estimated prior to the results presented in this chapter \citep{2018ApJ...868...79C}.

One of the most-commonly analysed (and sometimes only) parameter in observational studies is the apparent location of the radio sources (i.e. the centroid location).  In the following section, an analytical estimation for the scattering-induced shift in source locations is presented, derived by assuming isotropic density fluctuations.  So far in this thesis, the necessity to account for the anisotropy in the density fluctuations has been demonstrated, by illustrating that multiple source properties can be simultaneously reproduced only if anisotropy is considered.  However, isotropic scattering simulations have been able to reproduce individual properties, like the source locations (as discussed in Section~\ref{sec:ray-tracing_simulations_motivation}).  Nevertheless, in many of the observational studies focusing on source positions, the effects of scattering have not been examined.  A reason for this could be that radio-wave propagation simulations are complex and often computationally intensive (like those accounting for anisotropy; described in Chapter~\ref{chap:scattering}).  Hence, the upcoming section provides an alternative, analytical method for estimating the displacement in the observed sources, allowing for the scattering-induced shift to be considered when the use of simulations is not desired, a preferred approach to entirely neglecting this effect.
This chapter also demonstrates that the interpretation of the observations is altered once scattering is accounted for, highlighting the need for a consideration of the scattering-induced effects, even if isotropy is assumed.

\subsection{Analytical estimation of the scattering-induced shift} \label{sec:tau_deriviation}

Several simplifying assumptions were made in order to derive an analytical expression which estimates the radial shift induced by scattering that radio sources experience.  Following previous scattering estimations \citep{1952MNRAS.112..475C, 1968AJ.....73..972H, 1971A&A....10..362S, 1999A&A...351.1165A, 2007ApJ...671..894T}, homogeneous, isotropic, and stationary density fluctuations with a Gaussian correlation in an unmagnetised plasma environment are considered, described by a (Gaussian) spatial autocorrelation function
\begin{equation}\label{eqn:dn_dn}
  \langle\delta n(\vec{r}_1)\delta n(\vec{r}_2)\rangle = \langle\delta n^2\rangle \exp \left(-\dfrac{(\vec{r}_1-\vec{r}_2)^2}{h^2}\right)\, ,
\end{equation}
where $h$ is the characteristic density scale height (or ``radius'' of correlation), $n$ is the density, $r$ is the heliocentric distance, and $\langle ... \rangle$ denotes an ensemble average.
These isotropic density inhomogeneities cause photons of frequency $f$ to experience angular scattering.  Considering small steps $dr$ over which the photons' path is linear and the relative variation of the refractive index $\delta\mu / \mu$ over a single step is small ($\delta\mu / \mu < 0.1$), the expression for small scattering angles ($\Delta \theta < 0.1$~radians) in the coronal medium is given by \citep{1971A&A....10..362S}:
\begin{equation} \label{eqn:delta_theta_Steinberg}
	\langle\Delta \theta^2\rangle = \dfrac{2\sqrt{\pi}}{h} \, \dfrac{\langle\delta \mu^2\rangle}{\mu^2} \, dr \, .
\end{equation}
The quantity $\langle\delta \mu^2\rangle$ is the mean square fluctuation of the refractive index which can be related to $\langle\delta n^2\rangle$ via \citep{1971A&A....10..362S}:
\begin{equation} \label{eqn:delta_mu_Steinberg}
	\langle\delta \mu^2\rangle = \dfrac{1}{4\mu^2} \, \dfrac{f_{pe}^4(r)}{f^4} \, \dfrac{\langle\delta n^2 \rangle}{n^2} \, ,
\end{equation}
where $\langle\delta n \rangle / n \equiv \epsilon$ represents the relative density fluctuations, $f_{pe}$ is the electron plasma frequency (Equation~\ref{eqn:fpe_vs_ne}), and $f$ is the observed frequency.  The mean scattering rate per unit radial distance (obtained by combining Equations~(\ref{eqn:delta_theta_Steinberg}) and (\ref{eqn:delta_mu_Steinberg})) is given as: 
\begin{equation} \label{eqn:dtheta_dr_Steinberg}
	\dfrac{d\langle\Delta \theta^2\rangle}{dr} = \dfrac{\sqrt{\pi}}{2} \, \dfrac{f_{pe}^4(r)}{f^4} \, \dfrac{1}{\mu^4} \, \dfrac{\epsilon^2}{h} \, .
\end{equation}
The refractive index is defined as $\mu = 1 - f_{pe} / f$ (see Equation~(\ref{eqn:refractive_index})), leading to the following expression for angular scattering:
\begin{equation} \label{eqn:dtheta_dr}
  \dfrac{d\langle\Delta \theta^2\rangle}{dr} = \dfrac{\sqrt{\pi}}{2} \, \dfrac{f_{pe}^4(r)}{(f^2-f^2_{pe}(r))^2} \, \dfrac{\epsilon^2}{h} \, .
\end{equation}
It can be seen that the scattering rate $d\langle\Delta \theta^2\rangle / dt$ depends on the level of density fluctuations $\epsilon$, as well as the characteristic density scale height $h$, given that
\begin{equation} \label{eqn:dtheta_dt}
	\dfrac{d\langle\Delta \theta^2\rangle}{dt} = \dfrac{1}{v_{g}} \, \dfrac{d\langle\Delta \theta^2\rangle}{dr} \, ,
\end{equation}
where $v_{g}$ is the group velocity of the photons.  The scattering rate is also a decreasing function of the radial distance $r$ from the Sun, meaning that the scattering is frequent when $f \gtrsim f_{pe}$ (close to the source's emission location), but at larger heliocentric distances where $f \gg f_{pe}$ the scattering becomes negligible (see Section~\ref{sec:prop_effs}).
The effect of radio-wave scattering is quantitatively characterised through the optical depth (with respect to scattering) $\tau$:
\begin{equation}\label{eqn:tau}
   \tau(r)=\int_{r}^{1 \, au} \dfrac{d\langle\Delta \theta^2\rangle}{dr} \, dr=
   \int_{r}^{1 \, au} \dfrac{\sqrt{\pi}}{2} \, \dfrac{f_{pe}^4(r)}{(f^2-f^2_{pe}(r))^2} \, \dfrac{\epsilon^2}{h} \, dr \, .
\end{equation}

The apparent location of the source is taken to be the distance at which the radio-wave optical depth $\tau=1$.  This is assumed to be the heliocentric distance at which the transition between a region of strong scattering to a region of weak or no scattering occurs (referred to as the ``scattering screen'' in Chapter~\ref{chap:scattering}).  In other words, at distances for which $\tau < 1$, there occurs less than one scattering event on average, meaning that the contribution of scattering at those distances is negligible.

Equation~(\ref{eqn:tau}) requires the observed frequency $f$ to be defined as $\gtrsim f_{pe}$ in order for a physical result to exist, unlike other estimations that allow for $f=f_{pe}$ to be assumed (see, e.g., Section~\ref{sec:f_vs_R_relation} or \ref{sec:bandsplitting_proj_effs_estimation}).  Thus, Equation~(\ref{eqn:tau}) can be solved for all values of $f_{pe}$ that are less than $f$, and their corresponding $r$ values (defined by the chosen density model).  An approximation for harmonic emissions can also be obtained simply by taking $f \gtrsim 2 f_{pe}$.

It should also be emphasised that the expression is depended on the ratio of $\epsilon^2/h$ and requires it to be a fixed constant over $r$.  This implies that $\epsilon$ and $h$ can be assumed to vary with heliocentric distance without invalidating the results obtained via this calculation, as long as their ratio remains the same (see, e.g., \cite{1971A&A....10..362S, 1974SoPh...35..153R}).

While not as physically accurate as computationally-intensive radio-wave propagation simulations (like those discussed in Chapters~\ref{chap:scattering} and \ref{chap:observation_simulations}), the expression derived in this section provides a simple, fast, and analytical method of estimating the scattering-induced shift of radio sources, a preferred approach to entirely neglecting the contribution of this important effect.  This analytical approximation is not limited to a specific type of radio emissions (i.e. it is not dependent on the emissions' exciter), since the required inputs are merely the observed frequency $f$, a constant representing the ratio of $\epsilon^2/h$, and a coronal density model (used to relate $f$ to the radial distance $r$).  Depending on the input values chosen, this analytical approach can provide outputs similar to those from simulations, like the ones presented in Chapter~\ref{chap:scattering}.

\subsection{Application to split-band Type II emissions} \label{sec:split-band_shift_estimation}
\begin{figure}[ht!]
    \centering
	\includegraphics[width=0.7\textwidth, keepaspectratio=true]{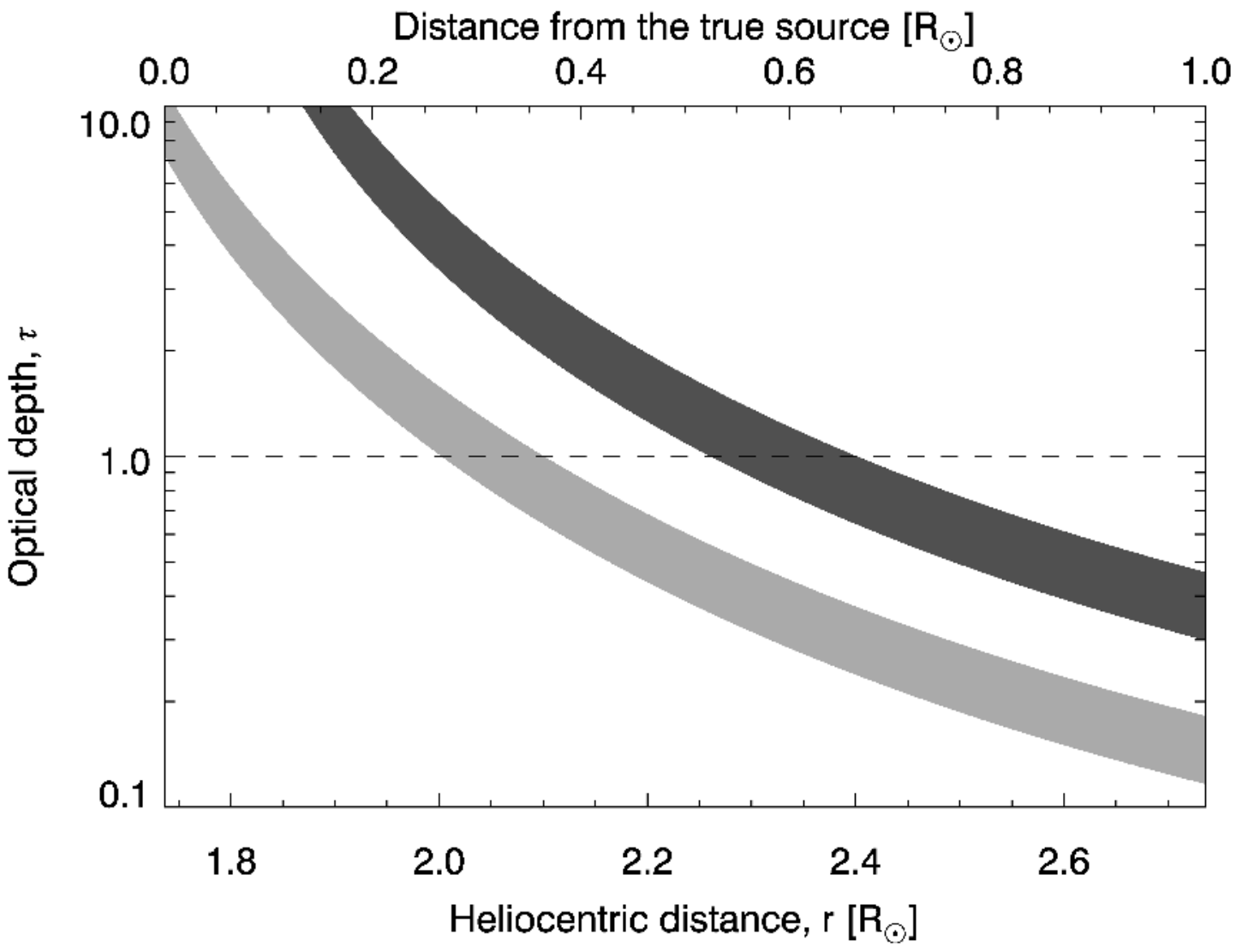}
    \caption[Radio-wave optical depth with respect to scattering as a function of heliocentric distance.]
    {Radio-wave optical depth (with respect to scattering) $\tau$ as a function of heliocentric distance $r$.  Results for two frequencies representing the subbands of a split-band Type II burst are shown, assuming that the subband emissions originate from the same location.  The area shaded in light grey represents the solutions for a radio source emitted at $f_U = 40$~MHz and for values of $\epsilon^2/h$ ranging from $4.5 \times 10^{-5} \, \mathrm{km^{-1}}$ (left boundary) to $7 \times 10^{-5} \, \mathrm{km^{-1}}$ (right boundary), whereas the dark grey area represents the solutions for emissions at $f_L = 32$~MHz for the same range of $\epsilon^2/h$ values.  The dashed line indicates the point at which $\tau=1$, taken to represent the heliocentric distance at which the radio emissions are observed.  The bottom $x$-axis illustrates the heliocentric distance of the sources given the 1$\times$Newkirk model, whereas the top $x$-axis illustrates the amount of scattering-induced radial shift of the observed sources from their true locations.
    Figure taken from \cite{2018ApJ...868...79C}.
	}
    \label{fig:2018_tau}
\end{figure}

Utilising the analytical method derived in Section~\ref{sec:tau_deriviation},
the separation between two sources of different frequencies can be estimated and compared to the imaged separation between split-band Type II sources.  The characteristic optical depth $\tau$ (Equation~(\ref{eqn:tau})) was solved for two frequencies and for a range of $\epsilon^2/h$ values---for comparison---assuming the 1$\times$Newkirk density model (see Equation~(\ref{eqn:r_Newkrik})) and fundamental emissions.  The values for the ratio of $\epsilon^2/h$ that were considered varied from $4.5 \times 10^{-5}$ to $7 \times 10^{-5} \, \mathrm{km^{-1}}$ \citep{1971A&A....10..362S, 1974SoPh...35..153R}.  The two frequencies used for the computation of $\tau$ (shown in Figure~\ref{fig:2018_tau}) are $f_U = 40$~MHz and $f_L = 32$~MHz, representing the upper- and lower-frequency subbands of the split-band Type II burst, respectively (see Figure~\ref{fig:2018_dynspec}).  To apply the calculations to a scenario in which the $f_U$ and $f_L$ sources are virtually co-spatial, like in the \cite{1974IAUS...57..389S, 1975ApL....16R..23S} model, values of the heliocentric distance $r$ corresponding to $f \gtrsim f_{pe}$ where $f_{pe} < f_L$ were considered in Equation~(\ref{eqn:tau}), both for the 40 and 32~MHz sources.  This implies that both sources are emitted at a heliocentric distance at which the local plasma density corresponds to a frequency just below $f_L$ (see Section~\ref{sec:f_vs_R_relation}).  According to the 1$\times$Newkirk model, the distance corresponding to $f_{pe} \lesssim f_L$ is expected at $\sim$1.74~$\Rs$, which is taken as the true heliocentric location for both the 40 and 32~MHz sources in the split-band Type II scenario imitated (cf. inset of Figure~\ref{fig:2018_cartoon}).

Figure~\ref{fig:2018_tau} shows the optical depth calculations for the two sources.  The area shaded in light grey represents the results for the 40~MHz source, calculated for $\epsilon^2/h$ values from $4.5 \times 10^{-5}$ (left side of shaded area) to $7 \times 10^{-5} \, \mathrm{km^{-1}}$ (right side).  The dark grey shaded area represents the equivalent results for a 32~MHz source.
The distance at which the sources are expected to be observed---once scattering effects are considered---is illustrated in the bottom $x$-axis, where the result must be taken at $\tau = 1$, indicated by the horizontal dashed line (i.e. the scattering screen).

It can be seen that both sources are shifted radially away from their true location ($\sim$1.74~$\Rs$), with the lower-frequency source experiencing a higher degree of shift compared to the higher-frequency source (as expected).  Given the taken $\epsilon^2/h$ values, the average estimated observed location for the 40~MHz source is $\sim$2.05~$\Rs$, whereas for the 32~MHz source it is $\sim$2.34~$\Rs$, i.e. a good agreement with the imaged heliocentric distances of the Type II sources at those frequencies (see Figure~\ref{fig:2018_locations_vs_model}).
As evident, the 40~MHz source shifts by $\sim$0.3~$\Rs$ away from its true location and the 32~MHz source shifts by $\sim$0.6~$\Rs$, indicated by the top $x$-axis of Figure~\ref{fig:2018_tau}, where the result is again taken at $\tau = 1$.  These estimations suggest that when imaged, a separation of $\sim$0.3~$\Rs$ is expected between sources emitted at 40 and 32~MHz, which are otherwise intrinsically co-spatial, purely because of the radio-wave scattering effects and their stronger impact on lower-frequency emissions.

\subsection{Consequences on the observation's interpretation} \label{sec:scattering_TypeII_consequences}

Radio-wave scattering effects can account for the large separation of $\sim$\nolinebreak$0.2 \pm 0.05~\Rs$ observed between the split-band Type II sources imaged by LOFAR, as discussed in Section~\ref{sec:split-band_shift_estimation} where co-spatial intrinsic locations were assumed for the calculations.  This outcome implies that the imaged separation can no longer be considered as supporting evidence for the \cite{1983ApJ...267..837H} band-splitting model, but instead supports models requiring the intrinsic emission sources to be co-spatial, like the \cite{1974IAUS...57..389S, 1975ApL....16R..23S} model.

In order to evaluate this result, the maximum separation between the upper- and lower-frequency subbands was taken, given the calculated uncertainties on the observed locations (see Figure~\ref{fig:2018_locations_vs_model}).  A total of $0.3$~$\Rs$ and $0.6$~$\Rs$ was subtracted from the upper- and lower-frequency source locations, respectively.  On average, the remaining physical separation between the two subbands was found to be $\lesssim$~0.02~$\Rs$.  If the subbands do not originate from regions upstream and downstream of a shock front, the remaining physical separation cannot sufficiently account for the observed spectral separation of $\sim$8~MHz between the two subbands, even if higher coronal densities are assumed.  In other words,
\begin{equation*}
	\Delta f_s = f_U - f_L = \kappa \, \sqrt{n_U(r)} - \kappa \, \sqrt{n_L(r+0.02)} \ll 8 \, \mathrm{MHz} \, ,
\end{equation*}
where r is taken to be $\sim$1.74~$\Rs$ for this observation, $\kappa$ is defined as in Equation~(\ref{eqn:fpe_vs_ne}), and $n_U$ and $n_L$ represent the densities at the upper- and lower-frequency subband sources, respectively.  This means that the \cite{1983ApJ...267..837H} model cannot be used to describe the studied split-band Type II observation once scattering effects are taken into account.

The radial shift of the sources away from their true location also affects the inferred coronal density.  The coronal density will be overestimated when the \textit{observed} source locations are used to infer the coronal density model best describing the local environment of the emissions, as done in Figure~\ref{fig:2018_locations_vs_model}.  The impact on the apparent density model is schematically illustrated in the inset of Figure~\ref{fig:2018_cartoon}.  Assuming that the intrinsic location $R_i$ of a 32~MHz source is at $\sim$1.74~$\Rs$, as given by the 1$\times$Newkirk density model, the density at that location is given by $n_e(R_i)$ (defined in Equation~(\ref{eqn:n_Newkirk})).  When scattering effects displace the source radially away from the Sun by $\sim$0.6~$\Rs$, its apparent position is $R_s = 2.34$~$\Rs$, which corresponds to an increased density given by $n_e(R_s)$.  The ratio between these two densities (assuming the 1$\times$Newkirk density model) is $n_e(R_i) / n_e(R_s) = n_e(1.75 \, \Rs) / n_e(2.34 \, \Rs) \simeq 4.3$.  In other words, in order for a 32~MHz source to appear at a location of $\sim$2.34~$\Rs$, a corona described by the 4.3$\times$Newkirk model would need to exist at the time and location of the observation.  This means that, in the studied event, scattering leads to the apparent coronal density (deduced from observations near 32~MHz) to be overestimated by a factor of $\sim$4.3.

Equivalently, the upper- and lower-frequency subband source locations can be corrected to obtain a description of the coronal density after the effects of scattering have been considered.  When the correction is applied on the observed locations, the upper-frequency subband is best described by the 1.9$\times$Newkirk model, whereas the lower-frequency subband is best described by the 1.3$\times$Newkirk model.  This means that there is a factor of $\sim$1.46 difference between the two density models, i.e. there is a density jump of $\sim$1.46 between the upper- and lower-frequency subbands.  The two subbands can no longer be described by the same density model, meaning that the argument for the \cite{1983ApJ...267..837H} model has weakened further.  It is worth mentioning that the density jump deduced from the dynamic spectrum by invoking the \cite{1974IAUS...57..389S, 1975ApL....16R..23S} model and the Rankine-Hugoniot conditions (see Section~\ref{sec:bandsplitting_models} and Equation~(\ref{eqn:shock_density_jump_n2/n1})), also has a value of $\sim$1.46, matching the value obtained from the corrected source locations.

\section{Discussion and Conclusions} \label{sec:bandsplitting_conclusions}
This chapter introduced, in the context of the \cite{1974IAUS...57..389S, 1975ApL....16R..23S} and \cite{1983ApJ...267..837H} models, the decades-long debate over the interpretation of imaging observations of split-band Type II bursts which depicted large physical separations between the sources of the two subbands.  A LOFAR observation of a split-band Type II burst was studied and used to address the shortcomings in the imaging capabilities of previously-available radio detectors.  LOFAR enabled imaging of the two subbands at the exact same time (see Section~\ref{sec:radio_obs}), comparing the relative positions of the emission sources without any time-delay ambiguities in the observations.

A large separation of $\sim$\nolinebreak$0.2 \pm 0.05~\Rs$ was observed between the sources of the upper- and lower-frequency subbands of the split-band Type II burst studied.  However, due to the limitations of the presented LOFAR observation (see Section~\ref{sec:imaged_bandsplitting}) and projection effects that may have been at play (see Section~\ref{sec:bandsplitting_proj_effs_estimation}), the observed separation of $\sim$\nolinebreak$0.2 \pm 0.05~\Rs$ can only be considered as a lower limit.
Moreover, when the observed sources were compared to the locations predicted by a coronal density model, no density jump was observed between the two subbands (see Figure~\ref{fig:2018_locations_vs_model}).  These observations---i.e. subband sources that are well separated and emitted in the same atmosphere---are in agreement with the expectations of the \cite{1983ApJ...267..837H} band-splitting model.  Specifically, the apparent heliocentric locations of the subbands were both described by the 4.5$\times$Newkirk density model.

The importance of appreciating the limitations of imaging observations when the information available does not represent the full 3D nature of the emissions was also illustrated in this chapter.  A mathematical model that allows for the estimation of the extent of projection effects through the calculation of the out-of-plane locations of split-band Type II sources was presented in Section~\ref{sec:bandsplitting_proj_effs_estimation}.  The model is applicable to all split-band Type II observations, whether they support models that expect co-spatial or physically-separated intrinsic sources.

\begin{figure}[ht!]
    \centering
	\includegraphics[width=0.9\textwidth, keepaspectratio=true]{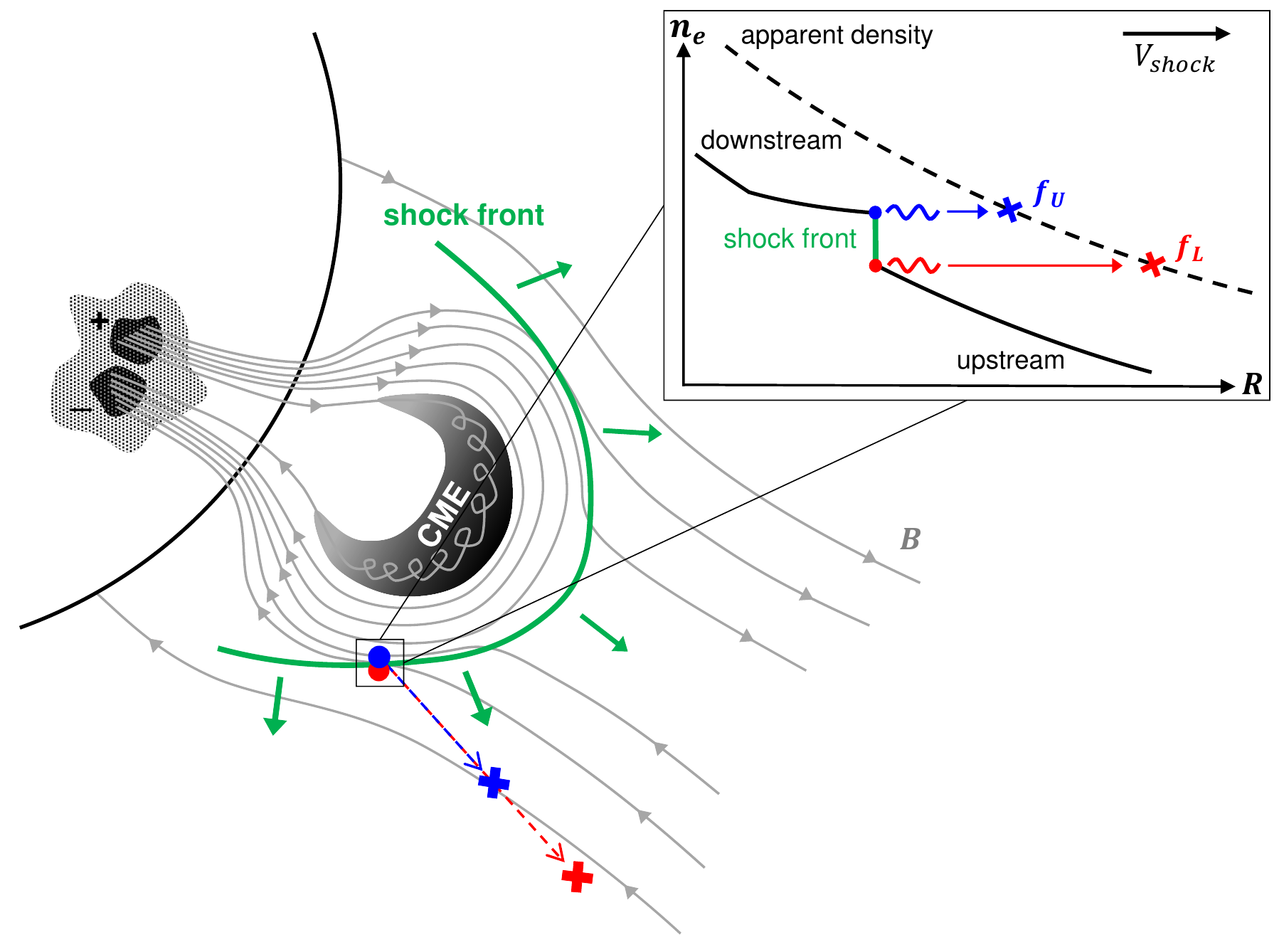}
    \caption[Schematic illustration of scattering effects on the observed split-band Type II source positions and the inferred coronal density model.]
    {
    Schematic illustration of the impact of scattering effects on the observed positions of split-band Type II sources and the apparent coronal density model.  A CME propagates away from an active region on the solar surface, disturbing the local magnetic field $B$ (grey lines) and driving a shock wave (green line).  The shock wave excites split-band Type II radio emissions whose intrinsic subband sources are virtually co-spatial, with the upper-frequency source ($f_U$, blue disk) located downstream of the shock front and the lower-frequency source ($f_L$, red disk) located upstream of the shock front.  Radio-wave scattering effects impact the propagation of the emitted radio waves, causing the sources to shift radially away from their true location, with the lower-frequency source affected the most, resulting in an apparent spatial separation between the two sources when imaged (blue and red crosses).  The radial shift and perceived separation lead to an overestimation of the coronal density model when deduced from the emission images, as indicated by the black dashed line in the inset.    
    Figure taken from \cite{2018ApJ...868...79C}.
	}
    \label{fig:2018_cartoon}
\end{figure}

The scattering-induced separation between two split-band Type II sources was quantitatively estimated for the first time.  Two sources emitting at 40 and 32~MHz were used in the quantitative estimation to represent the subband emissions of the Type II burst observed by LOFAR.  The calculations were set so that both sources are intrinsically co-spatial, as expected by the \cite{1974IAUS...57..389S, 1975ApL....16R..23S} model.
Figure~\ref{fig:2018_cartoon} is a schematic presentation of the consequences that radio-wave scattering effects have on the emissions of a split-band Type II burst, where the source emitted upstream of the shock (32~MHz) is indicated in red and the one emitted downstream (40~MHz) is indicated in blue.
It was found that the 40~MHz source shifts radially away from its true location by $\sim$0.6~$\Rs$, whereas the 32~MHz source shifts by $\sim$0.3~$\Rs$.  As such, it was shown that radio-wave scattering effects can induce a separation of $\sim$0.3~$\Rs$ between a 40~MHz and a 32~MHz source, even if they are emitted from the same location.  The estimated shift also implies that if the density model is deduced from a comparison with the observed source locations without a correction for the scattering-induced shifts---an approach frequently encountered in the literature---it will lead to an overestimation of the coronal density model, calculated to be up to a factor of $\sim$4.3 for the event presented in this chapter.  If that density model is then used to infer further values describing the behaviour of the exciter (see Section~\ref{sec:f_vs_R_relation}) or the local coronal environment (see Section~\ref{sec:bandsplitting_models}), they will also be deceptive.

While scattering is the dominant radio-wave propagation effect, other effects that alter the true properties of the radio emissions during observations also need to be considered in analyses.  Even though refraction is weaker than scattering, it will shift the perceived source locations radially closer to the Sun, partially counteracting the impact of scattering.  It has been shown, however, that for the split-band Type II emissions studied in this chapter, neglecting the effects of scattering would lead to a misleading interpretation of the observation.  Without accounting for the impact of scattering, the observations favour band-splitting models that require the intrinsic subband sources to be physically separated, like the \cite{1983ApJ...267..837H}.  When scattering effects are considered, the initial interpretation of the observation changes.  The observations are found to support models that expect the intrinsic subband sources to be virtually co-spatial, like the \cite{1974IAUS...57..389S, 1975ApL....16R..23S} model.  The necessity to consider radio-wave scattering effects in analyses of radio observations has therefore been illustrated.

The presented analytical estimation of scattering-induced shifts is not limited to the observed Type II event discussed in this chapter.  It can be applied to any type of radio burst emitted through the plasma emission mechanism.  The ability to apply this analytical method on any event is an advantage, but comes with a caveat that---depending on the scope of the study---can be restrictive.  It fails to reflect the fluctuating local conditions of the highly-variable coronal environment that can define the properties of an observation, some of which are considerably complex (see Chapter~\ref{chap:scattering} and the computationally-intensive ray-tracing simulations described therein).  As a simple example, the solar corona is non-static, meaning that the density at a specific location can considerably deviate from the ``idealised'' density models like that of \cite{1961ApJ...133..983N} (such that, for example, a 32~MHz source is not emitted at $\sim$1.74~$\Rs$), affecting any parameters obtained by assuming a certain density model.  The local coronal conditions of each event are unique, hence the variation in observed properties.  Nevertheless, attempting an estimation of the scattering effects---even through a simplified and generalised method---is preferred to neglecting any such contributions in radio observations, which can potentially lead to misleading interpretations.

\cleardoublepage
\chapter{A Transitioning Type II Burst} \label{chap:transitioning_typeII}

\textit{The work in this chapter has been published in \cite{2020ApJ...893..115C}.}

\section{First Observation of a Transitioning Type II Burst} \label{sec:transitioning_typeII}
Type II solar radio bursts can be separated into two categories---drifting and stationary Type II bursts---according to whether they drift with frequency or not (see Section~\ref{sec:typeIIs}).  However, a transition between a drifting and stationary state within a single Type II burst has not been reported, prior to the observation presented in this chapter.  This morphology has been termed a ``transitioning'' Type II burst, introducing a new sub-class of Type II solar radio bursts \citep{2020ApJ...893..115C}.

The focus of this chapter is to identify the sequence of events related to the transitioning Type II emissions by analysing multi-wavelength observations, and to understand the mechanisms that led to this morphology.

\subsection{Overview of the observations} \label{sec:2020obs_overview}
A Type II burst that transitions between a stationary and drifting state was observed on 15 July 2017 by LOFAR between $\sim$11:02 and 11:05~UT (see Section~\ref{sec:radio_spectroscopy}).  A Type III burst that intersected the stationary part of the Type II emissions was also recorded.  The LOFAR observation was conducted between 30--80~MHz with 24 core stations in the outer LBA configuration.  The coherent Stokes beam-formed mode was utilised, recording  only the Stokes I information, which formed a tied-array beam of 217 individual beams that covered a hexagonal area of approximately 2.8~$\Rs$ from the centre of the Sun (see Section~\ref{sec:tied-array}).  This configuration provided a temporal resolution of $\sim$0.01~s, a spectral resolution of $\sim$12.2~kHz, a sensitivity of $\lesssim$~0.03~sfu per beam, an average separation between beam centres of $\sim$6$\arcmin$, and synthesised beams with a FWHM of $\sim$10$\arcmin$ at 30~MHz.  The flux calibration of the recorded radio emissions was achieved by observing both Tau A and the ``empty sky'' before and after the observation (see Section~\ref{sec:flux_cal}).  For the analysis and presentation of the radio observations in this chapter, the temporal and spectral resolutions were rebinned and decreased to $\sim$0.21~s and $\sim$73.2~kHz, respectively.

A coronal jet eruption (a long and thin transient feature; see Section~\ref{sec:solar_activities}) was observed in temporal and spatial proximity to the radio emissions.  The jet originated at $\sim$10:51~UT from the edge of an active region located on the west side of the Sun.  The jet was imaged in EUV wavelengths by the AIA instrument onboard \textit{SDO} with a $\sim$12~s cadence.  The eruption was also recorded in X-ray wavelengths by the \textit{GOES}-15 XRS.

Following the jet eruption, the LASCO/C2 white-light coronagraph onboard \textit{SOHO} recorded ejecta which first appeared in its FoV at $\sim$11:12~UT from the west side of the Sun.  The C2 coronagraph imaged the ejecta---identified as two CME fronts---between the (plane-of-sky) heliocentric distances of $\sim$2.2 and 6~$\Rs$, with a $\sim$12 minutes cadence.  Given that Type II bursts are often related to CME-driven shocks, and that the observed radio emissions are consistent with the apparent location and timing of the imaged CME fronts, these CMEs ejections are believed to be associated with the excitation of the transitioning Type II burst.

\begin{figure}[ht!]
    \centering
	\includegraphics[width=1.0\textwidth, keepaspectratio=true]{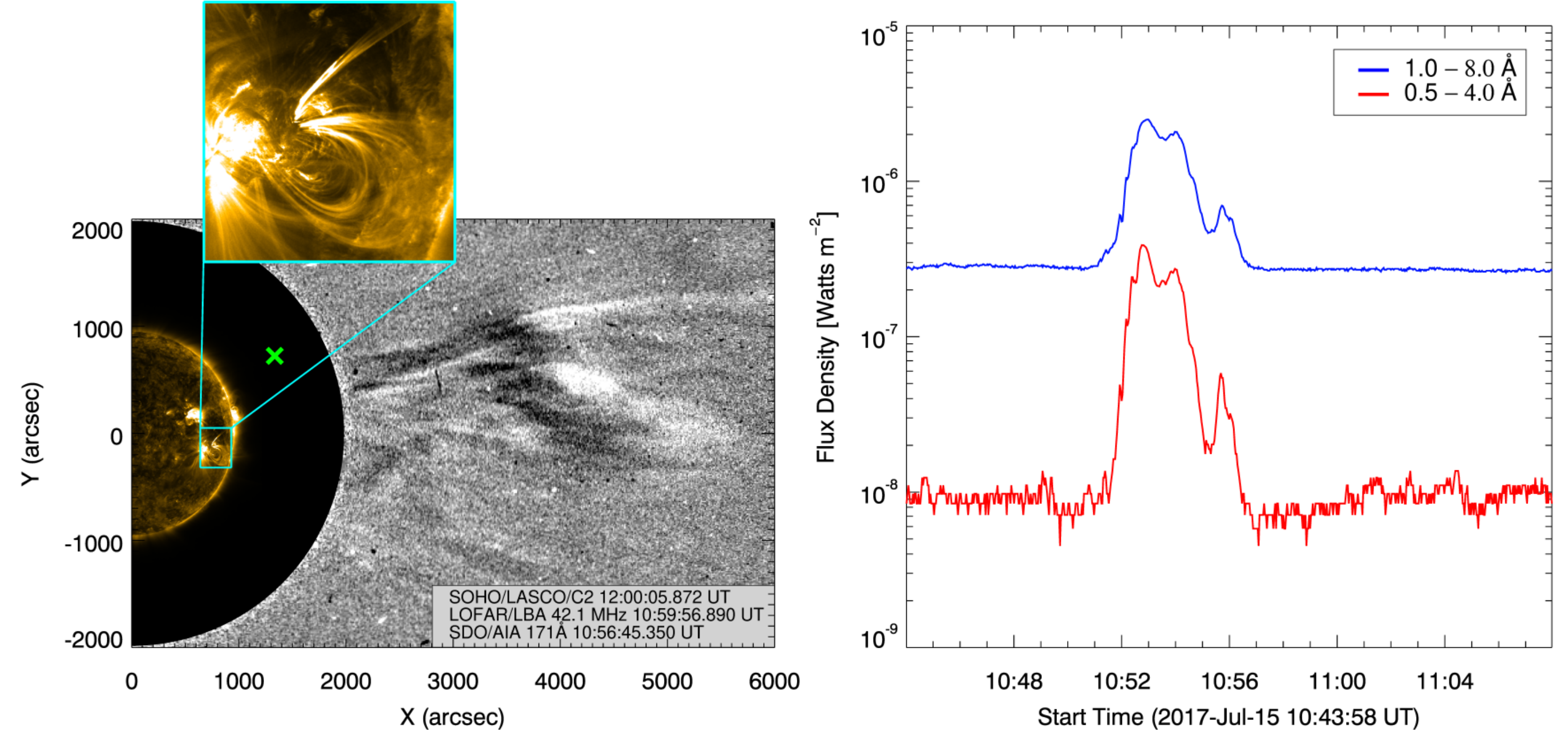}
    \caption[Radio, white-light, EUV, and X-ray observations from 15 July 2017.]
    {The left panel is a combination of LOFAR data, \textit{SOHO}/LASCO/C2 running-difference data, and \textit{SDO}/AIA (171~\AA) data.  The green cross illustrates the apparent location of the Type II burst emissions (at a given frequency and time) and the black disk represents the occulting disk of the C2 coronagraph.  The two CME fronts can also be distinguished, with brighter structures in LASCO's FoV reflecting relative increases in intensity and darker structures reflecting relative decreases in intensity.  The inset shows the coronal jet emerging from the northern edge of the active region.  The right panel shows the true X-ray flux density measured by the \textit{GOES}-15 XRS instrument during the jet's eruption at 0.5--4.0~\AA\ (red curve) and 1.0--8.0~\AA\ (blue curve).
    Figure taken from \cite{2020ApJ...893..115C}.
	}
    \label{fig:2020_inset_GOES}
\end{figure}

The left panel of Figure~\ref{fig:2020_inset_GOES} illustrates the temporal and spatial relation of the jet, the radio emissions, and the two CME fronts.  The location of the radio emissions is indicated by the green cross.  The radio source locations were estimated by fitting a 2D elliptical Gaussian on the 50\% maximum intensity of the LOFAR images, and calculating the centroids and the associated uncertainties, as described in Section~\ref{sec:centroid_calc}.  The inset highlights the active region of interest during the jet's eruption (seen on the northern edge).  The right panel of Figure~\ref{fig:2020_inset_GOES} shows the true X-ray flux density as recorded by the \textit{GOES}-15 XRS instrument, where a prominent peak can be identified near the jet's eruption time ($\sim$10:51 UT).

\subsection{Spectroscopic radio observations of the transitioning Type II burst} \label{sec:radio_spectroscopy}
\begin{figure}[ht!]
    \centering
	\includegraphics[width=1.0\textwidth, keepaspectratio=true]{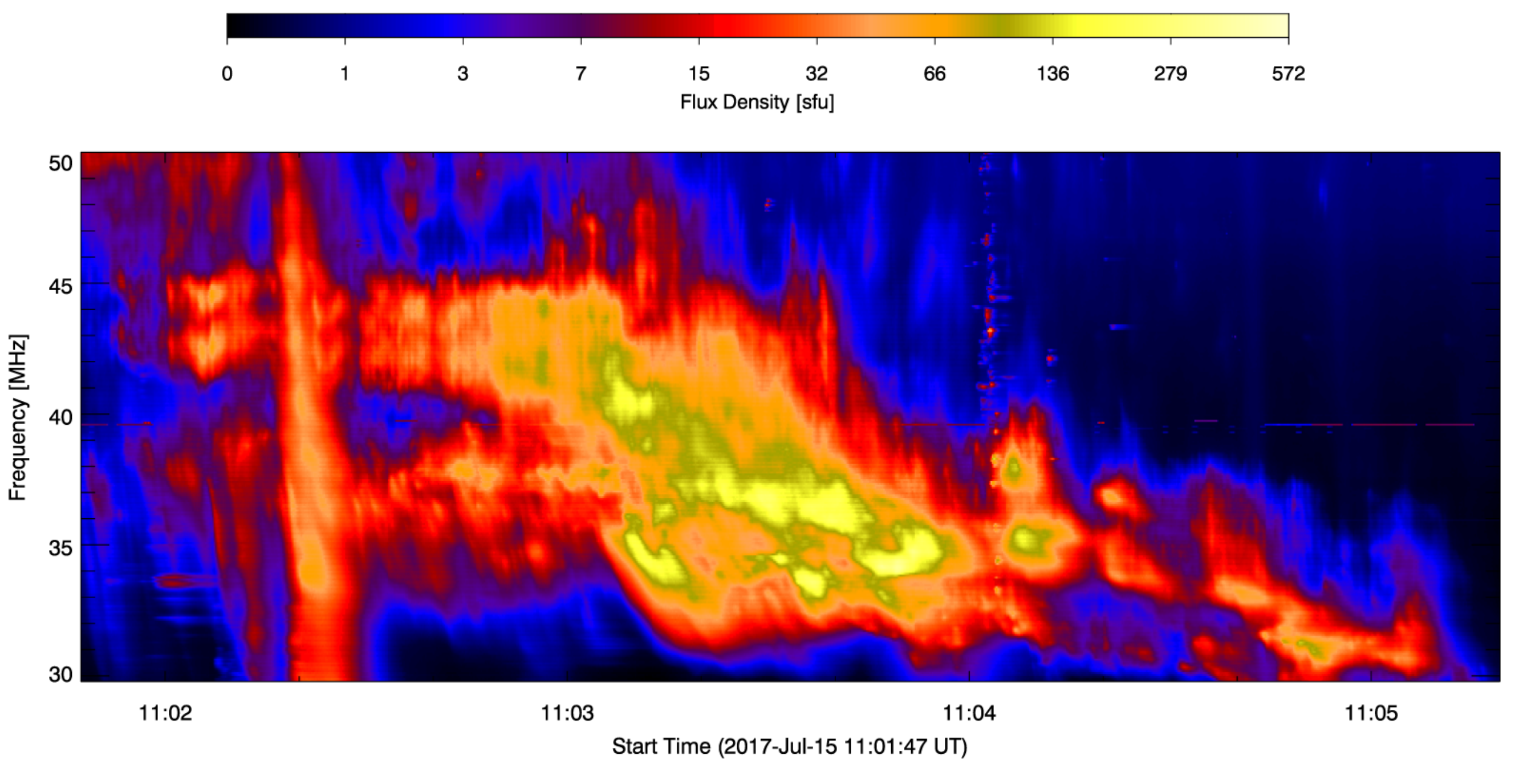}
    \caption[Dynamic spectrum of the transitioning Type II burst.]
    {Dynamic spectrum depicting a Type II solar radio burst transitioning between a stationary and drifting state, as observed on 15 July 2017 by LOFAR's LBA stations.  The stationary Type II part is observed between $\sim$11:02--11:03~UT and the drifting part between $\sim$11:03--11:05~UT.  A Type III burst is also observed (at $\sim$11:02:20~UT) during the stationary Type II emissions.  Prior to plotting, the spectral and temporal resolutions of the data were decreased from $\sim$12.2~kHz and $\sim$0.01~s to $\sim$73.2~kHz and $\sim$0.21~s, respectively.
    Figure taken from \cite{2020ApJ...893..115C}.
	}
    \label{fig:2020_dynspec}
\end{figure}

The dynamic spectrum obtained using LOFAR's LBA antennas is shown in Figure~\ref{fig:2020_dynspec}.  The transitioning Type II burst can be clearly distinguished, with the stationary part observed between $\sim$11:02 and 11:03~UT, and the drifting part observed between $\sim$11:03 and 11:05~UT.  The stationary Type II part consists of two bands where each of them experiences band-splitting.  The first pair of subbands appears between $\sim$41--45~MHz, whereas the second one appears between $\sim$35--39~MHz.  It is worth mentioning that the emissions of both subband pairs seem to start at the same time ($\sim$11:02~UT).

These pairs of subbands are not harmonically related, as the characteristic frequency ratio of 1:2 between fundamental and harmonic plasma emissions is not observed (see Section~\ref{sec:plasma_emmission}).  It is therefore possible that:
\begin{enumerate*}[label=(\roman*)]
	\item each pair of subbands is the result of an individual shock, or
	\item that both subband pairs are the result of a single shock and the Type II burst experiences simultaneous band splitting in two different locations.
\end{enumerate*}
The two pairs of subbands are not only emitted simultaneously, but also show similarities in morphology and in their temporal fluctuation of flux (see Figure~\ref{fig:2020_dynspec}).  Such resemblance in the emission patterns in the two pairs of subbands, implies that the regions of the shock (or shocks) which excite the Type II emissions propagate through the same coronal density region simultaneously (see, e.g., \cite{1974IAUS...57..389S, 1975ApL....16R..23S, 2001A&A...377..321V}).

The higher-frequency component of the first pair of subbands appears at around 44~MHz, whereas the lower-frequency component appears at $\sim$42~MHz.
The average relative frequency split ($\Delta f_s/f$; see Equation~(\ref{eqn:deltaf_over_f})) between these subbands is $\sim$0.05.  Equivalently, the higher- and lower-frequency components of the second pair of subbands appear at $\sim$37.5 and 36~MHz, respectively, corresponding to an average frequency split $\Delta f_s/f$ of $\sim$0.04.  These frequency-split values are somewhat lower than those often described as the typical range for Type II bursts which experience band splitting, i.e. $\Delta f_s /f =$ 0.1--0.5 (see Section~\ref{sec:typeIIs}).  However, these ``typical'' values are obtained from statistical analyses of drifting Type II bursts, not stationary.  The observed drifting Type II emissions, on the other hand, appear to drift in frequency at the rate of $\sim$\nolinebreak$-0.14$~$\MHzs$, which is within the typical range for Type II bursts (see Section~\ref{sec:typeIIs}), and is similar to that of the Type II burst presented in Chapter~\ref{chap:split-band_typeII}.

A Type III burst is also observed during the stationary Type II emissions (at $\sim$11:02:20~UT) and appears to intersect both pairs of subbands.  The Type III burst was found to have a frequency-drift rate $df/dt \approx -5$~$\MHzs$.

\begin{figure}[ht!]
    \centering
	\includegraphics[width=0.8\textwidth, keepaspectratio=true]{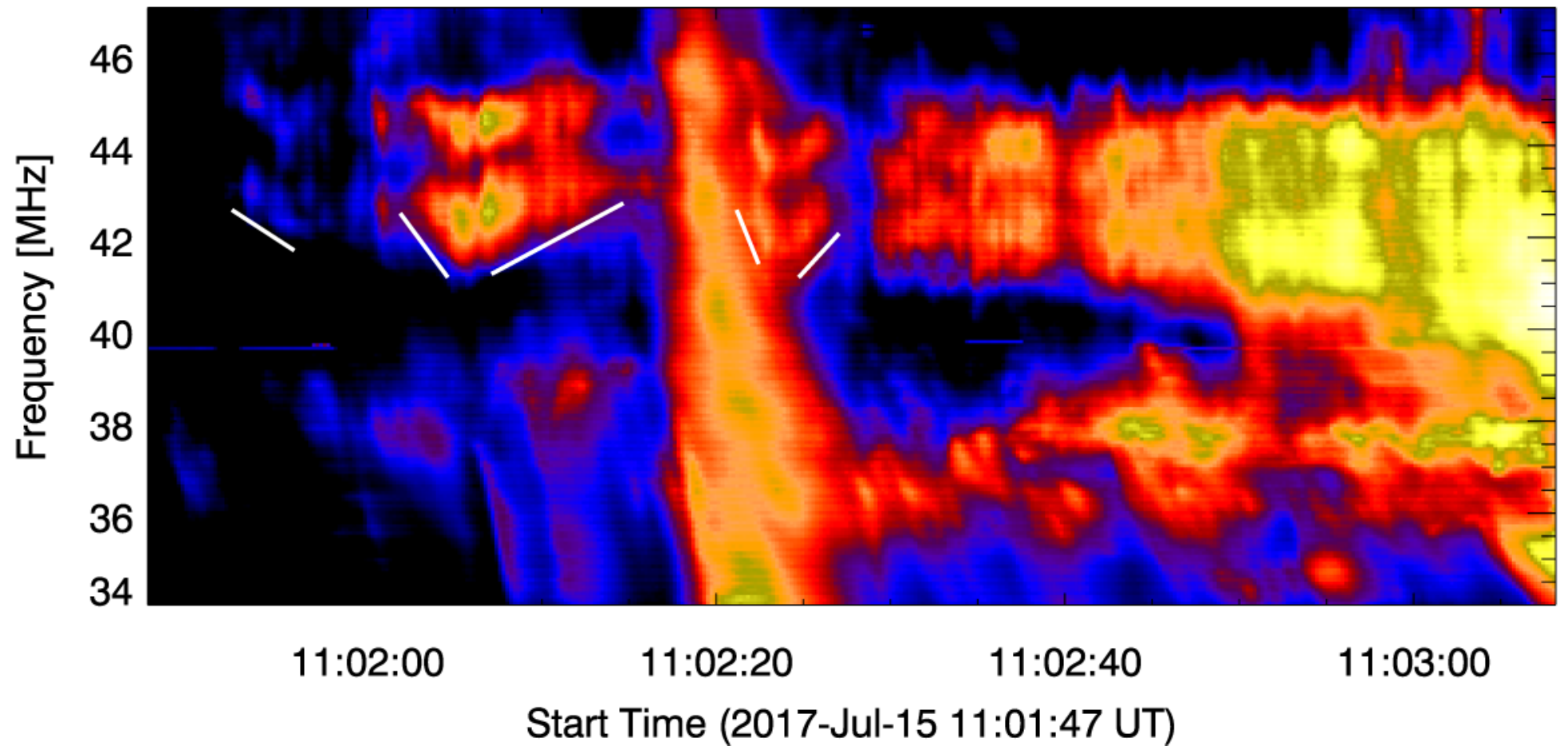}
    \caption[Fine structures within the stationary Type II emissions.]
    {A section of the dynamic spectrum shown in Figure~\ref{fig:2020_dynspec}, plotted with a different dynamic range to highlight the fine structures within the stationary part of the Type II burst.  The five white-line annotations emphasise the altering frequency-drift rates of some of the fine structures that are easily distinguishable.  As evident, some of the fine structures have positive drift rates and some negative.
    Figure taken from \cite{2020ApJ...893..115C}.
	}
    \label{fig:2020_fine_structures}
\end{figure}

Furthermore, a closer examination of the stationary Type II emissions revealed intriguing fine structures, illustrated in Figure~\ref{fig:2020_fine_structures}.  To highlight these structures, the dynamic spectrum in Figure~\ref{fig:2020_fine_structures} is plotted such that emissions below 1\% of the maximum intensity are omitted from the presentation.  White-line annotations have been used to emphasise the most prominent of these intriguing fine structures, which can be clearly seen to repeat in both components of the higher-frequency pair of subbands.  These structures are unusual due to the fact that they show both negative and positive frequency-drift rates.  They also do not resemble the well-known Type II structures referred to as ``herringbones'' which are often observed to emanate from a Type II burst's backbone, and have opposing drift rates on each side of the Type II backbone (see Section~\ref{sec:typeIIs}).  The frequency-drift rate of each of the fine structures annotated in Figure~\ref{fig:2020_fine_structures} was estimated to be---from left to right---approximately $-$0.25, $-$0.51, $+$0.21, $-$0.93, and $+$0.41~$\MHzs$.  The altering frequency drift of these fines structures could be interpreted as radio emissions signalling the presence of a pulsating exciter.  The observed behaviour is reminiscent of a Type II burst reported by \cite{2004SoPh..222..151M}, which had a ``waving backbone'' but on average showed no frequency drift.  It is notable, though, that the individual waving structures in that Type II lasted over several minutes, but for the case presented in this chapter, each fine structure lasts for only a few seconds.

\section{Complementary Observations} \label{sec:cme_and_jet}
\subsection{Examining the jet eruption} \label{sec:jet}
A solar flare of magnitude C1.4 was observed at $\sim$10:50~UT, just before the jet discussed in Section~\ref{sec:2020obs_overview} erupted.  The jet eruption lasted from $\sim$10:51 to 10:58~UT and originated from the same region on the Sun as the solar flare.  The jet's footpoint appears at the umbra-penumbra region of the active region identified as NOAA 12665, and above a light bridge on the sunspot (visible in the 1600 and 1700~\AA\ AIA channels) which has a magnetic configuration of Hale class $\beta$ (see Section~\ref{sec:solar_activities}).

\begin{figure}[ht!]
    \centering
	\includegraphics[width=0.496\textwidth, keepaspectratio=true]{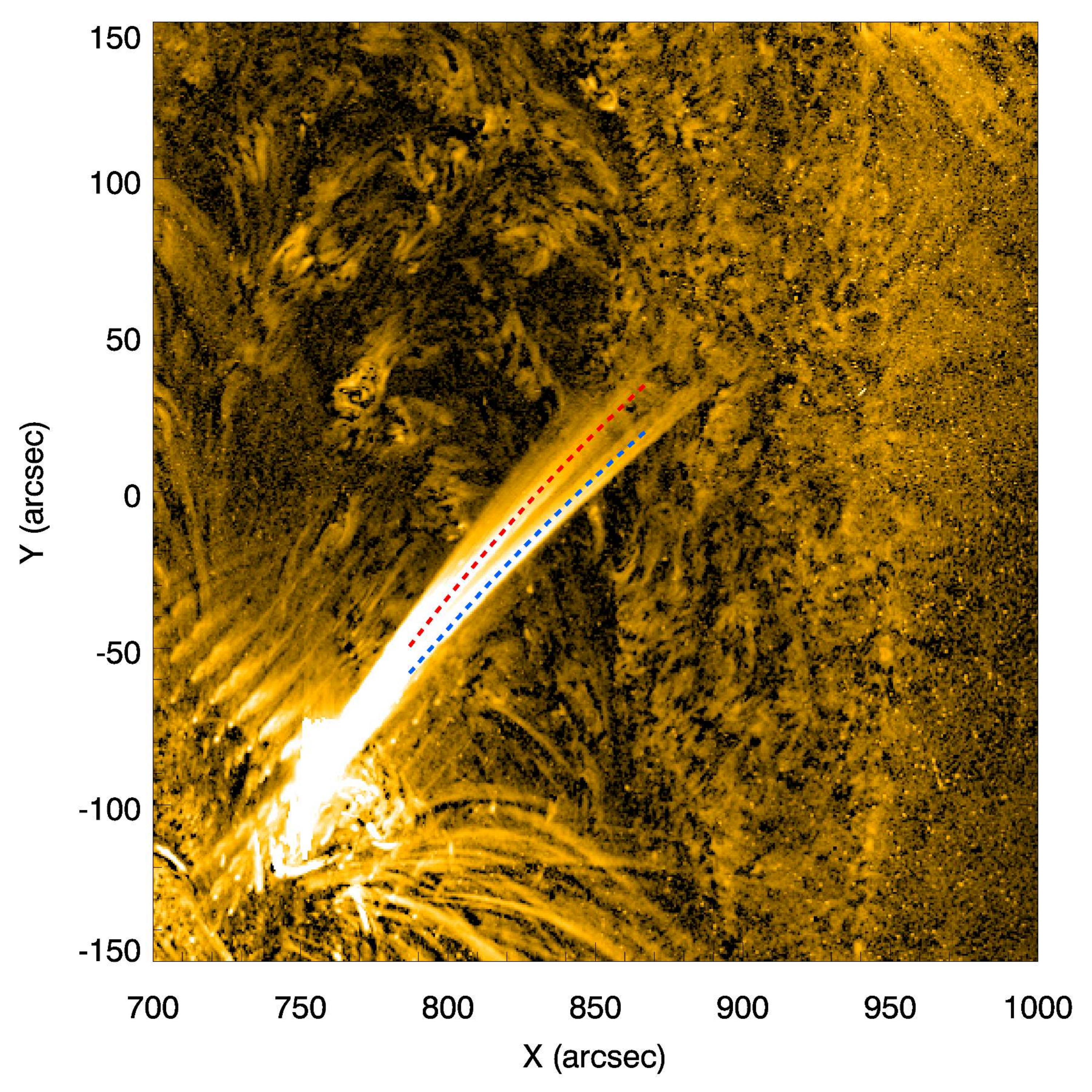}
	\includegraphics[width=0.496\textwidth, keepaspectratio=true]{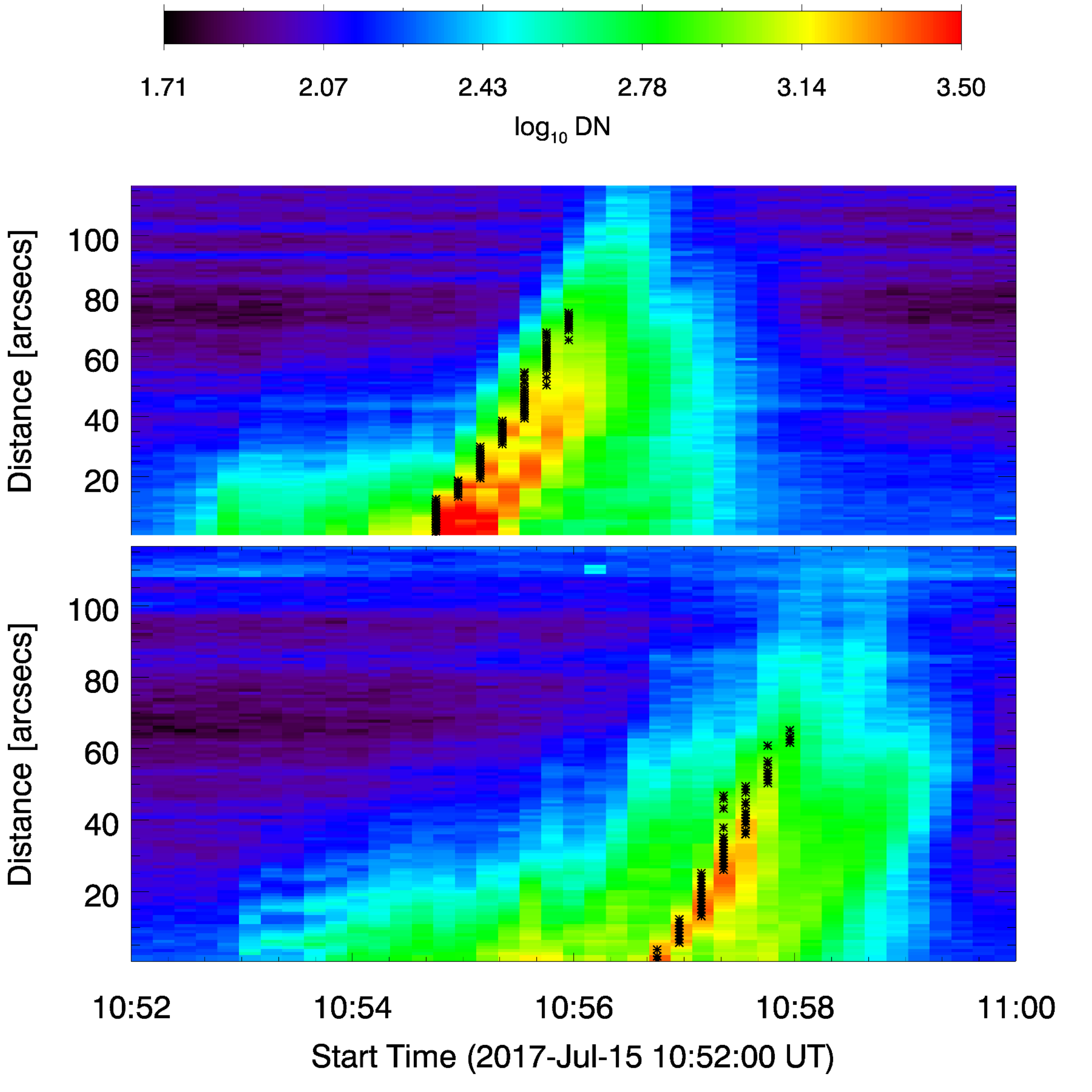}    
    \caption[Bifurcated jet spire and AIA stack plots.]
    {The left panel shows the jet (using AIA 171~\AA\ data) with the background subtracted.  Two artificial slits (red and blue dashed lines) highlight the two ejections of plasma.  The top right panel shows the AIA 171~\AA\ stack plot along the blue artificial slit, whereas the bottom right panel shows the stack plot along the red artificial slit, displaying the propagation of the jet as a function of time.  The black stars indicate the times where the intensity surpassed the background level by a factor of 10, used to find the speed of the bifurcated jet components.
    Figure taken from \cite{2020ApJ...893..115C}.
	}
    \label{fig:2020_jet}
\end{figure}

Figure~\ref{fig:2020_jet} depicts data obtained at 171~\AA\ by AIA.  The data represents the background-subtracted peak intensity of the jet, where reference time $\sim$10:45~UT (well before any eruption) was taken as the background.  The left panel of Figure~\ref{fig:2020_jet} shows the jet's spire which exhibits bifurcation, i.e. the spire erupts into two components \citep{2012ApJ...745..164S}.  The two components are marked with a red and blue dashed line.  These lines also represent the path along which two artificial slits were used in order to examine the propagation of each bifurcated component (see, e.g., \cite{2016A&A...589A..79M}), the results of which are shown by the stack plots in the right panel of Figure~\ref{fig:2020_jet}.  These are stack plots of distance against time, representing the erupting jet plasma from each of the bifurcated components.  The onset of the jet was estimated at the moment at which the intensity surpassed the background intensity level by a factor of 10.  The obtained onset times at each spatial point are illustrated in the stack plots with black crosses.

Using the onset times and the corresponding distance along the slits, the start time and plane-of-sky speed of each bifurcated component was calculated.  It was found that the southern component (indicated by the blue dashed line) occurred first, at $\sim$10:54:40~UT, and erupted with a plane-of-sky speed of $\sim$650~$\kms$.  The second, northern component (indicated by the red dashed line) occurred two minutes later, starting at $\sim$10:56:40~UT, and its speed was estimated to be $\sim$660~$\kms$.  The estimated onset time of each of the bifurcated components is strongly-correlated to the two peaks observed in the X-ray flux density measurements obtained by \textit{GOES}-15, depicted in the right panel of Figure~\ref{fig:2020_inset_GOES}.

It is worth mentioning that several other jets with similar characteristics to the one presented in this chapter were observed throughout the day.  Their footpoints seem to be located on the same edge of the active region of interest, and when ejecta emerge in the LASCO/C2 FoV, they trace the same streamer as the ejection presented in this chapter (discussed in Section~\ref{sec:cme}).  Easily recognisable examples of some of these other jets include those at $\sim$12:37, 14:43, 16:26, and 23:09~UT.

\subsection{Examining the CME eruption} \label{sec:cme}
\begin{figure}[htp!]
    \centering
	\includegraphics[width=1.0\textwidth, keepaspectratio=true]{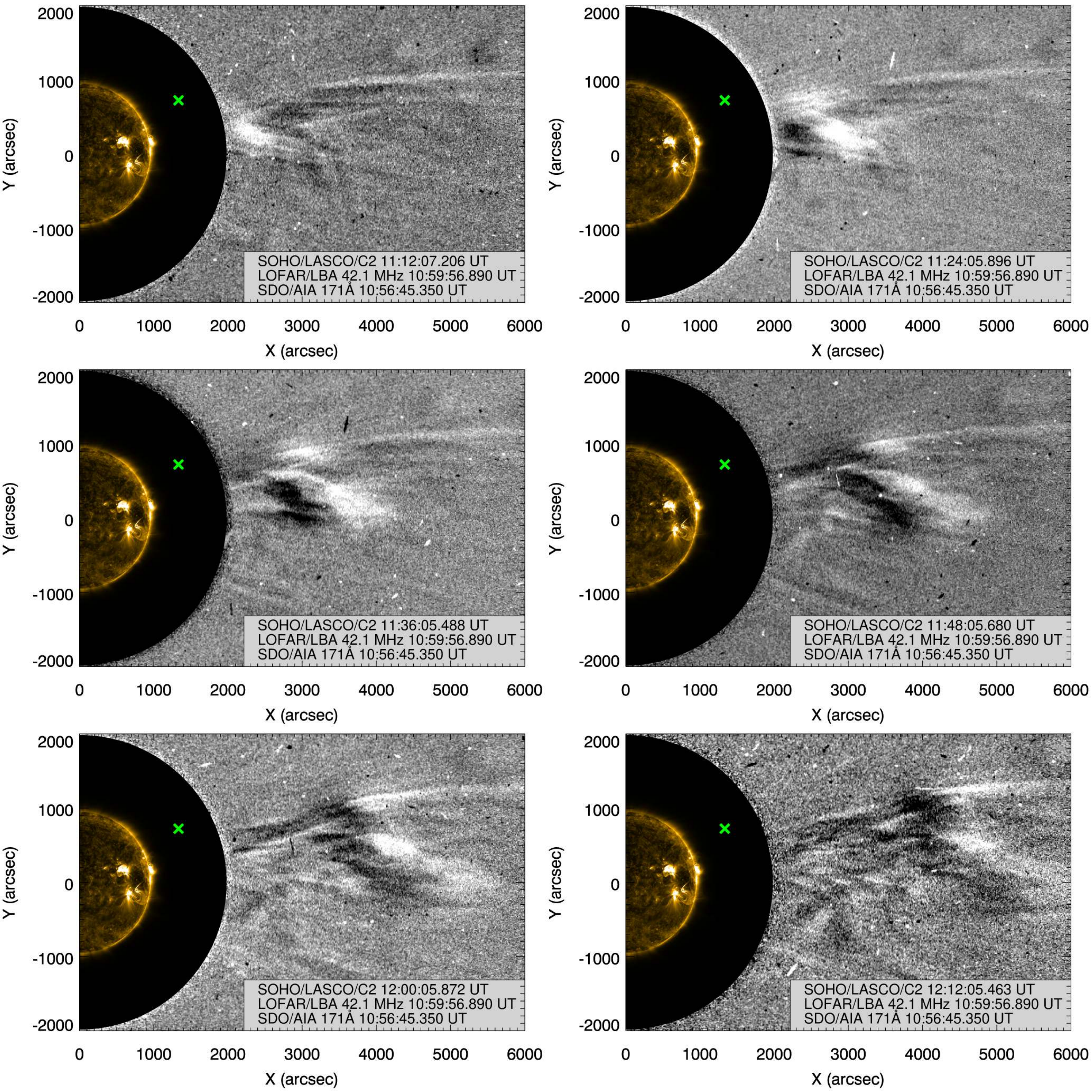}
    \caption[Consecutive white-light coronagraphic images of the CME fronts.]
    {Combination of multi-wavelength observations from 15 July 2017 that were temporally and spatially related to the Type II radio emissions.  Consecutive white-light running-difference images from the \textit{SOHO}/LASCO/C2 coronagraph show the temporal and spatial evolution of the CME fronts as they propagate away from the solar surface.  There is a $\sim$12 minute interval between each panel.  Brighter structures in LASCO's FoV reflect relative increases in intensity, whereas darker structures reflect relative decreases in intensity.  The black disk represents the occulting disk of the C2 coronagraph.  The solar surface at the time of the jet's eruption is shown in EUV using \textit{SDO}/AIA 171~\AA\ data.  The apparent position of the Type II radio burst observed by LOFAR is illustrated by the green cross.
    Figure taken from \cite{2020ApJ...893..115C}.
	}
    \label{fig:2020_cme_snapshots}
\end{figure}

The presence of CMEs at a time and location which coincide with those of the Type II burst, implies that the shock exciting the Type II emissions is driven by the CMEs.  Figure~\ref{fig:2020_cme_snapshots} illustrates the temporal and spatial evolution of the CMEs as captured in white-light by LASCO/C2, where the panels are taken $\sim$12 minutes apart.  To emphasise the CME fronts and coronal structures present, running-difference images of the white-light observations are presented.  The evolution of the CMEs is shown relative to both the solar surface and an approximated location of the Type II emissions (green cross).  The black disk depicts the occulting disk of the C2 coronagraph.

It can be seen from the top panels in Figure~\ref{fig:2020_cme_snapshots} that there are open magnetic fields forming a thin streamer, appearing to point towards the active region of interest.  This streamer was present long before the discussed eruptions and remained visible for several hours after these events.  It appears that the streamer formed during an eruption occurring from the same active region as the one associated with the jet discussed, but much earlier than the events studied in this chapter.

As mentioned in Section~\ref{sec:2020obs_overview}, two CME fronts were identified, as can be seen in Figure~\ref{fig:2020_cme_snapshots}.  The top left panel shows the front that appears in the C2 FoV first (at $\sim$11:12~UT), whereas the second front can be seen in the top right panel (imaged at $\sim$11:24~UT), emerging near the northern flank of the first front.  The two fronts can be distinguished through their apparent angular widths---a measure of their (widest) spatial span with respect to the solar centre (i.e. the angular width measured from the edge of one flank to the edge of the other flank of the CME front).  In this case, the spatial span of the CMEs was estimated using the C2 plane-of-sky images.  The average angular width of the first, southern front was found to be $\sim$14$\degr$, whereas the second, northern front had an average angular width of $\sim$5$\degr$.  As such, the first front will be referred to as the ``broader front'', and the second as the ``narrower front''.  It is worth noting, however, that both fronts are classified as narrow CMEs, since their apparent angular widths are $\lesssim$~15$\degr$ \citep{2001ApJ...550.1093G}.

Several structures from these fronts were manually tracked across the C2 FoV in order to obtain an estimate of the speed of the two CME fronts.  The heliocentric locations of these structures were calculated at every $\sim$12 minute interval (C2's imaging cadence), and the obtained distance-time data was linearly fitted in order to provide a rough estimate of the average plane-of-sky speed.
The average speed of the broader front was estimated to be $\sim$700~$\kms$, whereas the speed of the narrower front was found to be $\sim$560~$\kms$.  Features that were directly between the two fronts, i.e. near the southern flank of the narrower CME and near the northern flank of the broader CME, were also (manually) tracked.  Their average plane-of sky speed in the C2 FoV was estimated to be $\sim$470~$\kms$.  These parts are likely slowed down due to a partial interaction between the two fronts.

The CME fronts appear to evolve in a different manner from each other as they propagate away from the Sun.  As evident from Figure~\ref{fig:2020_cme_snapshots}, the narrower front appears to be confined by the streamer, the path of which it traces, and does not experience a significant expansion.  On the other hand, the broader front experiences a greater degree of expansion as it propagates away from the Sun.  It also seems to deflect towards the south, away from the path laid out by the streamer, and is the first to dissolve into the coronal background, completely dissipating by the time it reaches the edge of the C2 FoV.  The narrower front, however, maintains its shape even beyond the edge of the C2 FoV (i.e. $\gtrsim$~6~$\Rs$) and into the FoV of the C3 coronagraph (which images distances up to $\sim$30~$\Rs$).

The surface of the Sun was studied in order to identify the origin of the two CME eruptions and their relation to other activities on the Sun.  However, besides the eruption of the jet, there was no evidence of any erupting flux ropes or coronal dimming, often signifying the release of solar material into the corona (see Section~\ref{sec:TypeII-CME_relation}).  It is therefore believed that both CMEs were the result of material ejected during the eruption of the jet.  Specifically, it is believed that the bifurcation experienced by the jet's spire drove the two CME fronts, a behaviour similar to the findings of \cite{2012ApJ...745..164S} who studied the eruption of a bifurcated blowout jet believed to have caused two simultaneous CMEs.  The broader CME front---which appears first---is likely caused by the first bifurcated component, while the narrower CME front is likely caused by the component that erupted last (see Section~\ref{sec:jet}).

The characteristics attributed to the narrower CME front---i.e. that it traces the streamer but does not inflate it---along with the repetitive nature of jet eruptions from the specific area of the active region (see Section~\ref{sec:jet}), agree with the description of ``streamer-puff'' CMEs \citep{2005ApJ...635L.189B, 2016ApJ...822L..23P, 2018JPhCS1100a2024S}.  Streamer-puff CMEs were first identified as a new variety of CMEs by \cite{2005ApJ...635L.189B}.  There are described as narrow CMEs that ``move along the streamer, transiently inflating the streamer but leaving it intact'' \citep{2005ApJ...635L.189B}, and are driven by erupting jets \citep{2018JPhCS1100a2024S}.  As such, the narrower CME front discussed in this chapter is identified as a streamer-puff CME.

To summarise the sequence of events, a solar flare is observed at $\sim$10:50~UT, followed by a jet at $\sim$10:51~UT (imaged using \textit{SDO}/AIA data).  The spire of the jet bifurcates, with the first (and southern) component estimated to occur at $\sim$10:54:40~UT, and the second (and northern) component estimated to occur at $\sim$10:56:40~UT.  The approximated onset times of the jet's bifurcated components correlate with two peaks observed in X-ray data.  Type II radio emissions are observed from $\sim$11:02~UT, and a Type III burst that intersects the Type II emissions is recorded at $\sim$11:02:20~UT.  Two CMEs appear in the C2 FoV, with the broader (and southern) CME front first imaged at $\sim$11:12~UT, and the narrower (and northern) CME front first appearing in the C2 FoV at $\sim$11:24~UT.  The narrower CME is identified as a streamer-puff CME, as it traces a streamer which is present long before and after the listed events.

The projected heliocentric distances of the Type II emissions are smaller than $\sim$1.8~$\Rs$ and are thus outside the C2 FoV, as indicated in Figures~\ref{fig:2020_inset_GOES} and~\ref{fig:2020_cme_snapshots}.  This means that it is not possible to image both the Type II emissions and the CMEs at the same location.  Nevertheless, the Type II emissions appear to the north of both CME fronts (as seen in Figure~\ref{fig:2020_cme_snapshots}) and are seemingly related to the streamer-puff CME.  Although this is deduced from the 2D plane-of-sky depiction which lacks information on the 3rd dimension (as detailed in Section~\ref{sec:proj_effs_impact}), the apparent latitude of the Type II emissions above the ecliptic and above the streamer leads to the belief that the streamer-puff CME is the one driving the shock which excited the transitioning Type II burst.  Given the sources' apparent location and that the first CME structures appear in the C2 FoV $\sim$10 minutes after the Type II emissions are first observed, it is likely that the radio emissions are excited near the CME's flanks.

\section{LOFAR Imaging of the Radio Emissions} \label{sec:radio_imaging}
The motion of the exciter is reflected in the apparent motion of the radio sources in frequency and time.  It is therefore necessary to study the radio emission images in order to identify the mechanism resulting in the observed transitioning Type II burst.  Thanks to LOFAR's unprecedented observing capabilities, the behaviour of the emission sources before, during, and after the transition from a stationary to a drifting state has been examined.

\subsection{Imaging the transitioning Type II burst} \label{sec:2020_typeII_imaging}
\begin{figure}[ht!]
    \centering
	\includegraphics[width=1.0\textwidth, keepaspectratio=true]{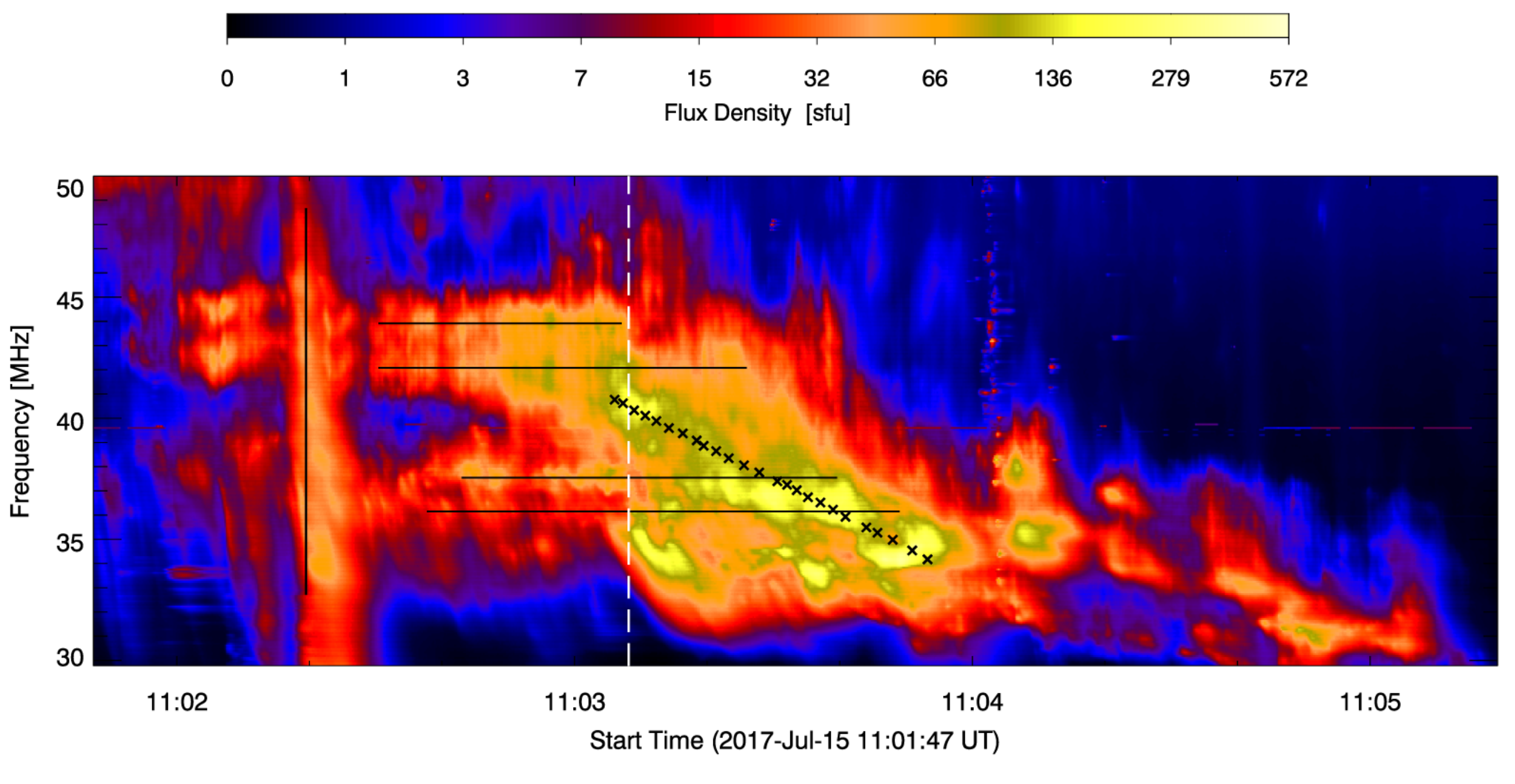}
    \caption[Annotated dynamic spectrum of the 15 July 2017 radio emissions.]
    {Annotated version of the dynamic spectrum shown in Figure~\ref{fig:2020_dynspec}.  The white dashed line indicates the defined time of transition from stationary to drifting Type II emissions, taken to be at 11:03:08.150~UT.  The black horizontal lines indicate the single-frequency slices taken for each subband in order to produce emission images before, during, and after the transition between the two states.  For the highest-frequency subband, no data was selected past the transition time.  The black crosses illustrate the points at which the drifting part of the Type II burst was imaged.  The black vertical line at $\sim$11:02:20~UT indicates the frequencies at which Type III sources were imaged.
    Figure taken from \cite{2020ApJ...893..115C}.
	}
    \label{fig:2020_annot_dynspec}
\end{figure}

The transition between the stationary and drifting Type II emissions occurs around 11:03:08~UT, as can be seen by the emissions depicted in the dynamic spectrum (Figures~\ref{fig:2020_dynspec} and \ref{fig:2020_annot_dynspec}).  This moment in time is indicated by the white dashed line in Figure~\ref{fig:2020_annot_dynspec}, and will be referred to as the ``transition time''.  A single and precise transition time is merely defined as a guiding point for the graphical illustrations in the forthcoming analysis of the emissions' source locations.  It is emphasised that the physical transition from a stationary to a drifting state likely lasts over a few seconds.

The Type II radio emissions are imaged at multiple moments in time, covering both the stationary and drifting parts, but a single frequency is used for each Type II subband.  This is done in order to examine how---if at all---the motion of the radio sources changes during the transition from a stationary to a drifting Type II burst.  Imaging each subband at a single frequency means that the frequency-dependent radio-wave propagation effects---which distort the intrinsic nature of the radio sources, like the scattering-induced radial shift (detailed in Section~\ref{sec:prop_effs} and Chapters~\ref{chap:scattering}--\ref{chap:split-band_typeII})---are entirely eliminated within each individual subband.  In other words, the relative motion observed within each subband is purely temporal and the inferred evolution is linked only to the motion of the exciter of the radio emissions.  However, the absolute heliocentric location of each of the subbands is affected by scattering effects, since all sources emitted at a specific frequency will shift away from the Sun, but by the same amount.  This implies that the apparent location of each single-frequency subband relative to the other subbands is distorted by scattering effects, with the lower-frequency subbands displaced the farthest from their true location.

The single-frequency slice used to image each of the four subbands observed during the stationary Type II part was selected roughly mid-way through each subband's emissions, i.e. it represents the subband's average frequency (see Figure~\ref{fig:2020_annot_dynspec}).
Emission images for the higher-frequency pair of subbands are produced at 43.9 and 42.1~MHz, for the higher- and lower-frequnecy components respectively.  Similarly, 37.5 and 36.2~MHz were selected for imaging the lower-frequency pair of subbands.  Data at these four frequencies was taken during the stationary emissions, as well as past the transition time and into the drifting Type II emissions, in order to image the behaviour of the sources before, during, and after the transition.  The temporal range of each of these four single-frequency slices is indicated by the black horizontal lines in Figure~\ref{fig:2020_annot_dynspec}.  As can be seen in Figure~\ref{fig:2020_annot_dynspec}, no data past the transition time was selected for the highest-frequency subband (imaged at 43.9~MHz), since none of the drifting emissions at that frequency could be confidently related to that subband.  As such, data at 43.9~MHz represents only the stationary Type II emissions.  Furthermore, to eliminate the possibility of imaging background noise emissions, data points whose flux did not exceed 1\% of the maximum flux value of the observation were omitted.

\begin{figure}[t!]
    \centering
	\includegraphics[width=1.0\textwidth, keepaspectratio=true]{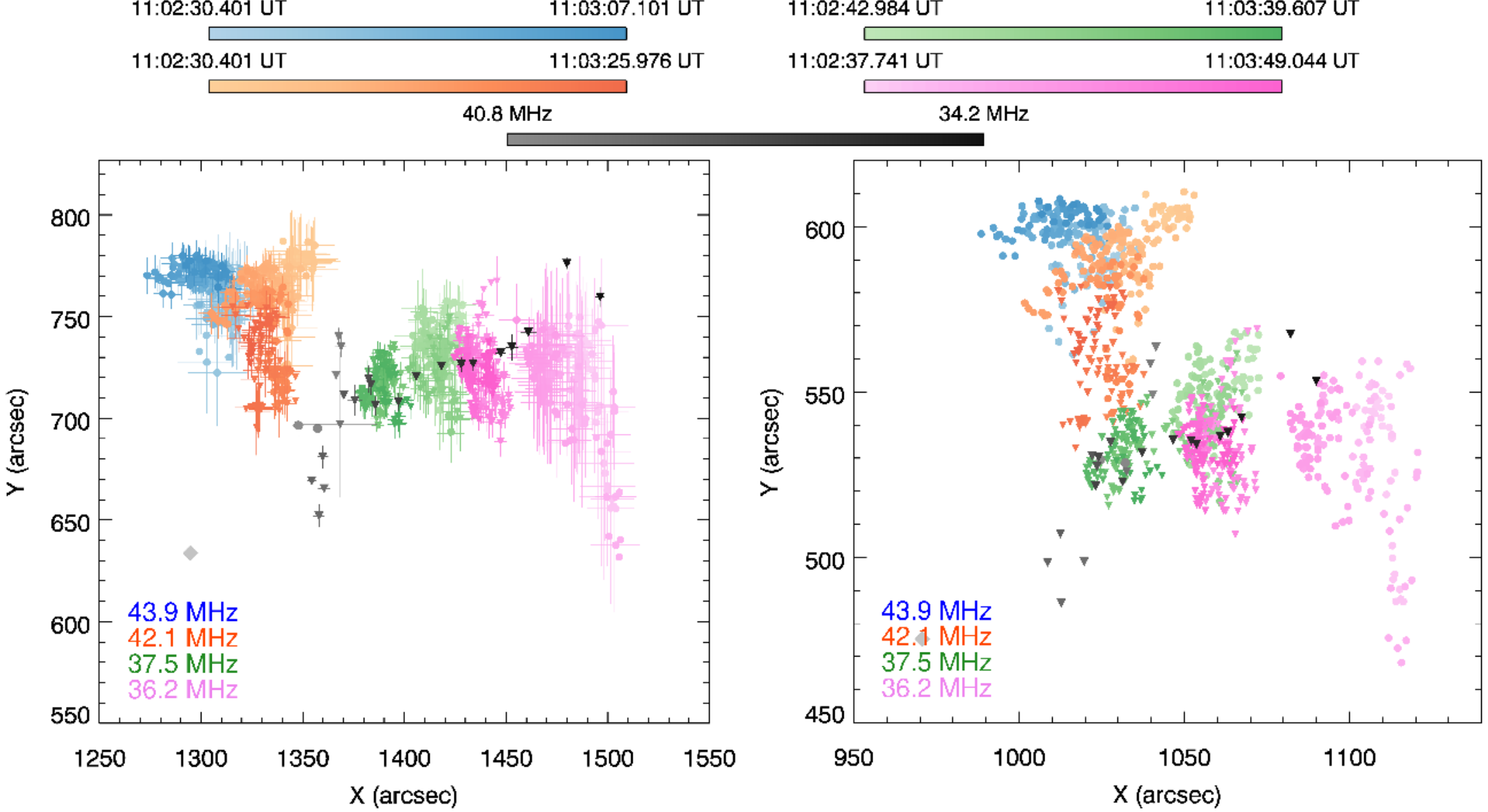}
    \caption[Transitioning Type II source locations.]
    {Estimated locations of the radio emission sources for the Type II structures annotated in Figure~\ref{fig:2020_annot_dynspec}.  The left panel displays the centroid locations with their associated errors obtained from the 2D elliptical Gaussian fits (see Section~\ref{sec:centroid_calc}), whereas the right panel displays the radially-corrected (for scattering-induced shifts) centroid locations without errors, in order to highlight the motion of the sources.  The subband imaged at 43.9~MHz is depicted in a blue colour scheme, the subband imaged at 42.1~MHz is depicted in an orange colour scheme, the subband imaged at 37.5~MHz is shown in a green colour scheme, and the subband imaged at 36.2~MHz is shown in a pink colour scheme.  The colour gradient represents a progression from earlier times (lighter) to later times (darker).  Grey centroids illustrate the motion of the drifting Type II emissions, starting from $\sim$40.8~MHz (light grey) until $\sim$34.2~MHz (dark grey).  Sources represented by a circle occurred before the defined transition time, whereas the ones represented by a downward-facing triangle occurred afterwards.  As indicated in Figure~\ref{fig:2020_annot_dynspec}, no sources past the transition time were imaged for the subband at 43.9~MHz (blue colour scheme).  Grey diamonds illustrate the central location of the LOFAR beams.
    Figure taken from \cite{2020ApJ...893..115C}.
	}
    \label{fig:2020_typeII_centroids}
\end{figure}

Figure~\ref{fig:2020_typeII_centroids} illustrates the obtained plane-of-sky source positions for the Type II sources imaged.  The centroids depicted by circles indicate that the sources occurred before the transition time, whereas the centroids depicted by downward-facing triangles occurred after the defined transition time.  The subband imaged at 43.9~MHz (and before the defined transition time) is presented in a blue colour scheme, the subband imaged at 42.1~MHz is presented in an orange colour scheme, the subband imaged at 37.5~MHz is shown in a green colour scheme, and the subband imaged at 36.2~MHz is shown in a pink colour scheme.  The colour progression corresponds to a temporal progression, with lighter colours used for earlier emission times.  Sources representing the drifting Type II emissions are also depicted for comparison.  These drifting emissions are imaged at decreasing frequencies with increasing time, as illustrated by the black crosses in Figure~\ref{fig:2020_annot_dynspec}.  The locations of the drifting Type II sources are presented in a grey colour scheme, where lighter shades represent higher-frequency sources and---consequently---earlier times.  The left panel of Figure~\ref{fig:2020_typeII_centroids} illustrates the centroids along with the associated uncertainties (see Section~\ref{sec:centroid_calc} and Equation~(\ref{eqn:gauss_centroid_error})).  The right panel shows the source locations following a correction for the scattering-induced radial shift, but without the error bars for a clearer illustration of the sources' spatial evolution.  The error bars, however, are not affected by the scattering correction and are thus the same as those in the left panel.  The observed heliocentric source locations ($R_{obs}$) were corrected for the scattering-induced shifts (obtaining $R_{true}$) by applying the analytical estimation described in Section~\ref{sec:tau_deriviation}.  Fundamental emissions, the 1$\times$Newkirk model, and $\epsilon^2/h = 4.5 \times 10^{-5} \, \mathrm{km^{-1}}$ were assumed.  The corresponding  plane-of-sky ($x$, $y$) locations---as illustrated in the left panel---were calculated using a simple trigonometric relation, given that the angle $\phi$ between the $x$-axis and the source remains constant during the radial correction for the scattering-induced shift:
\begin{equation*}
	\phi = \tan^{-1} \left( \dfrac{y_{obs}}{x_{obs}} \right) = \tan^{-1} \left( \dfrac{y_{true}}{x_{true}} \right) \, ,
\end{equation*}
so that
\begin{equation*}
	x_{true} = R_{true} \, \cos \phi
	\qquad \text{and} \qquad
	y_{true} = R_{true} \, \sin \phi \, ,
\end{equation*}
where parameters denoted with an \textit{obs} subscript represent the observed values, and those denoted with a \textit{true} subscript represent the corrected values.

The drifting Type II emission sources (shown in grey) appear to propagate away from the Sun, as expected for drifting Type II bursts (see, e.g., Section~\ref{sec:imaged_bandsplitting}).  The single-frequency centroids for each of the four subbands, however, do not appear to be gathered in a single location as expected for emissions excited by standing shocks (see Section~\ref{sec:typeIIs}).  This implies that the structure exciting the Type II emissions may not be completely stationary.  As can be seen by the colour progression, the sources appear to move towards the solar surface as time progresses.  The fact that sources which are imaged at a single frequency indicate a spatial evolution can be interpreted as the apparent motion towards the Sun of a structure with a constant density-to-background-density ratio ($n_{str}/n_{bg} = \text{constant}$), meaning that the excitation location of the emissions may change but not the emission frequency.  It should be emphasised that the apparent motion towards the Sun can be a mere side-effect of projection effects.  A source which moves away from the solar centre but at an angle to the observer's LoS can sometimes appear as if it is moving towards the solar centre in the plane of the sky, as described in Section~\ref{sec:proj_effs_impact} and illustrated in Figure~\ref{fig:2020_proj_effs_cartoon}.

Another intriguing aspect of the motion of the sources depicted in Figure~\ref{fig:2020_typeII_centroids} is that a ``jump'' can be observed in the collective position of the sources during the defined transition time.  This is highlighted by the use of circles for the pre-transition emission times and the downward-facing triangles for the post-transition emissions.
For example, when only the pink-coloured centroids in the left panel of Figure~\ref{fig:2020_typeII_centroids} are considered, it can be seen that the centroids could be grouped into two regions, with an easily-distinguished separation around axis $x=1455$~arcsec.  Similarly, the separation between the two green-coloured regions appears around $x=1400$~arcsec.

\subsubsection{Estimating the shock speed}
The speed of the shock exciting the Type II emissions was estimated using the imaged source locations.  The speed is given by the slope of the linear fit through the sources.  The corrected for scattering-induced shift
positions of the drifting Type II sources (grey data in the right panel of Figure~\ref{fig:2020_typeII_centroids}) and their respective emission times were used, resulting in an average plane-of-sky speed of $\sim$840~$\kms$.  It should be noted that if the correction for scattering-induced shifts is omitted, the resulting average plane-of-sky speed deduced from the apparent sources is $\sim$2220~$\kms$.  Such a shock speed would be unreasonably high, given the estimated CME speed ($\sim$560~$\kms$) within the C2 FoV.  The difference in the two estimations stems from the fact that the correction for the scattering-induced radial shift reduces the heliocentric heights of the lower-frequency sources to a larger extent, compared to that of the higher-frequency sources.  This decreases the collective spatial expansion of the sources over the given period of time, consequently decreasing the deduced shock speed.

For the sake of comparison, the shock speed was also estimated using Equation~(\ref{eqn:v_exc}), which requires a coronal density model and the drift rate of the Type II burst deduced from the dynamic spectrum ($\sim$-0.14~$\MHzs$).  The \cite{1961ApJ...133..983N} model was assumed, but the multiplicative factor $N$ (see Equation~(\ref{eqn:n_Newkirk})) was estimated through a comparison of the Newkirk model to the corrected sources, selecting the one that matched the corrected radial locations the best (similar to the method discussed in Chapter~\ref{chap:split-band_typeII} and illustrated in Figure~\ref{fig:2018_locations_vs_model}).  This approach resulted in a (radial) speed of $\sim$760~$\kms$, which is a reasonable value given the estimated CME speed ($\sim$560~$\kms$), and also agrees with the speed obtained using the corrected imaged locations ($\sim$840~$\kms$).

\subsection{Imaging the Type III burst} \label{sec:typeIII_imaging}
The Type III burst observed to intersect the stationary Type II emissions was also imaged.  Since Type III bursts are attributed to electrons tracing open magnetic fields (see Section~\ref{sec:typeIIIs}), the trajectory of the emission sources is expected to reflect the path laid out by the magnetic field.  As such, Type III sources tend to form a smooth curve, where the higher-frequency sources are found closer to the Sun than the lower-frequency sources (see Figure~\ref{fig:dyn_spec_cartoon} and, e.g., \cite{2014RAA....14..773R, 2019ApJ...885..140Z}).

\begin{figure}[ht!]
    \centering
	\includegraphics[width=1.0\textwidth, keepaspectratio=true]{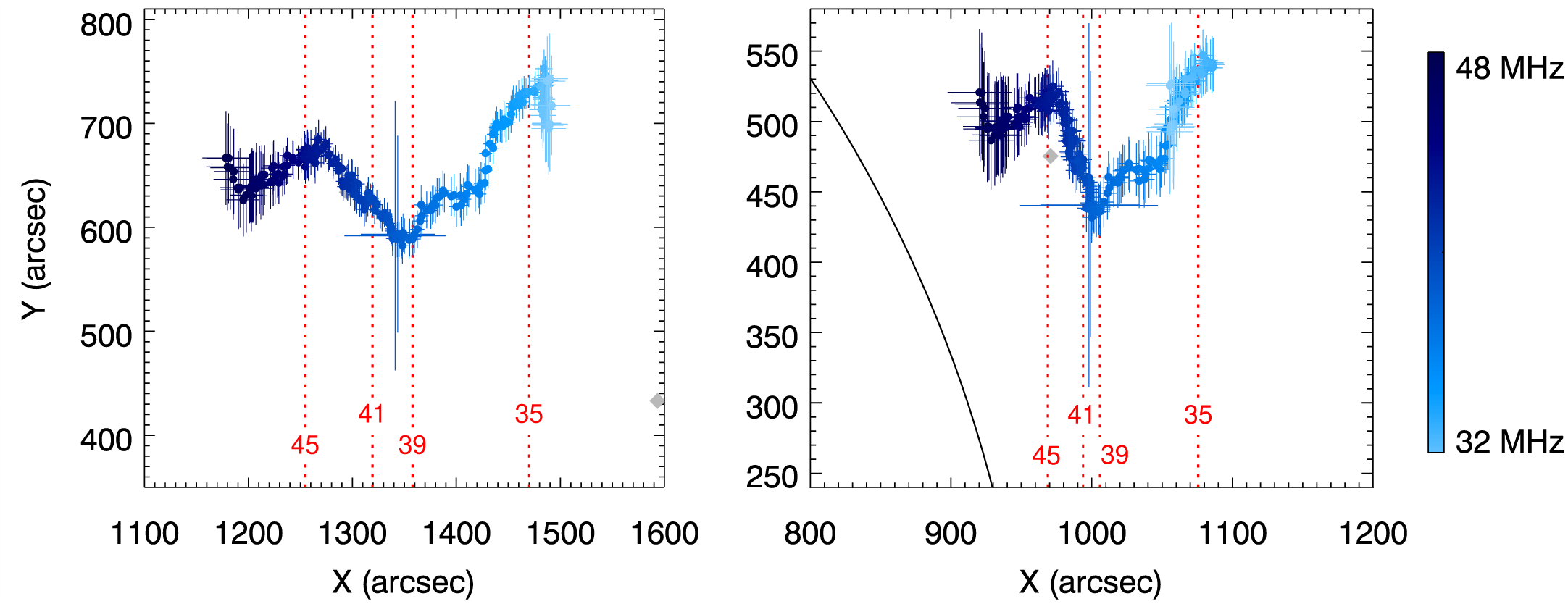}
    \caption[Source locations of the Type III burst intersecting the stationary Type II emissions.]
    {
    Estimated locations of the Type III sources with their associated errors, imaged at a single time ($\sim$11:02:20~UT) but multiple frequencies ($\sim$ 32--48~MHz), as indicated by the black vertical line in Figure~\ref{fig:2020_annot_dynspec}.  The red lines indicate the locations at which the four frequencies representing the bandwidths of the two pairs of Type II burst subbands are emitted (35--39~MHz and 41--45~MHz, see Section~\ref{sec:radio_spectroscopy}).  The left panel shows the apparent centroid locations, whereas the right panel shows the corrected (for the scattering-induced radial shift) locations.  Grey diamonds illustrate the central location of the LOFAR beams, while the solid black curve in the right panel represents the solar limb.
    Figure taken from \cite{2020ApJ...893..115C}.
	}
    \label{fig:2020_typeIII_centroids}
\end{figure}

Emission images for the observed Type III burst were taken at a single moment in time ($\sim$11:02:20~UT) but multiple frequencies ($\sim$32--48~MHz), as indicated by the black vertical line in Figure~\ref{fig:2020_annot_dynspec}.  The source positions are depicted in Figure~\ref{fig:2020_typeIII_centroids}, where the left panel shows the apparent centroid locations and the right panel shows the locations after a correction for the scattering-induced shifts was applied, as explained in Section~\ref{sec:2020_typeII_imaging}.
As can be seen, the positions of the Type III sources at the given moment in time form an unusual pattern.  Specifically, the snapshot does not depict the expected straight line or smooth arc-like shape that reflects the open magnetic field along which the electron beam propagates.  Instead, two striking changes in the positional progression with respect to frequency (not time) are observed.  The vector describing the position of the sources between $\sim$48--45~MHz points north-west (in solar coordinates), but an abrupt shift in the position of the sources occurs around 45~MHz (indicated by the red annotations), such that, for the sources between $\sim$45--39~MHz the vector points south-west.  After $\sim$39~MHz, a second abrupt shift occurs and causes the vector describing the sources between $\sim$39--32~MHz to once again point north-west.
The frequency range of 39--45~MHz coincides with the bandwidth of the three higher-frequency subbands of the stationary Type II burst (see Figure~\ref{fig:2020_annot_dynspec}).  The intriguing pattern of the Type III sources---imaged at a single moment in time---provides an insight into the shape of the magnetic field traced by the electron beam exciting this Type III event, at the given time and imaged heights, which coincide with the heights of the Type II emission sources (cf. Figure~\ref{fig:2020_typeII_centroids}).

\section{Proposed Generation Mechanism} \label{sec:gen_mech}
The multi-wavelength observations, discussed so far in this chapter, are combined in order to construct the complete picture of sequential events to which the excitation of the transitioning Type II emissions can be attributed.  Thus far, the Type II emissions were related to the streamer-puff CME, which was associated to the eruption of the jet.  However, it is not until the radio emission images are taken into account, that the way in which the transitioning Type II burst is generated can be assessed.

\begin{figure}[ht!]
    \centering
	\includegraphics[width=1.0\textwidth, keepaspectratio=true]{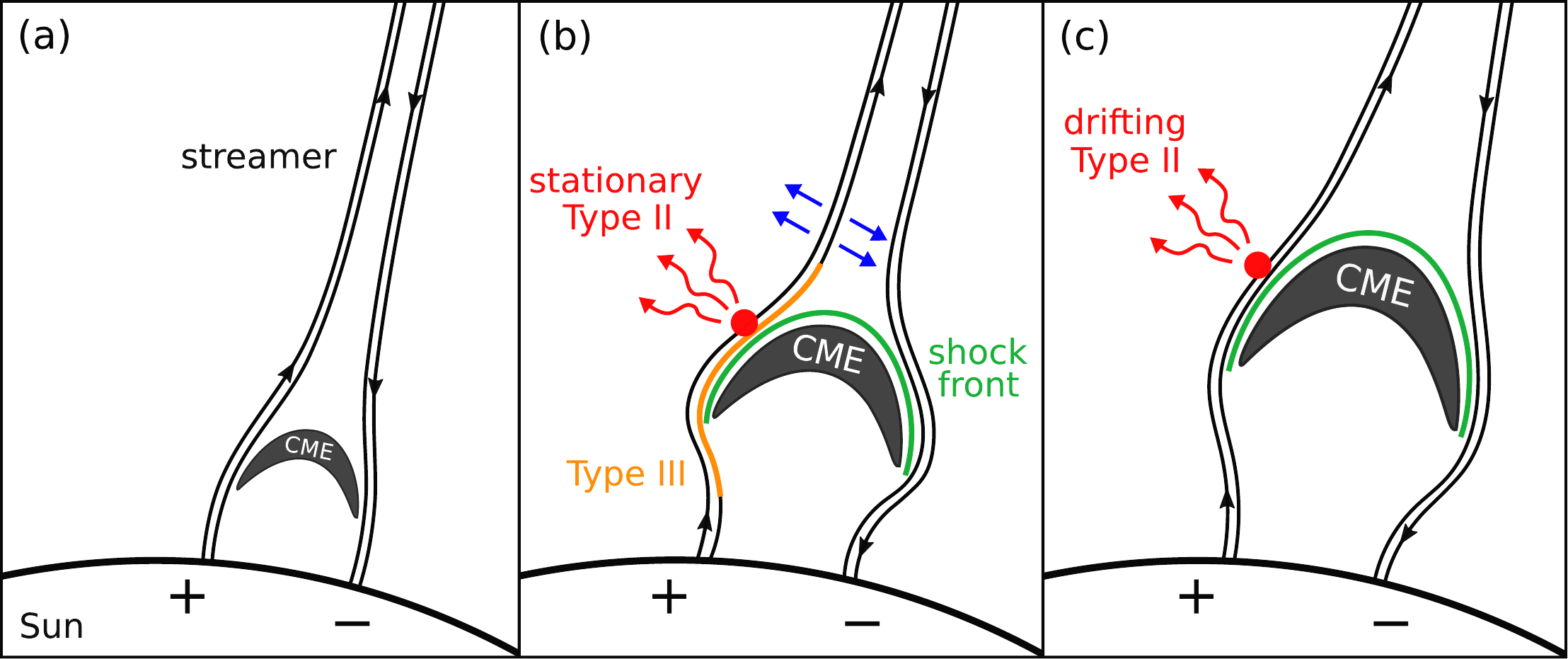}
    \caption[Schematic illustration of the mechanism generating the transitioning Type II burst.]
    {Schematic demonstration of the key phases of the mechanism generating the observed radio emissions.  Panel (a) illustrates the streamer-puff CME that was formed following the jet's eruption.  Panel (b) illustrates the CME as it propagates along the streamer and expands, as well as the shock front forming ahead of it (green curve).  The streamer undergoes an abrupt local expansion and the consequent compression by the shock results in the stationary Type II emissions (shown in red), as regions of the shock front are halted by the streamer.  The interplay between the streamer and the CME causes the streamer to pulsate (blue arrows), which is reflected in the negative and positive frequency-drift fine structures observed during the stationary Type II emissions (see Figure~\ref{fig:2020_fine_structures}).  An electron beam traces the curved magnetic fields confining the streamer and results in Type III emissions (orange curve).  Panel (c) shows the moment that the streamer succumbs to the CME's expansion and allows it to smoothly propagate away from the Sun, while the compression between the streamer and the moving shock excites the drifting Type II emissions (shown in red).
    Figure taken from \cite{2020ApJ...893..115C}.
	}
    \label{fig:2020_cartoon}
\end{figure}

Figure~\ref{fig:2020_cartoon} illustrates the three key phases of the sequence of events that are believed to have excited the various radio emissions captured in LOFAR's dynamic spectrum (Figure~\ref{fig:2020_dynspec}).  Panel \hyperref[fig:2020_cartoon]{(a)} depicts the streamer-puff CME which formed thanks to the presence of the streamer, following the eruption of the jet.  Once the CME gains a sufficient speed, such that the local Alfv\'en speed (which decreases with heliocentric distance) is exceeded, a shock front is formed ahead of the CME, as indicated by the green curve in panel \hyperref[fig:2020_cartoon]{(b)}.  The shock wave presses against the open magnetic fields forming the streamer, causing the streamer to undergo a localised expansion near the flanks of the CME, but not yet near the nose of the CME.  During the localised expansion, regions of the shock (on the CME's flank) are halted by the interplay with the streamer, effectively behaving as a standing shock.  It is believed that at this stage (see panel \hyperref[fig:2020_cartoon]{(b)}), three different---but nearly simultaneous---actions take place:
\begin{enumerate}
	\item The compression resulting from the interaction between the shock and the streamer excites the stationary Type II emissions (shown in red).  In other words, the stationary emissions are excited when the CME causes the streamer to expand, but before the undisturbed parts of the streamer (near the CME's nose) expand enough to allow for the smooth transition of the CME front.
	\item The interplay between the CME-driven shock and the streamer causes the streamer to pulsate (blue arrows).  These pulsations arise from the restoring force exerted by the magnetic fields confining the streamer, acting as a means of resisting the streamer's (global) expansion and keeping it intact during the CME's passage.  The magnetic field oscillations excite the negative and positive frequency-drift fine structures observed within the stationary Type II emissions (highlighted in Figure~\ref{fig:2020_fine_structures}).
	\item An electron beam traces the open magnetic fields confining the locally-expanded streamer, exciting the Type III burst (orange curve).  Consequently, the locations of the Type III sources (Figure~\ref{fig:2020_typeIII_centroids}) reflect the curvature exhibited by the magnetic fields due to the local inflation of the streamer.
\end{enumerate}
%
%
The final stage is schematically demonstrated in panel \hyperref[fig:2020_cartoon]{(c)}.  At this stage, the CME forces the streamer to succumb to its expansion, even at the regions near the nose of the CME.  This allows the shock to travel away from the Sun, without any of its parts being interrupted by the streamer to the extent that they are momentarily halted.  This is the instant at which the region of the shock exciting the radio emissions transitions from a standing shock to a drifting shock, and thus the Type II emissions no longer appear to be stationary.  However, the continuing compression between the flanks of the constantly-expanding CME and the streamer excite the drifting Type II emissions (shown in red in panel \hyperref[fig:2020_cartoon]{(c)}).  Furthermore, when the streamer expands to allow for the smooth propagation of the CME, the structure that was pulsating can no longer resist its displacement and abruptly ``jumps'' to a new location. This is believed to be the cause of the jump observed in the imaged Type II source locations (Figure~\ref{fig:2020_typeII_centroids}) at the time of the transition from stationary to drifting emissions.  It is likely that the Type II sources appear to be moving towards the Sun in the 2D plane-of-sky depiction, when in reality, they could be moving away from the Sun in the $z$-direction, as illustrated in Figure~\ref{fig:2020_proj_effs_cartoon}.
In other words, the inclination of the streamer could have changed in such a way that a propagation towards the solar centre is perceived in the 2D plane-of-sky observations (see Section~\ref{sec:proj_effs_impact}).

\section{Summary} \label{sec:transitioning_conclusions}
A Type II solar radio burst that transitions between a stationary and drifting state was reported for the first time, introducing a new subclass of Type II solar radio bursts: ``transitioning'' Type II bursts.  The aim in this chapter was to identify the sequence of events that are related to the radio emissions and understand what mechanisms are responsible for the generation of such a Type II morphology.

Ejected material appeared in the LASCO/C2 FoV close to the time and location of the radio emissions.  Two CME fronts with different spatial evolutions were identified.  One of them seemed to be confined by a streamer that was present both before and after the studied eruptions, and was only transiently inflated by the CME.  The driver of the CMEs was found to be a jet that erupted from the active region from which the streamer appeared to originate.  The spire of the jet bifurcated into two components, each of which is believed to have caused one of the two CME fronts.  The CME that traced the streamer was identified as a streamer-puff CME.  Due to its spatial relation to the radio source locations, this is the front believed to have excited the transitioning Type II emissions.

Besides the transition from stationary to drifting states, the presented Type II observation revealed other interesting features that can be useful in understanding the way that Type II radio bursts are formed.  One of these is the eye-catching band splitting experienced during the stationary Type II emissions.  Two bands, both of which experience band splitting, appear on the dynamic spectrum at approximately the same time.  The average relative frequency split $\Delta f_s/f$ of the two subband pairs is $\sim$0.04 for the upper-frequency pair and $\sim$0.05 for the lower-frequency one.  Both of these frequency splits are lower than the typical range reported for split-band Type II bursts that drift with frequency.  It is unclear, at this stage, whether these two pairs of subbands are the result of a single shock which simultaneously excites split-band emissions at two locations, or if they are the result of two individual shocks that happen to excite radio radiation at the same time.  Whilst attempting to answer this question was beyond the scope of the presented study, a careful analysis of this observation could potentially shed some light on the debated interpretation of band splitting and the associated locations of the emission sources (see Section~\ref{sec:bandsplitting_models} and Chapter~\ref{chap:split-band_typeII}).

The other intriguing aspect of this Type II burst is the fine structures observed within the stationary emissions, which exhibit both negative and positive frequency-drift rates.  They have a very small duration (a few seconds), differentiating them from previously-reported "waving" Type II emissions that alter between positive and negative frequency drifts that last over several minutes.  The magnitude of the frequency-drift rates of these fine structures varied significantly from one to another, suggesting that the speed of their exciter may be rapidly changing.  Additionally, the altering frequency drifts can be interpreted as signals of a pulsating exciter.

A Type III burst which intersects the stationary Type II emissions in the dynamic spectrum was also observed.  The Type III burst was imaged across several frequencies but at a single moment in time.  The snapshot of the source locations reveals a distinct pattern that does not resemble the expected one, i.e. a straight line or a smooth arc-like shape.  Instead, the positional progression (with respect to frequency) of the Type III sources changes abruptly at two locations.  The emission frequencies corresponding to the locations at which the two changes occur coincide with the bandwidth of the stationary Type II subbands.

It was of interest to probe the behaviour of the Type II sources before, during, and after the time of the transition, in order to examine their evolution and any significant changes that may be reflective of the spectral transition recorded by LOFAR.  To do so, the Type II emissions were imaged both before and after the transition from stationary to drifting states.  Four single-frequency slices which correspond to the average frequencies of the four subbands were chosen in order to image the transitioning behaviour.  The radio images revealed a ``jump'' in the source positions within each subband at the time of the transition.  Given that single frequencies were used to image each subband, this jump can only be related to the temporal evolution of the radio exciter and not to a spectral evolution or radio-wave propagation effects.  Furthermore, the sources of the stationary Type II emissions do not appear to be gathered at a single location, but instead seem to propagate towards the Sun, contrary to expectations for emissions related to stationary exciters.  This is another indication that the stationary Type II emissions are unlikely to be the result of a stationary exciter, like termination shocks (standing shocks) which have been related to some stationary Type II bursts (see Section~\ref{sec:typeIIs}).

The combination of multi-wavelength information and LOFAR's high-resolution spectroscopic and imaging observations of the radio emissions, allowed for the identification of the generation mechanism of the transitioning Type II emissions, as detailed in Section~\ref{sec:gen_mech}.  The observations suggest that the streamer-puff CME is the driver of the shock wave which excited the transitioning Type II burst.

The eruption of the jet leads to coronal ejections that---thanks to the existence of the streamer---form a streamer-puff CME.  As the CME expands and accelerates, a shock forms and interacts with the magnetic fields confining the streamer.  The streamer expands locally, halting the motion of parts of the shock front, causing them to act as a standing shock and exciting the stationary Type II emissions.  The interplay between the shock front and the streamer causes the magnetic fields to pulsate, exciting the negative and positive frequency-drift fine structures within the stationary Type II part.  At the same time, an electron beam traces the streamer's magnetic fields exciting Type III radio emission.  Subsequently, the source locations of the Type III burst reflect the local expansion of the streamer, resulting in the apparent abrupt positional changes and thus a pattern distinct from that of other ordinary Type III bursts.  Once the streamer succumbs to the constantly-expanding CME, it "jumps" to a less stressed position and expands near the nose of the CME as well, resulting in the observed jump in Type II source positions at the time of the spectral transition.  This expansion is sufficient for allowing the CME and all parts of the shock to proceed with their propagation away from the Sun undisturbed, such that the stationary Type II emissions cease to be excited.  Instead, the continuing compression between the streamer and the confined CME-driven shock excites the drifting Type II emissions, which have a frequency-drift rate that is characteristic of the typical drifting Type II bursts.

The validity of the proposed generation mechanism, beyond the reported observation, would have to be evaluated when other transitioning Type II bursts are recorded in the future.  Transitioning Type II bursts are expected to be observed very rarely compared to other Type II bursts---even as observing capabilities dramatically improve---given their apparent association to CMEs that travel along streamers (which are infrequent in comparison to standard CMEs) and the probability that a sequence of individual events (like those described) can occur.  Nevertheless, as previously mentioned, transitioning Type II bursts could provide unparalleled insight into their excitation mechanism, crucial for understanding the difference between the generation of Type II bursts that experience band splitting and those that do not, as well as understanding where the relevant emission source regions are located with respect to the shock front.

\cleardoublepage
\chapter{Conclusions and Final Remarks} \label{chap:conclusions}

The aim of this thesis was to combine state-of-the-art observations and simulations to investigate the intrinsic properties of solar radio emissions and how they are affected by their propagation through the turbulent coronal medium.  To enable such an investigation, observations were conducted with LOFAR so that the fine sub-second structures of radio bursts could be imaged.
Specifically, emissions between 30--80~MHz were observed using LOFAR's tied-array beam mode, which produced high temporal ($\sim$0.01~s) and spectral ($\sim$12.2~kHz) resolutions, as well as spatially-resolved sources imaged with high sensitivities ($\lesssim 0.03$~sfu per beam).
Recently-developed 3D radio-wave propagation simulations that consider anisotropic scattering from small-scale density fluctuations were utilised in order to compare their outputs to several of the observed source properties.  These were the first attempts at \textit{simultaneously reproducing} multiple observed properties, an approach deemed crucial for the appropriate evaluation of the mathematical framework characterising photon propagation.

In Chapter~\ref{chap:scattering}, the simulations were compared to a large collection of observed Type III burst source sizes and decay times, spanning a large range of frequencies.  When isotropic scattering was assumed, the simulations failed to simultaneously describe both the source sizes and decay times, indicating that an anisotropic scattering must be considered.  Therefore, the anisotropic scattering description was employed to produce time profiles, emission images, and directivity patterns for different levels of density fluctuations and anisotropy.  The input parameters that best matched the observed properties of a typical Type IIIb burst were identified.  It was demonstrated that scattering which is $\sim$3 times stronger in the perpendicular direction is required, consistent with observations of galactic point sources, which appear elongated along the perpendicular direction when observed through the upper solar corona.  Notably, it was found that strong scattering produced directivity patterns that were predominantly in the radial direction, even though the source was assumed to emit isotropically, addressing some of the early arguments against scattering from small-scale density fluctuations.  Moreover, the dependency of the imaged properties on the source-polar angle was investigated, showing that the sources near the limb appear to have smaller areas than those near the solar centre, since the $x$-size decreases with increasing angle.

In Chapter~\ref{chap:observation_simulations}, the sub-second properties of fine radio bursts were investigated across a single frequency and compared to the anisotropic scattering simulations.
Both the fundamental and harmonic components of a Type IIIb burst were considered, focusing on the temporal evolution of their flux, source location, and areal expansion.  Similar to Chapter~\ref{chap:scattering}, the isotropic scattering description failed to describe the observed features simultaneously, so the anisotropic scattering description was employed.  It was shown that while the fundamental properties can be reproduced when an intrinsic point source that instantly injects photons into the corona is assumed, the harmonic properties could not.  Instead, an intrinsic harmonic source with a finite size and finite emission duration was required.  The intrinsic size and emission duration were estimated through comparisons of the simulated properties to the observed.  Estimations of the level of density fluctuations, level of anisotropy, and the source's location were also obtained.
The sub-second properties and temporal evolution of a Drift-pair burst were also examined.
The inadequacy of reproducing the characteristic properties, assuming propagation in a medium where scattering is insignificant, was demonstrated.
Instead, the level of anisotropy was found to play a key role in the reproduction of the key Drift-pair properties and characteristic source evolution.  When strongly-anisotropic density fluctuations were assumed, the signal reflected from large-scale density inhomogeneities was sufficiently strong to form a second component (i.e. trailing Drift-pair component) with comparable intensities to the direct signal.  This result provided supporting evidence for the radio echo hypothesis, a once highly-criticised suggestion.  Unlike initial expectations, it was demonstrated that direct and reflected signals of fundamental plasma emissions can be observed.  Predominantly-perpendicular scattering was also necessary to produce the characteristic delay between the two components.  Both the time delay and relative intensity of the two components were found to decrease with increasing emission frequency, but were also found to depend on the assumed emission-to-plasma frequency ratio.  Specifically, lower frequency ratios resulted in lower time delays and higher intensity ratios.  Most importantly, the anisotropic scattering description reproduced the observed source evolution, i.e. an identical motion for both components, separated only in time.  Finally, it was found that sources with projected locations near the solar disk are more likely to produce Drift-pair bursts, and the narrow frequency range within which these bursts are observed was also justified.

In Chapter~\ref{chap:split-band_typeII}, simultaneous imaging of both subband sources of a split-band Type II burst was presented for the first time.  The imaged source locations implied that the emission sources originate from two spatially-separated locations on the shock front.  However, it was shown that once the frequency-dependent scattering shift is quantitatively accounted for, the split-band Type II sources become co-spatial, providing supporting evidence for band-splitting models that predict co-spatiality.  A further consequence of the scattering correction was that both sources move closer to the Sun, corresponding to lower coronal densities than the apparent locations.  To perform this scattering correction, an analytical expression allowing for the estimation of the scattering-induced radial shift was derived.  Moreover, the importance of projection effects on the perceived radio positions was also discussed.  A model that can be used to infer the out-of-plane location of split-band Type II sources from 2D images---as long as they are simultaneously imaged---was presented.

In Chapter~\ref{chap:transitioning_typeII}, a new sub-class of Type II solar radio bursts was reported.  A Type II burst that transitions between a stationary and drifting state---termed as a transitioning Type II burst---was observed for the first time.  Double band-splitting was also observed during the stationary Type II emissions, along with intriguing negative and positive frequency-drift fine structures.  The evolution of the Type II sources before, during, and after the transition time was investigated across four separate frequencies, representing each of the subbands.  A jump in the source locations at the time of the transition was observed.  Moreover, a Type III burst that intersected the stationary Type II emissions was also imaged, displaying a surprising source behaviour.  Sudden, abrupt changes in the location of the Type III sources was related to the stationary Type II emissions.  The objective of this study was to identify the mechanisms that generated such morphologies and source evolution.  A jet eruption and a CME were spatially and temporally related to the radio emissions.  It was found that upon eruption, the spire of the jet bifurcated, with each component driving a CME.  One of the CME fronts was confined by a streamer, forming a streamer-puff CME.  This front and its interaction with the magnetic fields of the streamer were associated to the radio emissions.  Parts of the streamer locally intercepting the CME's expansion and propagation induced the stationary Type II emissions.  An accelerated electron beam excited the Type III emissions, whose source locations reflected the local expansion of the streamer.  The negative and positive frequency-drift fine structures were related to the pulsation of the streamer and its interaction with the CME and its associated shock.  The drifting Type II emissions were attributed to the compression between the CME and the streamer, once the streamer succumbed to its expansion and allowed the CME front to propagate away uninterrupted.

As a whole, through the combination of imaging spectroscopy observations with very high resolutions and advanced radio-wave propagation simulations, this thesis emphasised the importance of considering radio-wave propagation effects---with a particular emphasis on anisotropic scattering---in analyses and subsequent interpretations of solar radio emissions.
The high degree to which density fluctuations in the solar corona alter the intrinsic emission properties and dictate what is received at the observer was demonstrated.  To advance the understanding and description of these effects, statistically large studies---similar to the ones presented in this thesis---need to be conducted.  It should be emphasised that several observed properties need to be simultaneously reproduced in order for any model of propagation effects to be deemed trustworthy.

Much of the work in this thesis would not be possible without the unprecedented imaging abilities of LOFAR, or the development of radio-wave propagation simulations that account for anisotropic density fluctuations.
This simultaneous characterisation of observed properties, enabled estimations of the level of anisotropy, the level of density fluctuations, and the source-polar angle.  These properties, however, could also be determined through the use of multi-vantage observations, by combining ground-based instruments like LOFAR with spaced-based ones.  Currently, such studies would be timely, given that the \textit{Solar Orbiter} and \textit{Parker Solar Probe} spacecraft (the two most recently-launched radio instruments) offer both in-situ and remote-sensing radio data, and will travel closer to the Sun than ever before.  As an example, simultaneous measurements of a single radio source from several vantage points will enable (through comparison with simulations) the estimation of the directivity of the emitted radiation, and thus the level of anisotropy.
It is, however, clear, that in order for significant progress to be made in the understanding of radio-wave propagation effects, ever-more-complex simulations need to be applied to increasingly-detailed observations of the finest radio emissions.
It is expected that future large interferometers with imaging capabilities that surpass that of LOFAR---like the upcoming Square Kilometre Array (SKA; e.g., \cite{2019AdSpR..63.1404N})---will further improve the radio-wave propagation understanding by capturing finer radio-burst structures with higher resolutions and sensitivity.

\cleardoublepage

\titleformat{\chapter}[display]
{\filleft \bfseries \color{myblue} \raggedright}
{\raggedleft \filleft \chapnumfont \textcolor{myblue!40} {\thechapter} \raggedleft}
{-24pt}
{\Huge}
\titlespacing{\chapter}{0pt}{0pt}{30pt}[0pt]

\bibliography{bibliography}
\bibliographystyle{aasjournal}

\titleformat{\chapter}[display]
{\filleft \bfseries \color{myblue} \raggedleft}
{\raggedleft \filleft \chapnumfont \textcolor{myblue!100} {\thechapter}}
{-40pt}
{\Huge}
\titlespacing{\chapter}{0pt}{-20pt}{60pt}[0pt]

\appendix
\chapter{Additional Simulation Equations} \label{chap:appendix}

\section{Fokker-Planck Equation and Diffusion Tensor} \label{sec:appendix_fokker_planck_eqn}
The spectral number density of photons (or photon number) $N(\vec{k}, \vec{r}, t)$ is described in the geometric optics approximation using a Fokker-Planck equation:
\begin{equation} \label{eqn:appendix_fokker_planck}
	\dfrac{\partial N}{\partial t} + 	\dfrac{d \vec{r}}{d t} \cdot 	\dfrac{\partial N}{\partial \vec{r}} + 	\dfrac{d \vec{k}}{d t} \cdot 	\dfrac{\partial N}{\partial \vec{k}} = 	\dfrac{\partial }{\partial k_i} \, D_{ij} \, \dfrac{\partial N}{\partial k_j} - \gamma N \, .
\end{equation}
Here, $\gamma$ is the collisional absorption coefficient (Equation~(\ref{eqn:appendix_gamma})) and $k_i$ describes the Cartesian coordinates of the photon wavevector $\vec{k}$, where the summation is performed over a repeated index $i, j =$1, 2, 3.  The number density of photons $N_0(\vec{r}) = \int N(\vec{k}, \vec{r}) \, d^3 \vec{k}$, and $d\vec{r}/dt$ and $d\vec{k}/dt$ are given by the Hamilton equations corresponding to the dispersion relation of electromagnetic waves in an unmagnetised plasma:
\begin{equation} \label{eqn:appendix_Hamilton_eq_drdt}
	\dfrac{d \vec{r}}{d t} = \vec{v}_g = \dfrac{\partial \omega}{\partial \vec{k}} = \dfrac{c^2}{\omega} \, \vec{k} \, ,
\end{equation}
\begin{equation} \label{eqn:appendix_Hamilton_eq_dkdt}
	\dfrac{d \vec{k}}{dt} = -\dfrac{\partial \omega}{\partial \vec{r}} = -\dfrac{\omega_{pe}}{\omega} \, \dfrac{\partial \omega_{pe}} {\partial \vec{r}} \, .
\end{equation}
The photon packet frequency $\omega$ is found from Equation~(\ref{eqn:EM_disp_rel}).
The diffusion tensor $D_{ij}$ appropriate to (anisotropic) scattering (given in terms of $\vec{k}$) is defined as
\begin{equation} \label{eqn:appendix_diffusion_tensor}
	D_{ij} = \left[ \dfrac{A_{ij}^{-2}}{\left(\vec{k} \, \mathbf{A}^{-2} \, \vec{k}\right)^{1/2}} - \dfrac{\left(\mathbf{A}^{-2} \, \vec{k}\right)_i \left(\mathbf{A}^{-2} \, \vec{k}\right)_j}{\left(\vec{k} \, \mathbf{A}^{-2} \, \vec{k}\right)^{3/2}}\right] D_A \, ,
\end{equation}
where $D_A$ is a $k$-independent coefficient given as
\begin{equation} \label{eqn:appendix_Da}
	D_A = \dfrac{\omega_{pe}^4}{32 \pi \, \omega \, c^2} \, \alpha \int^{\infty}_0 \tilde{q}^3 \, S \left( \tilde{q} \right) \, d \tilde{q} \, ,
\end{equation}
and $\mathbf{A}$ is the anisotropy matrix:
\begin{equation} \label{eqn:appendix_anis_matrix}
	\mathbf{A} = \begin{pmatrix}
		1 & 0 & 0\\
		0 & 1 & 0\\
		0 & 0 & \alpha^{-1}
	\end{pmatrix} \, ,
\end{equation}
where $\alpha$ is the anisotropy (see Equation~(\ref{eqn:2019_anisotropy_def})).

If isotropic scattering is assumed (such that $\alpha=1$), the diffusion tensor reduces to
\begin{equation} \label{eqn:appendix_diffusion_tensor_isotropic}
	D_{ij} = \dfrac{\nu_s \, k^2}{2} \, \left( \delta_{ij} - \dfrac{k_i \, k_j}{k^2} \right) \, ,
\end{equation}
where $\delta_{ij}$ is the Kronecker delta and $\nu_s$ is the scattering frequency, defined as
\begin{equation} \label{eqn:appendix_scatt_freq_isotropic}
	\nu_s = \dfrac{\pi}{4} \, \dfrac{\omega_{pe}^4}{\omega \, c^2 \, k^3} \, \bar{q} \, \epsilon^2 \, .
\end{equation}

\section{Stochastic Differential Equations} \label{sec:appendix_stoc_eqs}
Stochastic differential equations enable a numerical modelling of the radio-wave scattering effects, necessary for the simulations.
A form of the Fokker-Planck equation (i.e. Equation~(\ref{eqn:appendix_fokker_planck})) that is suitable for numerical computation can be obtained by writing the scattering term in Equation~(\ref{eqn:appendix_fokker_planck}) in the following way:
\begin{equation} \label{eqn:appendix_scatt_term}
	\dfrac{d N}{dt} =  \dfrac{\partial}{\partial k_i} \left(-N \, \dfrac{\partial D_{ij}}{\partial k_j} + \dfrac{\partial}{\partial k_j} \, \dfrac{1}{2} \, B_{im} \, B_{jm}^\text{T} \, N\right) \, ,
\end{equation} \label{eqn:appendix_Dij_Bim_Bjm}
where $\mathbf{B}$ is a positive semi-definite matrix with matrix elements determined by matrix $\mathbf{D}$, so that
\begin{equation}
	D_{ij} = \dfrac{1}{2} \, B_{im} \, B_{jm}^\text{T} \, .
\end{equation}

The non-linear Langevin equation for $\vec{k}(t)$ corresponding to the Fokker-Planck equation is
\begin{equation} \label{eqn:appendix_langevin_eq}
	\dfrac{dk_i}{dt} = \dfrac{\partial D_{ij}}{\partial k_j} + B_{ij} \xi_j \, .
\end{equation}
This equation is the definition of the stochastic integral in It\^{o}'s sense (adopted in the theory of random processes).  The presence of the It\^{o} drift (first right-hand-side term) conserves the value of $\vert \vec{k} \vert$ in elastic scattering events.

The effects of the large-scale refraction caused by the gradual variation of the ambient coronal density $n(\vec{r})$ are described via
\begin{equation} \label{eqn:appendix_dki_dt_eq}
	\dfrac{d k_i}{dt} = - \dfrac{\omega_{pe}}{\omega} \, \dfrac{\partial \omega_{pe}}{\partial r} \, \dfrac{r_i}{r} + \dfrac{\partial D_{ij}}{\partial k_j} + B_{ij} \, \xi_j \, .
\end{equation}
The radio-wave transport equation (Equation~(\ref{eqn:appendix_Hamilton_eq_drdt})) can be written in a similar manner:
\begin{equation} \label{eqn:appendix_dri_dt_eq}
	\dfrac{dr_i}{d t} = \dfrac{c^2}{\omega} \, k_i \, .
\end{equation}
The combination of the Langevin Equations~(\ref{eqn:appendix_dki_dt_eq}) and (\ref{eqn:appendix_dri_dt_eq}), describes the propagation, refraction, and scattering of radio-wave packets in an inhomogeneous plasma.

The stepping equations are used to describe the photons' wavevector ($k_i$) and position ($r_i$) at the next simulated time step of the stochastic process.  They are given by:
\begin{equation} \label{eqn:appendix_k_steps}
	k_i (t + \Delta t) = k_i(t) - \dfrac{\omega_{pe}(r(t))}{\omega} \, \dfrac{\partial \omega_{pe}}{\partial r} (t) \, \dfrac{r_i (t)}{r(t)} \, \Delta t + \dfrac{\partial D_{ij}}{\partial k_j} \Delta_t + B_{ij} \, \xi_j \, \sqrt{\Delta t}
\end{equation}
and
\begin{equation} \label{eqn:appendix_r_steps}
	r_i (t + \Delta t) = r_i(t) + \dfrac{c^2}{\omega} \, k_i(t) \, \Delta t \, .
\end{equation}
The vector $\vec{\xi}(t)$ describes a Gaussian white noise with properties $\langle \vec{\xi}(t) \rangle = 0$ and $\langle \xi_i(0) \, \xi_j(t) \rangle = \delta_{ij} \, \delta(t)$, where $\delta(t)$ is the Dirac delta function.  Random number $\xi_i$ are drawn from the normal distribution $N(0, 1)$ with zero mean and unit variance.  The time step $\Delta t$ is chosen such that it is shorter than the characteristic times of scattering and refraction.  Between the scattering and refraction time-scales, the mean scattering time ($1/\nu_s$) tends to be the shortest, so the chosen time step for the simulations is chosen to be $\Delta t = 0.1/\nu_s$.  Since the scattering frequency $\nu_s$ (Equation~(\ref{eqn:appendix_scatt_freq_isotropic})) decreases with increasing distance $r$, the time steps are shorter near the emission location of the radio waves, and become larger as the photons propagate away from the source (cf. Figure~\ref{fig:prop_effects_cartoon}).

In the case where anisotropic scattering is simulated, the values of $B_{ij}$ and $\partial D_{ij} / \partial k_j$ are defined as
\begin{equation} \label{eqn:appendix_Bij_anis}
	B_{ij} = \sqrt{ \dfrac{2 \, D_A}{\tilde{k}}} \left[ A_{ij}^{-1} - \dfrac{\left( \mathbf{A^{-2}} \, \vec{k} \right)_i \, \left( \mathbf{A^{-1}} \, \vec{k} \right)_j}{\tilde{k}^2} \right] 
\end{equation}
and
\begin{equation} \label{eqn:appendix_dDijdk_anis}
	\dfrac{\partial D_{ij}}{\partial k_j} = \dfrac{D_A}{\tilde{k}^5} \left[ -2 \tilde{k}^2 \, \left( \mathbf{A^{-4}} \, \vec{k} \right)_i + \left( \mathbf{A^{-2}} \, \vec{k} \right)_i \, \left( 3 \, \left( \vec{k} \, \mathbf{A^{-4}} \, \vec{k} \right) - \left(2 + \alpha^2 \right) \, \tilde{k}^2 \right) \right] \, .
\end{equation}

In the case of isotropic scattering,
\begin{equation} \label{eqn:appendix_Bij_dDijdk_isotropic}
	B_{ij} = \sqrt{\nu_s \, k^2} \, \left( \delta_{ij} - \dfrac{k_i \, k_j}{k^2} \right)
	\qquad \text{and} \qquad
	\dfrac{\partial D_{ij}}{\partial k_j} = - \nu_s \, k_i \, .
\end{equation}

\section{Characteristic Absorption Rate} \label{sec:appendix_abs_rate}
The collisional absorption coefficient $\gamma$ of radio waves (or characteristic rate of absorption) for binary collisions in a plasma is defined as
\begin{equation} \label{eqn:appendix_gamma}
	\gamma = \dfrac{\omega_{pe}^2}{\omega^2} \, \gamma_c \, ,
\end{equation}
where $\gamma_c$ is given as
\begin{equation} \label{eqn:appendix_gamma_c}
	\gamma_c = \dfrac{4}{3} \, \sqrt{\dfrac{2}{\pi}} \, \dfrac{e^4 \, n(\vec{r}) \, \ln \Lambda}{m \, v_{T_e}^3} \, .
\end{equation}
Here, $e$ is the electron charge and $n(\vec{r})$ is the density.  The thermal speed $v_{T_e} = \sqrt{T_e/m_e}$, with the electron temperature $T_e$ given in energy units.  It is assumed that the constant Coulomb logarithm $\ln (\Lambda) \simeq 20$, as per \cite{2014A&A...572A.111R}.
The temperature of the corona (which affects the collisional damping) is assumed to be isothermal, taken to be $\sim$86~eV ($\sim$1~MK).

\backmatter
\end{document}